\documentclass[12pt]{iopart}
\usepackage{iopams,graphicx}
\usepackage{color}

\newcommand{\JVec}{\boldsymbol{J}}

\newcommand{\SVec}{\boldsymbol{S}}

\newcommand{\qVec}{\boldsymbol{q}}
\newcommand{\pVec}{\boldsymbol{p}}

\newcommand{\TCollExpr}{
 - \frac{\hbar^2}{2} \displaystyle \sum_{\alpha\beta} \frac{\partial }{\partial q_{\alpha}} B_{\alpha\beta}(\qVec) \frac {\partial }{\partial q_{\beta}}
}
\newcommand{\VCollExpr}{V(\qVec)}
\newcommand{\HCollExpr}{\TCollExpr + \VCollExpr}

\newcommand{\gras}[1]{\boldsymbol{#1}}
\newcommand{\tensor}[1]{\mathsf{#1}}
\newcommand{\hfbtho}{\sc{hfbtho}}
\newcommand{\hfodd}{\sc{hfodd}}
\newcommand{\hfbtri}{\sc{hfbtri}}
\newcommand{\hfbaxial}{\sc{hfbaxial}}

\newcommand{\ev}{\sc{ev}}

\begin{document}

\review[Microscopic Theory of Nuclear Fission: A Review]{Microscopic Theory of Nuclear Fission: A Review}

\author{N Schunck$^{1}$, L M Robledo$^{2}$}

\address{$^{1}$ Nuclear and Chemical Science Division, Lawrence Livermore National Laboratory, Livermore, CA 94551, USA}
\ead{schunck1@llnl.gov}
\address{$^{2}$ Departamento de F\'{\i}sica Te\'{o}rica, Universidad Aut\'{o}noma de Madrid, E-28049 Madrid, Spain}

\vspace{10pt}
\begin{indented}
\item[]November 2015
\end{indented}

\begin{abstract}
This article reviews how nuclear fission is described within nuclear density
functional theory. A distinction should be made between spontaneous fission, 
where half-lives are the main observables and quantum tunnelling the essential 
concept, and induced fission, where the focus is on fragment properties and 
explicitly time-dependent approaches are often invoked. Overall, the cornerstone of 
the density functional theory approach to fission is the energy density 
functional formalism. The basic tenets of this method, including some well-known 
tools such as the Hartree-Fock-Bogoliubov (HFB) theory, effective two-body  
nuclear potentials such as the Skyrme and Gogny force, finite-temperature 
extensions and beyond mean-field corrections, are presented succinctly. The 
energy density functional approach is often combined with the hypothesis that 
the time-scale of the large amplitude collective motion driving the system to 
fission is slow compared to typical time-scales of nucleons inside the nucleus. 
In practice, this hypothesis of adiabaticity is implemented by introducing (a 
few) collective variables and mapping out the many-body Schr\"odinger equation 
into a collective Schr\"odinger-like equation for the nuclear wave-packet. The 
region of the collective space where the system transitions from one nucleus 
to two (or more) fragments defines what are called the scission configurations.
The inertia tensor 
that enters the kinetic energy term of the collective Schr\"odinger-like equation is 
one of the most essential ingredient of the theory, since it includes the 
response of the system to small changes in the collective variables. 
For this reason, the two main approximations used to compute 
this inertia tensor, the adiabatic time-dependent Hartree-Fock-Bogoliubov and 
the generator coordinate method, are presented in detail, both in their general 
formulation and in their most common approximations. The collective inertia tensor 
enters also the WKB formula used to extract spontaneous fission half-lives from 
multi-dimensional quantum tunnelling probabilities (For the sake of completeness, 
other approaches to tunnelling based on functional integrals are also briefly 
discussed, although there are very few applications.) It is also an important 
component of some of the time-dependent methods that have been used in fission 
studies. Concerning the latter, both the semi-classical approaches to 
time-dependent nuclear dynamics as well as more microscopic theories involving 
explicit quantum-many-body methods are presented. One of the trademarks of the 
microscopic theory of fission is the tremendous amount of computing needed for 
practical applications. In particular, the successful implementation of the 
theories presented in this article requires a very precise numerical resolution 
of the HFB equations for large values of the collective variables. This aspect 
is often overlooked, and several sections are devoted to discussing the 
resolution of the HFB equations, especially in the context of very deformed 
nuclear shapes. In particular, the numerical precision and iterative methods 
employed to obtain the HFB solution are documented in detail. Finally, a 
selection of the most recent and representative results obtained for both 
spontaneous and induced fission is presented with the goal of emphasizing the 
coherence of the microscopic approaches employed. Although impressive progress 
has been achieved over the last two decades to understand fission microscopically, 
much work remains to be done. Several possible lines of research are outlined 
in the conclusion.
\end{abstract}

\pacs{25.85.-w,25.85.Ca,25.85.Ec,21.60.Jz,21.60.Ev}
%
\vspace{2pc}
\noindent{\it Keywords}: Fission, Density functional theory, Hartree-Fock-Bogoliubov, 
Generator Coordinate Method, Scission, Adiabaticity, Large amplitude collective 
motion, Time-dependent density functional theory, Fission product yields, TKE, TXE.

\submitto{\RPP}
%
\maketitle
%
%

%
%

\section{Introduction}
\label{sec:intro} 

Experiments on the bombardment of Uranium atoms (charge number $Z=92$)
with neutrons performed by O. Hahn and F. Strassmann in 1938-1939 and 
published in \cite{hahn1938,hahn1939} showed that lighter elements akin 
to Barium ($Z=56$) were formed in the reaction. In February 1939, this 
observation was explained qualitatively by L. Meitner and O.R. Frisch in 
\cite{meitner1939} as caused by the disintegration of the heavy Uranium 
element into lighter fragments. This tentative explanation was based on the 
liquid drop model of the nucleus that had been introduced a few years earlier 
in \cite{bohr1936} by N. Bohr. A few months after these results, N. Bohr 
himself, together with J.A. Wheeler formalized and quantified Meitner's 
arguments in their seminal paper \cite{bohr1939}. They described fission as 
the process during which an atomic nucleus can deform itself up to the 
splitting point as a result of the competition between 
the nuclear surface tension that favours compact spherical shapes 
and the Coulomb repulsion among protons that favour very elongated
shapes to decrease the repulsion energy. They introduced the concepts of compound nucleus, 
saddle point (the critical deformation beyond which the nuclear liquid drop 
is unstable against fission) and fissility (which captures the ability of a 
given nucleus to undergo fission), provided estimates of the energy release 
during the process, of the dependence of the fission cross-section on the 
energy of incident particles, etc.. Although tremendous progress has been 
made since 1939 in our understanding of nuclear fission, many of the concepts 
introduced by Bohr and Wheeler remain very pertinent even today.


\subsection{Fission in Science and Applications}
\label{subsec:fission}

In simple terms, nuclear fission is the process during which an atomic 
nucleus made of $Z$ protons and $N$ neutrons ($A=N+Z$) may split into two or 
more lighter elements. The nuclear ``fissility'' parameter, given by $x \approx 
Z^{2}/50.88A(1-\eta I^2)$ with $I = (N-Z)/A$ and $\eta = 1.7826$, is a convenient quantity to 
characterize the ability of a nucleus to fission as suggested in \cite{bell1967,nix1969,bolsterli1972}. In the liquid drop model, 
the fissility is related to the ratio between the Coulomb and surface energy 
of the drop. For values of $x > 1$, the drop is unstable against fission, and  
nuclear fission can then occur spontaneously. This is the case, for instance, in heavy nuclei 
with large $Z$ values such as actinides or transactinides. The process is  
characterized by the spontaneous fission half-life $\tau^{\mathrm{SF}}_{1/2}$, 
which is the time it takes for half the population of a sample to undergo 
fission. Fission can also be induced through a nuclear reaction of a target 
nucleus with projectiles such as neutrons, protons, alpha particles or gamma 
rays (``photofission''). There are four well known cases ($^{239,241}$Pu and 
$^{233,235}$U) where the absorption of a neutron in thermal equilibrium 
with the environment -- with kinetic energy of a few tens of meV -- is 
sufficient to trigger fission (``fissile elements''). Note that $^{235}$U 
is the only such fissile element that is naturally occurring on earth.

Since the fissility parameter increases quadratically with $Z$, spontaneous fission is 
one of the most important limiting mechanisms to the existence of 
superheavy elements and is discussed in specialized review articles such as \cite{brack1972,hofmann2000}, as well as in 
several textbooks \cite{nilsson1995,krappe2012}. 
Theoretical studies of the location of the next island of stability beyond 
Lead are, therefore, largely focused on the accurate determination of 
spontaneous fission half-lives; see for instance the following papers  
\cite{bender2000,buervenich2004,staszczak2005,pei2009,sheikh2009,abusara2012,
warda2012,staszczak2013}. 

Nuclear fission also plays an important role in the formation of elements 
in the rapid neutron capture process ($r$-process) of nucleosynthesis in 
stellar environments. Modern simulations of neutron star-neutron star or 
neutron star-black hole mergers performed in \cite{goriely2011,korobkin2012,
just2015} suggest a vigorous $r$-process with fission recycling. In a fission 
recycling scenario, the nucleosynthesis flow proceeds beyond the $N=126$ 
closed shell to the region where neutron-induced, $\beta$-delayed, or 
spontaneous fission becomes likely. The fission products rejoin the 
$r$-process flow at lower $A$, continuing to capture neutrons until again 
reaching the fission region. Fission recycling is thus expected to contribute 
to the abundance pattern between the second ($A\sim 130$) and third 
$r$-process ($A\sim 190$) peaks. This appears consistent with observations 
of the $r$-process-enhanced halo stars -- unusual old, metal-poor stars at 
the edges of our galaxy that contain $r$-process elements in quantities 
measurable via high-resolution spectroscopy. Most of the relatively complete 
$56<Z<82$ patterns observed in these stars are strikingly similar and a good 
match to the solar $r$-process residuals within this element range as 
reported in \cite{sneden2008,roederer2014}. 

Because of the strong nuclear binding, the energy released during fission 
is very large (compared to other energy production sources), typically of 
the order of 200 MeV per fission event. Most of it is kinetic energy of the 
fission fragments while about 10--20\% of it is excitation energy. In a 
typical fission process, the formation of the fission fragments will thus 
be accompanied by neutron and gamma emission. In a nuclear reactor, these 
neutrons can be reabsorbed by the fissile elements present in the core, 
thereby triggering new fission events. The magnitude of this chain reaction 
can be controlled by neutron absorbers. Clearly, all characteristics 
of the fission process are essential to understand the physics behind 
the technology required for efficient, reliable and 
safe  nuclear technology applications. For instance, understanding the 
mechanism of induced fission at the level of making reliable prediction 
in regions where no experimental data is available will
be of paramount importance in technological applications.


\subsection{Defining a Microscopic Theory of Nuclear Fission}
\label{subsec:definition}

In this review, the term ``microscopic theory'' should be understood as 
of pertaining to the methods and concepts of nuclear density functional theory (DFT) -- in its broadest sense. 
In particular, we will use interchangeably the expression of DFT and 
self-consistent mean field (SCMF) theory. Beyond mean field extensions 
such as, e.g., projection or configuration mixing techniques, will also 
be generously included under the DFT label. The choice of this loose 
definition stems from the need to clearly differentiate current microscopic approaches 
to fission from configuration interaction methods such as {\it ab initio}
techniques or the nuclear shell model. It is also a reminder that the basic 
concepts and approximations essential in the theoretical description of 
fission are to a large extent independent of the formal differences between 
the SCMF and DFT.

\begin{figure}[ht]
\begin{center}
\includegraphics[width=0.65\linewidth]{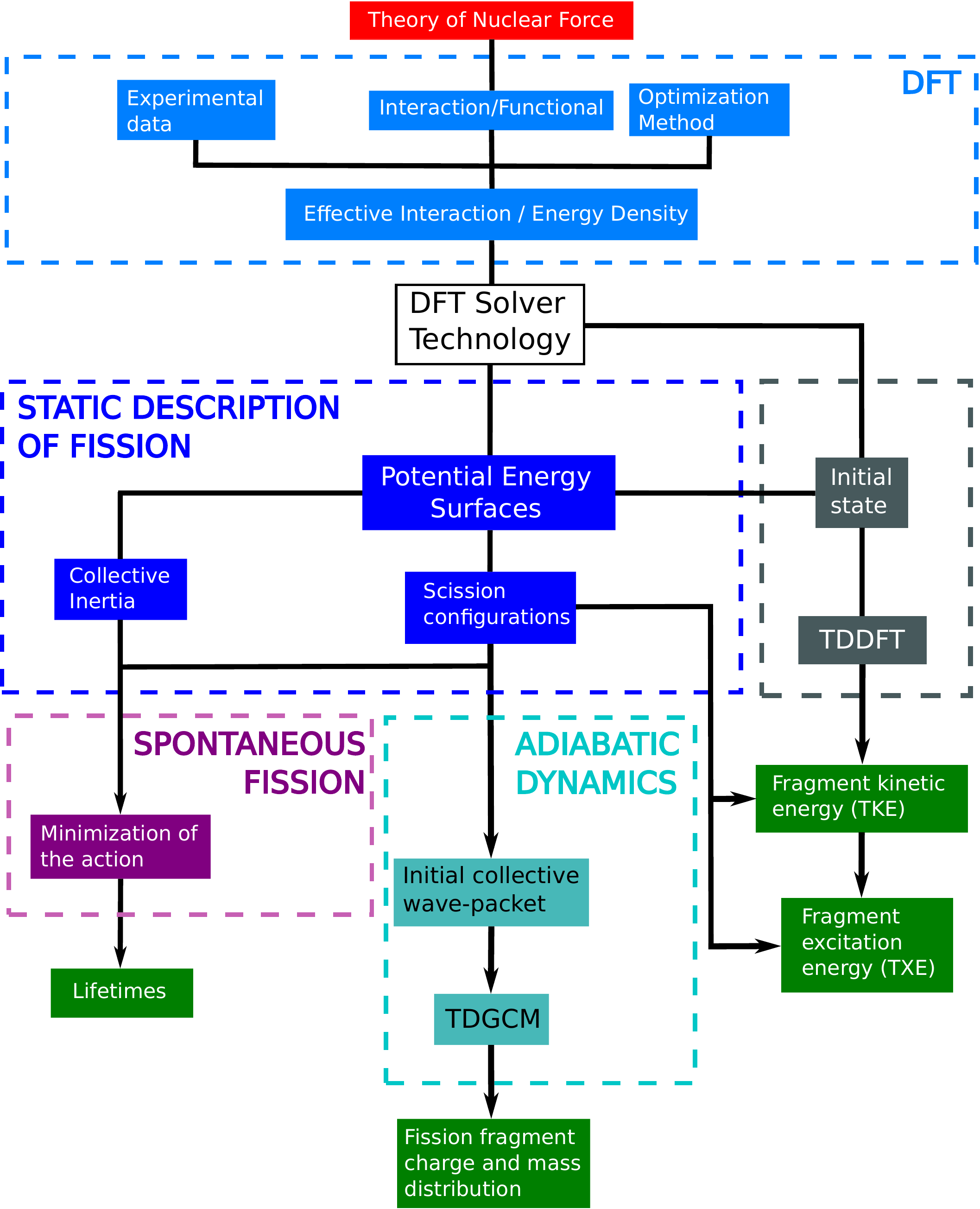}
\caption{Schematic workflow of the microscopic theory of fission based  
on nuclear density functional theory.}
\label{fig:diagramm}
\end{center}
\end{figure}

The overall framework of the microscopic theory of fission based on DFT/SCMF 
is summarized in figure \ref{fig:diagramm}. The starting point is to view 
the nucleus as a system of protons and neutrons, treated as point-like 
particles, in interaction. We further assume that the ground-state of the 
system can be well approximated by a symmetry-breaking quasiparticle vacuum. 
As a result of this approximation, the two basic degrees of freedom of the 
theory are the one-body density matrix $\rho$ and the pairing tensor 
$\kappa$. These objects are determined by solving the Hartree-Fock-Bogoliubov 
(HFB) equations, also known as Bogoliubov-de Gennes equations in condensed 
matter. The Wick theorem allows computing the expectation value of any 
operator based solely on the knowledge of the density matrix and pairing 
tensor. These concepts are briefly summarized in section \ref{subsubsec:hfb}. 

At the HFB approximation, the energy of the nucleus is a  functional of 
$\rho$ and $\kappa$. In the particular case of the self-consistent mean field, 
this functional is in fact derived from the expectation value of an effective 
nuclear Hamiltonian $\hat{H}$ on a quasiparticle vacuum; in the DFT picture, 
the energy density functional (EDF) is not necessarily related to any 
Hamiltonian. In both cases, the form of the EDF should in principle be 
constrained by our knowledge of nuclear forces. However, the various 
parameters that enter the definition of the EDF are typically readjusted to 
data in nuclear matter and finite nuclei to account for the HFB approximation. 
The Skyrme and Gogny forces are among the two most popular effective 
nuclear potentials that have been used, among others, in fission studies. 
The characteristics of the nuclear EDF, including a short discussion of 
how their parameters are adjusted to data, are presented in section 
\ref{subsubsec:edf}.

Allowing for spontaneous breaking of the symmetries of the nuclear force 
at the level of the quasiparticle vacuum is reminiscent of the historical picture 
of fission: if the one-body density matrix is not rotationally invariant, 
its spatial distribution represents a deformed nuclear shape. Fission can 
then be viewed as a process during which the deformation becomes so large 
that two separate fragments appear. This viewpoint can be formalized by 
introducing a set of collective variables that represent the motion of 
the nucleus as a whole and control the fission process. The characteristics 
of the resulting potential energy surface (PES), i.e., the energy as a function of 
the chosen collective variables, determines fission properties. For example, 
differences in the characteristics of potential energy barriers of nuclei 
qualitatively explains the range of spontaneous fission half-lives observed 
experimentally. In neutron-induced fission, the time ``from saddle to 
scission'', which is the time it takes for the nucleus to go from the top of 
the highest barrier to a configuration with two separated fragments, is also 
strongly dependent on the characteristics of the PES.

In the adiabatic approximation, it is further assumed that there is a perfect 
decoupling between the motion in collective space and the intrinsic motion of 
individual nucleons. This hypothesis originates from the observation that the 
time from saddle to scission is typically of the order of 10$^{-19}$ s. 
This is about two orders of magnitude slower than the time scale that can be 
inferred from the average binding energy per nucleon ($ B\approx 8 $ MeV). As 
a result, one can explore the collective space by seeking solutions to the 
HFB equations that yield specific values of the collective variables: this 
is how potential energy surfaces are computed in practice. Section 
\ref{subsec:collective} discusses the role of various collective variables in 
fission; see also section \ref{subsubsec:surface} for geometrical parametrizations 
of the nuclear surface used in semi-phenomenological approaches.

For specific values of the collective variables, the density of nucleons 
$\rho(\gras{r})$ may exhibit two disconnected regions of space with a high 
density of particles. Such a configuration corresponds to separated 
fragments. The frontier between configurations associated to the whole 
compound nucleus and those associated with the fission fragments is called 
the scission line; in $N$-dimensional collective spaces, it is in fact a 
$N-1$ hyper-surface. The actual definition of scission is ambiguous and is 
discussed in details in section \ref{subsec:scission}. In non-adiabatic 
approaches to fission such as time-dependent density functional theory 
(TDDFT), scission occurs ``naturally'' from the competition between 
short-range nuclear forces and Coulomb repulsion as the system is being 
evolved in time from some initial state (usually some specific quasiparticle vacuum). 
In the adiabatic approximation, however, there is no such ``dynamical'' 
mechanism to simulate the evolution of the system. 


\subsection{Challenges for a Predictive Theory of Fission}
\label{subsec:challenges}

As briefly mentioned in the previous section, there are different forms 
of nuclear fission, and the relevant observables that theorists need to 
compute thus differ depending on the mechanism under study. 
\begin{itemize}
\item Spontaneous fission (SF) is mostly characterized by the fission 
half-lives $\tau^{\mathrm{SF}}_{1/2}$, which range from 4.2 $\mu$s in 
$^{250}$No to over 1.4 $10^{10}$ years for $^{232}$Th, or a range of over 35 
orders of magnitude. The characteristics of the fission fragments are  
also important for astrophysical applications, or for establishing 
benchmarks data for nuclear material counting techniques used, e.g., in 
international safeguards, see for instance \cite{nichols2008}.
\item In neutron-induced or photofission reactions, the focus is more on 
the fission fragments. The relevant observables are thus the charge and mass 
distributions of the fragments; their total kinetic (TKE) and excitation 
energy (TXE) distributions, the average number of neutrons 
$\bar{\nu}$ emitted from each fragment and the average neutron energy; the 
average gamma multiplicity (i.e. number of gamma rays emitted) per fragment 
and the average gamma ray energy, etc. All these observables should be 
computed as a function of the energy of the incident particle (neutron or 
photon). Applications of induced fission in energy production typically 
require accuracy of less than a 1\% on these quantities.
\item In $\beta$-delayed fission, the compound nucleus has been itself 
produced by $\beta$ decay. The probability and characteristics of the 
subsequent fission will be dependent on the excited spectrum of the 
compound nucleus, and of the $\beta$-decay process itself. Calculating 
$\beta$-decay rates is, in itself, particularly challenging for nuclear 
DFT, since it is dependent on the detailed knowledge of the nuclear 
wave-function. Until now, there have been no attempt at computing 
$\beta$-delayed fission rates in a fully microscopic setting.
\end{itemize}
This variety of observables imposes various constraints on theory. For 
example, the description of neutron-induced fission with fast neutrons 
($E_{n} \approx 14$ MeV) requires the accurate modelling of the compound 
nucleus up to excitation energies of 20 MeV or so; predictions of 
spontaneous fission half-lives in superheavy nuclei are contingent on 
the predictive power of nuclear EDF far of stability; $\beta$-delayed 
fission rates depend on a quantitatively accurate description of weak 
processes in nuclei, etc.

Over the past decade, nuclear DFT has made great strides toward becoming 
a predictive tool for nuclear structure; see \cite{bogner2013} for a 
description of a comprehensive effort to build a universal energy functional. 
In particular, there is little doubt today that the theoretical framework 
outlined in the previous section and explained in greater details in sections 
\ref{sec:pes} and \ref{sec:dynamics} is sufficient to explain, at the very 
least semi-quantitatively, multiple aspects of the fission process, from 
the trend of fission half-lives along isotopic lines, to the shape of 
fission fragment mass distributions to the overall properties of fission 
fragments; see section \ref{sec:results} for an overview of recent results. 
The main question, therefore, is whether this framework can deliver the 
accuracy and precision required in applications -- be it to answer 
fundamental science questions or provide parameter-free input for nuclear 
technology.

Reaching such a goal will require progress on at least three fronts:
\begin{itemize}
\item At the most fundamental level, DFT predictions always depend on the 
intrinsic quality of the nuclear EDF. Currently, all EDFs used so far in 
fission studies have been based on phenomenological potentials such as the 
Skyrme and Gogny force. While these potentials have been essential in 
demonstrating the validity of the microscopic approach, there is a large 
consensus in the nuclear theory community that potentials with a much 
sounder connection to the theory of nuclear forces should be developed. 
How exactly to do this remains an open question.
\item On a more technical note, applications of DFT to fission are still 
limited by a number of approximations. In current adiabatic approaches, 
the solution to the nuclear many-body problem is obtained only for the 
lowest energy state and the number of collective variables is rather small; 
in current non-adiabatic time-dependent approaches, tunnelling is not 
possible, important correlations such as pairing are often modelled approximately 
and only one-body dissipation is taken into account. In both cases, open 
channels (particle evaporation, gamma emission) are rarely taken into 
account at the same level of detail.
\item Finally, one should emphasize that the physics of scission remains 
poorly known. How fission fragments acquire their identity, and the 
connection of this process with the physics of quantum entanglement, has 
not been studied. 
\end{itemize}

Although this article tries to provides as complete as possible a review 
of the current microscopic theory of fission, choices had to be made. First, 
we only mention topics where results were obtained with a DFT approach, 
and voluntarily leave out all the other areas where this is not the case. For 
example, we do not discuss the phenomenon of ternary fission observed 
in the spontaneous fission of actinides such as in \cite{ramayya1998}; the 
generation of angular momentum in the fission fragments discussed, e.g., in \cite{bonneau2007} 
and references therein; or the very complex problem 
of the fission spectrum (identifying the characteristics of the neutron, 
$\gamma$ and $\beta$-decay emitted during fission); see 
\cite{capote2016} for a recent discussion of the state of the art. 
In view of these considerations, our most 
controversial choice is certainly to leave out results obtained with the 
relativistic formulation of DFT (called relativistic mean-field or covariant 
density functional theory) reviewed in
\cite{ring1996,nikvsic2011}. This was motivated partly by the need to keep the 
length of this article reasonable, but also by the fact that most of the 
methods discussed in the non-relativistic framework can easily be ported to the 
relativistic one. There are excellent articles on fission within various 
versions of covariant 
density functional theory, and we refer, e.g., to \cite{rutz1995,bender1998,
buervenich2004,karatzikos2010,abusara2010,
abusara2012,lu2012,lu2014,zhao2015,zhao2016} for a short sample of the existing 
literature.

%
%

\newpage
\section{Potential Energy Surfaces}
\label{sec:pes}

As mentioned in the introduction, the hypothesis of adiabaticity has played 
a special role in the theory of fission since its introduction by Niels 
Bohr back in his 1939 paper \cite{bohr1939}. The notion that a small set 
of collective degrees of freedom was sufficient to describe most of the 
physics of fission has proved extraordinarily fruitful in 
semi-phenomenological approaches \cite{nix1965,brack1972,wilkins1976}. It 
is no surprise, therefore, that the same concept has been adapted to a more 
microscopic theory based on nuclear density functional theory (DFT).

The cornerstone of the implementation of the adiabatic approximation in 
fission theory is the definition of a potential energy surface (PES) in an 
arbitrary collective space. This PES is the analogue of the classical phase 
space of Lagrangian and Hamiltonian mechanics. In all the current approaches 
to fission that rely on the adiabatic approximation, the first step is, 
therefore, to define the most relevant collective variables and compute the 
PES.

In section \ref{subsec:micmac}, we briefly recall how this is done in 
macroscopic-microscopic methods, which are often an invaluable source of 
inspiration for density functional theory. In the latter, the PES is 
computed by solving the DFT equations, which take the form of the 
Hartree-Fock-Bogoliubov (HFB) equations. Section \ref{subsec:dft} gives a 
modern presentation of the energy density functional (EDF) implementation of 
DFT. This includes a reminder about the HFB theory in section 
\ref{subsubsec:hfb} and of its BCS approximation in section \ref{subsubsec:bcs}; 
an overview of the main components of standard energy 
functionals in section \ref{subsubsec:edf}; a foray into how to compute PES 
at high excitation energies in section \ref{subsubsec:excitation}; a list 
of the most important ``dynamical'' corrections to the PES in section 
\ref{subsubsec:beyond}. The large choice of possible collective variables, 
including either geometric or non-geometric quantities, is discussed in 
section \ref{subsec:collective}. Finally, we review the various criteria 
that have been introduced in the literature to define the scission 
configurations in section \ref{subsec:scission}.


\subsection{The Macroscopic-Microscopic Approach}
\label{subsec:micmac}

Starting with the pioneering work of Bohr and Wheeler, many theoretical 
studies of fission have been performed with empirical models derived from 
the liquid drop formula; see \cite{nix1965,brack1972} for comprehensive 
reviews. The introduction of the shell correction method by Strutinsky and 
collaborators in the nineteen sixties was essential to provide more 
microscopic insight to this approach and account for the role of nucleon 
degrees of freedom and of (some) features of nuclear forces. In the early 
nineteen seventies, the macroscopic-microscopic (MM) method was already well 
established and had been successfully applied to the problem of nuclear 
fission as exemplified in \cite{brack1972,bolsterli1972}. Since then, it 
has remained a popular and effective way to perform large scale studies of 
nuclear properties in general (\cite{moeller1995,werner1995,moeller1997,
moeller2006}) and fission in particular (\cite{moeller1989,moeller2001,
moeller2009}). Although this review is devoted to the microscopic theory 
of fission, it is not out of place to recall some of the most important 
features of the MM approach, since DFT has borrowed many of its most 
successful concepts.

In simple terms, the MM approach consists in viewing the nucleus as a 
finite chunk of nuclear matter, the energy of which is parametrized as a 
function of the charge, mass and deformations $\gras{q}$ of the nucleus. 
The total energy of the nucleus is decomposed as a sum of three terms: 
(i) a macroscopic energy $E_{\mathrm{mac}}$ that is often 
approximated by a deformed liquid drop or droplet formula and represents 
bulk nuclear properties, see the work by Myers and Swiatecki in 
\cite{myers1966,myers1969,myers1974} for a complete description of this 
term. In the language of second quantization, this is essentially a 
zero-body term; (ii) a shell correction $E_{\mathrm{shell}}$ that accounts 
for the distribution of single particle levels in the average nuclear 
potential and thus has a one-body origin; (iii) a pairing correction 
$E_{\mathrm{pair}}$, which has a two-body origin. The total energy thus 
reads formally
\begin{equation}
E(\gras{q}) = 
E_{\mathrm{mac}}(\gras{q}) + 
E_{\mathrm{shell}}(\gras{q}) +
E_{\mathrm{pair}}(\gras{q}),
\label{eq:mm}
\end{equation}
where $\gras{q}$ represents the set of all deformations characterizing the 
nuclear surface. The formalism has also been extended to account for finite 
angular momentum, e.g. in  \cite{werner1995,nilsson1995}, and intrinsic 
excitation energy, for instance in \cite{dudek1988,schunck2007}. 


\subsubsection{Parametrization of the Nuclear Surface}
\label{subsubsec:surface}

In the MM approach, the energy (\ref{eq:mm}) is a function of all the 
parameters needed to describe the nuclear shape. In other words, 
the nuclear surface must be parametrized explicitly. Numerous 
parametrizations have been introduced over the years; see \cite{hasse1988} 
for a comprehensive review. In this section, we wish to recall some of the 
most common parametrizations, especially those that have been introduced to 
describe extremely deformed nuclear shapes near scission. 

\begin{figure}[!ht]
\begin{center}
\includegraphics[width=0.75\linewidth]{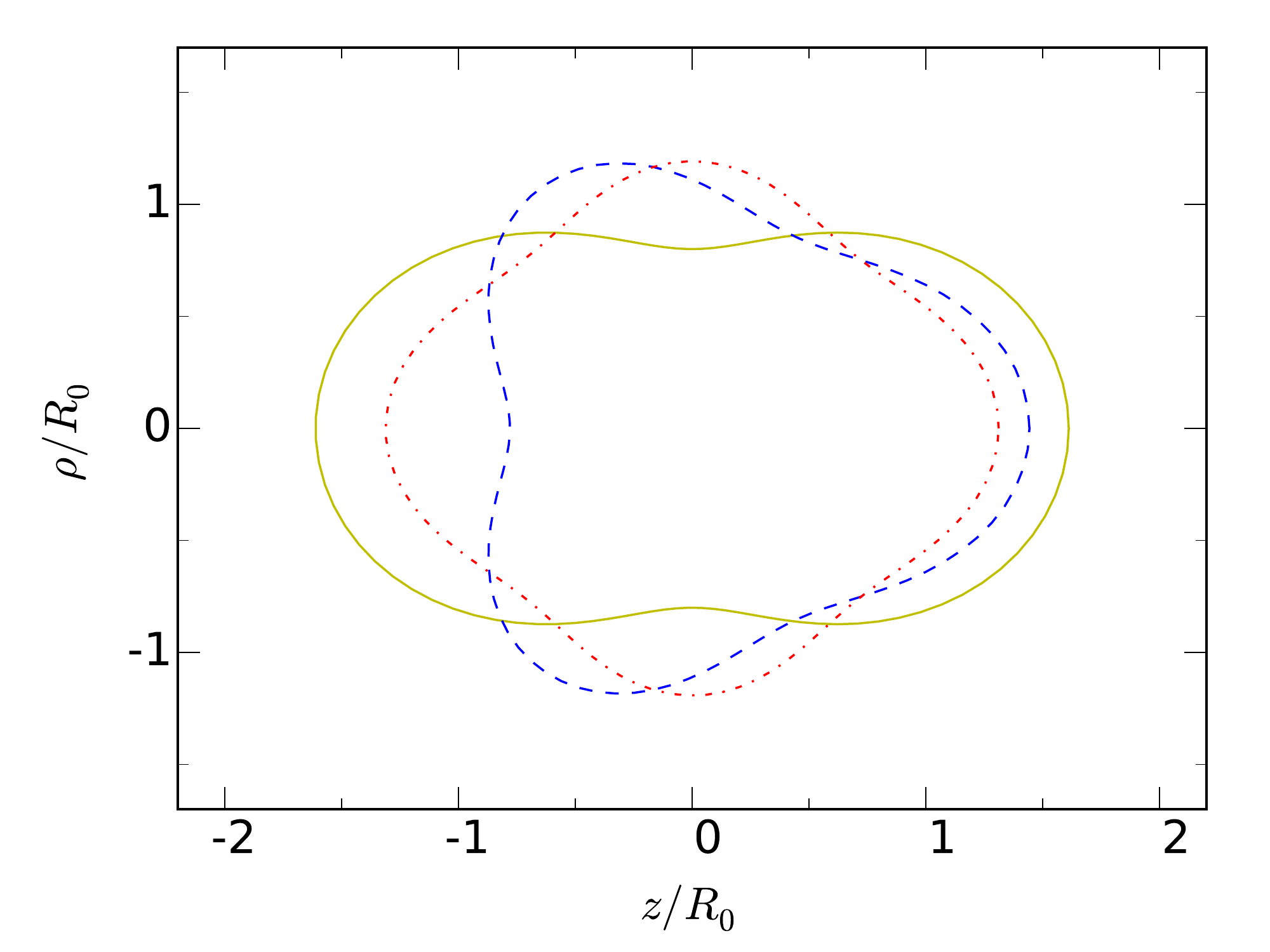}
\caption{Axial shapes obtained with the expansion (\ref{eq:multipole_exp}) of
the nuclear radius for $\alpha_{20}=0.8$ 
(plain curve), $\alpha_{30}=0.4$ (dashed curve) and  $\alpha_{40}=0.2$ (dashed-dotted 
curve). For each curve, all other deformation parameters are 0.}
\label{fig:spherical_harmonics}
\end{center}
\end{figure}

The simplest and most common parametrization is based on the multipole
expansion of the nuclear radius \cite{bohr1975},
\begin{equation}
R(\theta,\varphi) = R_{0}c(\alpha) 
\left[
1 + \sum_{\lambda\geq 2}\sum_{\mu=-\lambda}^{+\lambda} 
\alpha_{\lambda\mu}Y_{\lambda\mu}(\theta,\varphi) 
\right],
\label{eq:multipole_exp}
\end{equation}
where $R_{0} = r_{0} A^{1/3}$ is a parametrization of the nuclear radius 
for a spherical nucleus of mass $A$ ($r_{0} \approx 1.2 $ fm); $c(\alpha)$ 
is a factor accounting for the conservation of nuclear volume with 
deformation; the $\alpha_{\lambda\mu}$ are the deformation parameters; and 
$Y_{\lambda\mu}(\theta,\varphi)$ are the usual spherical harmonics. This 
parametrization of the nuclear surface is ideal for small to moderate 
deformations. However, a well-known problem of this formulation discussed 
in \cite{rohoziski1981,rohoziski1997} is that it does not provide a unique 
representation of the nuclear surface. In addition, the number of active 
deformation parameters needed to characterize very elongated shapes can 
become large. These difficulties also manifest themselves in the DFT 
framework since, as we will discuss in section \ref{subsubsec:multipole}, 
the collective variables most commonly used are the mass multipole moments 
of the nucleus, which are closely related to the spherical harmonics of 
(\ref{eq:multipole_exp}). For illustration, we show in figure \ref{fig:spherical_harmonics} 
a cross-section of the shapes obtained with the expansion (\ref{eq:multipole_exp})
for the nucleus $^{240}$Pu with $\lambda=2,3,4$ and $\mu=0$. 

\begin{figure}[!ht]
\begin{center}
\includegraphics[width=0.75\linewidth]{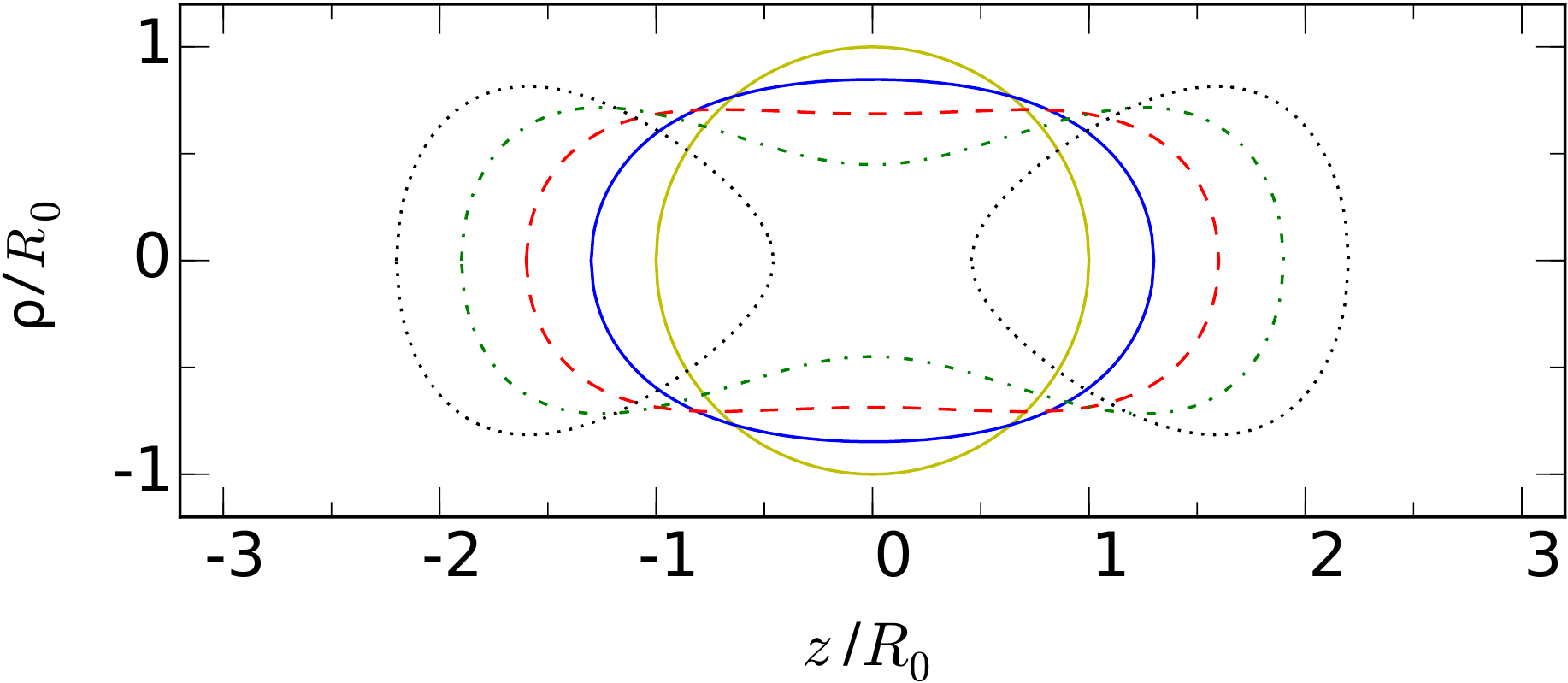}
\includegraphics[width=0.75\linewidth]{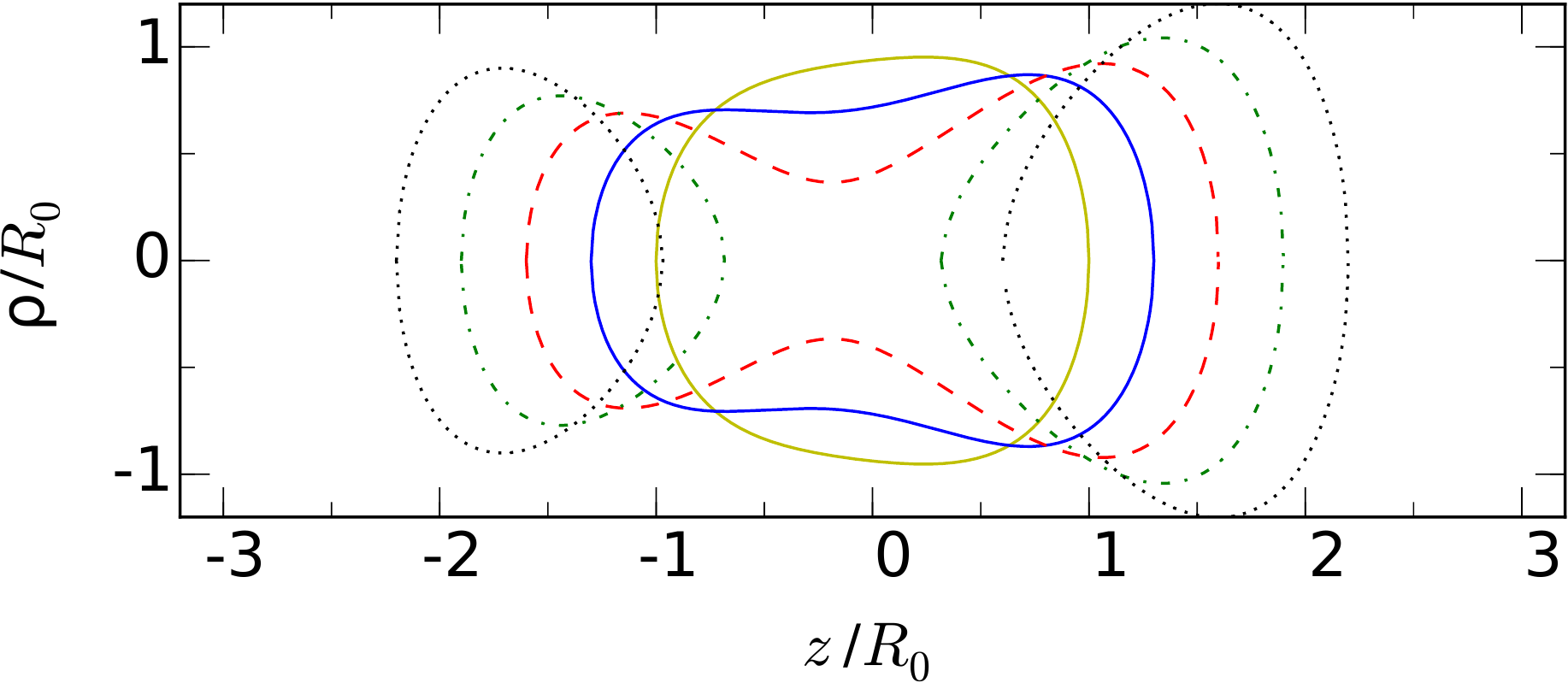}
\caption{Funny-Hills parametrization of nuclear shapes for 
$(h,\alpha) = (0,0)$ (top panel) and $(h,\alpha) = (0.2,0.3)$ (bottom 
panel). $c$ varies from 1.0 to 2.2 by step of 0.3}
\label{fig:funnyhills}
\end{center}
\end{figure}

Because of the limitations of expansion (\ref{eq:multipole_exp}), 
alternative parametrizations of the nuclear surface 
have been advocated over the years. One of the most popular is the Funny 
Hills shapes of \cite{brack1972}. Assuming axial symmetry along the 
$z$-axis of the intrinsic reference frame, a point on the surface is 
characterized by the usual cylindrical coordinates $(\rho,z,\varphi)$. In 
this case, the distance $\rho$ from the nuclear surface to the axis of 
symmetry is given by
\begin{eqnarray}
\rho^{2} = R_{0}^{2}c^{2}(1-\xi^{2})\left[ A + \alpha\xi + B\xi^{2} \right], & B\geq 0,\\
\rho^{2} = R_{0}^{2}c^{2}(1-\xi^{2})\left[ A + \alpha\xi\right] e^{Bc^{3}\xi^{2}}, & B<0,
\end{eqnarray}
where $A$, $B$, and $\alpha$ are the parameters characterizing the shape; 
$c$ is related to the elongation of the system, and defines the 
dimensionless variable $\xi= z/c$; in practice, $A$ is obtained from the 
volume conservation condition, and the parameter $B$ is substituted by 
another, $h = \frac{1}{2}B - \frac{1}{4}(c-1)$, which is related to the 
thickness of the neck. Therefore, the Funny Hills parametrizations is 
most commonly characterized by the set $(c,h,\alpha)$. The figure 
\ref{fig:funnyhills} illustrates the shapes obtained for $(h,\alpha) = (0,0)$ 
(top panel) and $(h,\alpha) = (0.2,0.3)$ (bottom panel), each case with 
$c$ varying between 1.0 and 2.2 by step of 0.3.

An alternative to the Funny-Hills parametrization is the one proposed by 
the Los Alamos group in \cite{nix1969}. It also applies to axially-symmetry 
shapes. In cylindrical coordinates, the nuclear surface is described by 
three quadratic surfaces of revolution
\begin{equation}
\rho^{2} = 
\left\{
\begin{array}{ll}
\displaystyle
a_{1}^{2} - \frac{a_{1}^{2}}{c_{1}^{2}}(z-l_{1}^{2})^{2}, & l_{1} - c_{1} \leq z \leq z_{1}, \medskip\\
\displaystyle
a_{2}^{2} - \frac{a_{2}^{2}}{c_{2}^{2}}(z-l_{2}^{2})^{2}, & z_{2} \leq z \leq l_{2} + c_{2}, \medskip\\
\displaystyle
a_{3}^{2} - \frac{a_{3}^{2}}{c_{3}^{2}}(z-l_{3}^{2})^{2}, & z_{1} \leq z \leq z_{2}.
\end{array}
\right.
\end{equation}
The nine parameters $(a_{i}, l_{i}, c_{i})_{i=1,2,3}$ are not all 
independent: the condition of volume conservation eliminates one 
parameter; the condition of continuity and differentiability of the 
surface at the two interfaces $z_{1}$ and $z_{2}$ eliminate two more 
parameters; fixing the centre of mass of the shape can also eliminate 
an additional parameter, so that only 5 independent parameters are 
ultimately needed. They are noted $(Q_{2}, \alpha_{g}, 
\varepsilon_{f1}, \varepsilon_{f2}, d)$ and correspond, respectively, 
to the elongation of the whole nucleus, the mass asymmetry of the two 
nascent fragments, the axial quadrupole deformation of the left and 
right fragment, and the thickness of the neck; see \cite{moeller2009} 
for details and derivations.

\begin{figure}[!ht]
\begin{center}
\includegraphics[width=0.75\linewidth]{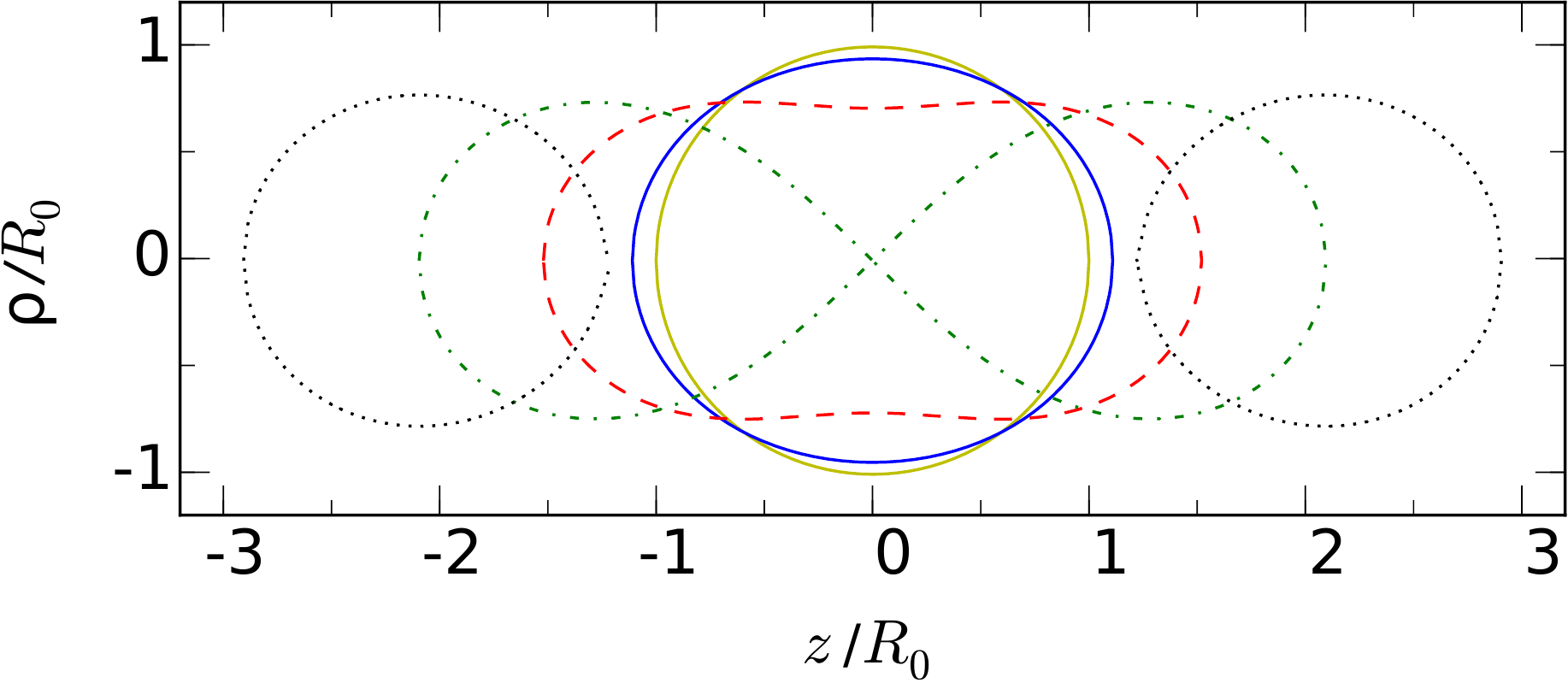}
\caption{Axial shapes obtained with the parametrization (\ref{eq:cassini}) for 
$u = 0.0, 0.4, 0.8, 1.0, 1.2$. The case $u=0$ corresponds to the sphere, 
the case $u=1.0$ to the scission point and the case $u=1.2$ to the two 
separated fragments}
\label{fig:cassini}
\end{center}
\end{figure}

The last category of nuclear shape parametrization originally introduced 
in \cite{pashkevich1971} is based on Cassini ovals. The general expression 
for Cassini ovals is
\begin{equation}
(a^{2} + z^{2} + \rho^{2})^{2} = 4a^{2}z^{2} + b^{4},
\label{eq:cassini}
\end{equation}
with $a$ and $b$ the only parameters. If $a=0$, then the shape reduces to 
a sphere of radius $b$; if $a>b$ (\ref{eq:cassini}) represents two separate 
fragments. When using this parametrization to represent the nuclear shape, 
the volume conservation condition allows eliminating one of the two 
parameters. Therefore, Cassini ovals represent a single-parameter 
parametrization of the nucleus that can reproduce shapes from the spherical 
point to two distinct fragments separated by an arbitrary distance. To 
account for mass asymmetry, Cassini ovals can be distorted by substituting 
$a^{2} \rightarrow a_{2} - \epsilon^{2}$ in the left-hand side of 
(\ref{eq:cassini}). The figure \ref{fig:cassini} illustrates some of the 
typical shapes obtained in the symmetric case of (\ref{eq:cassini}).

Once a parametrization of the nuclear surface has been established, the 
macroscopic part of the energy $E_{\mathrm{mac}}(\gras{q})$ can be computed. 
In the most advanced parametrization of the macroscopic energy, the surface, 
surface-symmetry, Coulomb and Wigner terms depend on the deformation via 
geometrical form factors that can be computed in a straightforward manner 
when the nuclear shape is known. The other terms are deformation-independent 
and adjusted to global nuclear properties \cite{myers1966,myers1969,
myers1974}.


\subsubsection{Quantum Corrections}
\label{subsubsec:corrections}

In addition to the macroscopic energy, the MM approach relies on various 
quantum corrections. Here, the adjective ``quantum'' recalls the fact that 
these corrections can only be computed by solving the Schr\"odinger equation 
of quantum mechanics. 

The shell correction is the most important correction. It was introduced in 
\cite{strutinsky1963} in order to account for the single particle shell 
structure in the calculation of the total energy. The starting point of the 
method is a phenomenological nuclear potential $U(\gras{r})$ including a central 
part such as e.g., the Nilsson, Woods-Saxon or Folded-Yukawa potential, a 
spin-orbit potential and the Coulomb potential for protons. The geometry of 
these potentials should be consistent with that of the nuclear surface: in 
the case of the Woods-Saxon or Folded-Yukawa potential, this implies that 
the various terms depend on the distance of the current point to the 
nuclear surface, parametrized as discussed in the previous section. Given 
the potential and its deformation, one solves the one-body Schr\"odinger 
equation in order to obtain the (deformed) single particle energies $e_{n}$ 
and wave functions $\varphi_{n}(\gras{r})$. The last step is to approximate 
the discrete sequence of states $\{ e_{n} \}_{n}$ by a continuous 
distribution $\tilde{g}(e)$ following the Strutinsky averaging procedure. 
The shell correction is then defined as
\begin{equation}
E_{\mathrm{shell}} = \sum_{n} e_{n} - \int \tilde{g}(e) ede.
\end{equation}
In the last expression, the shell correction should be computed for both 
protons and neutrons; the summation extends either over the $Z$ and $N$ 
lowest orbitals (ground-state) or over a subset of orbitals (excited 
states). In addition to providing a more realistic estimate of binding 
energies, the introduction of a one-body potential has given birth to a 
very powerful phenomenology based on single particle orbitals as basic 
degrees of freedom of nuclei; see \cite{nilsson1995} for a reference 
textbook on this topic.

Pairing correlations are incorporated into the macroscopic-microscopic 
approach in the form of an average pairing energy in the macroscopic energy $E_{\mathrm{mac}}$ 
supplemented by a pairing energy correction. The average pairing energy 
does not contribute to fission (it is independent of the deformation), 
while the pairing energy correction is treated in a very similar manner 
as shell effects. The solution to the BCS equations 
define the pairing correlation energy $E_{\mathrm{pair}}$. Assuming again 
a continuous level distribution, one can define a smooth pairing energy 
$\tilde{E}_{\mathrm{pair}}$. The difference $\delta E_{\mathrm{pair}} = 
E_{\mathrm{pair}} - \tilde{E}_{\mathrm{pair}}$ defines the pairing 
correction energy. In most applications to fission, pairing is described 
within the BCS theory with schematic interactions such as a constant 
seniority pairing force. Particle number conservation can be accounted 
for exactly by projection techniques as described in \cite{ring2000} or 
approximately through the Lipkin-Nogami prescription, see application in 
\cite{bolsterli1972,moeller1995}.Such a scheme was applied extensively in the fission 
studies by the LANL/LBNL and Warsaw groups, see, e.g., \cite{moeller2001,
moeller2009,kowal2010,randrup2011,randrup2011-a,jachimowicz2012,
jachimowicz2013} and references therein for some recent work.


\subsection{The Energy Density Functional Formalism}
\label{subsec:dft}

The energy density functional (EDF) formalism at the heart of the current 
microscopic theory of fission is the implementation of DFT in the particular context of atomic nuclei. Let us briefly 
recall that the DFT used in electronic structure theory relies on the 
rigorous existence theorems of Hohenberg and Kohn \cite{hohenberg1964}. 
It begins with the expression 
of the energy of the system as a functional of the local electron density 
$n(\gras{r})$. The density is determined in practice by solving the 
Kohn-Sham equations -- formally analogue to the Hartree equations 
\cite{kohn1965}. Note that various extensions of the Hohenberg-Kohn 
theorem to handle exchange terms exactly, excited states, systems at 
finite temperature, and superfluid correlations are also available; see
\cite{parr1989,dreizler1990} for a detailed presentation. In these 
extensions, the Kohn-Sham scheme is reformulated starting, e.g., from the 
full one-body density matrix for the exact treatment of exchange terms 
\cite{eschrig1996}. 

In nuclear physics, the true Hamiltonian is not known, nuclei are 
self-bound, and, as emphasized in \cite{drut2010}, correlation effects 
are much stronger than in electron systems. Also, pairing correlations 
play a special role, which is reflected by the importance of the Bogoliubov 
transformation in nuclear structure. There have been various attempts to 
extend the Hohenberg-Kohn theorem to self-bound systems characterized by 
their intrinsic density (defined in the centre of mass reference frame), 
see discussions in \cite{engel2007,barnea2007,giraud2008,messud2009,messud2011,
lesinski2014}. However, the resulting Kohn-Sham scheme does not seem to 
be as simple as in electronic DFT as shown in \cite{engel2007,barnea2007}. 
In addition, how to rigorously include symmetry-breaking effects (and 
subsequently restore these symmetries) in such schemes remain an open 
question.

Historically, most nuclear energy 
functionals have been explicitly derived from the expectation value of 
an effective nuclear Hamiltonian on a quasiparticle vacuum leading 
to the notion of self-consistent nuclear mean field reviewed in details 
in \cite{bender2003}. In the spirit of the Hohenberg-Kohn theorem, the 
SCMF is equivalent to expressing the energy as a functional of the 
intrinsic, one-body, non-local density matrix $\rho$ and non-local pairing 
tensor $\kappa$. In addition, these objects may break many of the 
symmetries of realistic nuclear forces such as translational or rotational 
invariance, parity, time-reversal invariance, and particle number. This 
spontaneous symmetry breaking is essential for introducing long-range 
correlations in the nuclear wave function as discussed in \cite{ring2000,
scheidenberger2014}. In the next sections, we recall some of the basic 
features of the nuclear EDF approach. We begin with the Hartree-Fock-Bogoliubov (HFB) 
theory in section \ref{subsubsec:hfb}, followed by its BCS approximation in 
section \ref{subsubsec:bcs}; in section \ref{subsubsec:edf}, 
we give a brief survey of standard energy functionals; we then introduce 
the finite temperature formalism in section \ref{subsubsec:excitation} as 
a method to describe excited states; finally, we list in section 
\ref{subsubsec:beyond} the various corrections to the energy that have a 
beyond mean-field origin.


\subsubsection{Self-Consistent Hartree-Fock-Bogoliubov Theory}
\label{subsubsec:hfb}

The HFB approximation to the energy is centred on the Bogoliubov 
transformation defining quasiparticle creation and annihilation operators 
in terms of a given single particle basis $(c_{i}, c_{i}^{\dagger})$ of a 
(restricted) Fock space
\begin{eqnarray}
\beta_{\mu}           & = & 
\sum_{i}U_{i\mu}^{*}c_{i} + \sum_{i}V_{i\mu}^{*}c_{i}^{\dagger},\\
\beta_{\mu}^{\dagger} & = & 
\sum_{i}U_{i\mu} c_{i}^{\dagger}+\sum_{i}V_{i\mu} c_{i}.
\end{eqnarray}
While the Hilbert space of single particle wave functions is infinite, 
in practice summations are truncated up to a maximum basis state $N$. It 
is convenient to introduce a block matrix notation and matrices of double 
dimension by writing the equation above as 
\begin{equation}
\left(\begin{array}{c}
\beta\\
\beta^{\dagger}
\end{array}\right)=\left(\begin{array}{cc}
U^{+} & V^{+}\\
V^{T} & U^{T}
\end{array}\right)\left(\begin{array}{c}
c\\
c^{\dagger}
\end{array}\right)=W^{+}\left(\begin{array}{c}
c\\
c^{\dagger}
\end{array}\right),
\end{equation}
which defines the matrix $W$ of the Bogoliubov transformation
\begin{equation}
W=\left(\begin{array}{cc}
U & V^{*}\\
V & U^{*}
\end{array}\right).
\label{eq:pes_W}
\end{equation}
The quasiparticle operators must satisfy canonical fermion commutation
relations, which can be summarized by the condition 
\begin{equation}
W^{+}\sigma W=\sigma, \ \ \ 
\sigma=\left(\begin{array}{cc}
0 & I_{N}\\
I_{N} & 0
\end{array}\right),
\end{equation}
with $I_{N}$ the $N$-dimensional identity matrix. The HFB 
wave function $|\Phi\rangle$ is defined as the vacuum of the 
quasiparticle annihilation operators, $\beta_{\mu}|\Phi\rangle=0$ for 
all $\mu$. This leads to writing $|\Phi\rangle=\prod_{\mu}\beta_{\mu}|0\rangle$ 
where $|0\rangle$ is the particle vacuum of Fock space and the product 
runs over all the $\mu$ indexes that render $|\Phi\rangle$ non zero. 

{\em Densities - }
If the ground-state wave function takes the form of a Bogoliubov vacuum, 
the Wick theorem guarantees that the basic degrees of freedom are the 
one-body density matrix $\rho$, the pairing tensor $\kappa$ and its 
complex conjugate $\kappa^{*}$ \cite{ring2000,blaizot1985}. Each of these objects 
can be expressed as a function of the Bogoliubov transformation,
\begin{equation}
\rho_{ij} 
= \left\langle \Phi\right|c_{j}^{\dagger}c_{i}\left|\Phi\right\rangle 
= (V^{*}V^{T})_{ij}, \ \ \ \ 
\kappa_{ij} 
= \left\langle \Phi\right|c_{i}c_{j}\left|\Phi\right\rangle 
= (V^{*}U^{T})_{ij}.
\label{eq:pes_densities}
\end{equation}
The notation can be further condensed by introducing the generalized 
density matrix,
\begin{equation}
{\cal R}=\left(\begin{array}{cc}
\rho & \kappa\\
-\kappa^{*} & 1-\rho^{*}
\end{array}\right),
\label{eq:pes_BigR}
\end{equation}
which is given in terms of the $W$ matrix of the Bogoliubov transformation
and the matrix of quasiparticle contractions
\begin{equation}
\mathbb{R}=\left(\begin{array}{cc}
\langle\beta_{\mu}^{\dagger}\beta_{\nu}\rangle           & \langle\beta_{\mu}\beta_{\nu}\rangle\\
\langle\beta_{\mu}^{\dagger}\beta_{\nu}^{\dagger}\rangle & \langle\beta_{\mu}\beta_{\nu}^{\dagger}\rangle
\end{array}\right)=\left(\begin{array}{cc}
0 & 0\\
0 & I_{N}
\end{array}\right) \label{eq:diagBigR}
\end{equation}
as $\mathcal{R}=W\mathbb{R}W^{\dagger}$. The generalized density matrix 
encapsulates in a single matrix all degrees of freedom of the theory and 
simplifies the formal manipulations required in the application of the 
variational least energy principle. 

{\em Variational principle - }
Since $\rho$, $\kappa$ and $\kappa^{*}$ are the only degrees of freedom, 
the energy can be expressed most generally as the functional $E[\rho,\kappa,
\kappa^{*}] \equiv E[\mathcal{R}]$. In practice, one often distinguishes between 
the particle-hole and particle-particle channel, 
\begin{equation}
E[\rho,\kappa,\kappa^{*} ] = E^{\mathrm{ph}}[\rho] + E^{\mathrm{pp}}[\kappa,\kappa^{*}]
\label{eq:fullEDF}
\end{equation}
although the distinction is a bit artificial since $\rho$ and
$\kappa$ are related through $\rho^2-\rho= - \kappa \kappa^+$.
The actual matrix $\mathcal{R}$ is obtained by minimizing the energy under
the condition that the HFB solution remains a quasiparticle vacuum, which 
is equivalent to $\mathcal{R}^{2} = \mathcal{R}$. We thus have to minimize
\begin{equation}
\mathcal{E} = E - \mathrm{tr} \left[ \Lambda(\mathcal{R}^{2} - \mathcal{R}) \right],
\end{equation}
where $\Lambda$ is a matrix of (undetermined) constraints. In this expression 
and the rest of this paper, ``tr'' refers to the trace in the single particle 
basis (possibly doubled). The variations of the energy are given by
\begin{equation}
\delta \mathcal{E} = 
\sum_{ab} \frac{\partial E}{\partial \mathcal{R}_{ab}}\delta \mathcal{R}_{ab} -
\sum_{ab} \frac{\partial}{\partial \mathcal{R}_{ab}}
\mathrm{tr} \left[ \Lambda(\mathcal{R}^{2} - \mathcal{R}) \right] \delta \mathcal{R}_{ab}.
\end{equation}
Since $\partial \mathcal{R}_{dc} /\partial \mathcal{R}_{ab} = \delta_{da}\delta_{cb}$, 
we find after some simple algebra,
\begin{equation}
\sum_{ab}\frac{\partial}{\partial \mathcal{R}_{ab}} \mathrm{tr} \left[ \Lambda(\mathcal{R}^{2} - \mathcal{R}) \right]
=
\sum_{ab} \left( \mathcal{R}\Lambda + \Lambda \mathcal{R} - \Lambda \right)_{ba}.
\end{equation}
Let us denote $\frac{1}{2}\mathcal{H}_{ba} = \partial E / \partial \mathcal{R}_{ab}$. 
In matrix form, the condition that the variations of the energy should be zero 
is expressed by the equation
\begin{equation}
\frac{1}{2}\mathcal{H} - \left( \mathcal{R}\Lambda + \Lambda \mathcal{R} - \Lambda \right) = 0.
\end{equation}
Pre- and post-multiplying this equation by $\mathcal{R}$ and subtracting the 
results after noticing that $\mathcal{R}^{2} = \mathcal{R}$, we find the final 
form of the HFB equation
\begin{equation}
[ \mathcal{H}, \mathcal{R} ] = 0.
\label{eq:pes_hfb}
\end{equation}
Since $\frac{1}{2}\mathcal{H}_{ba} = \partial E / \partial \mathcal{R}_{ab}$, we have 
by definition $\delta E = \frac{1}{2}\mathrm{tr}(\mathcal{H}\delta\mathcal{R})$. 
Considering the form (\ref{eq:pes_BigR}) of the generalized density, we find 
that the matrix of $\mathcal{H}$ reads
\begin{equation}
\mathcal{H} = \left(\begin{array}{cc}
 h  & \Delta \\
-\Delta^{*} & -h^{*}
\end{array}\right),
\label{eq:pes_BigH}
\end{equation}
with the mean field $h$ and pairing field $\Delta$ defined by 
\begin{equation}
\begin{array}{lcl}
\displaystyle h_{ij} = \frac{\partial E}{\partial \rho_{ji}},
& & \displaystyle \Delta_{ij} = \frac{\partial E}{\partial \kappa_{ij}^{*}},\\
\displaystyle \Delta_{ij}^{*} = \frac{\partial E}{\partial \kappa_{ij}}, 
& & \displaystyle h^{*}_{ij} =  \frac{\partial E}{\partial \rho_{ji}^{*}}.
\end{array}
\label{eq:pes_fields}
\end{equation}
The form of the HFB equation (\ref{eq:pes_hfb}) is not well suited for its
practical solution and, therefore, it is often reinterpreted by considering
that $[ \mathcal{H}, \mathcal{R} ] = 0$ implies that  $\mathcal{H}$ and
$\mathcal{R}$ can be diagonalized by the same Bogoliubov transformation. 
Since the generalized density matrix is diagonalized by $W$ as shown in 
(\ref{eq:diagBigR}), the same must hold
for $\mathcal{H}$,
\begin{equation}
\mathcal{H}W=W\mathcal{E}.
\label{eq:pes_HFBdiag}
\end{equation}
This represents a non-linear diagonalization problem (since $\mathcal{H}$
depends upon $W$ through the densities) with eigenvalues $\mathcal{E}$ and eigenvectors
$W$. It turns out that the matrix of eigenvalues can be written schematically
\begin{equation}
\mathcal{E}=\left(\begin{array}{cc}
 E_\mu  & 0 \\
 0 & -E_\mu
\end{array}\right),
\label{eq:pes_qp}
\end{equation}
that is, for each positive eigenvalue $E_{\mu}$, there is another one of 
opposite sign $-E_{\mu}$. The eigenvalues of the HFB matrix are referred to 
as ``quasiparticle energies''. In section \ref{subsec:solvers} page \pageref{subsec:solvers}, we 
discuss the technical aspects of solving such a non-linear eigenvalue problem.
The quasiparticle energies $E_\mu$ defined above should not be confused with
the single particle energies typical of the phenomenological mean field potentials 
of the macroscopic-microscopic method. The equivalent
to the single particle energies in the HFB case are the eigenvalues of the
Hartree-Fock (HF) Hamiltonian $h$ of (\ref{eq:pes_fields}). The study of 
those quantities allows to obtain the same degree of insight that can be 
obtained by the study of the Nilsson orbitals.

{\em Energy and fields - } In nuclear physics, the energy functional 
$E[\rho,\kappa,\kappa^{*}]$ is typically composed of two parts: one is 
derived from an effective two-body Hamiltonian,
\begin{equation}
\hat{H} = 
\sum_{ij} t_{ij} c_{i}^{\dagger}c_{j} 
+ 
\frac{1}{4} \sum_{ijkl} \bar{v}_{ijkl} c_{i}^{\dagger}c_{j}^{\dagger}c_{l}c_{k},
\end{equation}
where $t_{ij}$ refers to the matrix elements of the one-body kinetic energy 
operator and $\bar{v}_{ijkl}$ to the anti-symmetrized matrix elements of the 
two-body potential. The other part of the energy functional comes in the
form of various phenomenological density dependent terms that have to be handled 
with some care as they introduce additional terms in the HF Hamiltonian and
pairing fields that come in the form of rearrangement
potentials (see below). In the case of such an effective two-body Hamiltonian, 
the HFB energy is simply given by 
$E = \langle\Phi | \hat{H} | \Phi\rangle $, with $|\Phi\rangle = \prod_{\mu} 
\beta_{\mu}|0\rangle$ the quasiparticle vacuum introduced earlier. Considering
the definitions (\ref{eq:pes_densities}) for the density matrix and pairing 
tensor, the application of the Wick theorem gives
\begin{equation}
E= \mathrm{tr}(t\rho) + \frac{1}{2}\mathrm{tr}(\Gamma\rho) - \frac{1}{2}\mathrm{tr}(\Delta\kappa^{*}).
\label{eq:pes_EHFB_Veff}
\end{equation}
In this particular case, the mean field potential and the pairing field are 
given as a function of the two-body matrix elements by
\begin{equation}
\Gamma_{ij}=\sum_{kl}\overline{v}_{ijkl}\;\rho_{lk}, 
\ \ \ \ \ \ \ 
\Delta_{ij}=\frac{1}{2}\sum_{kl}\overline{v}_{ijkl}\;\kappa_{kl}.
\label{eq:pes_fields_Veff}
\end{equation}
and the HF Hamiltonian reads $h = t + \Gamma$. 
The notation can be further condensed by defining the generalized kinetic 
energy and mean field matrices by 
\begin{equation}
\mathcal{T}=\left(\begin{array}{cc}
t & 0\\
0 & -t^{*}
\end{array}\right), 
\ \ \ \ \ \ 
\mathcal{K}=\left(\begin{array}{cc}
\Gamma & \Delta\\
-\Delta^{*} & -\Gamma^{*}
\end{array}\right).
\label{eq:pes_BigT}
\end{equation}
With these notations, we note that $\mathcal{H}[\mathcal{R}] = \mathcal{T} + 
\mathcal{K}[\mathcal{R}]$. The total energy can then be written in the very 
compact form 
\begin{equation}
E = \frac{1}{4}\mathrm{tr}\left[\left(\mathcal{H}+\mathcal{T}\right)\mathcal{S}\right],
\end{equation}
with
\begin{equation}
\mathcal{S}=\left(\begin{array}{cc}
\rho & \kappa\\
-\kappa^{*} & -\rho^{*}
\end{array}\right)=\mathcal{R}-\left(\begin{array}{cc}
0 & 0\\
0 & I_{N}
\end{array}\right).
\label{eq:pes_BigS}
\end{equation}

{\em Density-dependent interactions - } 
In the practical implementation of the HFB theory in nuclear structure, 
special attention should be paid to the very common case of two-body, density-dependent 
effective ``forces'' $\hat{v}^{\mathrm{DD}}(\gras{r},\gras{r}') = 
\hat{v}^{\mathrm{DD}}[\rho(\gras{R})]\delta(\gras{r}-\gras{r}')$ with 
$\gras{R} = (\gras{r} + \gras{r}')/2$. Note that in this case, the potential 
$\hat{v}^{\mathrm{DD}}$ cannot be put into strict second quantized form as 
emphasized in \cite{erler2010}. However, one can still define an energy 
functional $E[\rho,\kappa,\kappa^{*}]$ by taking the  expectation value of 
$\hat{v}^{\mathrm{DD}}$ in coordinate space on a quasiparticle vacuum. Given the 
energy functional, the mean field and pairing field can then be defined from 
(\ref{eq:pes_fields}). When computing these derivatives with respect to 
the density matrix, additional contributions related to the matrix elements 
of $\partial\hat{v}^{\mathrm{DD}}/\partial\rho$ will arise (rearrangement
terms). They have the generic form
\begin{equation}
\Gamma_{ij} \rightarrow \Gamma_{ij} + \sum_{kl} \partial v^{\mathrm{DD}}_{ijkl}\rho_{lk} ,
\end{equation}
with
\begin{equation}
\partial v^{\mathrm{DD}}_{ijkl} 
=
\int d^{3}\gras{r}\int d^{3}\gras{r}'\; 
\varphi_{i}^{*}(\gras{r})\varphi_{j}^{*}(\gras{r}')
\frac{\partial\hat{v}^{\mathrm{DD}}}{\partial\rho}(\gras{R})
\varphi_{k}(\gras{r})\varphi_{l}(\gras{r}').
\end{equation}
Note that one could also consider pairing-tensor-dependent potentials in 
the same way. Recently, an extension of the above scheme as to consider
finite range density-dependent forces has been proposed \cite{chappert2015}.
In this case, the simple formulas above have to be adapted but the
same conceptual procedure remains valid. 

\label{par_thouless}
{\em Representation based on the Thouless theorem - } There is an alternative 
way to obtain and represent the HFB equation which is based on the Thouless 
theorem of \cite{thouless1960}. We recall that the Thouless theorem presented, 
e.g., in \cite{ring2000,blaizot1985}, establishes that two non-orthogonal 
quasiparticle vacuua $|\Phi_0\rangle$ and $|\Phi_1\rangle$ (corresponding to 
two different Bogoliubov transformations $W_0$ and $W_1$) are related through
\begin{equation}
|\Phi_1\rangle = \exp ( i \hat{Z} ) |\Phi_0\rangle,
\end{equation}
where $\hat{Z}$ is an hermitian one-body operator written in the quasiparticle
basis corresponding to $|\Phi_0\rangle$ as 
\begin{equation}
\hat{Z} = \sum_{\mu \nu} Z^{11}_{\mu \nu} \beta^\dagger_\mu \beta_\nu + 
\frac{1}{2} \sum_{\mu \nu} Z^{20}_{\mu \nu} \beta^\dagger_\mu \beta^\dagger_\nu + h.c.
\end{equation}
The complex matrices $Z^{11}$ (hermitian) and  $Z^{20}$ (skew-symmetric) are
determined by the requirement
\begin{equation}
W_1=W_0 \exp \left[ i \left(\begin{array}{cc}
Z^{11} & Z^{20}\\
-Z^{20\,*} & -Z^{11\,*}
\end{array}\right) \right]=W_0 \exp [i Z].
\end{equation}
The relation above can be used to generate a Bogoliubov transformation
$W_1$ given an initial one $W_0$ and a set of $Z^{11}$ and $Z^{20}$ coefficients.
Not surprisingly, the number of free parameters in $Z^{11}$ and $Z^{20}$ is the 
same as in a general Bogoliubov transformation $W$ after taking into account the 
condition $W^{+}\sigma W=\sigma$. Therefore, these matrices can be used as independent 
variational parameters. Starting from an initial Bogoliubov transformation $W$ 
and considering infinitesimal variations $\delta W$ characterized by the matrix 
$Z$ (that is, $Z^{11}$ and $Z^{20}$), we obtain to first order in $Z$ an explicit
expression for the variation of the Bogoliubov amplitude $W$ as $\delta W=iWZ$. 
It follows that the variation $\delta\mathcal{R}$ of the generalized density 
$\mathcal{R} = W \mathbb{R} W^{\dagger}$ under small variations $\delta W$ can 
be expressed in terms of the $Z$ matrix as
$\delta \mathcal{R} = i (\mathcal{ZR} - \mathcal{RZ}$) with $\mathcal{Z}=WZW^\dagger$.
From $\delta E = \frac{1}{2}\mathrm{tr}(\mathcal{H}\delta\mathcal{R})$
for the variation of the energy, we thus find
$\delta E= \frac{i}{2}\mathrm{tr}([\mathcal{R},\mathcal{H}]\mathcal{Z})$. 
The variational principle condition $\delta E=0$ yields the HFB equation 
$[\mathcal{R},\mathcal{H}]=0$. Finally, calculating the variation of the energy 
$\delta E$ to second order in $\mathcal{Z}$ tell us that, if $\mathcal{Z}$ is 
chosen as $i\eta[\mathcal{R},\mathcal{H}]^\dagger$, where $\eta$ is a 
sufficiently small step size, then $\delta E$ is negative. This result represents 
the essence of the gradient method commonly used in numerical implementations of 
the HFB equation and discussed in more details in section \ref{subsubsec:algorithms}; 
see also \cite{perez-martin2008} for a complete presentation.

\begin{figure}[!ht]
\begin{center}
\includegraphics[width=0.62\linewidth]{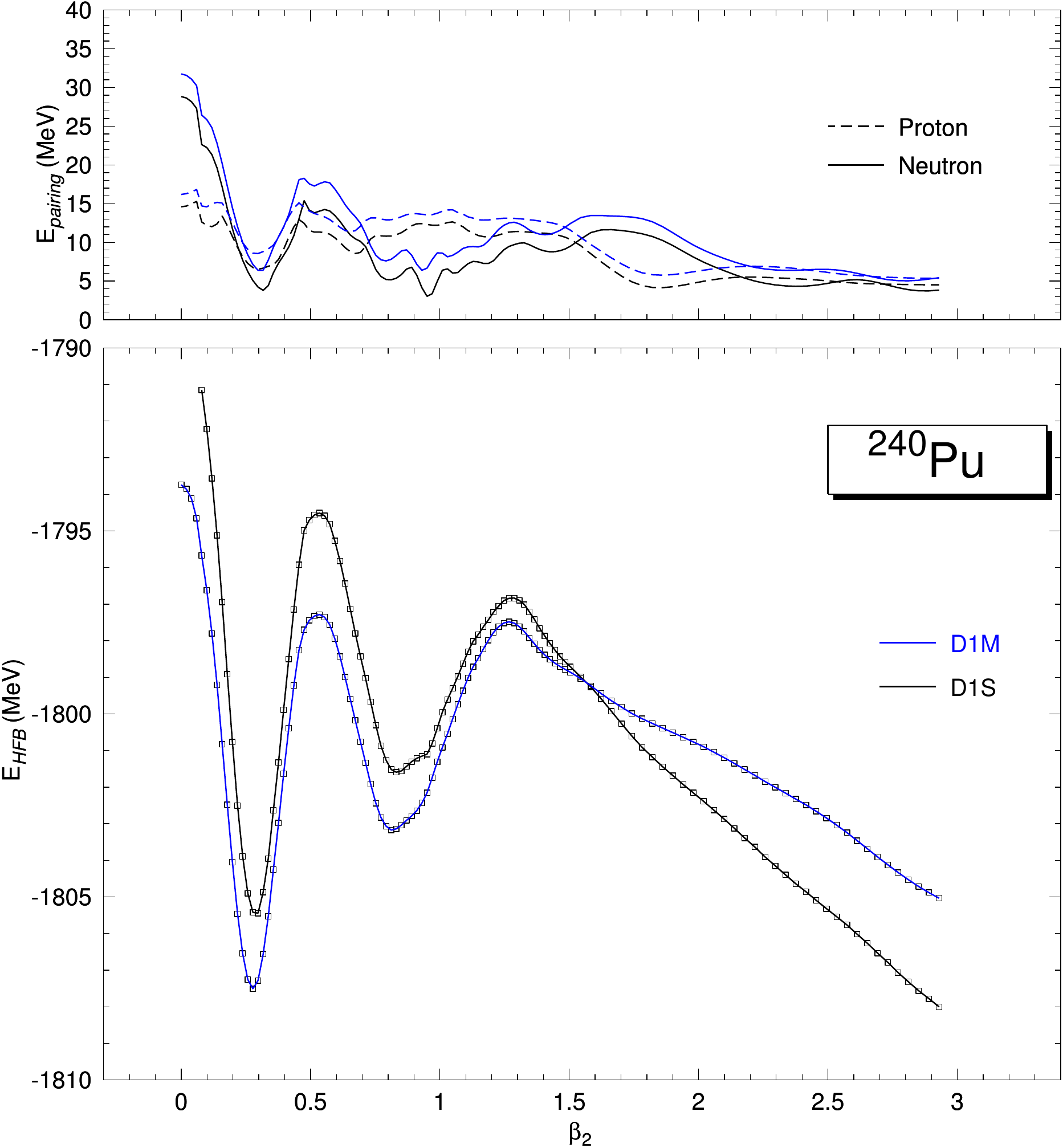}
\caption{Illustration of a constrained HFB calculation in the particular 
case of two parametrizations of the Gogny force, see \ref{subsubsec:edf}. 
Bottom panel: HFB energy (\ref{eq:pes_EHFB_Veff}) as a function of the 
quadrupole deformation. Top panel: pairing energy 
$\frac{1}{2}\mathrm{tr} (\Delta\kappa^{*})$ as a function of the quadrupole 
deformation.}
\label{fig:hfb}
\end{center}
\end{figure}

{\em Constraints - } By construction, the quasiparticle vacuum is not a eigenstate 
of the particle number operator. In practice, it is thus always necessary to impose 
a constraint on the average value of the particle number operator, both for protons 
and neutrons. Solving the HFB equation (\ref{eq:pes_hfb}) 
subject to constraints is also particularly important in fission studies. It is 
required in the evaluation of the potential energy surface as a function of the 
collective coordinates used to characterize the fission process. The 
handling of constraints in the formalism outline above is straightforward 
since it only requires replacing the HFB matrix by the auxiliary operator
$\mathcal{H}_{ij} \rightarrow \mathcal{H}_{ij} - 
\sum\lambda_{\alpha}\mathcal{O}_{\alpha,ij}$ where 
$\mathcal{O}_{\alpha,ij}$ are the matrix elements of the constraint 
operators $\hat{O}_{\alpha}$ in the doubled basis and $\lambda_{\alpha}$ 
are Lagrange multipliers. One-body operators such 
as, e.g., the particle number operator or multipole moments, have the same 
generic structure as in (\ref{eq:pes_BigT}). For example, solving the HFB 
equation with a constraint on particle number implies that the HFB matrix 
becomes
\begin{equation}
\mathcal{H} = \left(\begin{array}{cc}
h - \lambda & \Delta \\
-\Delta^{*} & -h^{*} + \lambda
\end{array}\right),
\label{eq:pes_BigH'}
\end{equation}
with $\lambda$ the Fermi energy. In the previous equation, $h-\lambda$ is a 
shorthand notation for $h_{ij} - \lambda\delta_{ij}$. Nearly all operators needed in fission are one-body 
operators, and their expectation value is thus given by 
$\langle \hat{O}_{\alpha}\rangle = \mathrm{tr}[\hat{O}_{\alpha}\rho]$. 
The figure \ref{fig:hfb} shows an example of constrained HFB solutions in 
the particular case of a single collective variable. In the figure, the 
quadrupole deformation $\beta_2$ is obtained from the axial quadrupole moment 
$\hat{Q}_{20}$ used as a constraint through  
$\beta_2 = \sqrt{5/(16\pi)} 4 \pi / (3 r_0^2 A^{5/3} ) \langle \hat{Q}_{20} \rangle$. 
Section \ref{subsec:collective} discusses in more details typical collective 
variables used in fission calculations.

To solve the non-linear HFB equation several approaches are used. A very
popular one is the iterative method, where the density matrix of the $n$-th 
iteration is used to compute the $\mathcal{H}$ matrix for the $(n+1)$-th one. 
Diagonalization of this matrix produces a new density matrix and the process 
is repeated until the generalized density matrix remains stationary up to 
the desired precision. Another popular method is based on the direct 
minimization of the energy using any of the variants of the  gradient 
method \cite{robledo2011}. Finally, the imaginary time method is 
also popular in the particular case of the HF+BCS equation. The advantages 
and disadvantages of the three of them will be discussed in more details in 
section \ref{subsubsec:algorithms}.


\subsubsection{BCS Approximation to the HFB Equation}
\label{subsubsec:bcs}

As immediately visible from the form (\ref{eq:pes_BigH'}) of the HFB matrix, 
solving the HFB equations require handling matrices twice larger than the 
simpler HF theory. In addition, the HFB spectrum is unbounded from below and 
from above, which leads to practical difficulties when trying to solve the 
HFB equation on a lattice, see section \ref{subsubsec:mesh}. These are some 
of the reasons why the BCS approximation to the HFB equation is sometimes 
preferred in applications.

The idea is to first write the HFB matrix $\mathcal{H}$ in a special basis -- 
the single particle basis where the HF Hamiltonian $h$ is diagonal with 
eigenvalues $e_{\mu}$ (the HF single particle energies). In other words, we 
find the transformation matrix $D$ such that $hD=De$ where $e$ is the matrix 
of eigenvalues $e_{\mu}$. The BCS approximation then consists in imposing  
that the skew-symmetric tensor $\Delta_{\mu\mu'}$ of (\ref{eq:pes_fields}) be diagonal in that basis. 
Specifically, we impose that it can only connect states with $\mu'=\bar{\mu}$, 
where the single particle state $\bar{\mu}$ is the partner of state $\mu$ 
under the time-reversal operator:
$\Delta_{\mu} \equiv \Delta_{\mu\mu'}\sim \Delta_{\mu\bar{\mu}} \delta_{\mu\mu'}$.
If we denote by $\Delta$ the diagonal matrix $\Delta_{\mu}$, we find that 
the HFB matrix (\ref{eq:pes_BigH'}) becomes
\begin{equation}
\mathcal{H}_{\mathrm{BCS}} = \left( \begin{array}{cccc}   
    e - \lambda    &     0     &     0   & \Delta \\
    0    &     e - \lambda     & -\Delta &    0   \\
    0    & -\Delta^* &    -e + \lambda  &    0   \\
\Delta^* &     0     &     0   &   -e + \lambda   \\
\end{array} \right).
\end{equation}
This matrix can easily be rearranged and block-diagonalized by a BCS 
transformation, which  is equivalent to solving the traditional ``gap 
equation'' of the BCS theory,
\begin{equation}
(e_{\mu} - \lambda)(u_{\mu}^{2} - v_{\mu}^{2}) + 2\Delta_{\mu}u_{\mu}v_{\mu} = 0.
\end{equation} 
The size of the matrix to be diagonalized, $h$, is reduced by a factor 
of 2 as compared to HFB, which in turns lowers the computational cost by a 
factor 8. Several test calculations, for example \cite{girod1983,samyn2005}, 
have shown that the HF+BCS approach is a reasonable approximation to the full 
HFB treatment for fission calculations. This conclusion does not hold any 
more for weakly-bound nuclei close to the neutron dripline, where the BCS 
approximation induces a coupling with an non-physical gas of neutrons as 
discussed in \cite{dobaczewski1996-a}. 

There is a relation between the quasiparticle energies $E_\mu$ of the HFB 
method and the single particle energies $e_\mu$, the Fermi energy  
$\lambda$ and the diagonal matrix element of the pairing field $\Delta_{\mu}$, 
$E_\mu =\sqrt{(e_\mu-\lambda)^2 + \Delta^2_{\mu}}$. It follows that the 
coefficients of the BCS transformation $u_\mu$ and $v_\mu$ acquire the very 
simple form
\begin{equation}
\left.\begin{array}{c}
u^2_\mu \\
v^2_\mu \\
\end{array}
\right\} =
\frac{1}{2} \left[1\pm \frac{e_\mu-\lambda}{\sqrt{(e_\mu-\lambda)^2+\Delta^2_{\mu}}}\right].
\end{equation}

In the case of a pure pairing force with constant matrix elements $-G$ 
the $\Delta_{\mu\mu'}$ matrix is already in diagonal form with 
constant matrix elements in the diagonal 
$\Delta = G \sum_{\mu>0} u_{\mu}v_{\mu}$. The $v_{\mu}$ and $u_{\mu}$ are 
again the coefficients of the BCS transformation. In this case, however, 
the attractive pairing interaction can only be active within a restricted 
window of single particle levels, being zero elsewhere.


\subsubsection{Energy Functionals}
\label{subsubsec:edf}

As already emphasized, most nuclear energy functionals used in fission 
studies are explicitly derived from the expectation value of effective, 
density-dependent, local two-body nuclear potentials at the HFB 
approximation. For the particle-hole channel, that is, the component 
$E^{\mathrm{ph}}[\rho]$ of the full functional (\ref{eq:fullEDF}), the 
two most popular functionals are based on the Skyrme and Gogny two-body 
effective potentials. Both EDFs are intended to be applicable throughout 
the nuclear chart using a limited set of parameters adjusted to global 
nuclear properties.

The Skyrme EDF is the energy functional $E[\rho]$ derived from the 
expectation value of the Skyrme potential introduced in \cite{skyrme1959} 
on a Slater determinant. The current standard form of the two-body
Skyrme interaction is
\begin{eqnarray}
\hat{V}_{12}(\gras{r}_{1},\gras{r}_{2}) 
& = & 
t_{0}(1 + x_{0}\hat{P}_{\sigma}) \delta(\gras{r}_{1}-\gras{r}_{2}) \\
& + &
\frac{1}{2}t_{1}(1 + x_{1}\hat{P}_{\sigma})
[\hat{\gras{k}'}^{2} \delta(\gras{r}_{1}-\gras{r}_{2}) +  \delta(\gras{r}_{1}-\gras{r}_{2}) \hat{\gras{k}}^{2}]  \medskip\nonumber \\
& + &
t_{2}(1 + x_{2}\hat{P}_{\sigma})
\hat{\gras{k}'}\cdot \delta(\gras{r}_{1}-\gras{r}_{2}) \hat{\gras{k}}  \medskip\nonumber \\
& + &
iW_{0}\hat{\gras{S}}\cdot [\hat{\gras{k}'}\delta(\gras{r}_{1}-\gras{r}_{2})\wedge\hat{\gras{k}}] 
\medskip\nonumber \\
& + &
\frac{1}{6}t_{3}(1 + x_{3}\hat{P}_{\sigma})
\rho^{\alpha}(\gras{R}_{12})
\delta(\gras{r}_{1}-\gras{r}_{2}),
\label{eq:pes_skyrme}
\end{eqnarray}
where 
\begin{equation}
\hat{\gras{k}} = \displaystyle-\frac{i}{2}(\gras{\nabla}_{1} - \gras{\nabla}_{2}),
\ \ \ 
\hat{\gras{k}'} = \displaystyle\frac{i}{2}(\overleftarrow{\gras{\nabla}}_{1} - \overleftarrow{\gras{\nabla}}_{2})
\end{equation}
are the relative momentum operator, $\gras{R}_{12} = (\gras{r}_1 + \gras{r}_2)/2$,  
$\hat{P}_{\sigma}$ is the spin exchange operator and 
$\hat{\gras{S}} = \gras{\sigma}_{1} + \gras{\sigma}_{2}$ is the total spin. 
By convention, $\hat{\gras{k}'}$ acts on the left. The Skyrme potential 
contains a phenomenological density-dependent term (term proportional 
to $t_{3}$). Originally this term was introduced with $\alpha=1$ and $x_3=1$ 
in \cite{vautherin1972} to simulate the effect of three-body forces, in 
particular with respect to the saturation of nuclear forces. Following 
\cite{koehler1976}, both $\alpha$ and $x_3$ are now usually taken as 
phenomenological adjustable parameters. The expectation value of the potential 
(\ref{eq:pes_skyrme}) on a Slater determinant can be expressed as a functional 
of the local density $\rho(\gras{r})$ and various other local densities 
derived from $\rho(\gras{r})$. The potential energy is then 
given by
\begin{equation}
E_{\mathrm{Skyrme}} = \int d^{3}\gras{r} 
\sum_{t=0,1} 
\left[ 
\mathcal{H}_{t}^{\mathrm{even}}(\gras{r}) + \mathcal{H}_{t}^{\mathrm{odd}}(\gras{r})
\right],
\label{eq:pes_skyrme_EDF}
\end{equation}
where $t=0$ refers to isoscalar energy densities, and $t=1$ to isovector 
ones. The contribution $\mathcal{H}_{t}^{\mathrm{even}}(\gras{r})$ to 
the total energy density depends only on time-even densities,
\begin{equation}
\mathcal{H}_{t}^{\mathrm{even}}(\gras{r}) =
C_{t}^{\rho}\rho_{t}^{2} 
+
C_{t}^{\Delta\rho}\rho_{t}\Delta\rho_{t} 
+
C_{t}^{\tau}\rho_{t}\tau_{t}
+
C_{t}^{J} \tensor{J}_{t}^{2}
+
C_{t}^{\nabla J} \rho_{t}\gras{\nabla}\cdot\gras{J}_{t},
\label{eq:pes_Heven}
\end{equation}
while the contribution $\mathcal{H}_{t}^{\mathrm{odd}}(\gras{r})$ depends 
only on time-odd densities,
\begin{equation}
\mathcal{H}_{t}^{\mathrm{odd}}(\gras{r}) =
C_{t}^{s}\gras{s}_{t}^{2} +
C_{t}^{\Delta s}\gras{s}_{t}\cdot\Delta\gras{s}_{t}
+
C_{t}^{T}\gras{s}_{t}\cdot\gras{T}_{t}
+
C_{t}^{j} \gras{j}_{t}^{2}
+
C_{t}^{\nabla j} \gras{s}_{t}\cdot\left(\gras{\nabla}\wedge\gras{j}_{t}\right).
\label{eq:pes_Hodd}
\end{equation}
The various densities involved in these expressions are: the kinetic energy 
density $\tau_t$; the spin current density $\tensor{J}_t$ (rank 2 tensor); the 
vector part of the spin current density $\gras{J}_{t}$ (which is obtained by 
the tensor contraction 
$\gras{J}_{\kappa,t} = \sum_{\mu\nu}\epsilon_{\mu\nu\kappa}\mathrm{J}_{\mu\nu,t}$); 
the spin density $\gras{s}_{t}$; the spin kinetic energy density $\gras{T}_{t}$ 
and the spin current density $\gras{j}_{t}$. Their actual expressions as a 
function of the one-body density matrix can be found in \cite{bender2003,engel1975,lesinski2007}. 
The Skyrme energy density is 
the most general scalar that can be formed by coupling the fields derived 
from the one-body density matrix up to second order in derivatives as 
discussed in \cite{perliska2004,dobaczewski1996}. There have been attempts 
to generalize the EDF (\ref{eq:pes_Heven})-(\ref{eq:pes_Hodd}) by going 
beyond the second order derivative \cite{carlsson2008,raimondi2011,
raimondi2011-a,becker2015} or by adding local, zero-range three-body forces to 
the potential given by (\ref{eq:pes_skyrme}) \cite{sadoudi2013}.

The Gogny force is the most-widely used non-relativistic effective 
finite-range nuclear potential. It can be viewed as a Skyrme potential 
where the zero range central potential is replaced by the sum of 
two Gaussians in the spatial part. In fact, the central part of the 
Skyrme force can be obtained by expanding a Gaussian central and 
spin-orbit potential up to second order in momentum space as shown in 
\cite{vautherin1972}. The main advantage of the finite range is that 
the matrix elements are free from ultraviolet divergences in the 
particle-particle channel, which allows using the same potential in 
both particle-hole and particle-particle channels without introducing
a window of active particles in the particle-particle channel. In its original formulation of 
\cite{gogny1975}, the Gogny force reads
\begin{eqnarray}
\hat{V}_{12}(\gras{r}_{1},\gras{r}_{2}) 
& = & 
\sum_{i=1}^2  \ (W_i + B_i\hat P_\sigma - H_i\hat P_\tau -M_i\hat P_{\sigma}\hat P_{\tau}) \ 
e^{-\frac{(\gras{r}_{1}- \gras{r}_{2})^{2}}{\mu_i^{2}}}
\medskip\nonumber\\
& + & 
iW_{\rm LS}\hat{\gras{S}}\cdot [\hat{\gras{k}'}\delta(\gras{r}_{1}-\gras{r}_{2})\wedge\hat{\gras{k}}]
\medskip\nonumber \\
& + & 
 \ t_3  (1+x_0 \hat P_\sigma) \
\rho^{\alpha}(\gras{R}_{12})
\delta(\gras{r}_{1}-\gras{r}_{2}).
\label{eq:pes_gogny}
\end{eqnarray}
Note that the energy of two-body finite-range potentials such as the Gogny could also 
be put in the form (\ref{eq:pes_skyrme_EDF}), the difference being that 
the energy densities $\mathcal{H}(\gras{r})$ would be expressed 
themselves not directly as functionals of the local density $\rho(\gras{r})$, 
but as integrals over $\gras{r}'$ involving the non-local density 
$\rho(\gras{r},\gras{r}')$.

For both the Skyrme and Gogny forces, the energy density functional is 
obtained from a central and spin orbit potential plus a phenomenological 
density-dependent term. Note that we have not discussed the inclusion 
of explicit tensor potentials in either the Skyrme or Gogny forces, see 
\cite{lesinski2007,bender2009,anguiano2012} for detailed discussions on
that topic. In the spirit of the 
Kohn-Sham theory, however, the EDF does not need to be derived from any 
potential and could be parametrized directly as a functional of the density. 
In the Barcelona-Catania-Paris-Madrid (BCPM) functional of \cite{baldo2008,
baldo2013}, this idea is only applied to the volume part of the functional, 
as it can easily be constrained by nuclear and neutron matter properties. 
The resulting functional of the density is supplemented with a term from 
a spin-orbit potential and another potential providing the surface term, 
leading to
\begin{equation} 
E[\rho] = T + E^{\mathrm{SO}}[\rho] + E_{\mathrm{int}}^{\infty}[\rho] + E_{\mathrm{int}}^{\mathrm{FR}}[\rho],
\label{BCPM} 
\end{equation}
where $T$ is the kinetic energy, $E^{\mathrm{SO}}$ is the spin-orbit
energy density obtained from a zero range spin-orbit potential, 
$E_{\mathrm{int}}^{\infty}$ is the bulk part given in terms of a fitting 
polynomial of the density with parameters adjusted to reproduce nuclear 
matter properties, and $E_{\mathrm{int}}^{\mathrm{FR}}$ is the surface 
energy obtained from a finite-range Gaussian potential. Similar ideas 
had also been pursued earlier by Fayans and collaborators in 
\cite{fayans1994,kroemer1995}. In their work, the parametrization of 
$E_{\mathrm{int}}^{\infty}$ was different from the BPCM functional, and 
the $E_{\mathrm{int}}^{\mathrm{FR}}$ term was derived from a Yukawa 
potential instead of a Gaussian.

In realistic calculations, the Coulomb potential must also be included 
for protons. The direct contribution of this potential to the energy 
does not pose any particular problem. The exchange term is usually 
treated in the Slater approximation although several studies such as 
\cite{anguiano2001} have shown the impact of both Coulomb exchange 
and the associated Coulomb anti-pairing effect in collective inertias.

As mentioned in section \ref{subsubsec:hfb}, the nuclear EDF comprises
two components, $E^{\mathrm{ph}}[\rho]$ in the p.h. channel and 
$E^{\mathrm{pp}}[\kappa,\kappa^{*}]$ in the p.p. channel. Once again, 
pairing functionals are most often obtained by taking the expectation 
value of an effective (two-body) potential on the quasiparticle vacuum. Users of 
the Gogny force often take the same potential for the pairing channel. 
In the case of the Skyrme EDF, it is customary to consider simple pairing 
forces that can be put into functionals of the local pairing density 
$\tilde{\rho}(\gras{r})$ introduced in \cite{dobaczewski1984}. The full 
one-body pairing density $\tilde{\rho}(\gras{r}\sigma,\gras{r}'\sigma')$,
depending on spatial ($\gras{r}$) and spin ($\sigma$) coordinates,
is related to the pairing tensor through $\kappa(\gras{r}\sigma,\gras{r}'\sigma')
= 2\sigma'\tilde{\rho}(\gras{r}\sigma,\gras{r}'-\sigma')$ ($\sigma, \sigma' = \pm 1/2$) 
and has similar symmetry properties as the one-body density matrix $\rho$. A 
commonly used pairing force is the density-dependent, zero range potential 
\begin{equation}
\hat{V}_{\mathrm{pair}}(\gras{r}_{1},\gras{r}_{2})
=
\sum_{\tau=n,p} V_{\tau}^{0} 
\left[ 1 - V_{\tau}^{1}\left(\frac{\rho_{0}(\gras{R}_{12})}{\rho_{c}}\right)^{\alpha}\right]
\delta(\gras{r}_{1}-\gras{r}_{2}),
\end{equation}
where $V_{\tau}^{0}$, $V_{\tau}^{1}$, $\rho_{c}$, $\alpha$ are adjustable 
parameters and $\rho_0$ the isoscalar one-body density. While $V_{\tau}^{0}$ 
controls the strength of the pairing interaction, $\rho_{c}$ represents 
the average density inside the nucleus (it is often set at 0.16 fm$^{-3}$) 
and, therefore, $V_{\tau}^{1}$ controls the surface or volume nature of 
the pairing interaction. 
Such a schematic pairing force was originally introduced in \cite{chasman1976}. 
Note that its zero-range requires introducing a cut-off in the number of 
active particles used to define the densities.
In both the BCPM and Fayans functionals, a similar zero-range density-dependent 
interaction was adopted \cite{fayans1994,PhysRevC.60.064312}. 

Irrespective of the mathematical form of the functional, nuclear EDFs 
contain free parameters that must be adjusted to experimental data. 
In the case of the Skyrme force, these are the $(t,x)$ parameters plus 
$W_0$ and $\alpha$; for Skyrme EDFs, the $C^{uu'}$ coupling constants 
(see \cite{kortelainen2010} for an alternative representation of the 
Skyrme EDF where volume coupling constants are expressed as a function 
of nuclear matter characteristics); in the case of the Gogny force, the 
parameters are the $W_i,B_i,H_i,M_i$ parameters of the central part, 
plus $W_{LS}$ and $x_0$. Fitting these parameters on select nuclear data 
is an example of an inverse problem in statistics, and several strategies 
have been explored in the past; see \cite{schunck2015-a,schunck2015-b} for 
a discussion.

In the case of the Skyrme EDF, there are hundreds of different 
parametrizations; see \cite{stone2007} for a review. However, not 
all of these parametrizations are suitable for fission studies, where 
the ability of the EDF to reproduce deformation properties is essential. 
The two most widely used Skyrme EDF to-date are the SkM* parametrization 
of Bartel {\it et al.} \cite{bartel1982} and the UNEDF1 parametrization of 
Kortelainen {\it et al.} \cite{kortelainen2012}. In both cases, the 
functionals were fitted by considering fission data in actinides, namely 
the height of the fission barrier in $^{240}$Pu for SkM* and the value of 
a few fission isomer excitation energies for UNEDF1. The case of the Gogny 
force is similar, although the number of Gogny parametrizations is much 
more limited since only 5 parametrizations have been published so far 
\cite{gogny1975,decharge1980,berger1984,chappert2008,goriely2009}. Just 
as for the Skyrme force, the original D1 and D1' parametrizations of the 
Gogny force could not reproduce accurately enough the height of the fission 
barrier in $^{240}$Pu. The D1S parametrization proposed by Berger in 
\cite{berger1989} was a slightly modified version of D1 where the surface 
energy coefficient in nuclear matter was modified to decrease the fission 
barrier. Since then, the D1S parametrization has been the most popular 
force for fission. Note that there are extensions of the Gogny force to 
include density dependencies in each term as in \cite{chappert2015} or to 
add a tensor force with a finite-range spatial part, see e.g., 
\cite{otsuka2006,anguiano2012}. 
In the BCPM case there is a single set of parameters, 
adjusted to nuclear matter properties (bulk part) and to the binding 
energies of even-even nuclei (surface term). Fission information has not 
been included in the fit.



\subsubsection{Excitation Energy}
\label{subsubsec:excitation}

As mentioned in the introduction, applications of the neutron-induced 
fission of actinides with fast neutrons (kinetic energies of the order 
of 14 MeV) involve excitation energies of the fissioning nucleus that can 
be as large as 20 MeV or more. At such excitation energies, the potential 
energy surface of the nucleus can be markedly different from what is 
obtained from constrained HF+BCS or HFB calculations. Similar and higher 
excitation energies can be reached during the formation of the compound 
nucleus in heavy ion fusion reactions, which are one of the primary 
mechanisms to synthesise heavy elements as discussed in \cite{hoffman2000}. 
As shown in \cite{bohr1975}, the level density increases exponentially 
with $E^{*}$, and at excitation energies beyond a few MeV, it becomes 
impossible to individually track excited states using direct methods. 
Finite-temperature density functional theory thus provides a convenient 
toolbox to quantify the evolution of nuclear deformation properties as a 
function of excitation energy. The inclusion of temperature in microscopic 
studies of fission has been done within the Hartree-Fock approach (FT-HF) 
in \cite{brack1974,bartel1985}, the HF+BCS approach in \cite{sheikh2009} 
and the fully-fledged HFB approach (FT-HFB) in \cite{martin2009,pei2009,sheikh2009,
schunck2015}. The reader can also refer to \cite{schunck2015} for additional 
references to the use of temperature in semi-microscopic methods. Note that 
in practical applications, the temperature must be related to the excitation 
energy of nucleus, which is not entirely trivial; see \cite{schunck2015} 
for a detailed discussion. 

Various elements of the FT-HF and FT-HFB theory can be found in \cite{blaizot1985,descloizeaux1968,lee1979,
goodman1981,tanabe1981,egido1993,egido2000,martin2003}. Let us recall that the 
nucleus is assumed to be in a state of thermal equilibrium at temperature $T$. 
It is then characterized by a statistical density operator $\hat{D}$, which contains all 
information on the system. In particular, once the density operator is known, 
the expectation value of any operator $\hat{F}$ can be computed by taking 
the trace $\mathrm{Tr}\, \hat{D}\hat{F}$ in any basis of the Fock space. Here, 
the notation ``Tr'' refers to the statistical trace by contrast to the trace ``tr'' 
used before with respect to a single particle basis. If we 
further assume the system can be described by a grand canonical ensemble (the 
average value of the energy and particle numbers are fixed), then the application 
of the principle of maximum entropy yields the following generic form of the 
density operator as 
\begin{equation}
\hat{D} = \frac{1}{Z} e^{-\beta(\hat{H} - \lambda\hat{N})},
\label{eq:density_operator}
\end{equation}
where $Z$ is the grand partition function, $Z = \mathrm{Tr}\, e^{-\beta(\hat{H} - \lambda\hat{N})}$, 
$\beta=1/kT$, $\hat{H}$ is the (true) Hamiltonian of the system, $\lambda$ the 
Fermi level and $\hat{N}$ the number operator. The demonstration of 
(\ref{eq:density_operator}) is given in \cite{blaizot1985,reichl1988}. 

In practice, the form (\ref{eq:density_operator}) is not very useful in nuclear 
physics, where the true Hamiltonian of the system is not known. The mean-field 
approximation provides a first simplification. It consists in replacing the true 
Hamiltonian $\hat{H}-\lambda\hat{N}$ by a simpler, quadratic form $\hat{K}$ of particle 
operators,
\begin{equation}
\hat{D}_{\mathrm{MF}} = \frac{1}{Z_{\mathrm{MF}}} e^{-\beta\hat{K}}, \ \ \ \
Z_{\mathrm{MF}} = \mathrm{Tr}\,e^{-\beta\hat{K}}.
\label{eq:D_MF}
\end{equation}
Given a generic basis $|i\rangle$ of the single particle space, with $c_{i}$ 
and $c_{i}^{\dagger}$ the corresponding single particle operators, the operator 
$\hat{K}$ can be written formally 
\begin{equation}
\hat{K} = \frac{1}{2}\sum_{ij} K_{ij}^{11} c_{i}^{\dagger}c_{j} +
\frac{1}{2}\sum_{ij} K_{ij}^{22} c_{i}c_{j}^{\dagger} \nonumber
+ \frac{1}{2}\sum_{ij} K_{ij}^{20} c_{i}^{\dagger}c_{j}^{\dagger}
+ \frac{1}{2}\sum_{ij} K_{ij}^{02} c_{i}c_{j}
\label{eq:D_HFB}.
\end{equation}
At this point, the matrices $K^{11}$, $K^{22}$, $K^{20}$ and $K^{02}$ in 
(\ref{eq:D_HFB}) are still unknown. 

The next step is to take advantage of the Wick theorem for ensemble averages \cite{GAUDIN196089}. This 
theorem establishes that when the density operator has the form (\ref{eq:D_MF}), 
statistical traces defining the expectation value of operators can be expressed 
uniquely in terms of the generalized density $\mathcal{R}$, which is defined by
\begin{equation}
\mathcal{R} = 
\left( \begin{array}{cc}
\langle c_{j}^{\dagger}c_{i} \rangle           & \langle c_{j}c_{i}           \rangle \medskip\\
\langle c_{i}^{\dagger}c_{j}^{\dagger} \rangle & \langle c_{j}c_{j}^{\dagger} \rangle,
\end{array}\right),
\label{eq:R_T}
\end{equation}
where $\langle \cdot \rangle$ refers to the statistical average: for instance, 
$\langle c_{j}^{\dagger}c_{i} \rangle = \mathrm{Tr}( \hat{D} c_{j}^{\dagger}c_{i} ) 
= \rho_{ij}$ is the one-body density matrix, $\langle c_{i}c_{j} \rangle = \mathrm{Tr}( \hat{D} c_{i}c_{j} ) 
= \kappa_{ij}$ is the pairing tensor, etc. The consequence of the Wick theorem is 
that once the generalized density $\mathcal{R}$ of (\ref{eq:R_T}) is known, the expectation 
value $\langle\hat{F}\rangle $ of an arbitrary operator $\hat{F}$ can be computed by 
\begin{equation}
\langle\hat{F}\rangle = \mathrm{Tr} \left( \hat{D}_{\mathrm{MF}} \hat{F} \right) 
\rightarrow 
\langle\hat{F}\rangle = \frac{1}{2}\mathrm{tr} \left( \mathcal{R} F \right). 
\end{equation}
In other words, the computation of the statistical trace over a many-body basis 
of the Fock space can be substituted by the much simpler operation of taking the 
trace within the quasiparticle space. In effect, this implies that the Wick theorem 
transfers the information content about the system from the density operator 
$\hat{D}_{\mathrm{MF}}$ into the generalized density $\mathcal{R}$.

The final step is thus to determine the generalized density. This is where we take 
advantage of the fact that the matrix elements of the operator $\hat{K}$ in 
(\ref{eq:D_HFB}) are arbitrary and can be taken as variational parameters of the 
theory. We thus determine them by requesting that the grand potential be minimum 
with respect to variations $\delta\hat{K}$. This leads to the identification of 
$K$ with the usual HFB matrix $\mathcal{H}$, $K = \mathcal{H}$
where $\mathcal{H}$ has the generic form (\ref{eq:pes_BigH}), and to the relation 
\begin{equation}
\mathcal{R} = \frac{1}{1 + \exp(\beta\mathcal{H})},
\label{eq:eq_HFB}
\end{equation}
where $\beta = 1/kT$. Equation (\ref{eq:eq_HFB}) is the FT-HFB equation; see 
\cite{blaizot1985,egido1993} for the demonstration. It establishes a self-consistency 
condition between the HFB matrix $\mathcal{H}$ and the generalized density 
$\mathcal{R}$. In practice, this self-consistency condition is most easily satisfied 
in the basis where $\mathcal{H}$ is diagonal. In this basis (the usual quasiparticle 
basis of the HFB equations), one easily shows the following relation
\begin{equation}
\mathrm{Tr} \left( \hat{D}\beta_{\mu}^{\dagger}\beta_{\mu} \right)
= \frac{1}{1 + e^{\beta E_{\mu}}} \delta_{\mu\nu} = f_{\mu\nu}\delta_{\mu\nu},
\end{equation}
with $E_{\mu}$ the quasiparticle energy, i.e., the eigenvalue $\mu$ of $\mathcal{H}$. This 
result allows one to extract the expression $\mathbb{R}$ of the generalized density in the 
quasiparticle basis, recall (\ref{eq:diagBigR}). By applying the Bogoliubov transformation, 
$\mathcal{R} = W\mathbb{R}W^{\dagger}$, one then finds the 
following expression for the density matrix and pairing tensor in the single particle basis,
\begin{eqnarray}
\rho_{ij}   & = & 
= \left( V^{*}(1-f)V^{T} \right)_{ij}
+ \left( UfU^{\dagger} \right)_{ij},
\label{eq:rho}\medskip\\
\kappa_{ij} & = & 
= \left( V^{*}(1-f)U^{T} \right)_{ij}
+ \left( UfV^{\dagger} \right)_{ij},
\label{eq:kappa}
\end{eqnarray}
where $U$ and $V$ are the matrices of the Bogoliubov transformation.

The finite-temperature extension of the HFB theory poses two difficulties.
While in the HFB theory at zero temperature the one-body density matrix 
and two-body pairing tensors are always localized for systems with 
negative Fermi energy \cite{dobaczewski1984,dobaczewski1996-a}, this is not 
the case at finite temperature. In particular quasiparticles with 
$E_{\mu} > -\lambda$ bring a non-localized contribution to the one-body 
density matrix through the $\left( UfU^{\dagger} \right)$ term. This effect 
is discussed in \cite{schunck2015,brack1974,bonche1984,bonche1985}. In 
addition, in the statistical description of the system by a grand canonical 
ensemble, only the average value of the energy and of the particle number 
(or any other constrained observable) are fixed. Statistical fluctuations 
are also present \cite{egido1988}. They increase with temperature and decrease 
with system size \cite{reichl1988}. The FT-HFB theory thus gives only the 
most probable solution within the grand-canonical ensemble, the one that 
corresponds to the lowest free energy. Mean values and deviations around 
the mean values of any observable $\hat{\mathcal{O}}$ should in principle 
be taken into account. In the classical limit, they are given by 
\begin{equation}
\bar{\mathcal{O}} =
\frac{\displaystyle \int d^{N}\gras{q}\; \mathcal{O}(\gras{q})
e^{-\beta F(T,\gras{q})}}{\displaystyle \int d^{N}\gras{q}\;  
e^{-\beta F(T,\gras{q})}}.
\end{equation}
Such integrals are computed across the whole collective space defined by 
the variables $\gras{q} = (q_{1}, \dots, q_{N})$ and require the knowledge 
of the volume element $d^{N}\gras{q}$. This was done for instance in 
\cite{martin2003}. Other possibilities involve functional integral methods 
as in \cite{levit1984}. 


\subsubsection{Beyond Mean-field}
\label{subsubsec:beyond}

Potential energy surfaces should in principle be corrected to account for
beyond mean field correlations. In particular, all broken symmetries 
(translational, rotational, parity, particle number) should be restored, 
which yields an additional correlation energy to the system. Moreover, the 
variational HFB equation should, in principle, be solved after the 
projection on good quantum numbers has been performed (variation after 
projection, VAP). 

The general problem of symmetry restoration is most naturally formulated 
in the framework of the SCMF model: given an effective Hamiltonian, one 
can associate a projector operator $\hat{P}_{g}$ to any broken symmetry 
and compute $E_{g} = \langle\Phi| \hat{H}\hat{P}_{g}|\Phi\rangle$. When 
$|\Phi\rangle$ is the symmetry-breaking HFB state minimizing the HFB energy, $E_g$ corresponds 
to the projection after variation (PAV) result; if $|\Phi\rangle$ is 
determined from the equations obtained after varying $E_{g}$, the energy 
is the VAP result. The correlation energy is defined as the difference 
$E_{g} - E_{0}$, where $E_{0} = \langle\Phi| \hat{H} |\Phi\rangle$. 
Unfortunately, projection techniques, especially the VAP, are 
computationally very expensive when multiple symmetries are broken (as 
in fission). In addition, the thorough analysis of \cite{stoitsov2007,
bender2009-a,duguet2009,lacroix2009} showed that they are, strictly speaking, 
ill-defined as soon as $\hat{H}$ contains density-dependent terms. Finally, 
in spite of the recent work of \cite{dobaczewski2009,hupin2011,hupin2012}, 
it is not really clear how such correlation energies can be rigorously 
computed in a strict EDF framework where no Hamiltonian is available.

In practice, the situation depends on the symmetry considered:
\begin{enumerate}
\item {\bf Translational invariance}: The kinetic energy of the centre of mass 
must be subtracted to account for the correlation energy due to restoration
of translational invariance \cite{ring2000}. The correlation energy is most 
often computed as $-\langle\hat{\gras{P}}^{2}_{\mathrm{cm}}\rangle/2mA$, 
which is a first-order approximation of the VAP result as shown in 
\cite{villars1971}. Here, $\hat{\gras{P}} = \sum_{i} \hat{\gras{p}}_{i}$ 
is the total linear momentum of the system and the expectation value is 
taken on the quasiparticle vacuum. In heavy nuclei, the study of 
\cite{bender2000} showed that its value can vary by about 1 MeV as a 
function of deformation.
\item {\bf Rotational invariance}: For the reasons mentioned above, 
the correlation energy induced by angular momentum projection is 
also often taken into account by approximate formulas. For large 
deformations of the intrinsic reference state, the rotational correction 
energy can be approximated by $\epsilon_{R} = -\langle \gras{J} \rangle^{2} 
/(2\mathcal{J})$, where $\mathcal{J}$ is the nuclear moment of inertia 
(which depends on the deformation) \cite{ring2000}. In this expression, 
the analyses of \cite{PhysRevC.2.892,villars1971,PhysRevC.62.054319} showed 
that one should in principle use the Peierls-Yoccoz moment of inertia  of 
\cite{peierls1957} for the denominator although many authors use the 
Thouless-Valatin one. Typical values of the rotational correction range
from zero MeV for spherical nuclei to 7-8 MeV for strongly quadrupole 
deformed configurations, as illustrated in the bottom panel of figure 
\ref{fig:corrections}.

\begin{figure}[!ht]
\begin{center}
\includegraphics[width=0.65\linewidth]{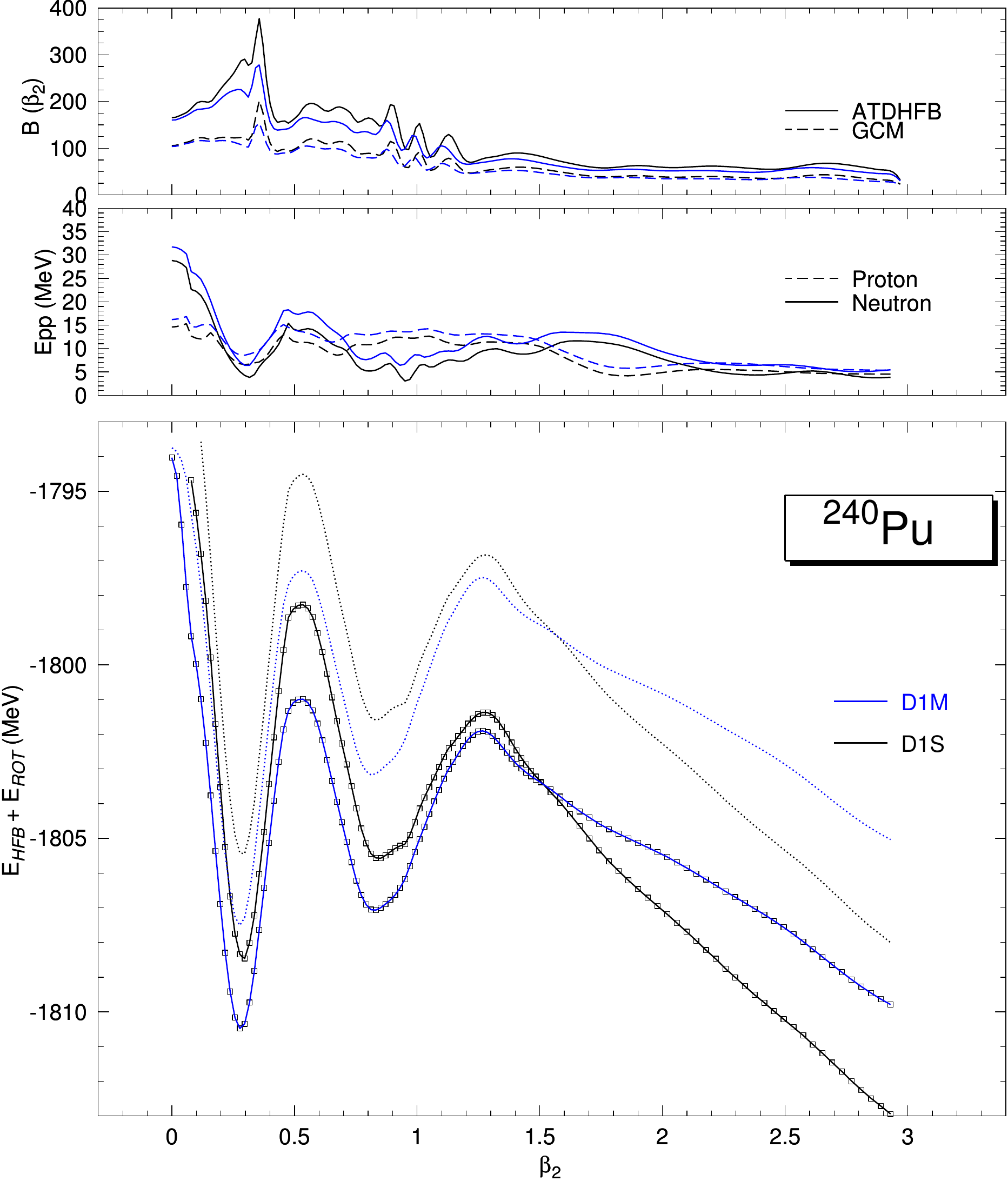}
\caption{Illustration of the impact of the rotational energy correction 
on the total HFB energy as a function of the quadrupole deformation $\beta_2$.
The HFB energy is represented by dotted lines and the one corrected with
the rotational correction by full lines with symbols.}
\label{fig:corrections}
\end{center}
\end{figure}

\item {\bf Reflection symmetry}: Asymmetric fission is explained by invoking 
potential energy surfaces where the nuclear shape breaks reflection 
symmetry. Parity projection is, therefore, required to restore left-right 
symmetry. Due to the discrete nature of the symmetry (only two states 
involved) the correlation energy is negligible for the typically large 
values of the octupole moment in fission as shown in \cite{hao2012}.
\item {\bf Particle number}: By definition, the quasiparticle vacuum does not 
conserve particle number and this symmetry should also be restored. There 
have been very few studies of the impact of this kind of correlation 
energy on the PES, and the few available results, for instance of 
\cite{Laft01}, are only based on the Lipkin-Nogami approximate particle 
number restoration scheme. Based on results obtained in other applications, 
this correlation energy could modify the values of the typical quantities 
characterizing the PES (barrier height, fission isomer excitation energy, 
etc.) by at most a couple of MeV. The impact on the collective inertia, 
however, could be significant. In addition to this, the particle number
breaking intrinsic states are averages of wave functions with different
numbers of protons and neutrons and therefore symmetry restoration is
critical to recover nuclear properties strongly dependent on particle 
number.
\end{enumerate}

The issue of how to describe those correlation energies in the transition 
from the one-fragment regime to the two-fragments one is also of paramount 
importance in order to determine the kinetic energy distribution of the 
fragments. The works of \cite{berger1980,urbano1981,goeke1983,skalski2007} 
are among the few attempts to describe such transitions in the HFB framework.


\subsection{The Collective Space}
\label{subsec:collective}

The success of the adiabatic approximation relies entirely on the small set 
of collective variables that are assumed to dominate the fission process. 
However, there is some degree of arbitrariness in the choice of these 
collective variables. The success of semi-phenomenological approaches to 
fission suggests using collective variables related to the shape of the 
nucleus (in its intrinsic frame). In DFT, this will be implemented by 
introducing the relevant operators and solving the HFB equations under 
constraints on the expectation values of said operators; see section 
\ref{subsubsec:multipole} for a discussion of the most typical choices.
Although these ``geometrical'' degrees of freedom are the most important for 
a realistic description of fission, recent studies have shown that fission 
dynamics can be very sensitive to additional collective degrees of freedom 
such as pairing correlations. Just as for nuclear deformations, there exist 
several possibilities to define collective variables associated to the 
pairing channel. These options are discussed in section 
\ref{subsubsec:other}.


\subsubsection{Parametrizations of the Nuclear Shape}
\label{subsubsec:multipole}

In nuclear DFT, the traditional parametrization of the nuclear shape is 
based on mass multipole moments $q_{\lambda\mu}$. These quantities are 
computed in the intrinsic frame of reference of the nucleus as expectation 
values of the operators
\begin{equation}
\hat{Q}_{\lambda\mu} = C_{\lambda\mu} \int d^{3}\gras{r}\; r^{\lambda}Y_{\lambda\mu}(\theta,\varphi),
\end{equation}
where $Y_{\lambda\mu}(\theta,\varphi)$ are the usual spherical harmonics, 
and $C_{\lambda\mu}$ are arbitrary coefficients introduced for convenience; 
see for example \cite{dobaczewski2004} for one particular choice. Since these 
operators are spin-independent, one-body operators, their expectation value 
is simply
\begin{equation}
q_{\lambda\mu} 
= \mathrm{tr} (\hat{Q}_{\lambda\mu}\rho)
= C_{\lambda\mu} \int d^{3}\gras{r}\; \rho(\gras{r})r^{\lambda}Y_{\lambda\mu}(\theta,\varphi).
\label{eq:pes_multipole}
\end{equation}
The mass multipole moments $q_{\lambda\mu}$ are the analogues to the
deformation parameters $\alpha_{\lambda\mu}$ of the nuclear surface 
parametrization of (\ref{eq:multipole_exp}) used in the 
macroscopic-microscopic approach. In fact, since multipole moments scale 
like the mass of the nucleus, it is sometimes advantageous to convert them 
to the $A$-independent deformation parameters $\alpha_{\lambda\mu}$. The 
most commonly used technique to do so, which is recalled in \cite{ring2000}, 
is to insert a constant density $\rho_0$ in (\ref{eq:pes_multipole}), compute 
the volume integral for a surface defined by (\ref{eq:multipole_exp}) and 
expand the result to first order in $\alpha_{\lambda\mu}$. The result is thus 
only valid for small deformations. At large deformations, other methods can 
be used, see, e.g., \cite{nikolov2011} and references therein for a discussion.
Note that there is an important difference between the explicit shape parametrization 
of the macroscopic-microscopic models and the choice of multipole moments as 
collective variables: in the nuclear DFT the multipole moments that are not 
constrained take a possibly non-zero value so that the energy is minimal, 
whereas in the macroscopic-microscopic approach they are zero.

\begin{figure}[!ht]
\begin{center}
\includegraphics[width=0.75\linewidth]{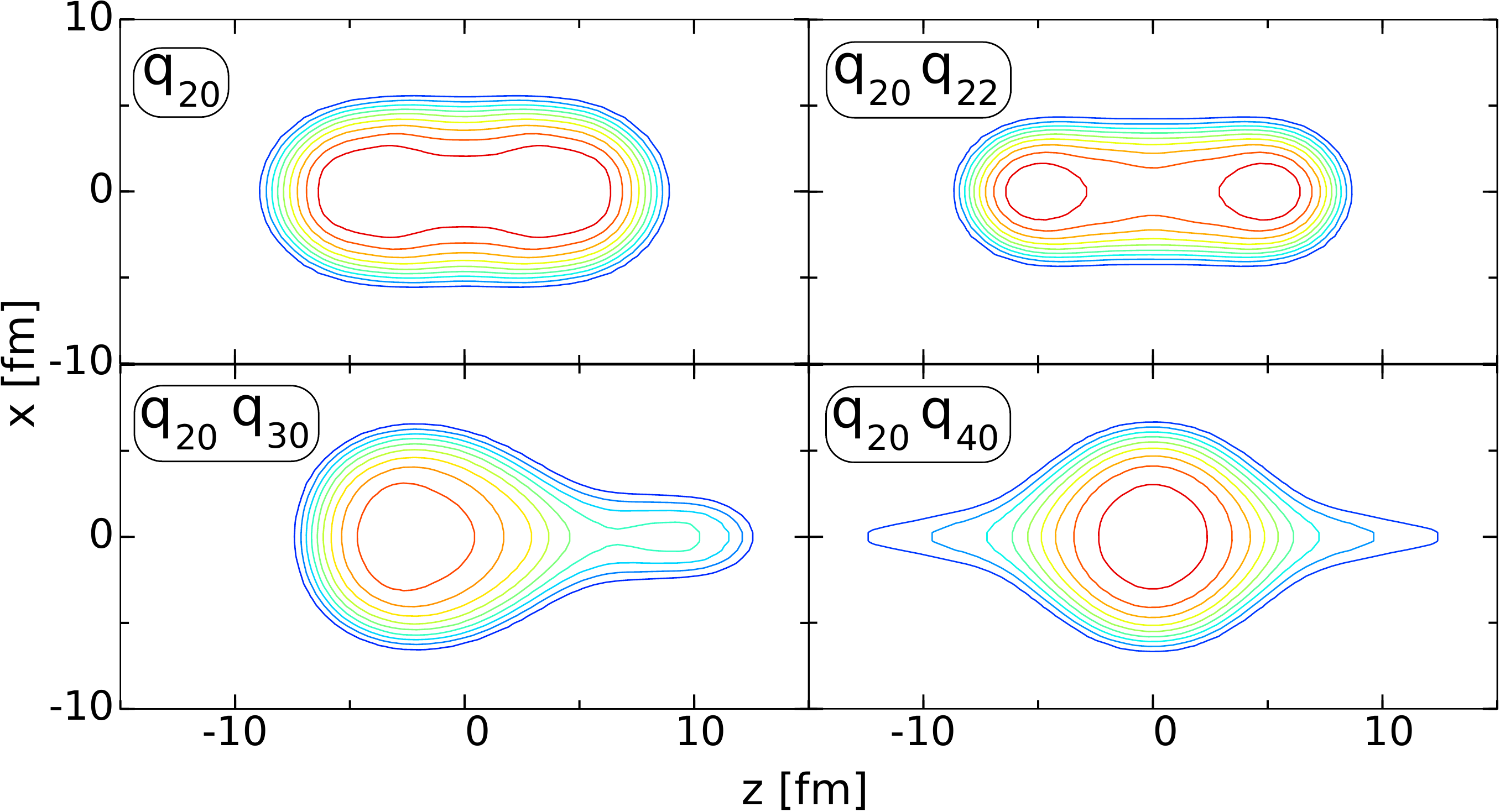}
\caption{Visual representation of the mass multipole moments. In each frame, the 
axial quadrupole moment is constrained to $q_{20} = 60$ b. In the frame labelled 
`$q_{20}$', there is no other constraint; in the frame `$q_{20}$ $q_{22}$', the 
triaxial quadrupole moment is fixed at $q_{22} = 30$ b (equivalent to $\gamma = 30^{\mathrm{o}}$); 
in the frame `$q_{20}$ $q_{30}$', the axial octupole moment is fixed at $q_{30} = 30$ b$^{3/2}$; 
in the frame `$q_{20}$ $q_{40}$', the axial hexadecapole moment is fixed at $q_{40} = 20$ b$^{2}$. 
All calculations are performed in $^{240}$Pu; see \cite{schunck2014} for 
technical details.}
\label{fig:pes_shapes}
\end{center}
\end{figure}

In the context of fission, by far the most important collective variable 
is the axial quadrupole moment $q_{20}$, which represents the elongation 
of the nucleus. The degree of triaxiality of nuclear shapes is captured 
by either $q_{22}$ or the ratio $\mathrm{tan} \gamma \propto q_{22}/q_{20}$ 
(the exact relation depends on the convention chosen for the normalization 
$C_{\lambda\mu}$ of the multipole moments).
Several studies, e.g., in \cite{staszczak2005,larsson1972,girod1983,schunck2014}, 
have confirmed that triaxiality is particularly 
important to lower the first fission barriers of actinides. 
The mass asymmetry of fission fragments, especially in actinides, 
can be well reproduced by introducing non-zero values of the axial mass 
octupole moment $q_{30}$. The hexadecapole moment $q_{40}$ has been mostly 
used as collective variable in the study of the neutron-induced fission of 
actinides: starting from the ground-state, the two-dimensional calculations 
of PES in the $(q_{20},q_{40})$ plane reported in \cite{berger1984,
berger1986,berger1989,schunck2014} showed the existence of two valleys, the 
fission (high $q_{40}$-values) and fusion (lower $q_{40}$ values) valleys; see 
also figure \ref{fig:pes_pu240} page \pageref{fig:pes_pu240}. 
The figure \ref{fig:pes_shapes} gives a visual representation of the impact 
of each of these multipole moments on the nuclear shape.

\begin{figure}[!ht]
\begin{center}
\includegraphics[width=0.35\linewidth,angle=-90]{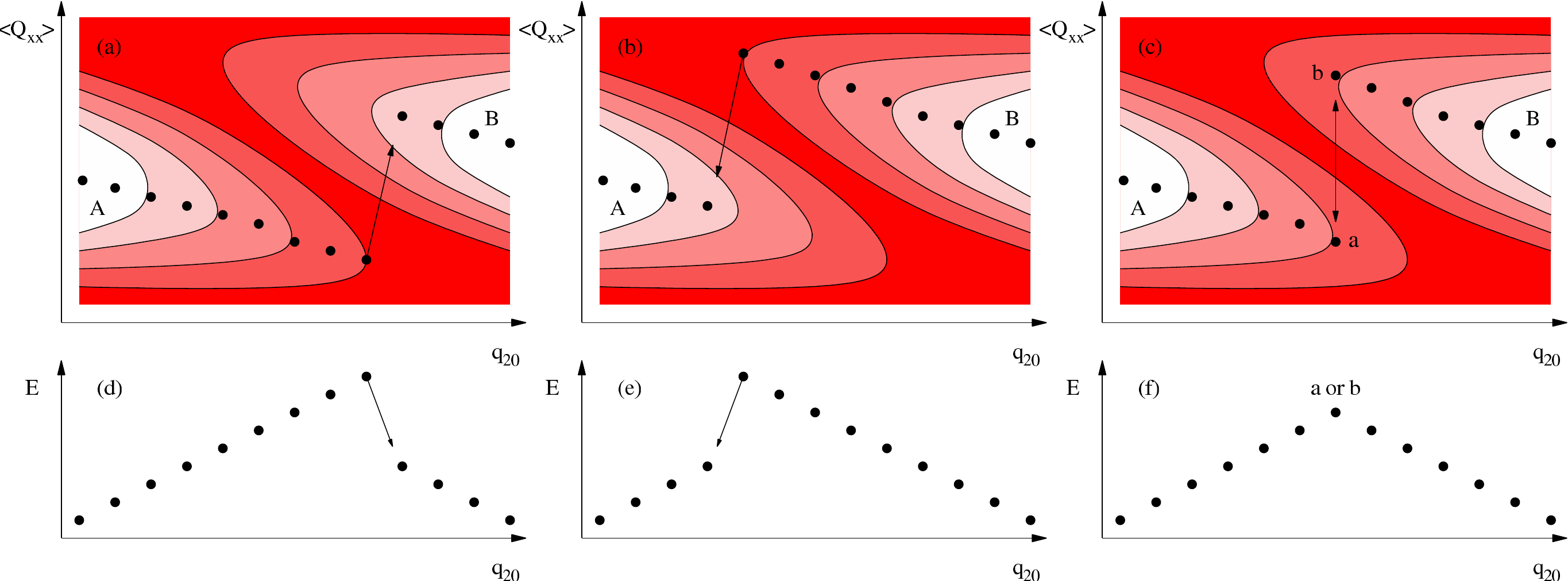}
\caption{Pedagogical illustration of discontinuities in PES calculations when 
the PES is generated by a step-by-step algorithm using neighbouring points. 
Figure taken from \cite{dubray2012}, courtesy of N. Dubray; 
copyright 2012 by IOP.}
\label{fig:continuity}
\end{center}
\end{figure}

Let us insist that in the HFB theory, the expectation value of {\em any} 
multipole moment can take non-zero values if symmetries allow it: for 
example, even when $q_{20}$ is the only collective variable (= the only
constraint on the HFB solution), all other multipole moments will vary 
along the $q_{20}$ path in such a way as to minimize locally the total 
energy. By taking advantage of the non-linear properties of the HFB 
equations and switching on/off constraints along the iterative process, 
it is therefore often possible to compute a $N$-dimensional PES that is 
guaranteed to be a {\it local} minimum in the full variational space. In 
practice, however, one often encounters situations where a HFB solution 
is a local minimum in the full variational space but not in the $N$-dimensional 
subspace defined by the collective variables $\gras{q}$. This point is 
illustrated in figure \ref{fig:continuity} for a toy-model two-dimensional 
collective space. In panel (a), calculations are initialized with the 
solution at point A and each value of $q_{20}$ is obtained by starting 
the calculation with the solution at a lower $q_{20}$ value. The resulting 
path follows the left valley in the ($q_{20}, Q_{xx})$ space and rejoins 
the right valley only when the barrier between the two vanishes; conversely, 
in panel (b) calculations are initialized at point B and the step-by-step 
process follows the right valley. The resulting trajectories are markedly 
different and show a discontinuity in energy. Panel (c) illustrates a 
possible continuous trajectory. These points were discussed extensively 
in \cite{dubray2012}.

\begin{figure}[!ht]
\begin{center}
\includegraphics[width=0.45\linewidth]{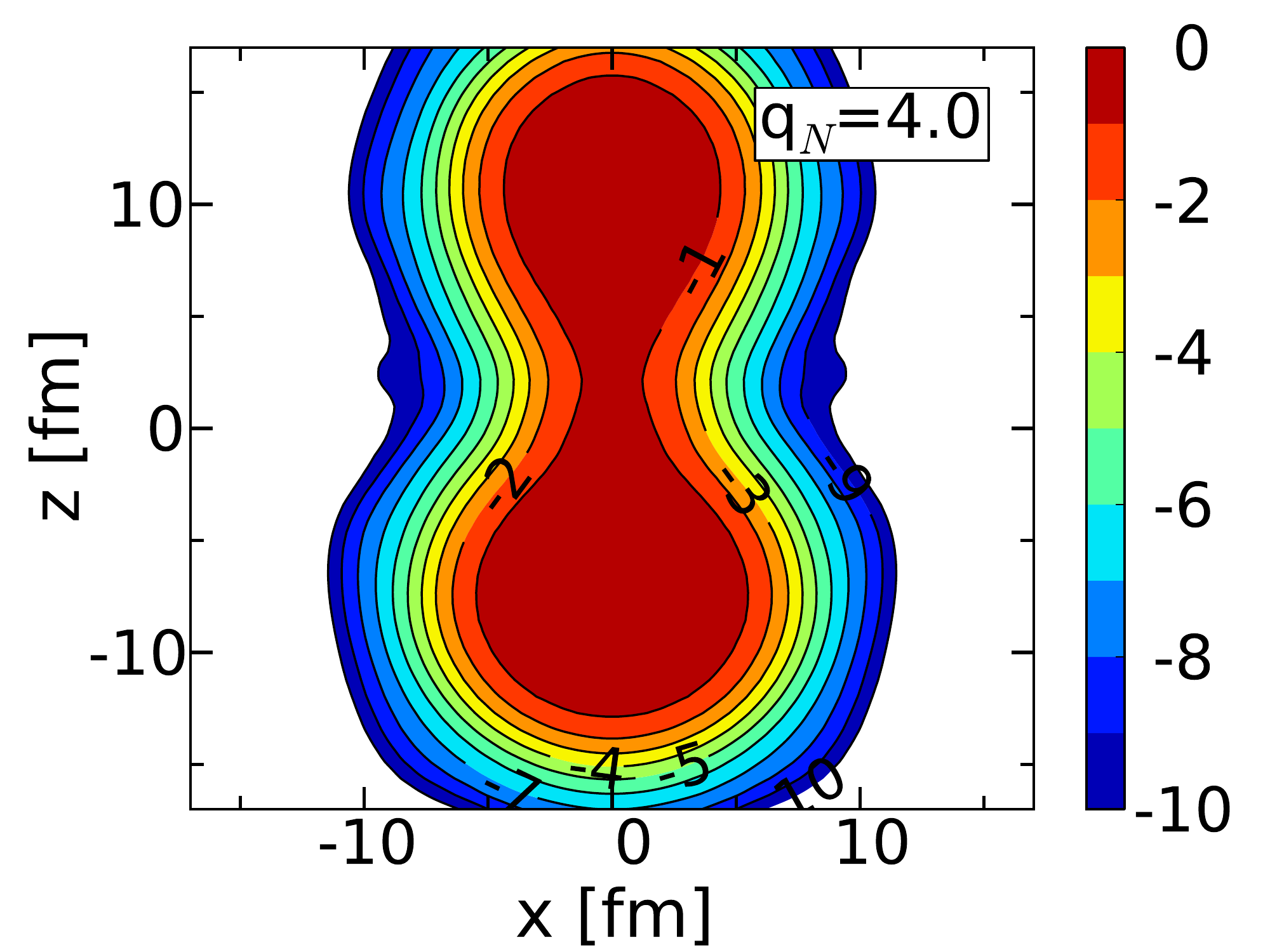}
\includegraphics[width=0.45\linewidth]{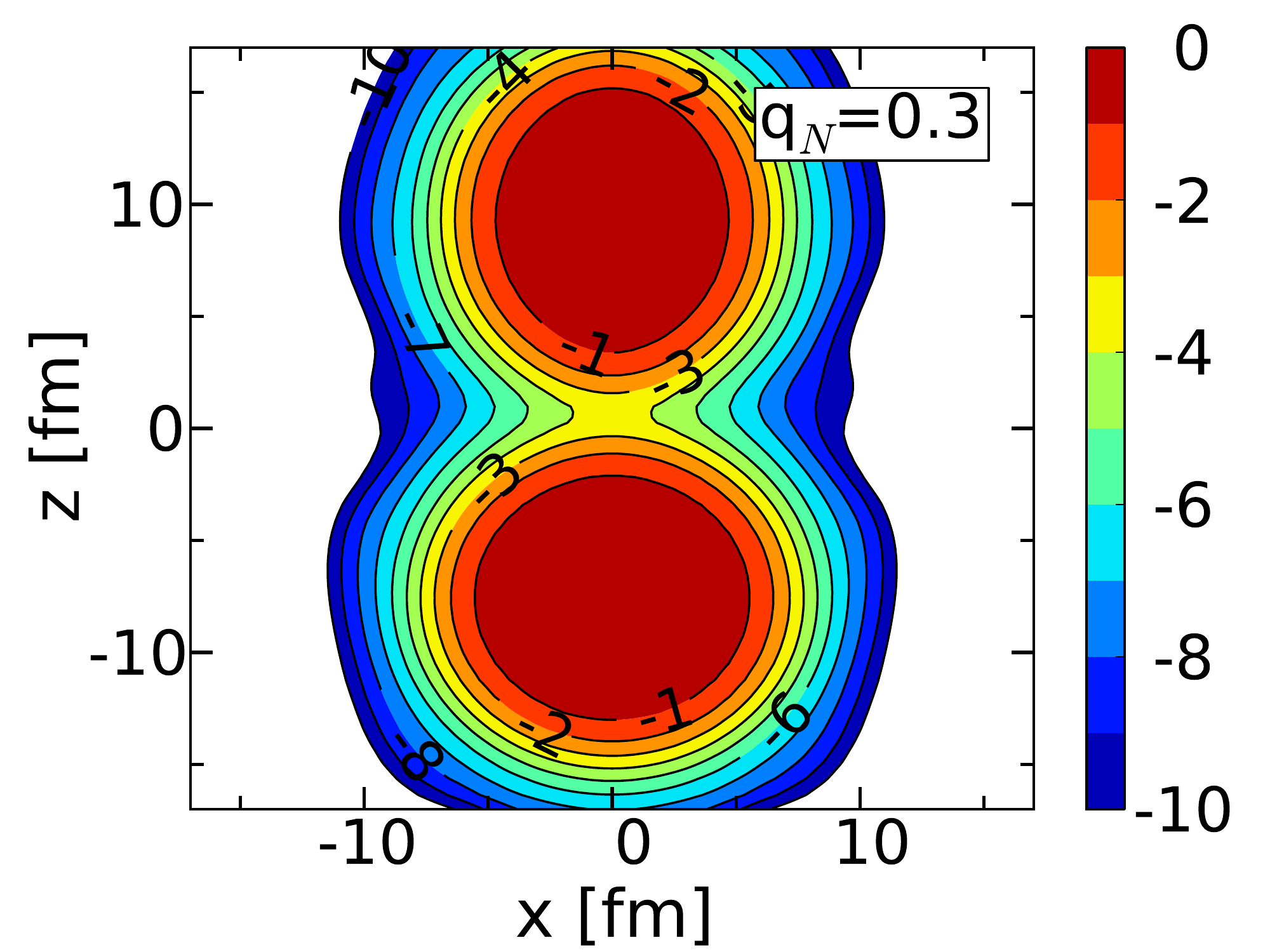}
\caption{Two-dimensional density profiles in the $(x,z)$ plane of the 
intrinsic reference frame for $^{240}$Pu in the scission region, with 
$q_{N} = 4.0$ (left) and $q_{N} = 0.3$ (right). In both cases, the value 
of the quadrupole moment is fixed at $345$ b.
Figures taken from \cite{schunck2014}, courtesy of N. Schunck; 
copyright 2014 by The American Physical Society.}
\label{fig:qN}
\end{center}
\end{figure}

In addition to the set $(q_{20}, q_{22}, q_{30}, q_{40})$ of multipole 
moments, a constraint on the size of the neck between the pre-fragments 
has often been used, in particular at very large elongations; see for 
instance \cite{schunck2015,schunck2014,berger1990,egido1997,warda2002,
younes2009-a}. The standard form of this operator is $\hat{Q}_{N} = \exp(-(\gras{r} - 
\gras{r}_{\mathrm{neck}})^{2}/a^{2})$, with $a$ an arbitrary range and 
$\gras{r}_{\mathrm{neck}}$ the position of the point with the lowest 
density between the two fragments. The figure \ref{fig:qN} shows the effect of decreasing the value of 
$q_N$ near the scission point of $^{240}$Pu. This constraint is often used 
as a way to continuously approach the final configuration of two separated 
fragments.

\begin{figure}[!ht]
\begin{center}
\includegraphics[width=0.45\linewidth]{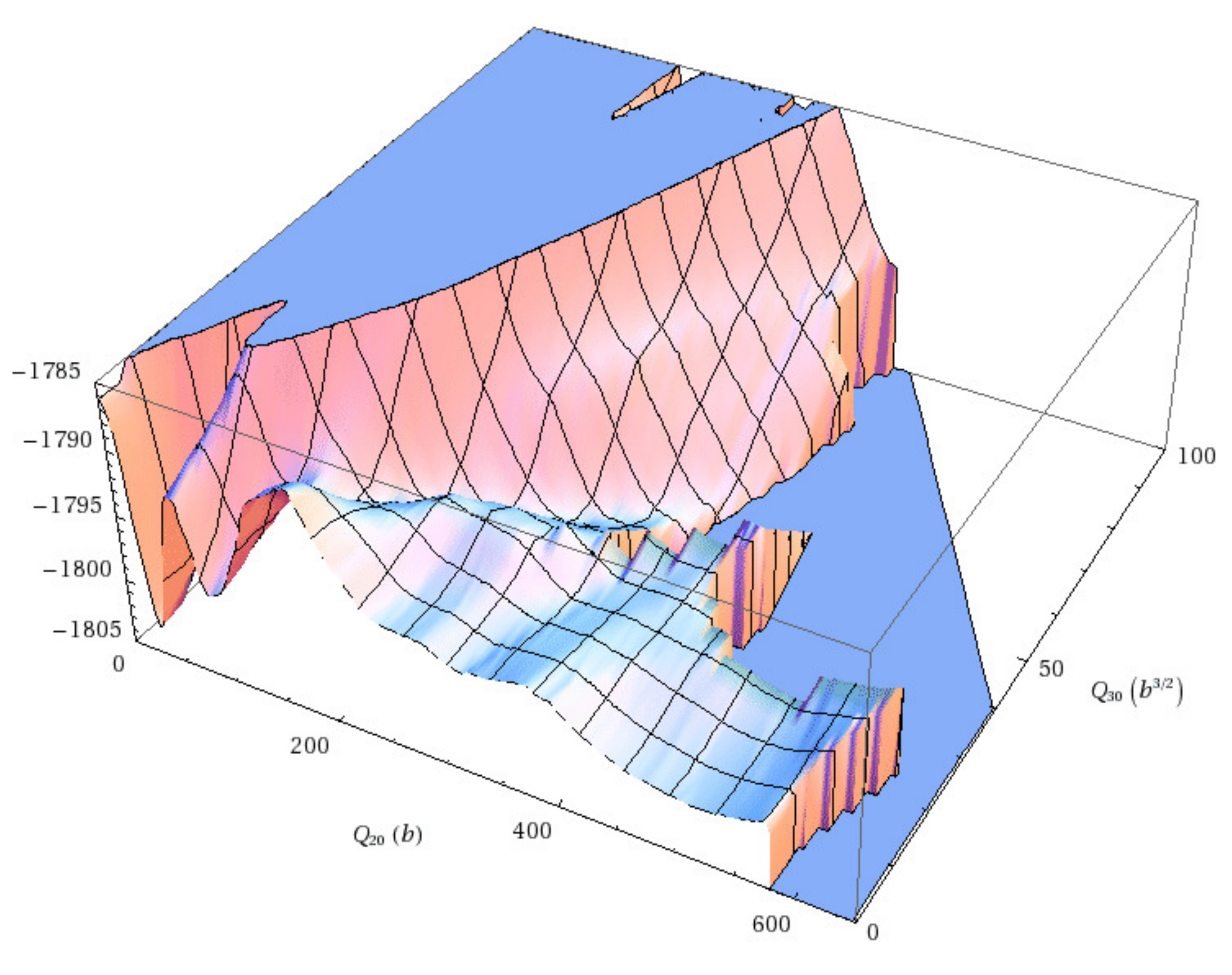}
\includegraphics[width=0.45\linewidth]{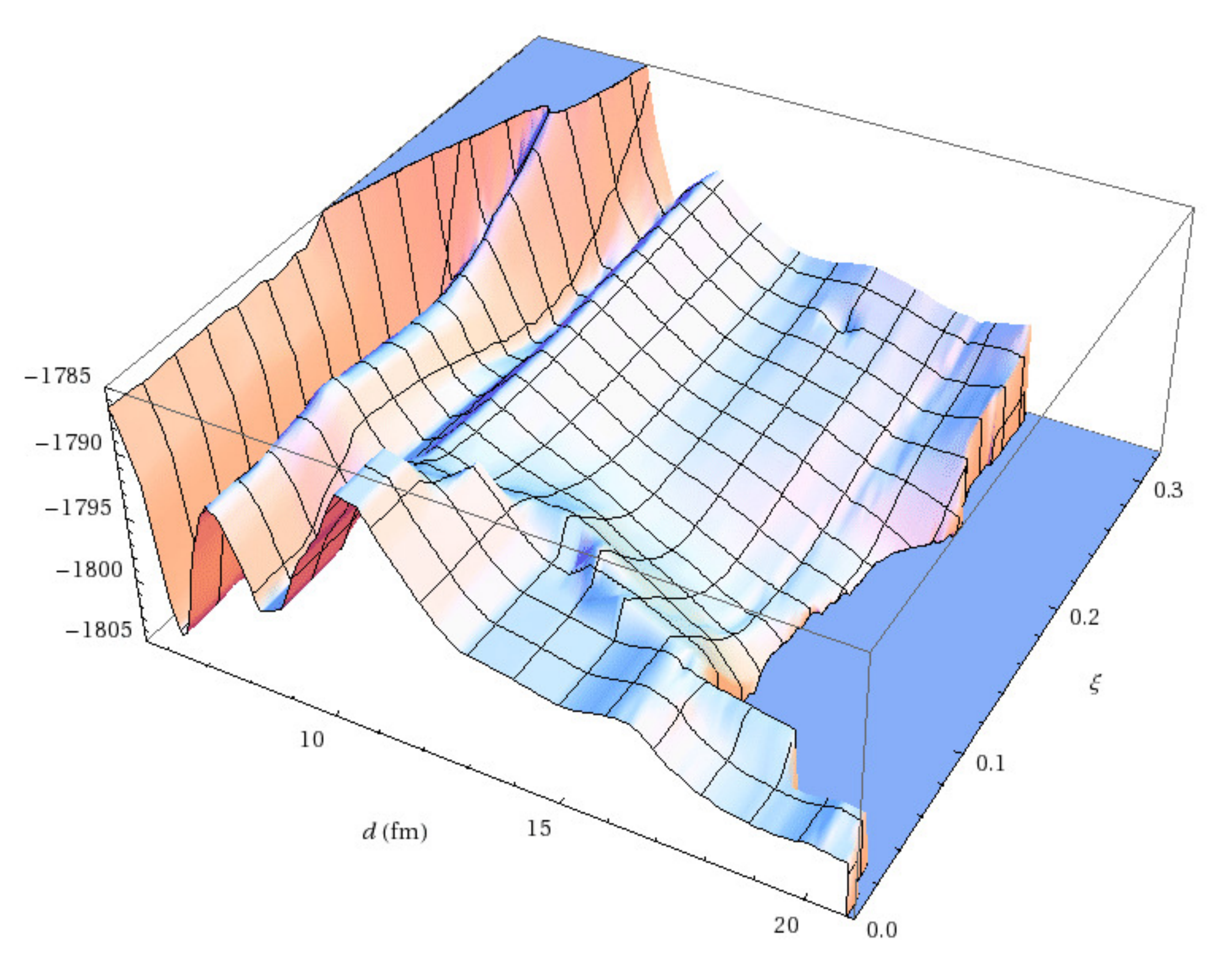}
\caption{Two-dimensional potential energy surface for $^{240}$Pu in the
$(q_{20},q_{30})$ collective space (left panel) and in the $(D,\xi)$
collective space (right panel). Calculations were performed with the
D1S parametrization of the Gogny force according to details given in
\cite{younes2012-a}.
Figure courtesy of W. Younes from \cite{younes2012-a}.}
\label{fig:pes_Dxi}
\end{center}
\end{figure}

Although multipole moments are widely used to characterize nuclear shapes 
in low-energy nuclear structure in general, and nuclear fission in 
particular, they are in fact not particularly well adapted to capture some 
of the most relevant features of fission. In particular, the charge 
and mass of the fission fragments computed at scission from PES generated 
with multipole moments are most of the time non-integers. Although particle 
number projection techniques were used in \cite{scamps2015-a}, they were 
not been applied on a large scale, as e.g., in the determination of fission 
product yields. In addition, Younes and Gogny noticed in their 
two-dimensional study of fission mass distributions for $^{239}$Pu(n,f) and $^{235}$U(n,f) 
presented in \cite{younes2012-a} that several fragmentations $(Z_{L},N_{L}), 
(Z_{R},N_{R})$ were missing along the scission line. This is a consequence of 
computing a potential energy surface in a restricted collective space: 
discontinuities of the surface, especially at scission, become critical, and 
the choice $(q_{20},q_{30})$ of collective variables does not guarantee that 
the proper fragmentations will be recovered. For these reasons, the authors 
suggested to adapt the practice of the macroscopic-microscopic approach by using as collective 
variables the distance between the two pre-fragments $d$ and the mass 
asymmetry between the fragments $\xi = (A_{R} - A_{L})/A$. These quantities 
can be computed by introducing the spatial operators
\begin{eqnarray}
\hat{d} & = &  
\frac{1}{A_{R}}zH(z-z_{\mathrm{neck}}) 
- 
\frac{1}{A_{L}}z[1 - H(z-z_{\mathrm{neck}})],\label{eq:D}\\ 
\hat{\xi} & = & 2H(z-z_{\mathrm{neck}}) - 1,\label{eq:xi}
\end{eqnarray}
where $H(x)$ is the Heaviside step function. As a result of this choice, 
they obtained a much more continuous potential energy surface.
The figure \ref{fig:pes_Dxi} shows the two PES, in the $(q_{20},q_{30})$ and
in the $(D,\xi)$ variables, side-by-side. Scission 
configurations are much better mapped out in the $(D,\xi)$ parametrization.

\subsubsection{Non-geometric Collective Variables}
\label{subsubsec:other}

In nuclear DFT, the amount of pairing correlations in the wave function is 
in principle automatically determined by the minimum energy principle. 
Altering the amount of pairing correlations in the wave function increases 
the energy by an amount that depends on the specific properties of the state 
considered (essentially, the level density) but is typically in the range of
1-2 MeV. Therefore, artificially 
modifying the amount of pairing correlations can modify the shape of the PES. 
Calculations in \cite{schunck2014,giuliani2013} have shown that increasing 
pairing correlations decreases the fission barrier and leads to scission occurring 
at lower elongations. Most important is the strong dependence of the collective 
inertia on the pairing gap already pointed out in \cite{urin1966,moretto1974}. 
The inertia is inversely proportional to the square of the pairing gap 
parameter and \cite{god1985-a,lazarev1987,PhysRevC.43.2200,sadhukhan2014} showed that larger 
pairing gaps imply a smaller inertia and therefore shorter fission half-lives.

There are several possibilities to define collective variables that 
characterize the amount of pairing correlations in the system:
\begin{enumerate}
\item The mean value of the number of particles fluctuation operator 
$\langle \Delta \hat{N}^{2} \rangle $ has been used in the recent calculations 
of \cite{giuliani2014}. The main drawback is that this is a two-body operator 
and therefore the computation of quantities of interest are more involved. 
Also its two-body character prevents the use of standard formulas used to 
compute the inertia. 
\item The average value of the pairing gap $\langle\Delta\rangle$ and the 
gauge angle $\phi$ associated with the particle number operator have been 
mostly studied within semi-microscopic approaches based on a phenomenological 
mean-field complemented by a BCS description of pairing, see, e.g., \cite{god1985,
staszczak1989}. The formalism could in principle be extended to a full HFB 
framework with realistic pairing forces by defining the average value of 
the pairing gap as the mean value of the Cooper pair creation operator 
$P^\dagger = \sum_k c^\dagger_k c^\dagger_{\bar{k}}$ 
times a convenient strength parameter $G$, i.e. $\langle\Delta\rangle = G \langle P^\dagger \rangle$.
In this way, the HFB theory with one-body constraints and all the subsequent
developments discussed below for one-body operators can be straightforwardly implemented.
However, including the gauge angle $\phi$ may cause 
problems when using density-dependent forces because of the singularities 
analysed in \cite{stoitsov2007,bender2009-a,duguet2009,lacroix2009}.
\item Another possibility is to use as starting point the Lipkin-Nogami (LN) 
method that adds to the HFB matrix a term $-\lambda_{2} \Delta \hat{N}^{2}$ 
with $\lambda_{2}$ in principle determined by the LN equation. Using instead 
$\lambda_{2}$ as a free parameter allows to change the mean value of 
$\Delta \hat{N}^{2}$ at will and therefore the strength of pairing 
correlations. This is the choice used in \cite{sadhukhan2014}.
\end{enumerate}


\subsection{Scission Configurations}
\label{subsec:scission}

\begin{figure}[!ht]
\begin{center}
\includegraphics[width=0.75\linewidth]{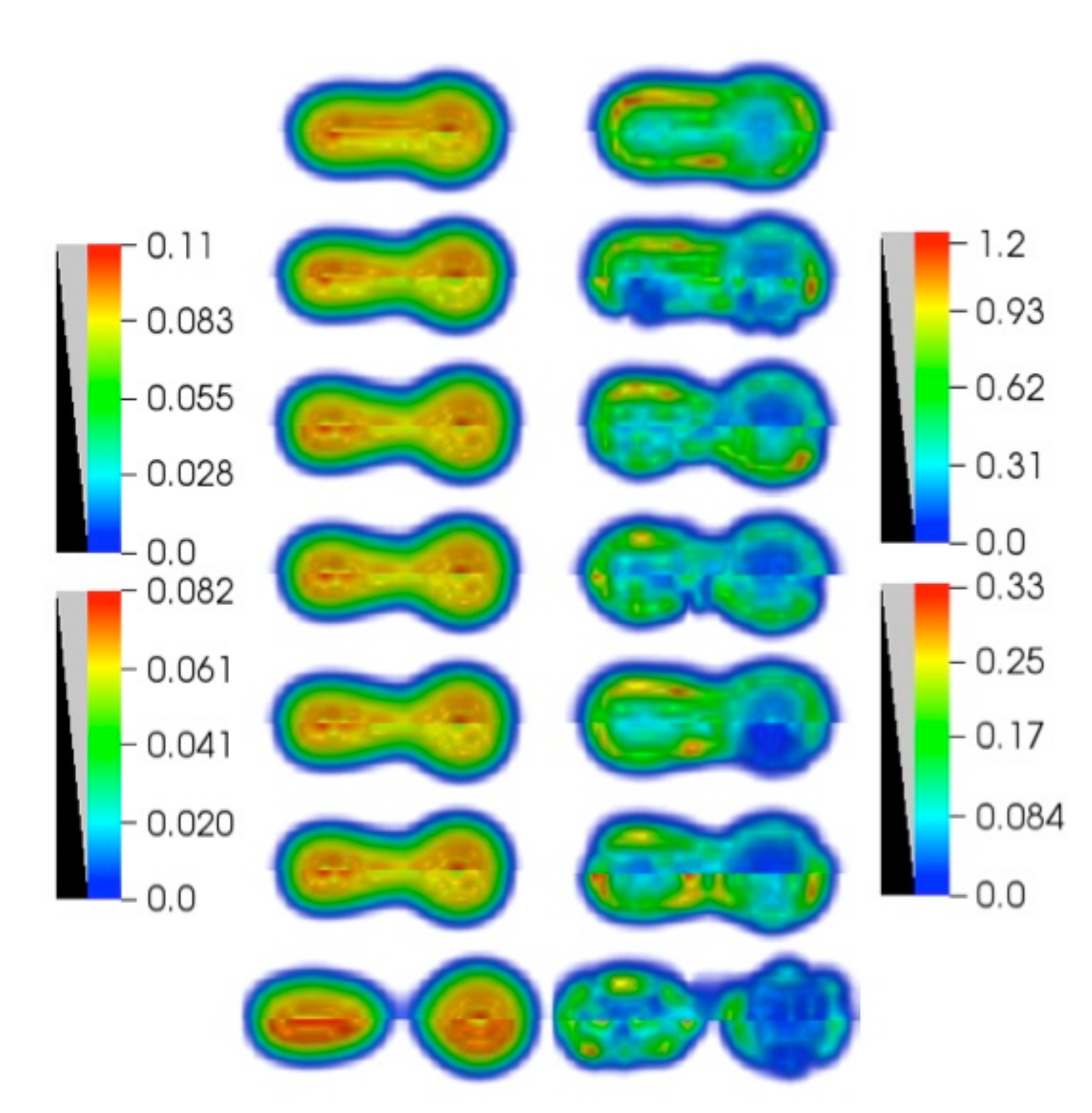}
\caption{Left panel: Neutron (proton) densities $\rho(\gras{r})$ in the top 
(bottom) half of each frame. Right panel: Neutron (proton) pairing field 
$\Delta(\gras{r})$ in the top (bottom) half of each frame. The time difference 
between frames is $\Delta t\approx 5. 10^{-21}$ s. The colour bar is in units of 
fm$^{-3}$ for the density and MeV for the pairing field.
Figure taken from \cite{bulgac2016}, courtesy of A. Bulgac; 
copyright 2016 by The American Physical Society.}
\label{fig:bulgac}
\end{center}
\end{figure}

As recalled in section \ref{subsec:definition}, scission is defined as the 
point where the nucleus splits into two or more fragments. In non-adiabatic 
time-dependent approaches to fission such as TDHF or TDHFB, scission automatically occurs at some 
time $\tau_{\mathrm{sc.}}$ of the time evolution of the compound nucleus as 
the result of the competition between nuclear and Coulomb forces. For example, 
the figure \ref{fig:bulgac} taken from the TDHFB calculation of \cite{bulgac2016} shows how the nuclear 
shape of $^{240}$Pu evolves as a function of time, from a compact deformed 
initial state to two separated fragments. Most importantly, owing to the 
conservation of total energy in TDHFB, these fragments are automatically in 
an excited state, see e.g. discussion in \cite{simenel2014}. The 
characteristics of the system before and after the split can thus easily be 
quantified and provide realistic estimates of fission fragment properties. 

In the adiabatic approximation, however, there is no scission mechanism: 
static potential energy surfaces are often pre-calculated for the compound 
nucleus within a given collective space and do not, strictly speaking, 
contain any information about the fragments. In these approaches, scission 
configurations must explicitly be {\it defined}. This definition happens 
to be essential to obtain sensible estimates of fission fragment properties. 
Indeed, the solutions to the HFB equation corresponding to very large 
elongations of the fissioning nucleus do not lead to excited fission 
fragments in contrast to non-adiabatic approaches. Because of the variational 
nature of the HFB equation, the fragments are essentially in their 
ground-state, which is counter to experimental evidence, see discussion in 
\cite{younes2011,younes2013}. In such adiabatic approaches, it is, 
therefore, necessary to invoke reasonable physics-based arguments to 
justify introducing scission configurations {\it before} the fragments are 
far apart from one another. Unfortunately, all quantities pertaining to 
fission fragments such as charge, mass, excitation energy, etc., happen to 
be extremely sensitive to the characteristics of scission configurations. 
In this section, we recall some of the definitions that have been introduced 
in the literature.


\subsubsection{Geometrical Definitions}
\label{subsubsec:geometric}

Historically, the concept of scission takes its origin in the liquid drop 
(LD) picture of the nucleus and reflects the fact that for very large 
deformations, the LD potential energy can be a multi-valued function of 
the deformation parameters, with at least one of the solutions corresponding 
to two separate fragments as exemplified in \cite{brack1972,nix1965,
strutinsky1963}. This is illustrated, for example, in the figure \ref{fig:cassini} page \pageref{fig:cassini} 
for the parametrization of the LD in terms of Cassini ovals. In the LD approach, these multi-valued regions originate 
from the finite number of collective variables (=deformations) and/or the 
lack of bijectivity between a set of parameters and a given geometrical shape. 

In DFT, such multi-modal potential energy surfaces are encountered when 
working in finite collective spaces as recalled in section 
\ref{subsubsec:multipole}. For example, studies of hot fission of actinide 
nuclei published in \cite{berger1984,berger1989,schunck2014} showed that the 
least energy fission pathway from ground-state to scission follows the fission 
valley, which lies higher in energy than the fusion valley. For given values 
of the axial quadrupole and octupole moments, these two valleys differ by the 
value of the hexadecapole moment $q_{40}$, and nuclear shapes in the fusion 
valley correspond to two well-separated fission fragments. Similar multi-modal 
potential energy surfaces have been observed in superheavy nuclei, for example 
in \cite{staszczak2009,bonneau2006}. Discontinuities in the energy (or any 
relevant collective variable) for large elongations of the fissionning 
nucleus are usually the first tell-tale signal of the transition to the 
scission region. 

\begin{figure}[!ht]
\begin{center}
\includegraphics[width=0.4\linewidth]{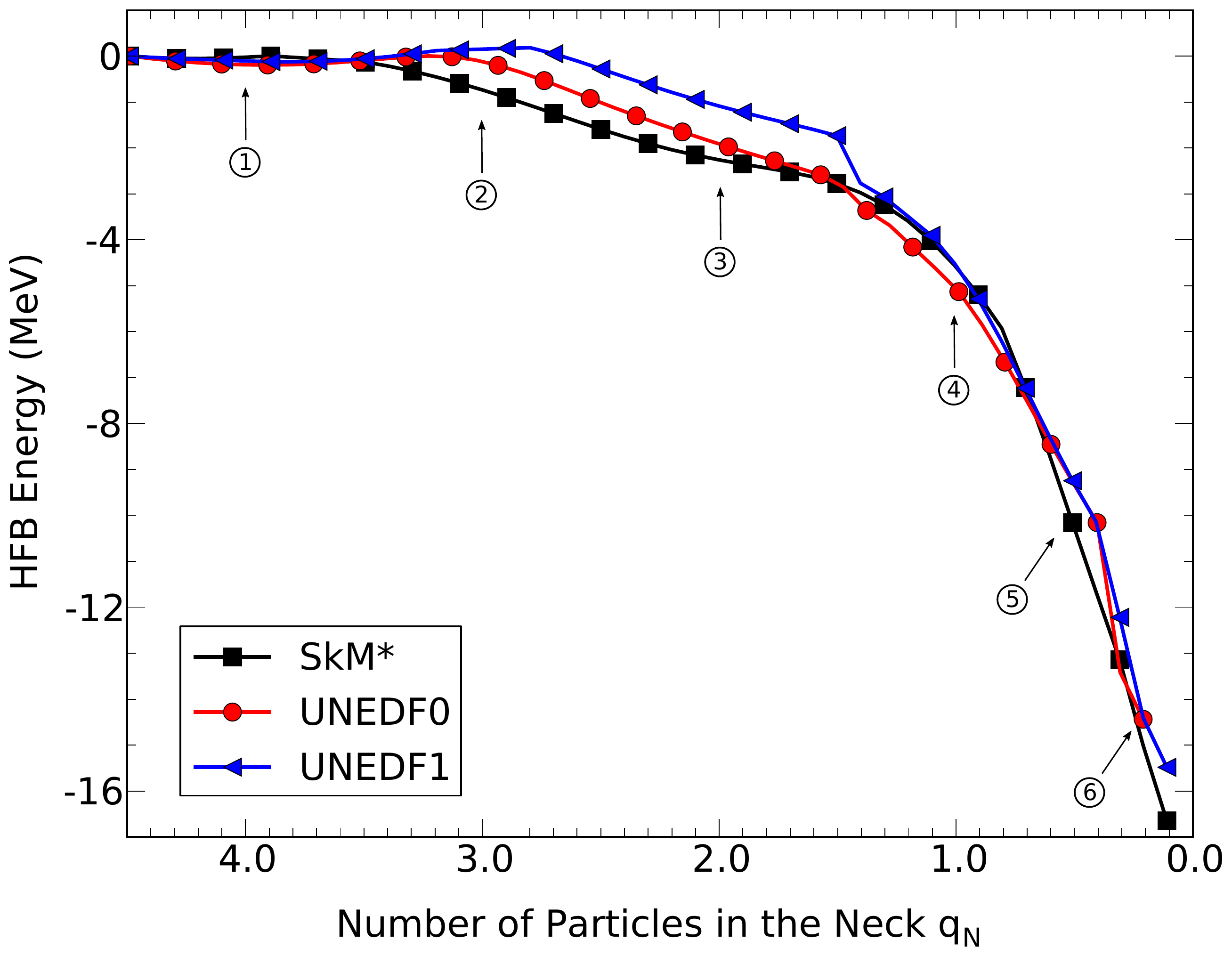}
\includegraphics[width=0.52\linewidth]{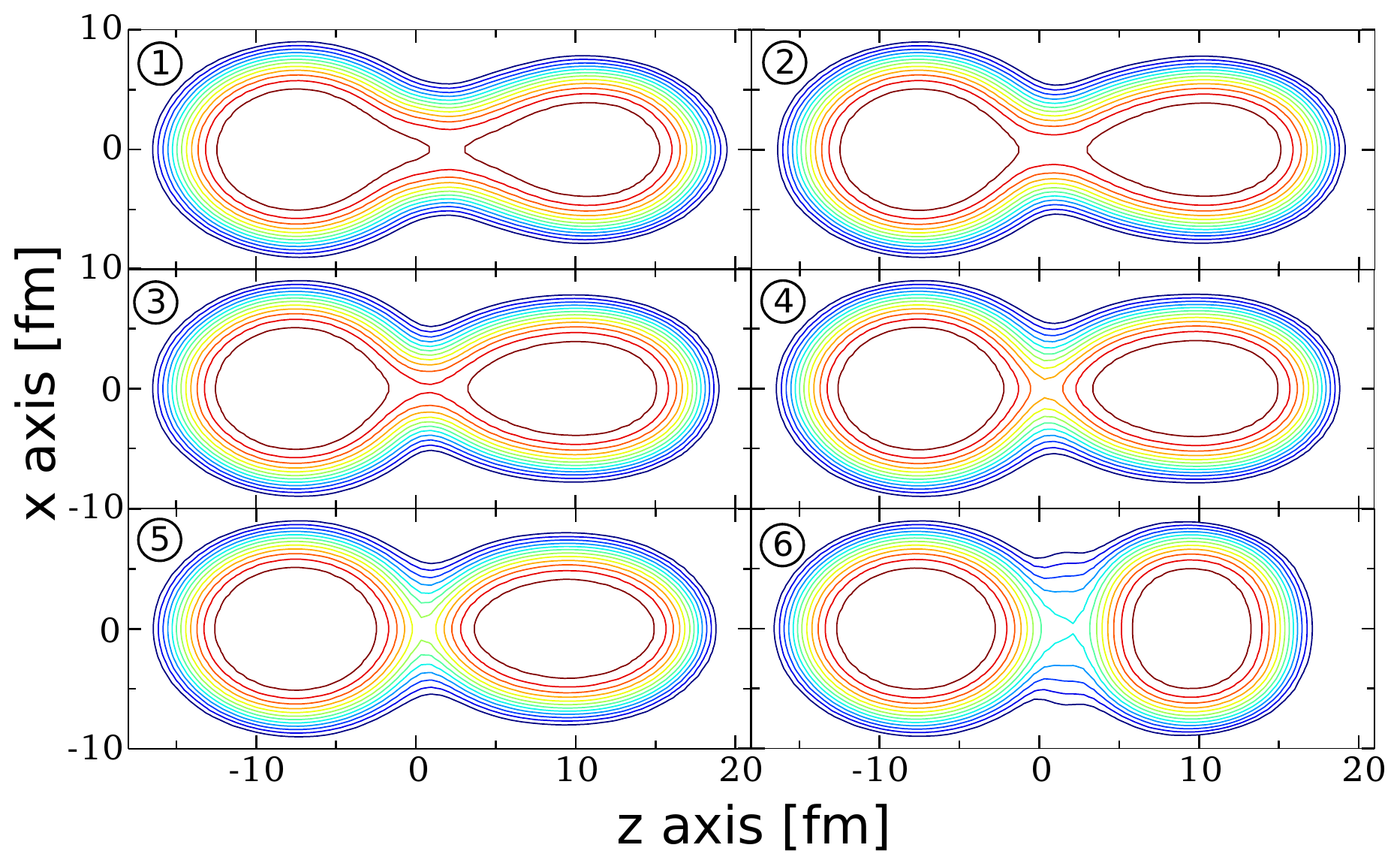}
\caption{Left panel: HFB energy as a function of $q_{N}$ in the scission region 
for the most likely fission path (point labelled 5 in figure \ref{fig:pes_pu240_1D} 
page \pageref{fig:pes_pu240_1D}) of $^{240}$Pu. The value of the quadrupole 
moment is fixed at $345$ b. Right panel: Corresponding two-dimensional density 
profiles in the $(x,z)$ plane of the intrinsic reference frame for six different 
values of $q_{N}$. 
}
\label{fig:scission}
\end{center}
\end{figure}

However, these discontinuities are also the manifestation of the truncated 
nature of the collective space: when additional collective variables such 
as, e.g., the size of the neck $q_{N}$, are used, they can disappear. The 
transition from a compact shape to very loosely joined fragments can thus be 
continuous. This point is illustrated in the top panel of figure 
\ref{fig:scission} adapted from \cite{schunck2014}. The graph shows the
energy of $^{240}$Pu as a function of $q_{N}$ at the point $q_{20} = 345$ b. 
This point is located just before the discontinuity in energy along the 
most likely fission path of figure \ref{fig:pes_pu240_1D} page 
\pageref{fig:pes_pu240_1D}. When the $q_N$ collective variable is used as a 
constraint, the system can go continuously to two separated 
fragments, as illustrated by the density contours of the bottom panel of 
the figure. 

In such cases, one needs a specific criterion to define when exactly one may 
{\it consider} the two fragments as fully separated. To this purpose, one can use 
the value of the density between the two fragments as in \cite{dubray2008}: 
the fragments are deemed separated if $||\rho||_{\mathrm{max}} \leq \rho_{\mathrm{sc.}}$. 
Alternatively, one may use the expectation value of the neck 
operator discussed in section \ref{subsubsec:multipole}. This quantity 
gives a measure of the number of particles in a slice of width $a$ 
centred on the neck position. The decision of considering the 
fragments as separated would be based on the condition 
$\langle\hat{q}_{N}\rangle \leq q_{\mathrm{sc.}}$. Whatever the criterion 
retained, however, there remains a part of arbitrariness in the definition 
of scission: how to choose the value of $q_{\mathrm{sc.}}$ (or 
$\rho_{\mathrm{sc.}}$)? 

Recently, there were attempts in  \cite{schunck2015,schunck2014,duke2012} 
to use topological methods to characterize scission in a less arbitrary 
manner. The main idea is to map the density fields $\rho_{\mathrm{n}}(\gras{r})$ 
and $\rho_{\mathrm{p}}(\gras{r})$ into an abstract contour net, and infer 
properties of those fields from the connectivity properties of these nets. 
Mathematically rigorous, the method can identify from the PES the regions 
where the variations of densities within the pre-fragments are commensurate 
with those in the compound nucleus. As a result of this work, the authors 
proposed to redefine scission as a region characterized by a range of values 
rather than by a fixed value of either $\langle\hat{q}_{N}\rangle$ or 
$||\rho||_{\mathrm{max}}$. 


\subsubsection{Dynamical Definitions}
\label{subsubsec:dynamical}

By design, geometrical definitions of scission only reflect static 
properties, and do not take into account the fact that the split is caused 
by a time-dependent competition between the repulsive Coulomb and the 
attractive nuclear force. As a result of this competition, however, scission 
may occur even when the two pre-fragments are separated by a sizeable neck. 
To mock up the dynamics of scission in static macroscopic-microscopic 
calculations, it was thus proposed already in \cite{davies1976} to use as a 
criterion for scission the ratio between the Coulomb and the nuclear forces 
in the neck region. These forces can be computed by taking the derivatives 
of the potential energy with respect to the relevant collective variable. 
Such techniques have been later extended in \cite{gherghescu2008} by adding 
an additional phenomenological neck potential. This ``dynamical'' definition 
of scission was adapted in the DFT framework by the authors \cite{bonneau2007,
younes2011}. 

\begin{figure}[!ht]
\begin{center}
\includegraphics[width=0.75\linewidth]{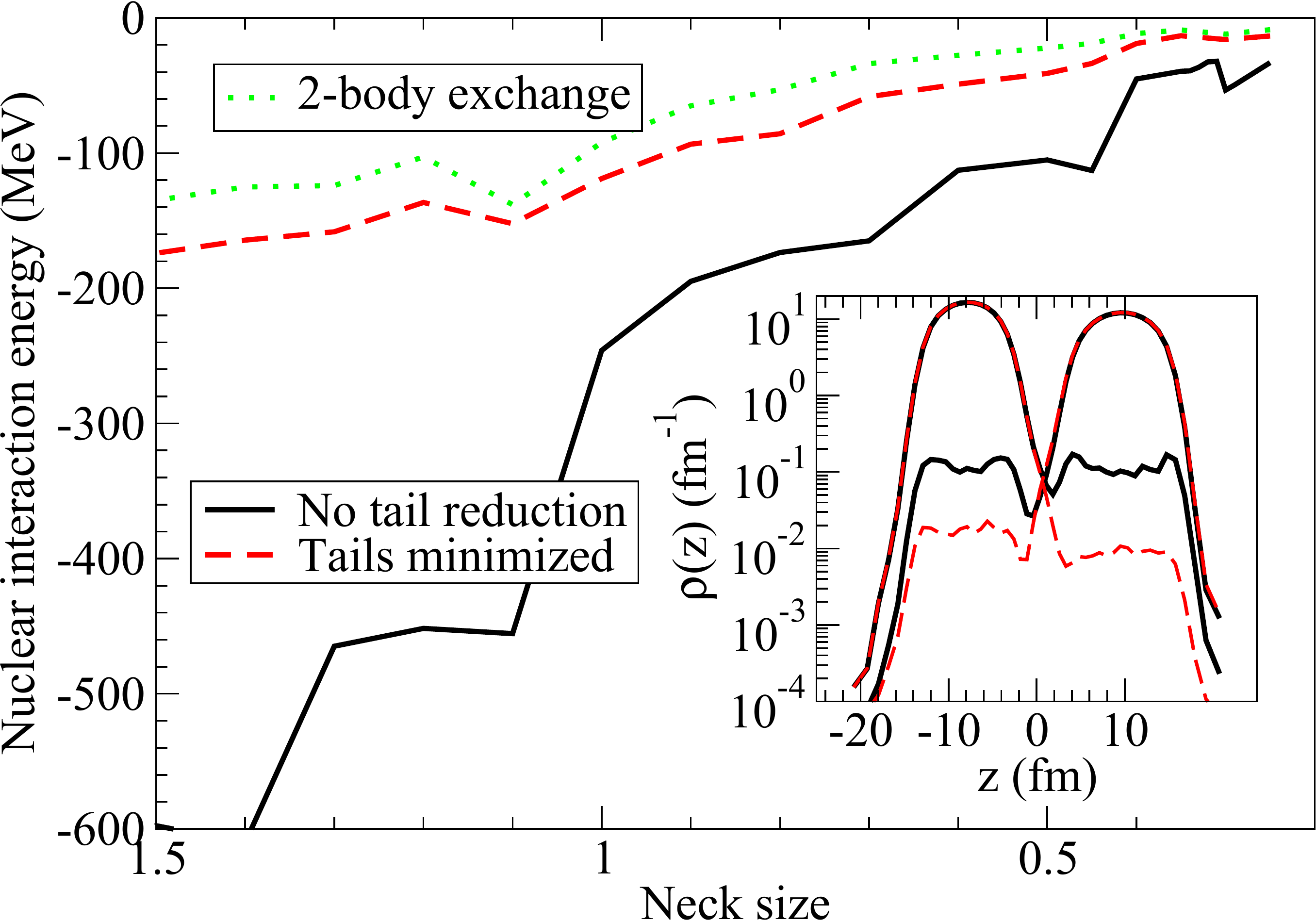}
\caption{Nuclear interaction energy between the two pre-fragments near the 
scission point of $^{240}$Pu as a function of the size of the neck. Calculations 
are done with the Gogny D1S effective force. Black: nuclear interaction 
energy before localization; dashed red: same after localization of the fragments 
by minimization of the tails; dotted green: two-body exchange contribution to 
the nuclear interaction energy. The insert shows the density profile along the 
symmetry axis of the nucleus.
Figure taken from \cite{younes2011}, courtesy of W. Younes; 
copyright 2011 by The American Physical Society.}
\label{fig:localization}
\end{center}
\end{figure}

Both the geometrical or dynamical definition of scission are, however, 
semi-classical, in the sense that they ignore quantum mechanical effects 
in the neck region. This limitation was highlighted by Younes and Gogny 
in a couple of papers \cite{younes2011,younes2009}. In the framework of 
DFT, the degrees of freedom of the fission fragments are the one-body 
density matrix and the pairing tensor, which are themselves obtained from 
the quasiparticle wave functions. Near scission, one can introduce a 
localization indicator (simply related to the spatial occupation of 
quasiparticle wave functions) to partition the whole set of quasiparticles 
into two subsets belonging to either one of the pre-fragments. The total 
one-body density matrix of the compound nucleus is then decomposed
\begin{eqnarray}
\rho(\gras{r}\sigma,\gras{r}'\sigma')
& =
\sum_{\mu\in\{ 1\}} V^{*}_{i\mu}V_{j\mu} \phi_{i}(\gras{r}\sigma)\phi^{*}_{j}(\gras{r}'\sigma')
+
\sum_{\mu\in\{ 2\}} V^{*}_{i\mu}V_{j\mu} \phi_{i}(\gras{r}\sigma)\phi^{*}_{j}(\gras{r}'\sigma'), \\
& = \rho_{1}(\gras{r}\sigma,\gras{r}'\sigma') + \rho_{2}(\gras{r}\sigma,\gras{r}'\sigma'). 
\label{eq:densities_fragments}
\end{eqnarray}
Similar expressions hold for the pairing tensor 
$\kappa(\gras{r}\sigma,\gras{r}'\sigma')$. With these definitions, it 
becomes possible to analyse fission fragment properties, including their 
intrinsic energy and their interaction energy within the DFT framework; see 
also \cite{schunck2014} for details. In \cite{younes2011}, Younes and Gogny 
made the crucial observation that the fission fragment density distributions 
thus extracted have large tails that extend into the other fragment and 
reflect the quantum entanglement between the two fragments as shown in 
figure \ref{fig:localization}. Because of these large tails, there was a 
substantial nuclear interaction energy between the fragments even when the 
size of the neck was very small. Similarly, the Coulomb interaction energy 
was much too high compared to its experimental value.

\begin{figure}[!ht]
\begin{center}
\includegraphics[width=0.75\linewidth]{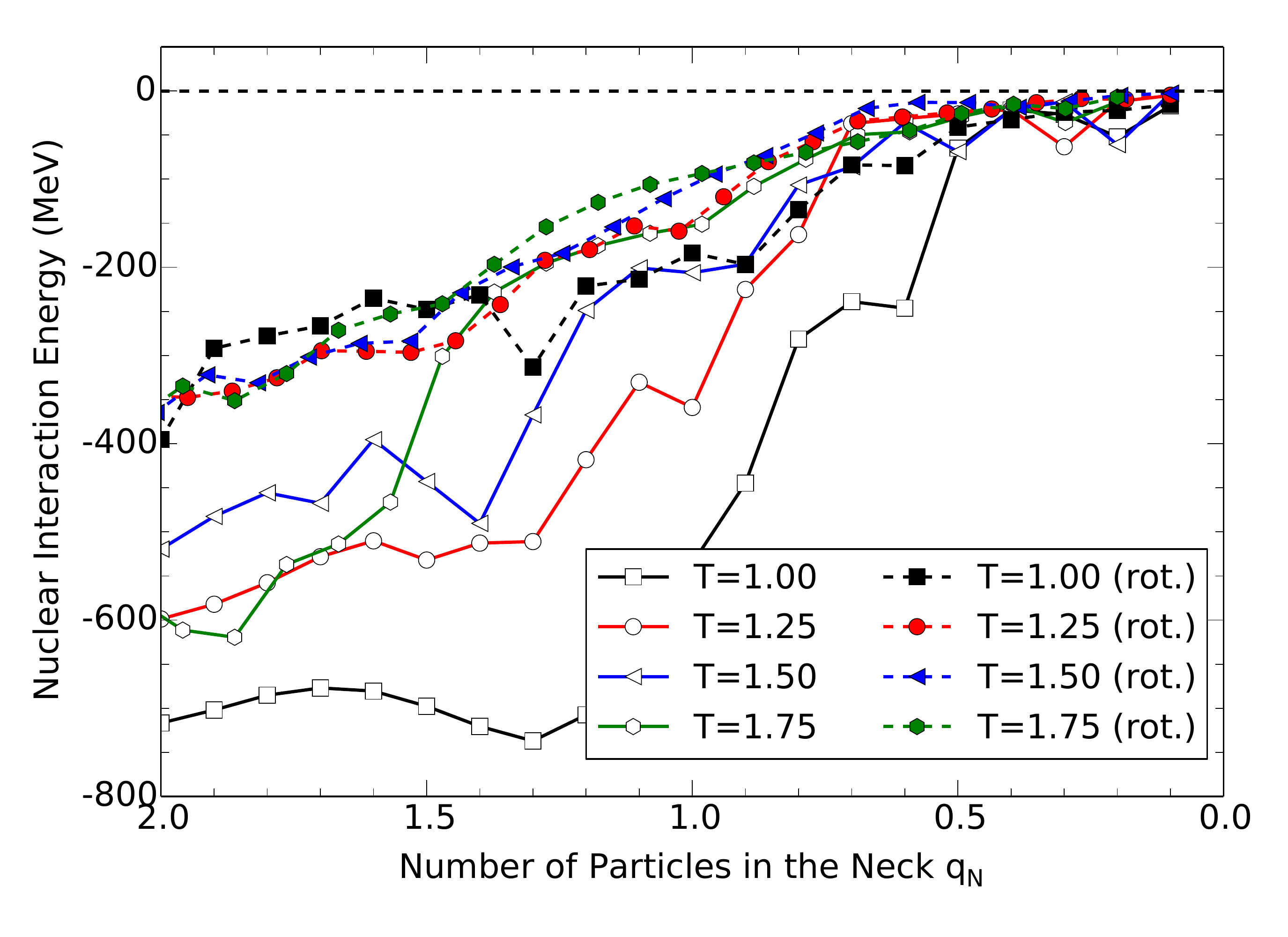}
\caption{Nuclear interaction energy between the two pre-fragments near the 
scission point (point labelled 5 in figure \ref{fig:pes_pu240_1D} page \pageref{fig:pes_pu240_1D}) of $^{240}$Pu as a function of the size of the neck. 
Calculations are done with the Skyrme SkM* effective force at finite 
temperatures. Plain curves with open symbols show the nuclear interaction 
energy before localization for $T=1.00, \dots, 1.75$ MeV; dashed curves 
with filled symbols show the energy after localization.
Figure taken from \cite{schunck2015}, courtesy of N. Schunck; 
copyright 2015 by The American Physical Society.}
\label{fig:localization_T}
\end{center}
\end{figure}

Most importantly, because the HFB solutions are invariant under a unitary 
transformation of the quasiparticle operators, it is possible to choose a 
representation in which this degree of entanglement is minimal, as discussed 
in \cite{schunck2014,younes2011,younes2009} leading to a ``quantum localization'' of the fission fragments. 
This freedom in choosing the representation of the HFB solutions is the analogue 
to localization techniques used in quantum chemistry and is discussed in some details in \cite{schunck2014}.
The implementation of the quantum localization procedure yields much more realistic 
estimates of fission fragment properties at scission, even for neck size 
up to $q_{N} \approx 0.3$. Figure \ref{fig:localization} illustrates the 
impact of choosing a representation that better reproduces the asymptotic 
conditions of two separated fragments: by reducing the tails by about an 
order of magnitude, the interaction energy is reduced by a factor 3 and 
is close to 0 for thin necks.

Recently, the localization method was extended at finite-temperature 
in \cite{schunck2015}. Figure \ref{fig:localization_T} illustrates the impact 
of minimizing the tails at different temperatures for the same case of 
$^{240}$Pu most likely fission, only with the SkM* Skyrme potential instead 
of the Gogny D1S. Although the practical implementation differs from the 
zero-temperature case because the generalized density matrix is not diagonal 
any longer after a rotation of the quasiparticles, it is still possible to reduce the 
tails of the fragment densities without changing the global properties of 
the compound nucleus. As the temperature increases, however, this procedure 
becomes more and more difficult, especially for well-entangled fragments, 
because of the coupling to the continuum; see section \ref{subsubsec:excitation}. 

%
%

\newpage
\section{Dynamics of Fission}
\label{sec:dynamics}

Fission is intrinsically a dynamical process where the quasi-static
ground state (or other quasi-static excited configurations) evolves
with time towards a two-fragment solution. Ideally, the probability
of such an event to occur could be computed using the rules of quantum
mechanics as 
\begin{equation}
\mathcal{P} = |\langle\Psi_{1}|\hat{U}(t_{1},t_{0})|\Psi_{0}\rangle|^{2},
\label{eq:Uoperator}
\end{equation}
where $|\Psi_{0}\rangle$ is the initial wave function, 
$|\Psi_{1}\rangle$ is the wave function of the two fragments and 
$\hat{U}(t_{1},t_{0})$ is the time evolution operator. In nuclear physics, 
computing any of the elements of the above expression represents a 
formidable task and therefore reasonable approximation schemes are in 
order. 

One of the most common such approximations is the hypothesis of 
adiabaticity already mentioned in the introduction and in section 
\ref{sec:pes}. Based on the related separation of scales between slow 
collective motion and fast intrinsic excitations, it is assumed that a 
small set of collective variables drives the fission process. Fission 
dynamics can then be studied in that reduced collective space using pure 
quantum mechanical methods based on configuration mixing. As we will 
show, this approach is particularly well adapted to compute spontaneous 
fission half-lives and fission product distributions. Formal aspects 
underpinning the hypothesis of adiabaticity are discussed in 
\cite{klein1991} in the context of the classical theory of collective 
motion.

A second, related, approximation consists in representing the nuclear 
wave function at each time $t$ by a mean field solution formally of the 
HF or HFB type. This leads to the concept of time-dependent density 
functional theory (TDDFT). This approach has recently gained ground with 
the development of supercomputers, and should, in principle, offer a more 
realistic description of fission fragment properties since it does not 
rely on adiabaticity.

A third approximation particularly relevant for spontaneous fission is the notion 
of tunnelling through a potential barrier, which is based on 
semi-classical concepts related to the least action principle of classical
dynamics. The least action principle establishes that the action, defined
as the integral of the Lagrangian from time $t_0$ to $t_1$, has to be 
stationary for the trajectories that satisfy the laws of motion of classical 
mechanics (Euler-Lagrange). The action thus defined is called Hamilton's action.
An alternative to Hamilton's action involves the integral of the momentum
as a function of the generalized coordinates $q$ (Maupertuis' action). Again, 
the physical trajectory of the system in phase space is the one for which the 
action is stationary. In quantum mechanics, where the concept of a trajectory 
does not apply, we usually want to compute the probability amplitude (\ref{eq:Uoperator}). 
If the initial and
final quantum states are eigenstates of the position operator, 
$|\Phi_0\rangle = |\vec{r}_0\rangle$, the probability amplitude can be written 
as the path integral over all possible trajectories connecting $\vec{r}_0$ at 
$t_0$ and $\vec{r}_1$ at $t_1$ of the exponential of 
$\frac{i}{\hbar} S_{\mathrm{cl}}$ where $S_{\mathrm{cl}}$ is the classical 
action. The main contribution to the path integral comes from the ``classical 
trajectories'', that is the ones that minimize the action \cite{feynman2012quantum}. 
The argument is still valid in the classically-forbidden regions where the 
action is an imaginary number \cite{COLEMAN1988178} -- see section 
\ref{subsubsec:instantons} below. 

Another approach is based on the semi-classical approximation to quantum 
tunnelling through a classically-forbidden region, which is at the heart of
the Wentzel, Kramers, Brillouin (WKB) approximation discussed in section 
\ref{subsubsec:wkb}; see also \cite{messiah1961} for a complete 
presentation. In this case, the idea is to write the wave function of the 
system as $\Psi(\gras{r}) = e^{\frac{i}{\hbar} W(\gras{r})}$ where $W(\gras{r})$ 
is the new unknown quantity. Inserting this expression for $\Psi(\gras{r})$ 
into the Schr\"odinger equation, one obtains a new equation for $W(\gras{r})$. 
The WKB approximation consists in expanding $W(\gras{r})$ in powers of $\hbar$ 
and keep only the zero-order term. In this limit, it turns out that $W(\gras{r})$ 
is the classical action. Minimizing it is again a way to increase the penetration
probability that can be extracted from the semi-classical wave function -- see
section \ref{subsubsec:wkb} below for practical formulas.

As can be concluded from this brief introduction, there is no quantum 
mechanics least action principle, but semi-classical arguments tend to point 
to favoured trajectories that minimize the classical action. In fission, only 
a few collective variables are considered as relevant quantities in the 
evolution of the system from the ground state to scission. Therefore, it is 
first necessary to define the classical action for a system characterized by 
these variables. The knowledge of the classical action in turns requires the 
knowledge of the inertia and the potential energy associated with the collective 
variables. 

In this section, we review these various approaches to fission dynamics. 
In the adiabatic approximation, collective inertia plays a special role 
as it contains the response of the nucleus to a change in the collective 
variables and effectively plays the role of the mass of the collective 
wave-packet. Section \ref{subsec:inertia} summarizes the various recipes 
to compute the collective inertia, from the generator coordinate method, 
section \ref{subsubsec:gcm}, to the adiabatic time-dependent 
Hartree-Fock-Bogoliubov approach, section \ref{subsubsec:atdhfb}. An 
accurate determination of collective inertia is essential both for 
spontaneous and induced fission. In spontaneous fission, it appears in 
the definition of the action and has, therefore, an exponential effect 
on the calculation of half-lives. This is discussed in section 
\ref{subsec:t12}. In induced fission, it naturally appears both in the 
classical and quantum treatment of dynamics in the collective space, 
which are based upon the Langevin and Kramers equations, and the 
time-dependent generator coordinate methods, respectively. Section 
\ref{subsec:induced} explores some of the similarities between these 
techniques, and compares them with non-adiabatic time-dependent density 
functional theory techniques. An excellent review covering in great detail 
some of this material can be found in Chapter 5 of \cite{krappe2012} where 
the reader is referred for further details.


\subsection{Collective Inertia}
\label{subsec:inertia}

In a phenomenological picture of fission, the collective inertia $B$ 
can be introduced when the dynamics is assumed to be restricted to a path in 
the manifold of collective variables with the associated classical action
\begin{equation}
S(s_1,s_0) = \int_{s_0}^{s_1} ds \sqrt{B(s)(V(s)-E_0)}.
\label{eq:classicalS}
\end{equation}
Here, $s$ is the parameter describing the path. For simplicity, it is 
assumed that all collective variables $\gras{q}$ (multipole moments, neck, 
pairing gap, etc), are smooth functions of $s$. The collective inertia along 
the path is
\begin{equation}
B(s) = \sum_{\alpha\beta} B_{\alpha\beta} \frac{dq_{\alpha}}{ds} \frac{dq_{\beta}}{ds}
\end{equation}
and is given in terms of the inertia tensor $\tensor{B} \equiv B_{\alpha\beta}$ defined for a 
pair of collective variables $q_{\alpha}$ and $q_{\beta}$. Note that the 
expression for the action and other related formulas like the penetrability
factor of the WKB formula for the spontaneous fission lifetime have not been
derived from first principles and only represent reasonable quantities 
inspired by semi-classical arguments to the tunnelling process \cite{landau2013quantum}. 

In nuclear physics, the notion of collective inertia also arises naturally 
in theories of large amplitude collective motion such as the generator 
coordinate method (GCM) or the adiabatic time-dependent Hartree-Fock-Bogoliubov 
(ATDHFB) theory \cite{ring2000}. Therefore these general approaches to the quantum 
many-body problem provide rigorous methods to compute the collective 
inertia needed in fission. Below we briefly review the derivation of both 
the GCM and ATDHFB masses and discuss some of the common approximations 
used to lessen the computational load.


\subsubsection{Generator Coordinate Method}
\label{subsubsec:gcm}

The generator coordinate method (GCM) is a general quantum many-body 
technique designed to encapsulate collective correlations in the wave 
function. It is based on the expansion of the unknown many-body wave 
function of the system on a basis of known many-body states. The 
variational principle is used to determine the set of expansion coefficients. 
The technique is closely related to the configuration interaction (CI) 
method popular in quantum chemistry and known in nuclear physics as the
"shell model". Usually, basis states are continuous 
functions of a finite set of coordinates (such as deformation parameters, 
Euler rotation angles, etc.) whereas in the CI method, they can be 
unrelated to each other (for instance a set of two quasiparticle 
excitations). In this section, we focus on the GCM as a tool to extract 
a collective inertia tensor. The time-dependent extension of the GCM also 
provides a powerful tool to extract fission fragment distributions, and 
will be presented separately in section \ref{subsubsec:tdgcm}.

In the GCM, the general ansatz for the wave function is
\begin{equation}
|\Psi\rangle = \int d \qVec\; f (\qVec) | \Phi (\qVec) \rangle,
\label{eq:gcm_ansatz}
\end{equation}
where $|\Phi (\qVec) \rangle$ represents a set of known wave functions
depending on a general label $\qVec$ that can include not only continuous
but also discrete variables. Also the integral has to be taken in a broad sense as
representing either sums over discrete values of $\qVec$, genuine integrals or
an admixture of the two.  In fission studies, the $| \Phi (\qVec) \rangle$
are usually quasiparticle vacuua obtained by solving the HFB equation with constraints 
on a set of $n$ operators $\hat{Q}_{\alpha},\alpha=1,\ldots,n$. As before, 
the boldface symbol $\qVec$ represents the set of collective variables such 
that $\langle \Phi (\qVec) | \hat{Q}_{\alpha} | \Phi (\qVec) \rangle = 
q_{\alpha}$. Applying the variational principle to the energy with the 
amplitudes $f(\qVec)$ as variational parameters leads to the 
Hill-Wheeler equations,
\begin{equation}
\int d \qVec'\; h(\qVec,\qVec') n(\qVec,\qVec') f (\qVec') = E 
\int d \qVec'\; n(\qVec,\qVec') f (\qVec').
\label{eqn:HW}
\end{equation}
It is an integral equation with the norm kernels,
\begin{equation}\label{gcm:norm}
n(\qVec,\qVec') = \langle \Phi (\qVec)  | \Phi (\qVec') \rangle ,
\end{equation}
and energy kernels,
\begin{equation}
h(\qVec,\qVec') = 
\frac { \langle \Phi (\qVec) | \hat{H} | \Phi (\qVec') \rangle }
{\langle \Phi (\qVec)  | \Phi (\qVec') \rangle  }.
\end{equation}
Since the set of wave functions $| \Phi (\qVec) \rangle$ is in general not
orthogonal, the $f ( \qVec )$ amplitudes can not be interpreted as probability
amplitudes, and $|f ( \qVec )|^{2}$ is not a probability density. In order to obtain
probability amplitudes, the set $| \Phi (\qVec) \rangle$ has to be orthogonalized
using standard techniques of linear algebra to obtain what are called ``natural
states'' $| \tilde{\Phi} (\qVec) \rangle$. These states are defined by folding $| \Phi (\qVec) \rangle$ with
the inverse of the square root of the norm kernel,
\begin{equation}\label{gcm:natural}
| \tilde{\Phi} (\qVec) \rangle = \int d \qVec' \left[n(\qVec,\qVec')\right]^{-1/2}  | \Phi (\qVec') \rangle.
\end{equation}
The square root of the norm kernel is defined as
\begin{equation}\label{gcm:norm_sqroot}
n(\qVec,\qVec') = \int d \qVec'' \left[n(\qVec,\qVec'')\right]^{1/2}  \left[n(\qVec'',\qVec')\right]^{1/2}
\end{equation}
which corresponds to a Cholesky decomposition of the positive-definite norm kernel.

{\it Collective Schr\"odinger Equation - }
The connection of the GCM with the theory of fission comes from the 
reduction of the Hill-Wheeler equation (\ref{eqn:HW}) to a collective Schr{\"o}dinger-like 
equation (CSE) of the collective coordinates, which naturally leads to the 
definition of a collective inertia. Following \cite{brink1968,onishi1975,une1976,krappe2012}, 
the CSE is derived after assuming that 
the norm overlap (\ref{gcm:norm}) is a sharply peaked function of the 
coordinate difference $\gras{s}=\qVec-\qVec'$ and smoothly depends upon 
the average $\bar{\qVec}=\frac{1}{2}(\qVec+\qVec')$ in such a way that 
the norm kernel is well approximated by a Gaussian
\begin{equation}
n(\qVec,\qVec') = \exp\left(- \frac{1}{2}\gras{s}\tensor{\Gamma} (\bar{\qVec}) \gras{s} \right).
\end{equation}
In this expression the width $\tensor{\Gamma}$ is to be understood as a rank 2 tensor
with components $\Gamma_{\alpha\beta}$
and the exponent is thus given explicitly by $- \frac{1}{2}\sum_{\alpha\beta} 
\Gamma_{\alpha\beta} s_{\alpha} s_{\beta}$. If we assume that the components $\Gamma_{\alpha\beta}$ 
of the width are slowly varying functions of the coordinate $\bar{\qVec}$, 
they can be related to the norm kernel by
$\Gamma_{\alpha\beta}=\frac{\partial}{\partial q_\alpha}\frac{\partial}{\partial q_\beta '}n(\qVec,\qVec')$. 
This expression can also 
be computed using standard linear response techniques from the alternative definition
\begin{equation}
\Gamma_{\alpha\beta}=\langle \Phi (\qVec) | \frac{\overleftarrow{\partial}}{\partial q_\alpha} 
\frac{\vec{\partial}}{\partial q_\beta}|\Phi (\qVec)\rangle.
\end{equation}
which involves the ``momentum operator'' 
$\frac{\vec{\partial}}{\partial q_\beta}$ - see below.
Another, simpler way, is to evaluate the overlap of the two HFB wave functions for
near $\qVec$ and $\qVec'$ values
using the Onishi formula \cite{ring2000} and make a local fit to a Gaussian. 
If the Gaussian overlap 
approximation (GOA) is valid, it does not make sense to accurately compute 
the Hamiltonian overlap for values of $\gras{s}$ greater than the inverse 
of the square root of the width $\tensor{\Gamma}^{-1/2}(\bar{\qVec})$
\footnote{The square root of $\tensor{\Gamma}$ has to be understood
in a matrix sense. The components  $\Gamma_{\alpha\beta}$ of $\tensor{\Gamma}$
are also the matrix elements of a symmetric positive-definite matrix. The square root is 
then understood as the Cholesky decomposition of $\tensor{\Gamma}$.
}. 
Therefore, a reasonable approximation is to expand $h(\qVec,\qVec')$ around 
$\gras{s} = 0$ (or $\gras{q}=\gras{q}'=\bar{\gras{q}}$) and keep quadratic 
terms only,
\begin{eqnarray}
h(\qVec,\qVec') 
& =  h(\bar{\qVec},\bar{\qVec}) \nonumber\\
& +  h_{\qVec} (\qVec-\bar{\qVec}) 
+ h_{\qVec'} (\qVec'-\bar{\qVec}) \nonumber\\
& 
+  \frac{1}{2} \left[(\qVec-\bar{\qVec}) \tensor{H}_{\qVec\qVec} (\qVec-\bar{\qVec}) 
+   (\qVec'-\bar{\qVec}) \tensor{H}_{\qVec'\qVec'} (\qVec'-\bar{\qVec}) \right.\nonumber \\
& \;\;\;\;\;\;\; \left. + 2 (\qVec-\bar{\qVec}) \tensor{H}_{\qVec\qVec'} (\qVec'-\bar{\qVec})
\right] + \cdots
\label{eq:GCM_quadratic}
\end{eqnarray}
In this expression $h_{\qVec}$ denotes the set of partial derivatives 
with respect to $\qVec$ of the energy kernels, 
\begin{equation}
h_{\qVec} \equiv (h_{q_{1}}, \dots, h_{q{N}} ),\ \ \ 
h_{q_{\alpha}} = \left. \frac{\partial h(\qVec,\qVec')}{\partial q_{\alpha}}\right|_{\gras{q}=\gras{q}'=\bar{\gras{q}}}
\end{equation}
This quantity is a vector and products like $h_{\qVec} (\qVec-\bar{\qVec})$ 
have to be understood as scalar products $\sum_{\alpha} h_{q_{\alpha}} 
(q_{\alpha} -\bar{q}_{\alpha})$. On the other hand, $\tensor{H}_{\qVec\qVec'}$ 
denotes the set of second partial derivatives with respect to $\qVec$ and 
$\qVec'$ and therefore it is a rank 2 tensor with components 
$H_{q_{\alpha}q'_{\beta}}$ 
\begin{equation}
\tensor{H}_{\qVec\qVec'} \equiv (H_{q_{\alpha}q'_{\beta}})_{\alpha,\beta=1,\dots,N},\ \ \ 
H_{q_{\alpha}q'_{\beta}} 
= 
\left. \frac{\partial^{2} h(\qVec,\qVec')}{\partial q_{\alpha}\partial q'_{\beta}}\right|_{\gras{q}=\gras{q}'=\bar{\gras{q}}},
\end{equation}
and products like $\tensor{H}_{\qVec\qVec'} (\qVec-\bar{\qVec})(\qVec'-\bar{\qVec})$ 
are to be understood as contractions of this tensor with the two vectors 
$\qVec-\bar{\qVec}$ and $\qVec'-\bar{\qVec}$, namely $\sum_{\alpha\beta} 
H_{q_{\alpha}q'_{\beta}}(q_{\alpha} -\bar{q}_{\alpha})(q'_{\beta} - 
\bar{q}_{\beta})$. Using the exponential form of the norm kernels, assuming 
a constant width $\tensor{\Gamma}$ and the quadratic expansion 
(\ref{eq:GCM_quadratic}) for the energy kernels, it is possible to reduce 
the Hill-Wheeler equation to the following CSE in the collective space 
defined by the variables $\qVec$,
\begin{equation}
\left( -\frac{\hbar^{2}}{2}  \frac{\partial}{\partial \qVec} \tensor{B}_{\mathrm{GCM}}(\qVec) \frac{\partial}{\partial \qVec}
+
V(\qVec) - 
\epsilon_\mathrm{zpe} (\qVec) \right) g_\sigma (\qVec) 
= \epsilon_\sigma g_\sigma (\qVec).\label{eq:CSE}
\end{equation}
The derivation is given in \cite{ring2000,onishi1975,krappe2012}. 
The collective mass $\tensor{M}_{\mathrm{GCM}}(\qVec)$ is a rank 2 tensor that 
depends on $\qVec$ and is the inverse to the collective inertia 
$\tensor{B}_{\mathrm{GCM}}(\qVec)$ (curvature) of the collective Hamiltonian 
when expanded to second order in the variable $\gras{s}$,
\begin{equation} \label{eq:GCMMASS}
\tensor{M}_{\mathrm{GCM}}^{-1}(\qVec) \equiv \tensor{B}_{\mathrm{GCM}}(\qVec) = 
\frac{1}{2}\tensor{\Gamma}^{-1}(\tensor{H}_{\qVec\qVec'}-\tensor{H}_{\qVec\qVec})\tensor{\Gamma}^{-1}.
\end{equation}
The potential energy $V(\qVec)$ is the HFB energy and 
$\epsilon_\mathrm{zpe} (\qVec)$ is a zero point energy correction, 
\begin{equation}
\epsilon_\mathrm{zpe}(\qVec)=\frac{1}{2}\tensor{H}_{\qVec\qVec'}
\tensor{\Gamma}^{-1}
\label{gcm:epsilon0}
\end{equation}
given in terms of the contraction of the two tensors to form a scalar. 
The zero point energy (zpe) correction represents a quantum correction to the
classical potential for the collective variables and given by the 
HFB energy. The zpe correction corresponds
to the energy of a Gaussian wave packet \cite{REINHARD1975120,ring2000}. The
interpretation of this term is similar to the rotational energy correction
but with the rotation angles replaced by the collective variables. 
The solution of (\ref{eq:CSE}) provides the energies $\epsilon_\sigma$ and
wave functions $g_\sigma(\qVec)$ of the collective modes. The wave functions $g_\sigma(\qVec)$
are related to the amplitudes $f(\qVec)$ of the Hill-Wheeler equation (\ref{eqn:HW})
through the relation 
\begin{equation}
f_\sigma (\qVec) = \int d\qVec' n(\qVec,\qVec')^{-1/2} g_\sigma (\qVec'),
\end{equation}
which can be used to compute, for instance, mean values of other observables
where the Gaussian approximation is not justified.
Apart from the simplification that the local reduction brings to the 
solution of problem, an interesting physical picture emerges: the 
collective dynamics is driven by the behaviour of the potential energy 
surface (PES) given by the HFB energy as a function of $\qVec$ with 
some coordinate-dependent quantal corrections $\epsilon_{\mathrm{zpe}}(\qVec)$ 
and inertia $\tensor{M}_{\mathrm{GCM}}(\qVec)$. If the $\qVec$ are chosen to be 
the collective variables driving the nucleus to fission (quadrupole 
moment, octupole moment, neck, etc.) then the probability of tunnelling 
through the fission barrier is given by the integral of the exponential 
of the action computed with the collective potential and the collective 
inertia of the CSE approximation to the Hill-Wheeler equation of the 
GCM. 

{\it Local approximation - }
The collective inertia can be evaluated from the energy kernels using 
numerical differentiation to obtain $h_{\qVec}$ and $\tensor{H}_{\qVec\qVec'}$. 
Various finite difference schemes can be used and the precision is 
controlled by the value of $\delta\qVec$ and $\delta\qVec'$. However, 
it is also common practice to use the explicit expression for the 
``momentum'' operator associated to the variable $\qVec$. Following 
\cite{une1976,reinhard1987}, we write
\begin{equation}
\frac{\partial}{\partial q_{\alpha}} | \Phi (\qVec)\rangle 
= 
\frac{i}{\hbar}\hat{P}_{q_{\alpha}} | \Phi (\qVec)\rangle.
\end{equation}
The action of the momentum operator $\hat{P}_{q{\alpha}}$ on the quasiparticle  
vacuum $|\Phi(\gras{q})\rangle$ can be obtained from the Ring and 
Schuck theorem of \cite{ring1977},
\begin{equation}
\hat{P}_{q_{\alpha}} | \Phi (\qVec)\rangle 
= 
\sum_{\mu < \nu} \left[ 
(P_{q_{\alpha}}^{20})_{\mu \nu } \beta^\dagger_\mu \beta^\dagger_\nu 
-
(P_{q_{\alpha}}^{20*})_{\mu \nu } \beta_\mu \beta_\nu 
\right] | \Phi (\qVec)\rangle.
\end{equation}
The quasiparticle matrix elements of the momentum operator 
$(P^{20}_{q_{\alpha}})_{\mu \nu }$ are obtained by expanding the 
HFB solution at point $\qVec+\delta\qVec$ to first order in $\delta\qVec$. 
They are related to the matrix of the derivatives of the generalized 
density with respect to the collective variables,
\begin{equation}
\frac{\partial\mathcal{R}}{\partial q_{\alpha}} = 
\frac{i}{\hbar}
\left(\begin{array}{cc}
0 & P_{q_{\alpha}}^{20} \\
-P_{q_{\alpha}}^{20*} & 0
\end{array}\right),
\label{eq:dRdq}
\end{equation}
and take the generic form
\begin{equation} \label{eq:P20}
\left( \begin{array}{c} 
\left( P^{20}_{q_{\alpha}} \right)_{\mu\nu} \\ 
-\left(P^{20\,*}_{q_{\alpha}} \right)_{\mu\nu} 
\end{array} \right) 
= \sum_{\beta}
[M^{(-1)}_{\alpha \beta}]^{-1} \sum_{\mu' < \nu'} \left(\mathcal{M}^{-1} \right)_{\mu\nu\mu'\nu'} 
\left( \begin{array}{c} 
\left(q^{20}_{\beta} \right)_{\mu'\nu'}\\ 
\left(q^{20\,*}_{\beta}\right)_{\mu'\nu'}
\end{array} \right),
\end{equation} 
where $\mathcal{M}$ is the linear response matrix in the quasiparticle 
basis,
\begin{equation}
\mathcal{M} = \left(\begin{array}{cc}
A & B \\ B^{*} & A^{*}
\end{array}\right),
\label{eq:qrpa_matrix}
\end{equation}
and $A$ and $B$ are given by (see, e.g., \cite{ring2000,blaizot1985})
\begin{equation}
A_{\mu \nu \mu' \nu' } = \langle \beta_\nu\beta_\mu \hat{H} \beta^\dagger_{\mu'}\beta^\dagger_{\nu'}  \rangle, \ \ \ 
B_{\mu \nu \mu' \nu' } = \langle \beta_\nu\beta_\mu\beta_{\nu'}\beta_{\mu'} \hat{H} \rangle. \label{eqn:AandB}
\end{equation}
In (\ref{eq:P20}), we have introduced the moments  
\begin{equation}
M^{(-n)}_{\alpha \beta} 
= 
\left( q^{20\,\dagger}_{\alpha},\  q^{20\,T}_{\alpha} \right) 
\mathcal{M}^{-n} 
\left( \begin{array}{c} 
q^{20}_{\beta} \\
q^{20\,*}_{\beta}    
\end{array} \right).
\label{eq:moments}
\end{equation}
where $\mathcal{M}^{-n} = \mathcal{M}^{-1}\times \dots \mathcal{M}^{-1}$ $n$ times. 
Using the explicit form of the momentum operator we can then express the width tensor 
in terms of the moments (which are also rank 2 tensors)
\begin{equation}
\tensor{\Gamma} =  \frac{1}{2} 
\left[ \tensor{M}^{(-1)}(\qVec) \right]^{-1} 
\tensor{M}^{(-2)}(\qVec) 
\left[\tensor{M}^{(-1)}(\qVec) \right]^{-1}.
\label{gcm:eps0crank}
\end{equation}

The linear response matrices of (\ref{eqn:AandB}) have been defined only
in the case of interactions deriving from a Hamiltonian $\hat{H}$. For
density-dependent interactions and generic EDFs that  
cannot be expressed as mean values of a Hamiltonian, the generalization of 
(\ref{eqn:AandB}) involves second derivatives of the energy with 
respect to the variational parameters. Typical expressions are given
in \cite{blaizot1985} in the most general case. The idea is to 
use the short version of Thouless theorem \cite{ring2000} 
$|\Phi (Z) \rangle = \sum_{\mu < \nu} Z_{\mu\nu} \beta^\dagger_\mu \beta^\dagger_\nu |\Phi (0) \rangle$ 
to define the density and pairing tensor entering the EDF as functions
of the independent variables $Z_{\mu\nu}$ and $Z^*_{\mu\nu}$ ( for instance, 
$\rho_{ji} (Z^*,Z) = \langle \Phi (Z) | c^\dagger_ic_j | \Phi (Z) \rangle / \langle \Phi (Z)  | \Phi (Z) \rangle$ )
and define 
\begin{equation}
A_{\mu \nu \mu' \nu' } = \frac{\partial E (Z^*,Z) }{\partial  Z_{\mu'\nu'} \partial Z^*_{\nu\mu}} \ \ \ 
B_{\mu \nu \mu' \nu' } = \frac{\partial E (Z^*,Z) }{\partial  Z^*_{\nu'\mu'} \partial Z^*_{\nu\mu}}. \label{eqn:AandB_Z}
\end{equation}

{\it Cranking Approximation - }
The evaluation (\ref{eq:moments}) of the moments $\tensor{M}^{(-n)}$ requires inverting the full 
linear response matrix $\mathcal{M}$. Computationally, this is a 
daunting task that is often alleviated by approximating $\mathcal{M}$ by a diagonal matrix, which simplifies enormously the 
inversion problem. This ``cranking approximation'' corresponds to 
neglecting the residual quasiparticle interaction. The diagonal matrix 
elements are simply the two-quasiparticle excitation energies. With this 
simplification, (\ref{eq:moments}) reduces to the more manageable form 
\begin{equation}
M_{\alpha\beta}^{(-n)} 
= 
\sum_{\mu < \nu } \frac{\langle \Phi | \hat{Q}^\dagger_{\alpha} | \mu \nu \rangle
\langle \mu \nu | \hat{Q}_{\beta} | \Phi \rangle}{(E_\mu+E_\nu)^n},
\label{eq:momentsapp}
\end{equation}
where $|\Phi\rangle $ stands for the quasiparticle vacuum at point $\qVec$ and $|\mu \nu \rangle $ 
represents a two quasiparticle excitation built on top of that vacuum, 
i.e. $|\mu \nu \rangle = \beta^{\dagger}_{\mu} \beta^{\dagger}_{\nu} |\Phi\rangle$. 
The combination of introducing the local momentum operator to substitute 
for exact numerical differentiations with the cranking approximation is 
also referred to as the ``perturbative cranking'' approximation. In this 
case, the curvature term $\tensor{H}_{\qVec \qVec}$ vanishes and some algebra leads to 
\begin{equation}
\tensor{H}_{\qVec \qVec'} = \frac{1}{2} \left[ \tensor{M}^{(-1)}\right]^{-1}.
\end{equation} 
Introducing this last result in (\ref{eq:GCMMASS}) while taking into account 
the form (\ref{gcm:eps0crank}) for the norm overlap, the collective 
mass tensor reduces to
\begin{equation}
\tensor{M}_\mathrm{GCM} (\qVec) 
= 
4 \tensor{\Gamma} \tensor{M}^{(-1)}  \tensor{\Gamma}.
\label{gcm:masscrank}
\end{equation}
The zero point energy correction becomes
\begin{equation}
\epsilon_\mathrm{zpe} (\qVec) = \frac{1}{2} \tensor{\Gamma} \tensor{M}_{\mathrm{GCM}}^{-1}(\qVec).
\label{gcm:epsilon0b}
\end{equation}
This last expression must again be understood as the contraction of the two 
tensors. Both expressions (\ref{gcm:masscrank}) and (\ref{gcm:epsilon0b}) are 
then used in the evaluation of the action $S(\qVec)$ that enters the 
WKB expression of the spontaneous fission half-life (\ref{eq:tSF}). 

{\it Variable Width - }
The above discussion has been restricted for simplicity to a constant 
Gaussian width $\tensor{\Gamma}$. In case this assumption is not strictly valid, a change of 
variables to a new set of coordinates $\gras{\eta}$ is required. The new 
variables are defined to make the width locally constant,
\cite{god1985-a}
\begin{equation}
\sum_{\alpha} (d\eta)_{\alpha}^2 
= 
\sum_{\alpha\beta}d(\qVec-\qVec')_{\alpha}\Gamma_{\alpha\beta}(\gras{q})d(\qVec-\qVec')_{\beta}.
\end{equation}
This last expression is reminiscent of the field of differential 
geometry where the new variables $\gras{\eta}$ have a constant metric 
$\delta_{\alpha\beta}$ and distances in the original variables $\qVec$ 
have to be measured with the new metric $\tensor{\Gamma}$. In this 
framework the expressions derived above are still valid if standard 
derivatives with respect to $q_{\alpha}$ are replaced by covariant 
derivatives $\mathcal{D} / \mathcal{D} q_{\alpha}$ that include in 
their definition the corresponding Christoffel symbols \cite{god1985-a,
GOZDZ1985281,krappe2012}. For an introduction to differential geometry, 
the reader may refer to chapter 5 of \cite{borisenko1968}. 
As the metric in the variables $\qVec$ is now 
coordinate-dependent, the volume element of integrals must be modified to 
account for an extra $\sqrt{\det\tensor{\Gamma}}$ which is the
determinant of $\tensor{\Gamma}$. Also the 
kinetic energy term has to incorporate the bell and whistles of 
differential geometry involving covariant derivatives. The final expression reads
\begin{equation}
\left( -\frac{\hbar^{2}}{2} 
\frac{1}{\sqrt{\det\tensor{\Gamma}}}\frac{\partial}{\partial \qVec}\sqrt{\det\tensor{\Gamma}}\;\tensor{B}_{\mathrm{GCM}}(\qVec) \frac{\partial}{\partial \qVec}
+ V(\qVec)
- \epsilon_\mathrm{zpe} (\qVec) \right) g_{\sigma} (\qVec) 
= \epsilon_{\sigma} g_{\sigma} (\qVec).
\end{equation}
The use of covariant derivatives is still required in the evaluation of the
inertia. As the calculation of the Christoffel symbols and its use in the covariant 
derivatives is rather cumbersome, it is often assumed that the width matrix $\tensor{\Gamma}$,
which is used also as the metric of the curved space,
varies slowly with the $\qVec$ coordinates. In this case, and given that the Christoffel 
symbols are defined in terms of partial derivatives of the metric, they can be neglected and
therefore the covariant derivative reduces to the standard partial derivative. 
However the term $\sqrt{\det\tensor{\Gamma}}$ is kept in its original form. 


\subsubsection{Adiabatic Time-Dependent Hartree-Fock-Bogoliubov}
\label{subsubsec:atdhfb}

For a given choice of collective variables $\{ \qVec \}$, the ansatz 
(\ref{eq:gcm_ansatz}) for the GCM wave function leads to a notorious 
underestimation of the collective inertia $\tensor{M}_\mathrm{GCM} (\qVec)$, even 
when computed with exact numerical derivatives. This is illustrated in the 
top panel of figure \ref{fig:corrections} page \pageref{fig:corrections}, 
which shows the collective inertia as a function of the quadrupole deformation 
for both the GCM and ATDHFB prescriptions (in the perturbative cranking approximation). 
Ring and Schuck recall, in
their textbook \cite{ring2000}, how one-dimensional GCM calculations using 
a shift of the centre of mass as collective variable $\qVec$ fail to 
reproduce the exact result in the exactly solvable case of translational 
large amplitude collective motion (where the collective mass is just the 
sum of the masses of all nucleons). Back in 1962, Peierls and Yoccoz had 
showed in \cite{peierls1962} that adding another collective variable 
corresponding to the momentum of the centre of mass was sufficient to 
reproduce the exact mass. Qualitatively, a naive implementation of the 
GCM where the only collective degrees of freedom $\qVec$  are time-even 
functions (such as, e.g., HFB states under various constraints on multipole 
moments) is bound to fail, since it does not contain enough information on 
the actual motion in the collective space, which is controlled by time-odd 
momenta. In fact, Reinhard and Goeke showed in their review 
\cite{goeke1980}, that a ``dynamic GCM'' (DGCM), where the set of 
collective variables $\qVec$ is expanded to include the associated momenta 
$\hat{\gras{P}}_{\qVec}$, was necessary to provide a more realistic 
description of collective dynamics. The implementation of the DGCM in 
practical applications is, however, a lot more involved than the usual GCM. 

The adiabatic time-dependent HFB (ATDHFB) approximation of the TDHFB equation 
thus provides an appealing alternative; see, e.g., 
\cite{brink1976,giannoni1976,villars1977,baranger1978,goeke1978,giannoni1980,
giannoni1980-a,giannoni1984,yuldashbaeva1999,baran2011} for a presentation 
of the theory. The ATDHFB is based on a small velocities expansion of the 
TDHFB equation,
\begin{equation}
i\hbar\dot{\mathcal{R}} = [\mathcal{H},\mathcal{R}],
\end{equation}
where the generalized density matrix $\mathcal{R}(t)$ is given by 
(\ref{eq:pes_BigR}) and the HFB matrix by (\ref{eq:pes_BigH}) -- both are 
now time-dependent. As we will show below, this expansion introduces a set 
of collective ``coordinates'', which are time-even generalized densities, 
and the related collective ``momenta''. Coordinates and momenta differ by 
their properties with respect to time-reversal symmetry. Once these 
quantities are defined,  the energy of the system can be expressed as the 
sum of a potential part and a kinetic part which is a quadratic function 
of the momenta, both parts being time-dependent functions. It is then
possible to assign a collective inertia associated to any point of the 
collective space. At this point the problem is reduced to a classical one 
and it is not possible to describe phenomena involving quantum tunnelling 
through the barrier. However, we can still resort to the semi-classical 
description of tunnelling based on the WKB method. Within this scheme it 
is thus possible to compute spontaneous fission lifetimes using quantities 
provided by the HFB theory and its time-dependent extension ATDHFB. 

Note that all applications of ATDHFB to fission so far have been 
performed in the particular case where the collective path, i.e., the trajectory 
of the system in the collective space, is not calculated dynamically by 
solving the ATDHFB equations, but predefined as a set of constrained 
HFB calculations. This is so even though (i) there are several possible 
prescriptions to compute the collective path, as can be seen, e.g. in the 
studies of \cite{goeke1981,mukherjee1981,mukherjee1982} and (ii) comparisons 
of dynamically-defined collective paths with constrained ones indicate 
that there can be significant differences in the collective inertia tensor 
and zero-point energy corrections; see the examples studied in 
\cite{goeke1983,yuldashbaeva1999}. The main reason for this approximation 
is that the ATDHFB theory is mostly used as a tool to compute a realistic 
collective inertia for WKB-types of calculations, and the actual evolution 
of the system in collective space is not needed. The adiabatic self-consistent 
collective model formulated in \cite{matsuo2000,hinohara2007} represents an 
alternative formulation of the adiabatic approximation to the TDHFB equations 
that solve many of the formal difficulties but has not been applied to the 
specific case of fission yet.

The starting point of the ATDHFB method is the expansion of the full 
time-dependent generalized density $\mathcal{R}(t)$ according to
\[
\mathcal{R}(t)=e^{-i\chi(t)}\mathcal{R}_{0}(t)e^{i\chi(t)}
\]
where $\mathcal{R}_{0}(t)$ is a time-dependent, time-even generalized density
matrix satisfying the standard relation $\mathcal{R}_{0}^{2}=\mathcal{R}_{0}$. 
$\chi(t)$ is also a time-even operator. The time-even density 
$\mathcal{R}_{0}(t)$ is considered as some kind of coordinate variable whereas 
$\chi(t)$ will be related to its conjugate momentum, which is assumed to be a 
small quantity. Expanding the density matrix in powers of $\chi(t)$ 
\begin{equation}
\mathcal{R}(t) = \mathcal{R}_{0}(t)+\mathcal{R}_{1}(t)+\mathcal{R}_{2}(t)+\cdots
\label{eq:Rexp}
\end{equation}
we obtain terms which are either time-odd, such as 
$\mathcal{R}_{1}=-i[\chi,\mathcal{R}_{0}]$, or time-even, such as 
$\mathcal{R}_{0}$ or $\mathcal{R}_{2}$. Introducing the above expansion 
in the TDHFB equation, we obtain a set of two equations, one for time-odd 
and the other for time-even quantities,
\begin{eqnarray}
i\hbar\dot{\mathcal{R}}_{0} & = & [\mathcal{H}_{0},\mathcal{R}_{1}]+[\mathcal{K}_{1},\mathcal{R}_{0}],
\label{eq:ATDHFB_I}\\
i\hbar\dot{\mathcal{R}}_{1} & =[\mathcal{H}_{0},\mathcal{R}_{0}]+ & [\mathcal{H}_{0},\mathcal{R}_{2}]+[\mathcal{K}_{1},\mathcal{R}_{1}]+[\mathcal{K}_{2},\mathcal{R}_{0}],
\label{eq:ATDHFB_II}
\end{eqnarray}
where $\mathcal{H}_{\mu}$ and $\mathcal{K}_{\mu}$ are obtained from
the expressions (\ref{eq:pes_BigH}) and (\ref{eq:pes_BigT}) of 
$\mathcal{H}$ and $\mathcal{K}$ with the densities $\mathcal{R}_{\mu}$ 
-- all being time-dependent quantities. In the next step, the TDHFB energy 
is expanded according to (\ref{eq:Rexp}) in order to obtain a zero-order 
term, which resembles the HFB energy for the $\mathcal{R}_{0}$ density, and 
a second order term reminiscent of a kinetic energy and given by 
\begin{equation}
\frac{1}{2}\textrm{tr}\left\{ \dot{\mathcal{R}}_{0}[\mathcal{R}_{0},\mathcal{R}_{1}]\right\}.
\label{eq:kinATDHFB}
\end{equation}
In this expression, the first term $\dot{\mathcal{R}}_{0}$ plays the 
role of a generalized velocity, whereas the second one is a momentum-like 
quantity. The identification of this term with a kinetic energy is at the 
origin of the definition of the ATDHFB mass. In order to obtain a more 
explicit definition of the mass, we use (\ref{eq:ATDHFB_I}) to express 
$\mathcal{R}_{1}$ in terms of $\dot{\mathcal{R}}_{0}$. To this end it is 
customary to introduce the ATDHFB basis, which diagonalizes $\mathcal{R}_{0}(t)$ 
at all times $t$ and has an analogous block structure as the traditional HFB 
basis. In the ATDHFB basis, the matrix of $\hat{\chi}$ is noted generically 
\begin{equation}
\chi = \left(\begin{array}{cc}
\chi^{11} & \chi^{12} \\
\chi^{21} & \chi^{22}\end{array}\right),
\end{equation}
and one can show that only the blocks $\chi^{12}$ and 
$\chi^{21}$ of the matrix of $\chi(t)$ are relevant for the dynamics 
(and they are related through $\chi^{21} = \chi^{12\,\dagger}$). After some 
manipulations, one finds
\begin{equation}
\left( \begin{array}{c}
\chi^{12} \\ 
\chi^{12 \,*}
\end{array} \right) 
= \mathcal{M}^{-1}
\left( \begin{array}{c} 
\dot{\mathcal{R}}_{0}^{12} \\ 
\dot{\mathcal{R}}_{0}^{12 \,*}
\end{array} \right),
\end{equation}
where $\mathcal{M}$ is the same linear response matrix as in (\ref{eq:qrpa_matrix}). 
Inserting this relationship in (\ref{eq:kinATDHFB}) we end up with an expression 
that is fully reminiscent of the kinetic energy
\begin{equation}
\mathcal{K} \equiv
\frac{1}{2} \left( \chi^{12\,\dagger},\  \chi^{12 \,T} \right) 
\mathcal{M} 
\left( \begin{array}{c}
\chi^{12} \\ \chi^{12 \,*} 
\end{array}\right)
=
\frac{1}{2} \left( \dot{\mathcal{R}}_{0}^{12,\dagger},\ \dot{\mathcal{R}}_{0}^{12 \,T} \right) 
\mathcal{M}^{-1}
\left( \begin{array}{c}
\dot{\mathcal{R}}_{0}^{12} \\ 
\dot{\mathcal{R}}_{0}^{12 \,*} 
\end{array} \right).
\label{eq:kinATDHFB_final}
\end{equation}
This expression shows that the linear response matrix is indeed the 
matrix of inertia. 

As for the GCM, the above expressions are used in a framework where it 
is assumed that just a few collective coordinates are responsible for 
the time evolution of the system. This assumption allows to express the 
time derivative of the density matrix as
\begin{equation}
\dot{\mathcal{R}}_{0} = 
\sum_{\alpha} \frac{\partial \mathcal{R}_{0}}{\partial q_{\alpha}} \dot{q}_{\alpha},
\end{equation}
where the $q_{\alpha}$ are the relevant collective coordinates and 
$\dot{q}_{\alpha}$ their time derivatives. The kinetic energy 
(\ref{eq:kinATDHFB_final}) becomes
\begin{equation}
\mathcal{K}  = \frac{1}{2} \sum_{\alpha\beta} M_{\alpha\beta}\dot{q}_{\alpha} \dot{q}_{\beta} ,
\end{equation}
with 
\begin{equation} 
M_{\alpha\beta} = 
\frac{\partial \mathcal{R}_{0}}{\partial q_{\alpha}} 
\mathcal{M}^{-1}
\frac{\partial \mathcal{R}_{0}}{\partial q_{\beta}}
\label{eq:ATDHFBMass}
\end{equation}
which constitutes the expression of the collective inertia for the relevant 
collective coordinates $q_{\alpha}$. 

As in the GCM case, the final expression of the ATDHFB mass involves the 
inverse of the linear response matrix. The only (very recent) attempt to 
invert this matrix explicitly has been reported in \cite{lechaftois2015}. 
Most often, the problem is simplified by resorting to the same two 
approximations encountered earlier and discussed in details in \cite{baran2011}: 
the ``cranking approximation'' uses the diagonal approximation to the linear 
response matrix with two quasiparticle energies as diagonal elements. The 
partial derivatives of the density $\mathcal{R}_{0}$ with respect to the 
collective variables $q_{\alpha}$ are evaluated numerically by obtaining 
the HFB wave function with the constraints $\qVec+\delta \qVec$ and using 
finite difference approximations to the partial derivatives 
\footnote{Obviously, higher order formulas are used to compute the 
derivatives that involve also multiples of $\delta q_i$}. However, the 
partial derivatives of the generalized density can also be obtained by 
applying linear response theory, which leads to an explicit expression of 
the partial derivatives involving again the inverse of the linear response 
matrix. If the same cranking approximation is used as before, we then 
obtain the ``perturbative cranking'' formula for the ATDHFB mass,
\begin{equation}
\tensor{M}_{\mathrm{ATDHFB}} = \hbar^{2} [\tensor{M}^{(-1)} ]^{-1} \tensor{M}^{(-3)} [ \tensor{M}^{(-1)} ]^{-1},
\label{eq:ATDHFBMassPC}
\end{equation}
with the moments defined in (\ref{eq:momentsapp}).

\begin{figure}[!ht]
\begin{center}
\includegraphics[width=0.7\linewidth]{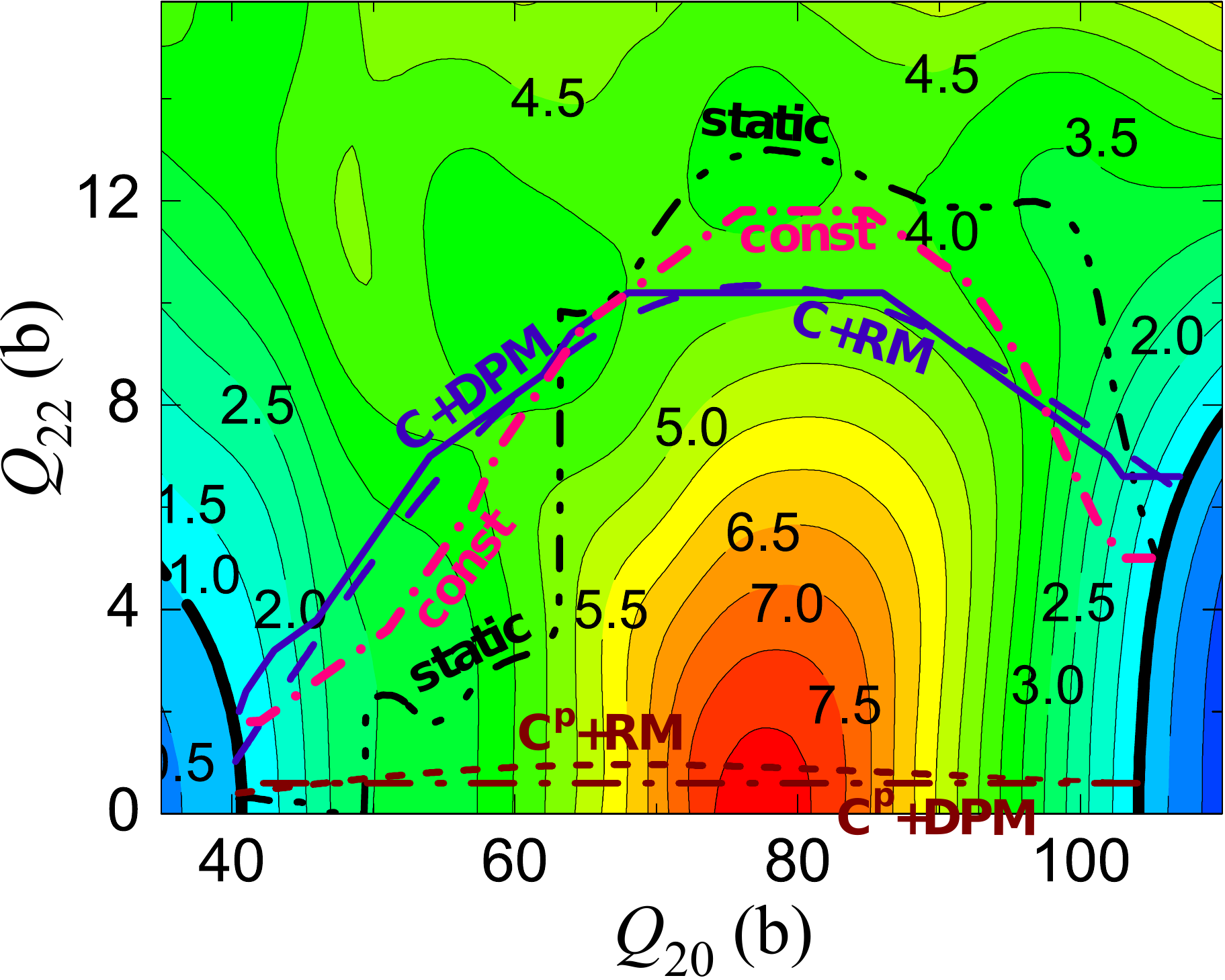}
\caption{Dynamic paths for spontaneous fission of $^{264}$Fm, calculated
for the non-perturbative $\mathcal{M}^{\mathrm{C}}$ (\ref{eq:ATDHFBMass}) and perturbative 
$\mathcal{M}^{\mathrm{CP}}$ (\ref{eq:ATDHFBMassPC}) versions of the ATDHFB cranking inertia using the dynamic programming 
method (DMP) and Ritz method (RM) to minimize the collective action 
integral. The static pathway (``static'') and that corresponding to a 
constant inertia (``const'') are also shown. The trajectories of turning 
points $s_{\mathrm{in}}$ and $s_{\mathrm{out}}$ are marked by thick solid 
lines. Figure taken from Ref. \cite{sadhukhan2013}, courtesy of J. 
Sadhukhan; copyright 2014 by The American Physical Society.}
\label{fig:cranking_perturbative}
\end{center}
\end{figure}

The validity of the perturbative cranking approximation (replacing exact 
derivatives by a linearization) has recently been tested in the 
calculation of fission pathways of superheavy elements in \cite{sadhukhan2013}. 
The figure \ref{fig:cranking_perturbative} shows the fission pathways in 
$^{264}$Fm in the $(q_{20},q_{22})$ collective space obtained under 
different approximations: the path obtained by only considering the lowest 
energy is marked ``static''; the path obtained by minimizing the action 
while taking a constant inertia is marked ``const''; the paths obtained 
by minimizing the action with the collective inertia computed at the 
cranking approximation either directly or perturbatively are denoted by 
``C+DPM'' and ``C$^{p}$+DPM'' and ``C+RM'' and ``C$^{p}$+RM'', respectively. 
There is a clear, qualitative difference between the perturbative and 
non-perturbative treatment of collective inertia.


\subsection{Tunnelling and Fission Half-Lives}
\label{subsec:t12}

Strictly speaking, the ground-state of many nuclei is not a stationary 
state since there are open channel through which the nucleus can decay. 
Nevertheless, it still makes sense to do this approximation since the
evolution towards those open channels must proceed through 
classically-forbidden regions with a tiny transmission probability 
coefficient. Similarly, in spontaneous fission the evolution of the nucleus 
from the ground-state to scission configurations and to a two-fragment final
state is achieved by tunnelling through classically-forbidden regions. Fission 
lifetimes are then proportional to the tunnelling transmission coefficient 
through multi-dimensional potential energy surfaces in the relevant 
collective space.   


\subsubsection{The WKB Approximation}
\label{subsubsec:wkb}

The Wentzel, Kramers, Brillouin (WKB) approximation is often used to get
an estimate of the transmission coefficient through the fission barrier \cite{landau2013quantum}. To
get an idea of how it works, let us consider first the one-dimensional case.
The basic idea is to substitute the classical expression for the momentum
$p=\sqrt{2m(E-V(x))}$ into the quantum mechanical identity
\begin{equation} 
\frac{d^2\Psi (x)}{dx^2}=-\frac{p^2}{\hbar^2}\Psi(x).
\label{eq:WKB1}
\end{equation}
This second-order differential equation is then solved under the assumption 
that the wave function can be expressed in the generic form 
$$
\Psi (x) = A(x) e^{i\phi (x)}.
$$
Inserting this ansatz in (\ref{eq:WKB1}) and assuming that the amplitude 
$A(x)$ varies slowly with $x$ one obtains 
$$
A(x) =\frac{C}{\sqrt{p(x)}},
$$ 
and 
$$\phi (x) = \pm \frac{1}{\hbar} \int dx\; p(x),
$$
which is valid in the ``classical'' region where $E\ge V(x)$. The extension
to classically-forbidden regions is straightforward, requiring the introduction
of a complex momentum $p(x)$ that leads to an exponential wave function in 
that region  
\begin{equation}
\Psi (x) = \frac{C_\pm}{\sqrt{|p(x)|}} 
\exp \left( \pm \frac{1}{\hbar} \int dx \sqrt{2m(V(x)-E)} \right).
\end{equation}
The positive sign in the exponent yields an exponentially increasing 
tunnelling probability and therefore the corresponding amplitude $C_+$ has 
to be very small. By keeping only the term with the minus sign for the wave 
function in the classically-forbidden region, we obtain for the transmission 
coefficient
\begin{equation}
T = |\Psi(b)|^2 / |\Psi(a)|^2 
= 
\exp \left( -  \frac{2}{\hbar} \int_a^b dx \sqrt{2m(V(x)-E)} \right),
\label{eq:WKB2}
\end{equation}
where $a$ and $b$ are the inner and outer turning points at the barrier 
corresponding to the energy $E$.

The extension of these ideas to fission is not entirely straightforward 
because of the infinite degrees of freedom of the nuclear many-body 
system. In the adiabatic approximation, however, the use of the WKB 
formula is somewhat justified owing to the reduction in the number of 
relevant collective variables. Clearly, the role of the mass of the 
particle in the one-dimensional problem recalled above should be played 
by the collective inertia tensor, and the potential energy 
should be replaced by the HFB energy along the collective path (possibly 
supplemented by quantum corrections such as the rotational energy 
correction, zero-point energies, etc.). The actual path used to compute 
the transmission coefficient is determined by invoking the last action 
principle mentioned in the introduction to this chapter: The physical path 
is the one for which the classical action (\ref{eq:classicalS}) is minimal. 
The spontaneous fission lifetime 
$\tau^{\mathrm{SF}}_{1/2}$ is then usually given by the inverse of the 
product of the transmission probability $T$ times the number of assaults 
to the barrier per unit time $\nu$
\begin{equation}
\tau^{\mathrm{SF}}_{1/2} = \frac{1}{\nu} \exp \left( \frac{2}{\hbar} \int_a^b ds \sqrt{B(s)(V(s)-E_0)} \right).
\label{eq:tSF}
\end{equation}
Usually $1/\nu$ is estimated assuming that the zero point energy correction of the
nucleus in its ground-state if of the order of one MeV, which leads to 
the value $1/\nu = 10^{-21}$s. Other authors (e.g. \cite{staszczak2013}) 
prefer to compute this parameter as the zero-point energy $\hbar \omega$ 
of the ground-state of the potential well corresponding to the collective 
variable $q$. For the mass, the two expressions for the collective 
inertia obtained previously in the framework of the GCM and ATDHFB 
methods are used, usually within the perturbative cranking approximation. 
Some results have recently been published in \cite{sadhukhan2013} without 
the perturbative treatment of the derivatives, but to our knowledge, 
there has been no calculations where the exact masses (involving the 
inverse of the full linear response matrix) have been used. 

As expected, the results for $\tau^{\mathrm{SF}}_{1/2}$ are very 
sensitive to the particular approximation used, especially for nuclei 
with large $\tau^{\mathrm{SF}}_{1/2}$ values. This is a straightforward 
consequence of the exponentiation of the action, leading to an exponential 
dependence on the potential, energy and inertia tensor. In applications 
of the WKB method, the collective potential $V(q)$ is usually supplemented 
by the various corrections discussed in details in section 
\ref{subsubsec:beyond}. Strictly speaking, zero-point energy corrections 
(ZPE) should only be computed when the GCM framework is used to compute 
the collective inertia. Some authors have argued that even in the ATDHFB 
case, where no ZPE correction is present by construction, some sort of 
ZPE can still be used by taking the GCM form and replacing the GCM by the 
ATDHFB mass, see for instance \cite{warda2012,sadhukhan2013,staszczak2013}. 
In any case, it is commonly assumed as in \cite{reinhard1987} that the ZPE corrections 
associated to the quadrupole moment 
vary slowly with the collective variable and therefore only represent a 
displacement of the energy origin.

Finally, the $E_0$ parameter is often taken as the HFB ground-state 
energy. However it has been argued, e.g. in \cite{PhysRevC.13.229,warda2002}, 
that the dynamics of the collective 
degree of freedom has an associated zero-point energy correction on 
top of the potential minimum (the HFB ground-state energy). This zero-point 
energy is often taken as a phenomenological parameter varying in the range 
0.5 - 1 MeV. Some authors prefer to compute it using the formulas 
obtained in the GCM formalism \cite{staszczak2013}. 


\subsubsection{Multidimensional Quantum Tunnelling}
\label{subsubsec:instantons}

Although the WKB method is the most popular choice to compute fission 
half-lives, several authors have considered alternative methods based on 
functional integrals, see for instance \cite{levit1980,levit1980-a,
negele1982,puddu1987,negele1989,bonasera1997,scamps2015}. Such techniques offer, at 
least in principle, the possibility to be extended to arbitrary many-body 
systems without relying on an explicit choice of collective coordinates 
(and the underlying adiabaticity hypothesis). Qualitatively, it can be 
thought of as an extension of time-dependent density functional theory 
in the classically-forbidden region ``below the barrier''. In this section, 
we only summarize some of the main features of the theory by following 
\cite{negele1982} in the simplified case of a one-dimensional problem.

\begin{figure}[!ht]
\begin{center}
\includegraphics[width=0.7\linewidth]{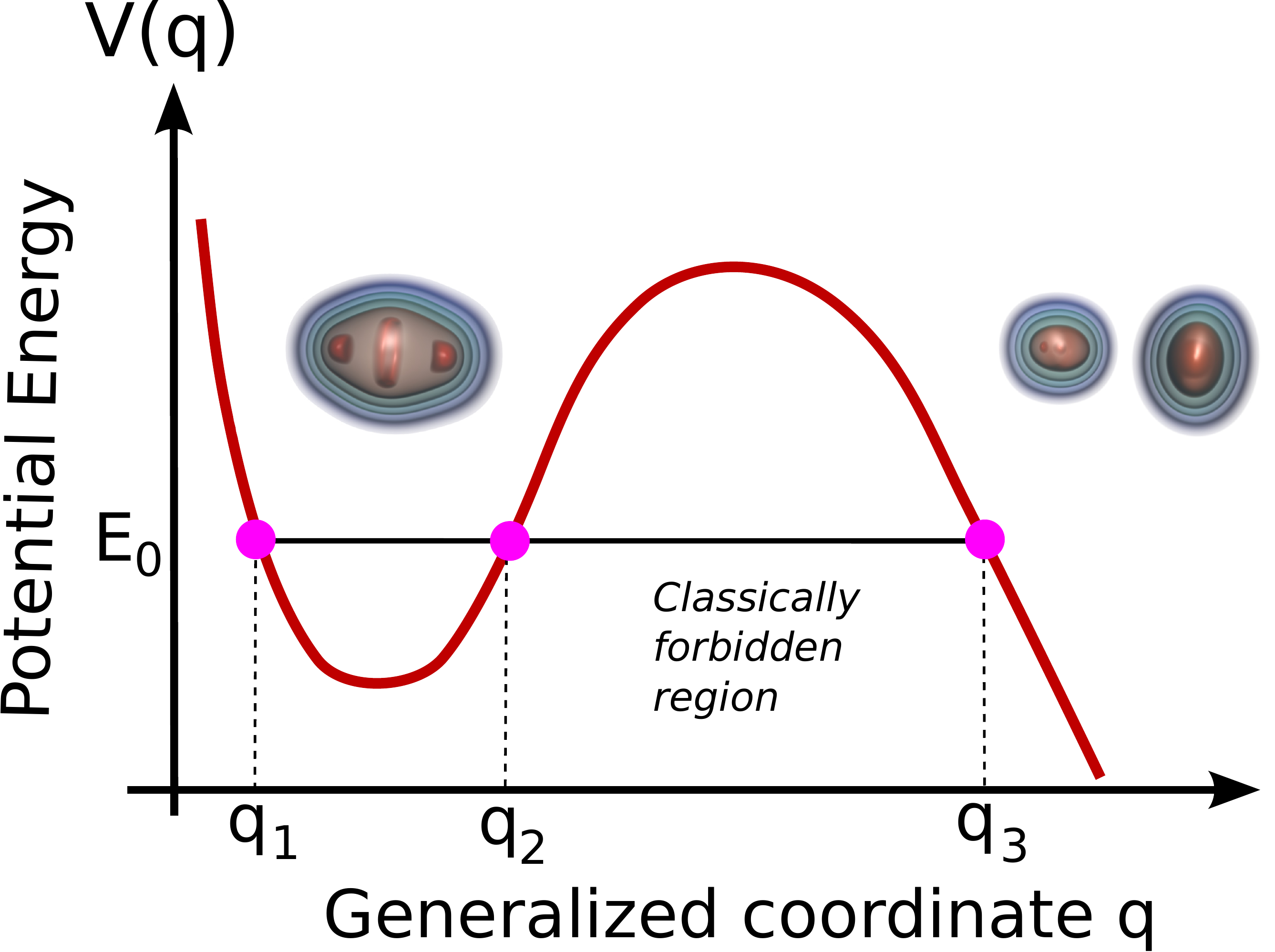}
\caption{Schematic representation of a quantum tunnelling problem in one 
dimension. The system is assumed to be described by a Hamiltonian of the 
type $H(q,p) = p^{2}/2m + V(q)$.}
\label{fig:barrier}
\end{center}
\end{figure}

We thus assume that the potential energy of the system has the typical 
fission-like features shown in figure \ref{fig:barrier}. We then adopt a 
functional integral representation of the quantum-mechanical evolution 
operator $\hat{U}$, and following \cite{negele1982} write
\begin{eqnarray}
\sum_{n} \frac{1}{E-E_{n} + i\epsilon}
& =
\textrm{Tr} \left[ \frac{1}{\hat{H}-E+i\epsilon} \right] \\
& =
-i\int_{0}^{\infty} dT\; e^{iET}\int dq\; \langle q|e^{-i\hat{H}T}|q\rangle
\end{eqnarray}
with $q$ the one dimensional coordinate, and the summation in the left-hand 
side extends over all eigenstates of the Hamiltonian. The poles of the 
resolvent as a function of $E$ give the energy of bound states and 
resonances. In the classically-forbidden region, no bound states are 
possible and the resonances appear as complex energies with an imaginary 
part providing the width $\Gamma$ of the state. 

In order to estimate the integral over $q$, the integrand is first converted 
to a Feynman path integral
\begin{equation}
\int dq \langle q|e^{-i\hat{H}T}|q\rangle = 
\int dq \int \mathcal{D}[q(t)]e^{iS[q(t)]},
\end{equation}  
where $S[q(t)]$ is the classical action. This Feynman path integral is then 
computed in the static path approximation (SPA). The SPA picks out only 
the path $q_{0}(t)$ that is periodic both with respect to the coordinate 
$q$ and the momentum $p$, $q(0) = q(T)$ and $p(0) = p(T)$, and corresponds 
to a minimization of the classical action. The remaining integral over $T$ 
is computed using again the SPA, which provides the relation 
$E=-\partial S/\partial T$ connecting $E$ with the classical energy. The 
contribution to the resolvent is proportional (the proportionality factor 
depends on second order corrections to the SPA that we do not discuss here) 
to $e^{iW(E)}$ where $W(E)=ET_{cl} + S_{cl} (T_{cl})$ and $T_{cl}$ is the 
classical period of motion for an energy $E$. Summing over all integer 
multiples of the period at an energy $E$ gives the contribution 
$e^{iW(E)}/(1-e^{iW(E)})$ to the resolvent for the orbits in the 
classically-allowed region around the local minimum. The poles of this 
quantity are at those energies satisfying $W(E)=2\pi n$, which is the WKB 
quantization energy formula up to a factor $\pi$. As argued in \cite{levit1980-a}, better treatment of the 
omitted factor in front of the phase provides the missing $\pi$. In the 
classically-forbidden region, we can repeat the same kind of arguments by 
going into imaginary time, $t \rightarrow i\tau$. Still neglecting the 
quadratic corrections to the SPA, the total contribution of all paths to 
the resolvent is then given in \cite{levit1980} by
\begin{equation}
\textrm{Tr} \left[ \frac{1}{\hat{H}-E+i\epsilon} \right]
=
\frac{e^{iW_{1}(E)}-e^{-W_{2}(E)}}{1-e^{iW_{1}(E)}-e^{-W_{2}(E)}},
\end{equation}
where $W_{1}(E)$ is the action in the classical allowed region and $W_{2}(E)$
is its generalization in the classically-forbidden region,
\begin{equation}
\begin{array}{l}
W_{1}(E) = \displaystyle 2 \int_{q_{1}}^{q_{2}} dq \sqrt{2m(E-V(q))}, \medskip\\
W_{2}(E) = \displaystyle 2 \int_{q_{2}}^{q_{3}} dq \sqrt{2m(V(q)-E)}.
\end{array}
\end{equation}
The expression of the trace is slightly different if corrections are taken 
into account and is given in \cite{negele1982}. The poles of the resolvent 
are now the solutions of $ 1-e^{iW_{1}(E)}-e^{-W_{2}(E)}=0$ and therefore lie 
in the complex plane. Assuming that the last term is small, the complex 
energy solutions are given by $E_{n}+\frac{i}{2}\Gamma_{n}$ where $E_{n}$ 
is the solution of the WKB energy quantization condition. The width of 
the resonance is given by
\begin{equation}
\Gamma_{n} = \frac{\omega_{cl}(E_{n})}{\pi} e^{-W_{2}(E_{n})},
\end{equation}
which is a factor of 2 smaller than the WKB formula in the large $W_{2}$ 
limit. Again, this factor is recovered when including quadratic corrections, 
as argued in \cite{negele1982}. It can also be shown that the region to 
the right of the barrier does not contribute significantly to the resolvent.

These ideas can in principle be extended to a quantum many-body system
as discussed in \cite{levit1980-a,negele1982,skalski2008}. The peculiarity 
of this approach is that the SPA to the Feynman path integral corresponds 
to the mean field solution instead of the classical trajectories. Aside 
from that, the method requires solving many TDHFB-like equations in 
imaginary time with periodic boundary conditions to describe the dynamics 
below the barrier. A non trivial and still unresolved issue is how to 
connect periodic trajectories in classically-allowed regions to the 
solutions under the barrier. In one dimension there is only one way to do 
so but in a multidimensional case the connection is far from trivial. Early 
studies in \cite{negele1989} point to the important role of symmetry-breaking. 
The connection to the standard WKB formula is still missing although the 
results of \cite{skalski2008} seem to suggest that the WKB formula might 
grasp some of the physics involved.

Although not directly connected with functional integral methods, the tunnelling through a 
multidimensional barrier has also been studied in a two-dimensional model 
using a semi-classical approximation with complex classical trajectories 
\cite{RING197850}. Finally, we also mention the recent attempt in 
\cite{scamps2015} to compute tunnelling probabilities by considering a 
complex absorbing potential: although the method was tested on simple 
fission model Hamiltonians, it could in principle be extended to more 
microscopic collective Hamiltonians of the form (\ref{eq:evolution0}).


\subsection{Time-Dependent Methods and Induced Fission}
\label{subsec:induced}

As mentioned in the introduction, the observables of interest 
in induced fission are essentially the distribution of fission fragment 
properties such as their charge, mass, total kinetic energy or total 
excitation energy. Such distributions emerge naturally from a 
time-dependent description of fission: if one can simulate the time 
evolution of the system from an initial state defining the compound nucleus 
to a final state characterized by two fragments, then repeating the 
calculation for several initial conditions would allow one to construct 
all the distributions of fission fragment observables.

Let us recall that the time evolution of a (non-relativistic) many-body 
quantum system is given by the time-dependent Schr\"{o}dinger equation. It 
originates from the requirement that the variations of the quantum 
mechanical action defined by 
\begin{equation}
\mathcal{S}[\Psi]
 =
\int_{t_{0}}^{t_{1}} 
\langle \Psi(t) \,|\, \left[ \hat{H} - i\hbar \frac{\partial}{\partial t}\right] \Psi(t) \rangle,
\label{eq:schrodinger_t}
\end{equation}
be zero with respect to variations of the many-body wave function 
$|\delta\Psi(t)\rangle$. Solving (\ref{eq:schrodinger_t}) for a realistic 
nuclear Hamiltonian is of course a formidable task. Just as in the case 
of spontaneous fission half-lives, the calculation of fission fragment 
observables is performed by making additional approximations.


\subsubsection{Classical Dynamics}
\label{subsubsec:langevin}

The simplest and most drastic of all approximations is to forgo the quantum 
nature of the nucleus and treat fission dynamics in a semi-classical way. 
We do not intend to give a comprehensive description of the various 
stochastic methods used to describe nuclear dynamics, and refer the reader 
to the review \cite{abe1996} by Abe {\it et al.} where this topic is discussed 
in great details. Here, our goal is simply to recall how some of these 
techniques have been applied to describe induced fission, especially fission 
fragment distributions, particle evaporation and fission probabilities.

We introduce the conjugate momenta $\pVec$ of the collective variables 
$\qVec$ driving fission. At any time $t$, the nucleus is represented by a 
point in phase space with coordinates $(\qVec(t), \pVec(t))$ where  
the potential energy is $V(\gras{q})$. If the total energy of the nucleus is 
$E$, then the local excitation energy is $E^{*}(\gras{q}) = E - V(\gras{q})$ 
(note that for such a classical treatment of nuclear dynamics, the excitation 
energy must be positive $E^{*}(\gras{q})\geq 0$ for all $\qVec$).
The dynamics of the system can be represented in several ways:
\begin{itemize}
\item The Langevin equations directly give the position of the system in 
phase space at any time $t$. They read
\begin{eqnarray}
\dot{q}_{\alpha} & = \sum_{\beta} B_{\alpha\beta}p_{\beta}, \\
\dot{p}_{\alpha} & = 
- \frac{1}{2}\sum_{\beta\gamma} \frac{\partial B_{\beta\gamma}}{\partial q_{\alpha}}p_{\beta}p_{\gamma}
- \frac{\partial V}{\partial q_{\alpha}} 
- \sum_{\beta\gamma} \Gamma_{\alpha\beta} B_{\beta\gamma}p_{\gamma}
+ \sum_{\beta} \Theta_{\alpha\beta}\xi_{\beta}(t),
\end{eqnarray}
with $\tensor{B}(\qVec)\equiv B_{\alpha\beta}(\qVec)$ the tensor of inertia, 
$\tensor{\Gamma}(\qVec) \equiv \Gamma_{\alpha\beta}(\qVec)$ 
the coordinate-dependent friction tensor (not to be confused with the width in the GCM) 
and $V(\qVec)$ the potential 
energy in the collective space. The Langevin equations are non-deterministic 
owing to the presence of the random force $\gras{\xi}(t)$. The strength of 
this random force is controlled by the parameter $\tensor{\Theta} \equiv\Theta_{\alpha\beta}$. In 
applications of the Langevin equation to fission such as, e.g. in 
\cite{nadtochy2007,sadhukhan2010,randrup2011}, this parameter is usually 
related to the friction tensor through the fluctuation-dissipation theorem, 
$\sum_{k}\Theta_{ik}\Theta_{kj} = \Gamma_{ij}T$, with $T\equiv T(\gras{q})$ a 
local nuclear temperature related to the excitation of the system $E^{*}(\gras{q})$ 
at point $q$ . Furthermore, it is often assumed that the random variable is 
a Gaussian white noise process characterized by
\begin{eqnarray}
\langle \xi_{\alpha} \rangle = 0 ,\\
\langle \xi_{\alpha}(t)\xi_{\beta}(t')\rangle = 2\delta_{\alpha\beta}\delta(t-t'),
\end{eqnarray}
where in this equation, $\langle . \rangle$ refer to statistical 
averaging. This absence of memory for $\gras{\xi}$ implies that the 
Langevin equations represent a Markovian stochastic process, i.e., the 
value of the random force at time $t$ does not depend on previous values 
at time $t'<t$; see \cite{reichl1988} for an introduction. It can be 
thought of as a random walk on the PES defined by $V(\qVec)$. Each such 
random walk from some initial state to a properly defined scission 
configuration defines a fission event; repeating the procedure, e.g., 
by Monte-Carlo sampling, allows one to reconstruct the full distribution 
of fission fragments.
\item The Kramers equation gives the probability distribution function 
for the nucleus to be at a given point in phase space. It reads
\begin{eqnarray}
\dot{f}(\qVec,\pVec,t)
& = \sum_{\alpha\beta}
\left[
- B_{\alpha\beta}p_{\beta}\frac{\partial}{\partial q_{\alpha}}
+ \left(  \frac{\partial V}{\partial q_{\alpha}} + \sum_{\delta}\frac{1}{2}\frac{\partial B_{\beta\gamma}}{\partial q_{\alpha}} p_{\beta}p_{\delta}
\right)\frac{\partial}{\partial p_{\alpha}} \right.\nonumber\\
& \left.
+ \sum_{\gamma}\Gamma_{\alpha\beta}B_{\beta\gamma}\frac{\partial}{\partial p_{\alpha}}p_{\delta}
+  D_{\alpha\beta}\frac{\partial^{2}}{\partial p_{\alpha}\partial p_{\beta}}
\right] f(\qVec,\pVec,t),
\label{eq:kramers}
\end{eqnarray}
where $D_{\alpha\beta} = \Gamma_{\alpha\beta}T$ and $T$ is the temperature. Since the 
Kramers equation does not contain an explicit random term, it lends 
itself to analytic approximations. On the other hand, its generalization 
to $N$-dimensional collective spaces is numerically more involved since 
it is a second-order differential equation in terms of collective 
variables.
\end{itemize}

Historically, the Kramers equation was essentially used to extract 
an analytic expression for the induced fission width $\Gamma_{f}$ by 
considering fission as a diffusion process over a potential barrier 
approximated by an $N$-dimensional quadratic surface (of the type 
$\frac{1}{2}\sum_{ij} q_i q_j$ where $q_i$ are the collective variables), for instance in 
\cite{grange1983,jing-shang1983,weidenmueller1984}. More recently, 
progress in computing capabilities has enabled solving the full Langevin 
equations in several dimensions. Results reported by various groups 
differ mostly in the number and type of collective degrees of freedom, 
as well as the prescription for the friction tensor $\tensor{\Gamma}$. Most 
studies have been performed in the high temperature regime, where the 
potential energy surface is approximated by a liquid-drop-like formula, 
for example \cite{nadtochy2005,nadtochy2007,sadhukhan2011,nadtochy2012}. 
The Langevin equations have also been solved for macroscopic-microscopic 
PES in the limiting case of strong friction (strongly damped Brownian 
motion) in \cite{randrup2011,randrup2011-a,randrup2013}. Note that both 
the Langevin and Kramers equations involve the collective potential energy 
$V(\gras{q})$ and the inertia tensor $\tensor{B}(\gras{q})$. In the 
recent work of \cite{sadhukhan2015}, these quantities were computed using 
the ATDHFB formula, and Langevin dynamics was solved from the outer 
turning point to scission in order to extract spontaneous fission fragment 
distributions.


\subsubsection{Time-dependent Density Functional Theory}
\label{subsubsec:tddft}

Time-dependent density functional theory (TDDFT) provides a fully microscopic 
approach to describe real time fission dynamics. It is a reformulation of the many-body 
time-dependent Schr{\"o}dinger equation as discussed in \cite{fiolhais2003,
marques2004}. If one considers an interacting electron gas in a time-dependent 
external potential, then the Runge-Gross existence theorem of \cite{runge1984} 
asserts that given an initial state, all properties of the system can be 
expressed as a functional of the (time-dependent) local one-body density 
providing the potential satisfies certain regularity conditions. Just as in the 
static case, the Kohn-Sham scheme can in principle be applied so that the 
TDDFT equations turn into simple time-dependent Hartree-like equations. However, 
again as in the static case, the form of the exchange-correlation 
time-dependent potential $\hat{v}_{xc}(\gras{r},t)$ is not known. What is called 
the adiabatic approximation in TDDFT (not to be confused with the adiabatic 
approximation in fission theory) consists in assuming that $\hat{v}_{xc}(\gras{r},t)$ 
has the same functional dependence on the density as at $t=0$, 
\begin{equation}
\hat{v}_{xc}^{\mathrm{adiabatic}}(\gras{r},t) = \hat{v}_{xc}[\rho](\gras{r},t).
\end{equation}
As for the Hohenberg-Kohn theorem of static DFT, there is no direct analogue of 
the Runge-Gross theorem for self-bound nuclear systems characterized by symmetry-breaking 
intrinsic densities; see also \cite{messud2009-a} for a discussion of self-bound systems 
with symmetry-conserving internal densities. In spite of this, the popular time-dependent Hartree-Fock 
(TDHF) and time-dependent Hartree-Fock-Bogoliubov (TDHFB) are {\it de facto} 
adaptations of adiabatic TDDFT in nuclear physics, and we give below a very 
brief presentation of each of these methods.

The TDHF equation
\begin{equation}
i\hbar \dot{\rho} = [h, \rho]
\end{equation}
can be obtained by enforcing that the many-body wave-function remains a 
Slater determinant at all times \cite{ring2000}. Given an initial density at time $t_0$, which is 
typically obtained by solving the static HF equations under a set of 
constraints, solving the TDHF equation provides the full time-evolution 
of the system. If the initial condition is such that the system has enough 
excitation energy -- a point discussed in details in \cite{goddard2015,tanimura2015} -- 
this time evolution may follow the system past the scission point and lead to 
two separated, excited fission fragments. In some way, TDHF is the microscopic 
analogue of the Langevin equation in that it simulates a single fission event in real time. 
One of the earliest applications of TDHF in \cite{negele1978} was made by 
Negele and collaborators to study the fission of actinides. At the time, a 
number of approximations were needed such as axial and reflection symmetry, 
no spin-orbit potential, and a coarse spatial grid. Progress in computing 
have enabled more realistic simulations including the full Skyrme potential 
and 3D geometries as in \cite{simenel2014}. In all cases, the initial point 
must have a deformation larger than the ``dynamical fission threshold'' 
introduced in \cite{tanimura2015} in order for the system to fission. The 
existence of such a threshold was explained by Bulgac and collaborators 
in \cite{bulgac2016} as the consequence of neglecting pairing correlations. 
Because TDHF does not rely on the hypothesis of adiabaticity, it is expected 
to give a much more realistic description of the scission point, in 
particular of fission fragment properties. The initial results on fragment 
total kinetic and excitation energies reported in \cite{simenel2014,bulgac2016} are 
very promising.

As was emphasized already several times, pairing correlations play a 
crucial role in fission. While, in principle, the TDDFT equations could 
provide the exact time-evolution of the system with only a functional of 
the density $\rho$, our ignorance of the form of this functional forces us 
to introduce explicitly a Kohn-Sham scheme based on symmetry-breaking 
reference states and a (time-dependent) pairing tensor $\kappa$. This 
problem is identical to the static case. Nuclear dynamics is then be
described by the TDHFB equation, 
\begin{equation}
i\hbar \dot{\mathcal{R}} = [\mathcal{H}, \mathcal{R}].
\end{equation}
While formally analogous to the TDHF equations, the TDHFB equation is  
substantially more involved numerically. Indeed, the number of (partially) 
occupied orbitals at each time $t$ is much larger than the number of 
nucleons. In spherical symmetry, the TDHFB equation has recently been 
solved without specific approximations \cite{avez2008}. In deformed nuclei, 
it has been solved in the canonical basis in \cite{ebata2010,
nakatsukasa2012,nakatsukasa2012-a}, but no application of this formalism 
to fission has been performed yet. The first pioneer calculation of 
neutron-induced fission with full TDHFB, which the authors refer to as 
time-dependent local density approximation, has also been reported in 
\cite{bulgac2016}. We note that, as for the static problem, the TDHFB 
equation can be approximated by the simpler TDHF+BCS limit as done, for 
instance in \cite{scamps2012,scamps2015-a}. However, the TDHF+BCS approach does not 
respect the continuity equation, contrary to TDHFB, which may lead to 
non-physical results in specific cases such as particle emission. The 
authors of \cite{scamps2012} advocated using a simplified version of 
TDHF+BCS where occupation numbers are frozen to their initial value and 
do not change with time.


\subsubsection{Collective Schr{\"o}dinger Equations}
\label{subsubsec:tdgcm}

If the TDHF equations are the analogue of the classical Langevin equations, 
the time-dependent generator coordinate method (TDGCM) can be viewed as the 
microscopic translation of the Kramers equation (\ref{eq:kramers}). The TDGCM is a 
straightforward extension of the static GCM introduced in section 
\ref{subsubsec:gcm}, where the ansatz for the solution to the time-dependent 
many-body Schr{\"o}dinger equation takes the form 
\begin{equation}
|\Psi(t)\rangle = \int d\qVec\; f(\qVec, t)|\Psi(\qVec) \rangle.
\label{eq:gcm_t_ansatz}
\end{equation}
As with (\ref{eq:gcm_ansatz}), the functions $|\Psi(\qVec)\rangle$ are 
known many-body states parametrized by a vector of collective variables 
$\qVec$, which most often are chosen as the solutions to the static HFB 
equations under a set of constraints $\qVec$, see also 
section \ref{subsec:collective} for a discussion of collective variables.

Inserting the ansatz (\ref{eq:gcm_t_ansatz}) in the variational principle 
(\ref{eq:schrodinger_t}) provides the time-dependent analogue of (\ref{eqn:HW}), 
where the only difference is that the functions $f(\qVec,t)$ are now 
time-dependent. All applications of the TDGCM method have been made by 
further assuming the Gaussian overlap approximation for the norm kernel. 
Generalizing the procedure given in section \ref{subsubsec:gcm} in the case 
where the overlap kernel does not depend on the collective variable (constant 
metric), we find the time-dependent collective Schr\"odinger equation
\begin{equation} 
i \hbar \frac{\partial}{\partial t} g(\qVec,t)= \left[ \HCollExpr\right] g(\qVec,t),
\label{eq:evolution0} 
\end{equation} 
where the function $g(\qVec,t)$ is related to the weight function 
$f(\qVec, t)$ of (\ref{eq:gcm_t_ansatz}) according to
\begin{equation}
g(\qVec,t) = \int d\qVec'\; f(\qVec',t) [n(\qVec,\qVec')]^{1/2},
\end{equation}
(see (\ref{gcm:natural}) for the definition of the square root of the norm)
and contains all the information 
about the dynamics of the system. As before, the rank 2 tensor  
$\tensor{B}(\qVec) \equiv B_{\alpha\beta}(\qVec)$ is the collective inertia of the system in the 
collective space, and $V(\qVec)$ is the potential energy. 

Equation (\ref{eq:evolution0}) implies a continuity equation for the 
quantity $|g(\qVec,t)|^2$,
\begin{equation}
\frac{\partial}{\partial t} |g(\qVec, t)|^2 = -\nabla \cdot \JVec(\qVec, t).
\label{eq:continuityEquation}
\end{equation}
This equation can be derived in perfect analogy with standard one-body 
quantum mechanics, see for example the derivation in 
\cite{messiah1961}. It suggests that $|g(\qVec,t)|^2$ can be interpreted 
as a probability amplitude for the system to be characterized by the collective 
variables $\qVec$ at time $t$. Consequently, the vector $\JVec(\qVec, t)$ 
is a current of probability in perfect analogy with the results of one-body  
quantum mechanics (see, e.g., equation IV.9 in \cite{messiah1961}),
\begin{equation}
\JVec(\qVec,t)= \frac{\hbar}{2i} \tensor{B}(\qVec) 
\left[ g^*(\qVec,t) \nabla g(\qVec,t) - g(\qVec,t) \nabla g^*(\qVec,t) \right].
\end{equation}
When more than one collective variable are involved, the coordinates of 
the current of probability are
\begin{equation}
J_{\alpha}(\qVec,t)= 
\frac{\hbar}{2i} \sum_{\beta=1}^{N} B_{\alpha\beta}(\qVec) 
\left[ 
g^*(\qVec,t) \frac{\partial g}{\partial q_{\beta}}(\qVec,t) 
- 
g(\qVec,t) \frac{\partial g^{*}}{\partial q_{\beta}}(\qVec,t) 
\right].
\label{eq:current}
\end{equation}
Therefore, like 
the classical Kramers equations, the TDGCM equation give the evolution of 
the flow of probability in the collective space. 
The probability amplitude $|g(\qVec,t)|^2$ and the current (\ref{eq:current}) 
are the key quantities to extract fission fragment distributions in the 
TDGCM+GOA approach to nuclear fission. Based on the adequate identification 
of scission configurations, see discussion in section \ref{subsec:scission} page \pageref{subsec:scission}, 
it is possible to estimate the probability of a given scission configuration 
at point $\qVec$ by simply calculating the integrated flux of the probability 
current through the scission hyper-surface at that same point $\qVec$. If we 
define the integrated flux $F(\xi,t)$ through an oriented surface element 
$\xi$ as
\begin{equation}
F(\xi,t) = \int_{t=0}^{t} dT
       \int_{\qVec\in\xi} 
       \JVec(\qVec,t)\cdot d\SVec.
\label{eq:fluxDef}
\end{equation}
then, following \cite{berger1991,goutte2005,younes2012-a}, the fission fragment mass yield 
for mass $A$ is
\begin{equation}
y(A) \propto \sum_{\xi\in\mathcal{A}} \lim_{t\rightarrow +\infty} F(\xi, t),
\label{eq:yield}
\end{equation}
where $\mathcal{A}$ is the set of all oriented hyper-surface elements $\xi$ belonging 
to the scission hyper-surface such that one of the fragments has mass $A$. In practice, 
fission fragment mass yields are normalized,
\begin{equation}
Y(A) = \frac{y(a)}{\sum_{a} y(a)}
\label{eq:yield_normalized}
\end{equation}

%
%

\newpage
\section{Numerical Methods}
\label{sec:numerics}

One of the reasons behind the resurgence of fission studies in a microscopic 
framework is the availability of high-performance computing facilities 
throughout the world. Computational aspects are very often overlooked in the 
discussion of fission theory. Yet, it is essential to bear in mind that all 
theories of fission share an inextinguishable thirst in computing power that 
even the largest supercomputers can barely quench. In fact, the microscopic 
theory of fission has been identified the two recent reports \cite{young2009,
bishop2009} by US agencies as a science justification for the construction of 
exascale computers.

Indeed, while the backbone of the microscopic theory of fission was for the 
most part formalized already at the beginning of the 1980ies, practical 
applications were very limited. As a simple example, consider the 
aforementioned pioneering work of Negele and collaborators in 1978 on the 
dynamics of induced fission with the time-dependent Hartree-Fock theory: 
calculations were performed in axial symmetry, neglecting the spin-orbit 
component of the Skyrme force and the exchange Coulomb force, using a 
constant gap approximation for pairing, a discretization of space with a 
mesh size of $h = 0.65 $ fm, and using 3-point finite differences for 
derivatives yielding an error of at least 2 MeV on the energy. Today, all 
of these approximations can be removed, but significant work on algorithms, 
code development and parallelization techniques is constantly needed.

The goal of this section is to give a comprehensive review of the various 
numerical methods needed to implement the microscopic theory of fission. 
In section \ref{subsec:solvers}, we review the technology of DFT solvers, 
which are essential tools to map out the potential energy surface of the 
nucleus in the adiabatic approximation. In particular, we offer a critical 
discussion of the advantages and drawbacks of basis expansion methods and 
lattice techniques. In section \ref{subsec:dynamics}, we present some of 
the methods and challenges related to the description of fission dynamics, 
in particular the implementation of time-dependent DFT solvers.


\subsection{DFT Solvers}
\label{subsec:solvers}

In the adiabatic approximation of nuclear fission, DFT solvers are used to 
compute the potential energy surface of the nucleus of interest within a 
given collective space or to provide the wave function for an initial state 
in time-dependent calculations. In practice, this requires solving the HFB 
equations for a set of constraints $\gras{q}$. As mentioned in section 
\ref{subsec:collective}, these collective variables may correspond to 
geometrical properties of the nucleus, such as the expectation value of 
multipole moments, or non-geometrical quantities such as the fluctuation of 
particle number. The success of the microscopic approach to fission as 
outlined in sections \ref{sec:pes}-\ref{sec:dynamics} depends to some extent 
on the ability of the chosen collective space to accurately capture the 
physics of fission. This implies that (i) there are enough collective 
variables (ii) the ``spatial'' resolution of the collective space is good 
enough, (iii) the numerical precision is good enough.

Today, there is a relative consensus that at least a handful of different 
collective variables are needed, see for instance the discussions in 
\cite{moeller2001,younes2009}. In particular, the elongation of the nucleus, 
the degree of mass asymmetry, triaxiality, the thickness and density of 
particles in the neck between the two pre-fragments are among the most 
fundamental quantities. If we assume for the sake of argument that the 
collective space is described by a $N=5$ dimensional hyper-cube and a 
grid of $n=100$ points per dimension, one finds that $10^{10}$ deformed 
HFB calculations must be performed to fully scan the potential energy 
landscape. To put this number in perspective, we recall that high-precision 
HFB solutions for triaxial, reflection-asymmetric shapes take up to a 
few hours on standard architectures. The computational challenge is thus 
formidable. 

In this section, we present the techniques used to numerically solve the 
HFB equation. While many of these techniques are well-known, we emphasize 
their advantages and drawbacks in the specific context of fission studies. 
We can distinguish between two main classes of DFT solvers: those based on 
the expansion of HFB solutions on a basis of the single particle Hilbert 
space, and those based on direct numerical integration.


\subsubsection{Basis Expansion Techniques}
\label{subsubsec:basis}

Basis expansion techniques are ubiquitous in practical applications of 
quantum mechanics and quantum many-body theory. Expanding the solutions to 
the Schr{\"o}dinger or Dirac equation on a basis of known functions yields a 
linear eigenvalue problem that can be solved very efficiently. In particular, 
the method is oblivious to the local or non-local character of the underlying 
nuclear potential. Moreover, the formulation of beyond mean-field extensions 
such as, e.g., the generator coordinate method or projection techniques is 
straightforward. 

In nuclear science, the eigenfunctions of the one-centre harmonic oscillator 
(HO) have historically played a special role. They are known analytically in 
spherical, cylindrical, Cartesian coordinates, among others. Talmi, Moshinski 
and Talman showed long ago in \cite{talmi1952,moshinsky1959,talman1970} that 
any product of two HO basis functions could be expanded into a sum of single 
HO functions, which allows for the exact separation of the centre of mass 
and relative motion for two-body potentials. Using the harmonic oscillator 
basis also greatly simplifies the calculation of matrix elements of Gaussian 
potentials, such as, e.g., the Gogny force as highlighted in \cite{gogny1975,
parrish2013}. Special care must be taken in the evaluation of matrix elements 
of states with large quantum numbers, e.g. by recurring to known properties 
of hypergeometric sums as in \cite{egido1997}; see also \cite{younes2009-b} 
for a recent account on the calculation of matrix elements of the Gogny force 
with axially symmetric harmonic oscillator wave functions. High-accuracy 
expansions of the Coulomb or Yukawa potentials onto a finite sum of Gaussians 
were also used in \cite{girod1983,quentin1972,dobaczewski2009-a} for the 
precise calculation of the Coulomb exchange contribution to the nuclear mean 
field. Last but not least, the nuclear mean field can be well approximated by 
a HO, which is one of the reasons for the spectacular success of the 
phenomenological Nilsson model of nuclear structure \cite{nilsson1995}. 

Several DFT solvers based on expanding the HFB solutions on the HO basis 
have been used in fission studies. Let us mention in particular the codes 
{\hfbtho} (\cite{stoitsov2005,stoitsov2013}) and {\hfodd} (\cite{dobaczewski2009-a,
dobaczewski1997,dobaczewski1997-a,dobaczewski2000,dobaczewski2004,dobaczewski2005,
schunck2012}), which have been released under open source 
license. Both codes implement generalized Skyrme-like functionals in the 
particl-hole channel and density-dependent delta-interaction in the particle-particle channel. 
The latest version of {\hfodd} also implements EDF based on finite-range 
pseudopotentials such as the Gogny force in both channels. 
Axial and time-reversal symmetries are built-in in {\hfbtho}, but 
reflection-asymmetric shapes are possible; by contrast, {\hfodd} breaks 
all possible symmetries of the nuclear mean-field. The two codes have been 
carefully benchmarked against one another, and the {\hfbtho} kernel is 
included as a module of {\hfodd}. These codes were used to study both 
spontaneous and induced fission; see the following papers  
\cite{staszczak2005,pei2009,sheikh2009,staszczak2013,schunck2015,schunck2014,
staszczak2009,baran2011,sadhukhan2013,mcdonnell2013,schunck2013,schunck2013-a}. 
In applications with the Gogny forces, the codes {\hfbaxial} and {\hfbtri} 
have been widely used to study spontaneous fission in actinides and superheavy 
elements \cite{warda2002,giuliani2013,martin2009,warda2011,rodriguez-guzman2014}.

Let us recall that the three-dimensional quantum HO is characterized by its 
frequency vector $\gras{\omega} = (\omega_x, \omega_y, \omega_z)$ (in 
Cartesian coordinates). The frequency, measured in MeV, is also related to 
the oscillator length $b_{\mu} = \sqrt{\hbar/m\omega_{\mu}}$ (in fm), where 
$m$ is the mass of a nucleon. While the single particle Hilbert space is 
of course infinite, practical implementations require truncating the HO 
basis. This is achieved in different ways. For a spherical basis with 
$\omega_x = \omega_y = \omega_z \equiv \omega_0$, one usually imposes a 
cut-off in the number $N_{\mathrm{shell}}$ of oscillator shells. Each $N$-shell 
contains $(N+1)(N+2)$ degenerate states \cite{bohr1975,nilsson1995}. 
Deformed or ``stretched'' HO bases are introduced to describe elongated 
geometries. In these cases, $\omega_x \neq \omega_y \neq \omega_z $, but 
the condition of volume conservation yields 
$\omega_x\omega_y\omega_z = \omega_0^3$ and accordingly $b_0^3 = b_x b_y b_z$. Since there is no degeneracy of the 
HO shells any more, one must introduce additional criteria to truncate the 
basis. Among the popular choices are the ratios $p = \omega_x/\omega_y$ and 
$q = \omega_x/\omega_z$, which define the deformation of the basis. 
Alternatively, this basis deformation can also be defined by 
introducing an ellipsoidal liquid drop characterized by the $(\beta,\gamma)$ 
Bohr deformations; see \cite{ring2000} for the relation between $(\beta,\gamma)$ 
and the $(\alpha_{20},\alpha_{22})$ parameters of the expansion (\ref{eq:multipole_exp}).
The quantities $p$ and $q$ can then be expressed as ratios of the radii of each 
of the principal axes of this ellipsoid, $p = R_y/R_x$ and $q = R_z/R_x$, each 
radius being a function of $(\beta,\gamma)$. This method was introduced in \cite{girod1983} 
and generalized in \cite{dobaczewski1997}.
An additional number of states $n_{\mathrm{max}}$ is sufficient to completely 
determine the basis states.

\begin{figure}[ht]
\begin{center}
\includegraphics[width=0.75\linewidth]{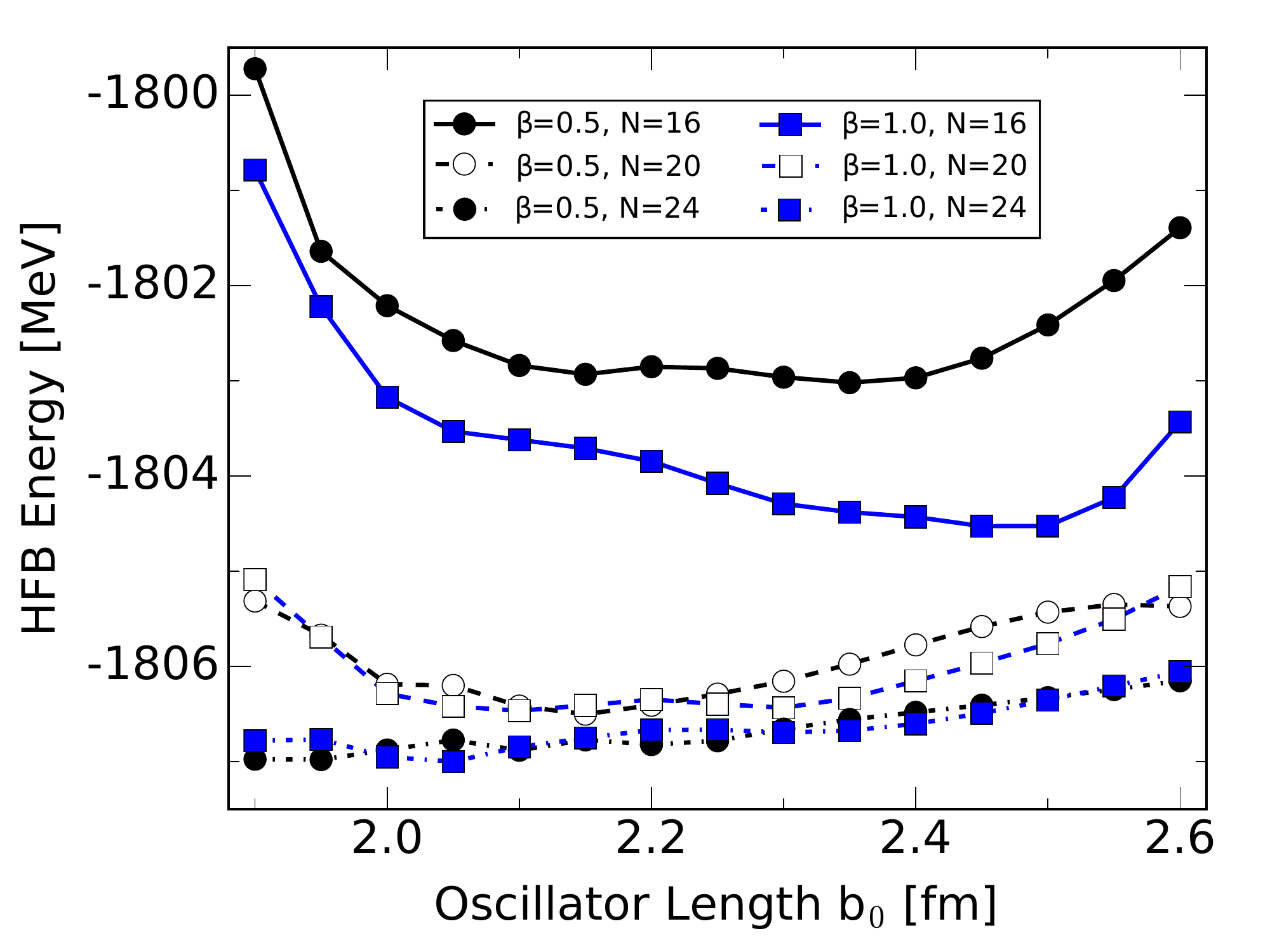}
\caption{Convergence of DFT calculations using HO expansions. The figure is 
obtained at the HFB approximation with the SkM* functional and a surface-volume 
pairing and shows the total energy of $^{240}$Pu as function of the oscillator 
length $b_0$ (in fm) for different number of oscillator shells $N_{\mathrm{shell}}$ 
for the configuration defined by $\langle \hat{Q}_{20}\rangle = 200$ b and
$\langle \hat{Q}_{40}\rangle = 50$ b$^{2}$. Stretched HO bases with different 
deformations $\beta = 0.5$ and $\beta = 1.0$ are used; adapted from 
\cite{schunck2015-a}.}
\label{fig:HO_1}
\end{center}
\end{figure}

As important as the truncation schemes are the choices for the oscillator
lengths used in the HO basis. Typically, these quantities are used
as  additional variational parameters that are adjusted to minimize
the energy. This search for optimal oscillator lengths is obviously less
important the bigger the size of the basis is, as illustrated in figure \ref{fig:HO_1}. Therefore, there is always
a compromise between using a large basis where the precise value of
the oscillator lengths is less relevant but calculations are expensive, or 
using a smaller basis at the cost of repeating, for the configuration of 
interest, the HFB calculation several times for different oscillator 
lengths. In this respect various phenomenological formulas relating the 
oscillator lengths to the imposed deformation parameters of the nucleus 
can be used as in \cite{schunck2014}. 
 
Because of the truncation of the basis, the solution of the HFB equation 
becomes dependent on the characteristics of the basis, namely $N_{\mathrm{shell}}$, 
$\omega_0$, and $q$ in the most common case of an axially-deformed basis. 
This dependence is clearly spurious and disappears in the limit of an 
infinite basis. In practical calculations, however, its effect must be 
properly quantified. The figure \ref{fig:HO_1} illustrates the expected size 
of truncation effects as a function of the oscillator frequencies and 
maximum spherical shell number. We show the energy of a deformed 
configuration along the fission path of $^{240}$Pu characterized by 
$\langle \hat{Q}_{20}\rangle = 200$ b and $\langle \hat{Q}_{40}\rangle = 50$ 
b$^{2}$. Even at $N_{\mathrm{shell}}=24$, the energy varies by several hundreds 
of keV over the range $b_0\in[1.9,2.6]$ fm.. 

\begin{figure}[ht]
\begin{center}
\includegraphics[width=0.75\linewidth]{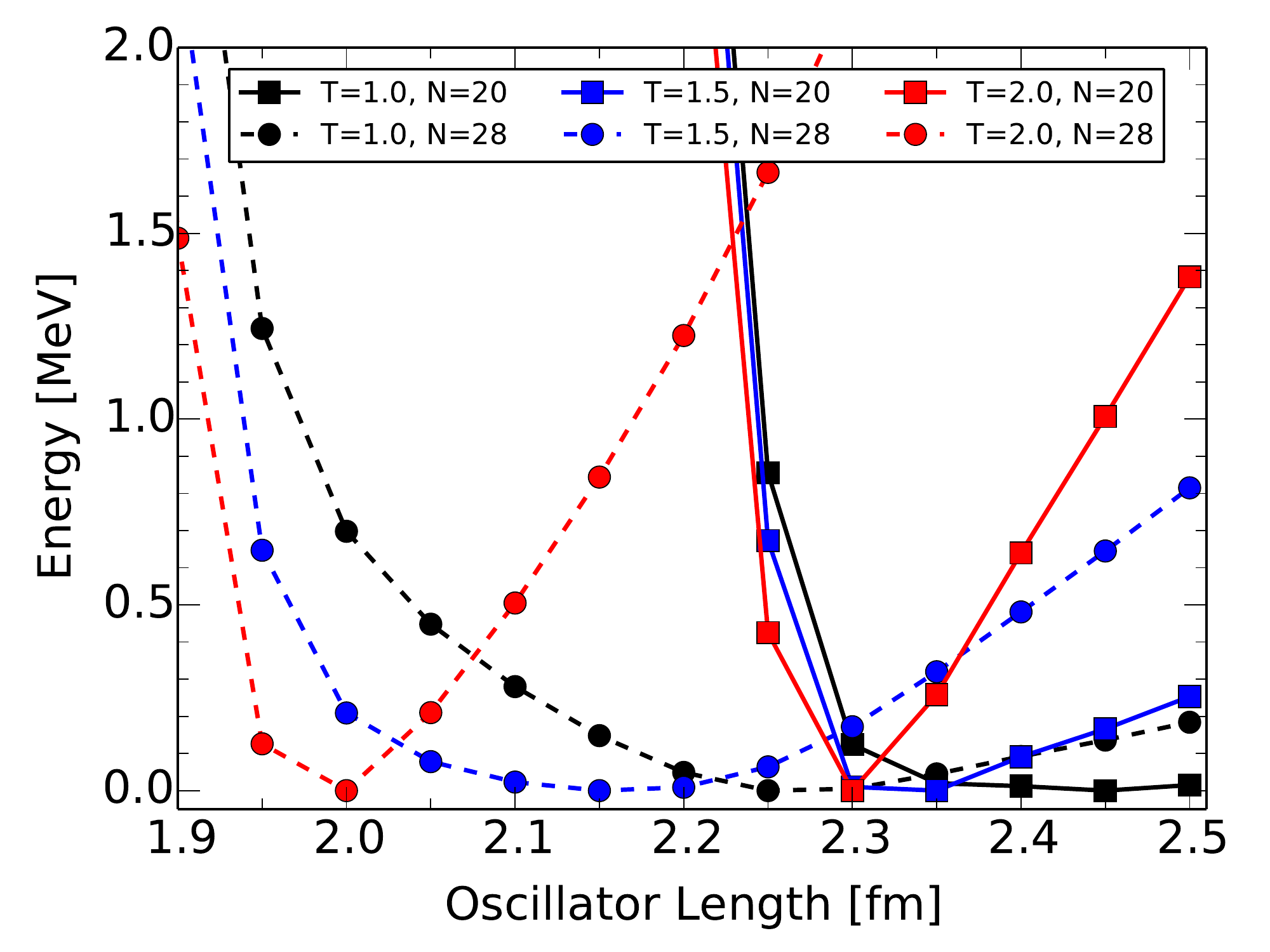}
\caption{Similar as figure \ref{fig:HO_1} for different nuclear temperatures. 
The basis deformation is fixed at $\beta = 1.0$. For better legibility, all 
curves have been normalized to their minimum value.}
\label{fig:HO_2}
\end{center}
\end{figure}

The work reported in \cite{warda2002,schunck2013-a,samyn2005,nikolov2011} 
showed the impact of basis truncation on fission properties, mostly on the 
static properties of the potential energy surface such as, e.g. the height 
of fission barriers . One should bear in mind that  
truncation errors typically amount to a few hundred keV at the top of the 
first fission barrier in actinides. Such errors can cause several orders of 
magnitude changes in spontaneous fission half-lives because of the 
exponential factor in (\ref{eq:tSF}). In addition, the truncation error 
increases with the mass of the nucleus, especially when pairing correlations 
are non-zero. Because of the Gaussian asymptotic behaviour of HO wave functions, 
the convergence of basis expansions in weakly bound nuclei near the dripline 
is also problematic, see possible alternatives for DFT solvers in, e.g., 
\cite{stoitsov1998,schunck2008-a}. This could have a major impact in fission 
fragment calculations of nuclei involved in the r-process discussed in 
\cite{grawe2007,martinez-pinedo2007}. Finally, we already mentioned that 
both spontaneous fission at high excitation energy and neutron-induced 
fission are often described in the finite-temperature HFB theory, where the 
density matrix contains a spatially non-localized component. This increases 
truncation errors accordingly. In figure \ref{fig:HO_2}, we show the evolution 
of the energy as a function of the oscillator length for FT-HFB calculations 
at $T=1.0, 1.5, 2.0$ MeV for large bases characterized by shell numbers 
$N_{\mathrm{shell}}=20$ and $N_{\mathrm{shell}}=28$. We note that the plateau 
condition for convergence degrades substantially as nuclear temperature 
increases: at $T=1.0$ MeV, the energy does not change by more than 50 keV 
over the range $b_0\in[2.2, 2.4]$ fm, while at $T=2.0$ MeV, even a small 
change of $\delta b = 0.05$ fm around the minimum induces variations of 
energy of 200 keV.

In the vast majority of DFT solvers based on the HO basis expansion, the 
basis functions are the eigenstates of the one-centre, three-dimensional 
quantum HO. In the nineteen eighties, the French group at CEA 
Bruy{\`e}res-le-Ch{\^a}tel developed an axial two-centre DFT solver, in 
which the basis functions are superpositions of the eigenfunctions of two 
one-centre HO shifted by a distance $d$ (adjustable). By construction, the 
set of all such functions is not orthogonal, which requires a Gram-Schmit 
orthogonalization procedure. This represents a disadvantage as the number 
of linearly independent states depends upon the distance $d$ and therefore 
wave functions at different elongations are expanded in different 
sub-spaces of the full Hilbert space. As a consequence, the evolution of 
observables in the transition from one subspace to the neighbouring ones 
is not necessarily smooth. On the other hand, this technology is especially 
advantageous to describe very deformed nuclear shapes and/or two fragments 
such as the ones encountered near scission. This code has been used to 
study induced fission in actinide nuclei, see \cite{berger1984,berger1986,
berger1989,dubray2008,goutte2004,goutte2005} for instance.


\subsubsection{Mesh Discretization and Lattice Techniques}
\label{subsubsec:mesh}

The coordinate space formulation of the HFB equation provides one of the 
simplest methods to remedy the limitations of basis expansion techniques. 
In this case, the HFB equation (\ref{eq:pes_HFBdiag}) becomes
\begin{eqnarray}
\int d^3\gras{r}' \sum_{\sigma'}\left(
\begin{array}{cc}
h(\gras{r}\sigma,\gras{r}\sigma') - \lambda\delta_{\sigma\sigma'} & \Delta(\gras{r}\sigma,\gras{r}\sigma') \medskip \\
-\Delta^{*}(\gras{r}\sigma,\gras{r}\sigma') &
-h^{*}(\gras{r}\sigma,\gras{r}\sigma') + \lambda\delta_{\sigma\sigma'}
\end{array}
\right)
\left( \begin{array}{c}
U_{\mu}(\gras{r}\sigma') \medskip\\
V_{\mu}(\gras{r}\sigma')
\end{array}
\right) \nonumber\medskip\\
=
E_{\mu}
\left( \begin{array}{c}
U_{\mu}(\gras{r}\sigma) \medskip\\
V_{\mu}(\gras{r}\sigma)
\end{array}
\right),
\label{eq:HFB_rspace}
\end{eqnarray}
with the mean field $h(\gras{r}\sigma,\gras{r}'\sigma')$ and pairing 
field $\Delta(\gras{r}\sigma,\gras{r}'\sigma')$ expressed in coordinate 
space, see for instance \cite{bender2003} for transformation rules between 
configuration and coordinate space. 
While the separation of the Schr{\"o}dinger equation makes this direct 
approach very straightforward if spherical symmetry is conserved (see 
applications in \cite{dobaczewski1984,dobaczewski1996-a}), extensions to 
deformed nuclei require introducing lattice techniques, as finite 
differences becomes either numerically too inaccurate or computationally 
impractical. There are several different examples of coordinate-space 
approaches to solving the HFB equation that have been applied to fission 
studies.

The HFB equation (\ref{eq:HFB_rspace}) is a particular example of an integro-differential equation. 
The B-spline collocation method (BSCM) was introduced 
in nuclear theory already more than two decades ago in \cite{umar1991,
kegleyjr.1996} to solve such equations. In practice, the BSCM has been 
successfully implemented in cylindrical coordinates for the particular case 
of axially-deformed nuclei. After discretizing the spatial  domain as a set 
of knots, one introduces a set of interpolating B-spline 
functions with order $M$ in each knot. Any arbitrary function or operator 
is then represented at the collocation points defined by the maximum of the 
splines functions at each knot. With this technique, the HFB equation takes 
the form of the standard non-linear eigenvalue problem, which can be solved 
iteratively by successive diagonalizations. This approach was implemented 
in \cite{teran2003,blazkiewicz2005,pei2008}. Vanishing boundary conditions 
are assumed at the boundaries of the domain (a cylinder of radius $R$ and 
length $2L$). 

Such lattice-based techniques achieve a very high numerical precision 
regardless of the underlying nuclear geometry. Their convergence is 
essentially characterized by the order $M$ of the B-spline and the 
maximum value $\Omega_{\mathrm{max}}$ of the $z$-projection of angular 
momentum [For Hartree-Fock calculations, $\Omega_{\mathrm{max}}$ is simply 
equal to the maximum $j$-value of occupied single particle orbitals]. They 
are especially suited to dealing with very deformed nuclei, weakly-bound 
systems or nuclei at high temperature. Comparisons with HO basis expansion 
techniques given in \cite{pei2008,nikolov2011,schunck2015-a,schunck2015-b} show that a 
prohibitively large number of basis states would be needed to reach similar 
precision. The downside of the BSCM techniques is the larger computational 
cost: explicit parallelization across a few dozen cores is needed to keep 
run time within a few hours. In addition, extending the codes to handle 
either finite-range nuclear potentials or fully triaxial geometries would 
be costly, and such extensions do not exist yet. 

A slightly different lattice technique relies on using a variational 
method with a set of Lagrange functions introduced in \cite{baye1986}. As 
in the BSCM, both functions and operators are represented on the resulting 
mesh, the functions by a vector and the operators by a matrix. This 
technique is implemented in three-dimensional Cartesian meshes in the code 
{\ev} of \cite{bonche2005,ryssens2015}. The use of Lagrange meshes to 
estimate derivatives delivers high and controllable numerical precision 
as demonstrated in \cite{ryssens2015-a}. In particular, it is possible to 
limit numerical errors to only a few dozens keV across an arbitrarily large 
deformation span with even relatively coarse meshes. Note that the 
implementation of finite-range potentials represents a significant 
increase in computations, see however \cite{hashimoto2013} for a recent 
example using the finite range Gogny force. 

Another popular technique used in DFT solvers based on the coordinate 
space representation consists in evaluating derivatives, which are needed 
to define the Laplacian in the kinetic energy, in momentum space. Consider 
a Cartesian grid of $N_x\times N_y\times N_z$ points defined in such a way 
that  $x_i = i dx$ for $i = 1, \dots N_x$ (and similarly for the coordinates 
$y$ and $z$). Then the momentum is discretized according to 
$p_i/\hbar = i(2\pi/L)$. As is well known, derivatives in momentum space 
are simple multiplications. The transformation back to coordinate space 
can be performed by Fast Fourier Transforms. This technique has been used 
in \cite{rutz1995,bender1998,bender2000,buervenich2004,bender2003,
bulgac2013,maruhn2014,bulgac2016}. 

We finish this section by mentioning two slightly alternative 
lattice-based techniques. Finite element analysis was introduced in the  
series of papers \cite{poeschl1996,poeschl1997,poeschl1997-a,poeschl1998} 
to solve the equations of the relativistic mean-field. Although this 
technique seemed promising, it was not disseminated further. Most recently, 
the NUCLEI SciDAC collaboration has published in \cite{pei2014} a new DFT 
solver based on multi-resolution analysis and wavelet expansions. The 
advantage of multi-resolution is the possibility to impose the numerical 
precision desired.  


\subsubsection{Algorithms to Solve Self-Consistent Equations}
\label{subsubsec:algorithms}

The HFB equation (\ref{eq:pes_hfb}) is a non-linear problem since the HFB 
matrix $\mathcal{H}$ depends on the generalized density $\mathcal{R}$, 
recall (\ref{eq:pes_BigH'}) and (\ref{eq:pes_fields}). Even in the BCS 
approximation, the non-linearity of the equation remains since the HF mean 
field $h$ is still a functional of the one-body density $\rho$. There are 
three main strategies to solve such problems: successive diagonalizations, 
gradient methods of imaginary time evolution. In each case, the algorithm 
must be initialized. This can be done by solving, e.g., the Schr\"odinger 
equation with a Nilsson potential followed by the BCS approximation: this
provides an initial one-body density matrix $\rho^{(0)}$ and pairing tensor 
$\kappa^{(0)}$, which define the HF or HFB matrix at the first iteration. 
Given this initialization, the three algorithms mentioned proceed as 
follows:
\begin{itemize}
\item {\bf Successive iterations}: In this method the HFB equation is 
written in the form of a diagonalization problem
\begin{equation}
   \mathcal{H} W = W\mathcal{E}
\end{equation} 
with $W$ having the structure of (\ref{eq:pes_W}) and $\mathcal{E}$ of 
(\ref{eq:pes_qp}). The $W^{(i)}$ matrix at iteration 
$i$ is used to compute the next iteration of the HFB matrix, 
$\mathcal{H}^{(i+1)}$, which is diagonalized to obtain $W^{(i+1)}$. 
The iterative process is repeated until convergence, i.e. , 
$||W^{(i)}-W^{(i+1)}|| \leq \varepsilon$ within a 
given accuracy $\epsilon$. The convergence of the method is not guaranteed 
and ``jumping'' solutions such that $W^{(i)}=W^{(i+q)}$
can occur. These deficiencies are usually overcome by annealing the density 
at iteration $i+1$ with the density at iteration $i$
$$
\mathcal{R}^{(i+1)} \rightarrow (1-\alpha)\mathcal{R}^{(i+1)}+\alpha\mathcal{R}^{(i)}
$$
with the annealing parameter $\alpha$. More advanced linear mixing schemes 
have been introduced in quantum chemistry and ported to nuclear structure 
in \cite{baran2008}. Constraints are handled very efficiently by using 
methods based on the augmented Lagrangian method or approximate 
linear response theory, see \cite{schunck2014,staszczak2009,younes2009}. 
In the special case of the HF+BCS approximation to the HFB equation, the 
same overall algorithm is applied to diagonalize the HF mean field $h$ instead 
of the HFB matrix $\mathcal{H}$.

\item {\bf Gradient methods}: The HFB equation is a direct consequence of 
the variational principle on the HFB energy. Therefore solving the HFB 
equation is equivalent to finding a minimum of the HFB energy. The gradient 
method was used as early as the late nineteen seventies in \cite{mang1976,
egido1980} to solve the HFB equations using a phenomenological approach 
to determine the gradient step and is based on the representation of the 
HFB equation in terms of the Thouless matrix $Z$, see page 
\pageref{par_thouless}. The efficient handling of constraints 
inherent to the method is especially convenient in fission calculations: 
during the iterative process one has to ensure that the descendant 
direction is orthogonal to the gradient of the constraint condition. The 
orthogonality is imposed by modifying the objective function by subtracting 
the mean value of the constraint operator multiplied by a Lagrange 
multiplier that is subsequently used to impose the orthogonality condition 
at every iteration. The procedure is straightforwardly extended to many 
constraints. The iteration count of the method is high although the cost 
of evaluating the gradient is just slightly higher than evaluating the 
$\mathcal{H}$ matrix. To reduce the number of iterations, the conjugate 
gradient method introduced in \cite{egido1995} chooses the steepest descent 
direction as the ``conjugate'' of the previous one. The method is very 
efficient for a quadratic function but the cost of each iteration increases 
because of the need to do a line minimization at each step. The Newton or 
second order method relies on the use of the Hessian matrix to modify the 
gradient direction in such a way that the minimum is reached in just one 
iteration for a quadratic function. The iteration count is severely reduced 
but the cost of evaluating the Hessian is very high (in the HFB case, the 
Hessian resembles the matrix of linear response theory), as is the evaluation
of its inverse. For not so many degrees of freedom the method is competitive 
as shown in \cite{Schmid1984205} but becomes prohibitively expensive in 
typical applications in nuclear physics with effective forces. An enormous 
simplification of the method is achieved if the Hessian matrix is assumed 
to be diagonal dominant as the most important contribution to the diagonal 
matrix elements is the sum of two quasiparticle energies. In this way the 
computation and inversion of the Hessian matrix is enormously simplified. 
This method was implemented in the {\hfbaxial} and {\hfbtri} DFT solvers for 
the Gogny force. For an early account of the different gradient method techniques applied
to the solution of the HF and HFB problems with an emphasis on the
election of the gradient step, see \cite{Reinhard1982418}

\item {\bf Imaginary time method}: In the particular case of the HF+BCS 
approximation, one can also use the imaginary time method. It is based 
on the general result that the eigenvalues of a Hamiltonian $\hat{H}$ 
evolve with time through an oscillatory phase factor $\exp(-i/\hbar E_n t)$ 
depending on the eigenvalue $E_n$. If time is replaced by a complex 
quantity $t\rightarrow -i\tau$ the complex phase in front of each 
eigenvalue becomes real and a decaying function with a decay rate that 
increases with excitation energy. Therefore, applying the time evolution 
operator with imaginary time to a general linear combination of eigenvalues 
of the Hamiltonian will converge to the lowest eigenvalue (ground-state 
energy) after a sufficiently long time. The problem with this method is 
that the time evolution operator involves the exponential of the true 
Hamiltonian, typically approximated by the two-body effective Hamiltonian 
(\ref{eq:pes_EHFB_Veff}), which is very expensive to compute. In practice, 
the full Hamiltonian $\hat{H}$ is therefore replaced by a simpler 
approximation: In nuclear physics, this method  was introduced in 
\cite{davies1980} by replacing $\hat{H}$ by the HF Hamiltonian $\hat{h}$. 
This is the technique implemented in the {\ev} suite of code of 
\cite{bonche2005,ryssens2015} to solve the HF+BCS equation. The direct 
extension of this algorithm to the case of the HFB equation is not trivial
because the HFB matrix is unbounded from below: the lowest eigenvalue is 
infinite resulting in the divergence of the algorithm at 
$\tau \rightarrow+\infty$. In \cite{gall1994}, the authors introduced the 
two-basis method, where the HFB matrix is diagonalized in a small 
discrete basis composed of the eigenstates of the HF Hamiltonian obtained 
from the imaginary time evolution.
\end{itemize}


\subsection{Dynamics}
\label{subsec:dynamics}

The various time-dependent methods used to describe fission dynamics, in 
particular induced fission, were discussed in section \ref{subsec:induced}. 
These methods present specific numerical challenges that we briefly address 
below in section \ref{subsubsec:tddft_num}. In addition, the calculation of 
the collective inertia tensor which has a critical impact on fission 
half-lives, has also been subject to several approximations, which are 
discussed in section \ref{subsubsec:inertia_num}.


\subsubsection{Time-dependent Approaches}
\label{subsubsec:tddft_num}

As emphasized in section \ref{subsec:induced}, time-dependent density 
functional theory (TDDFT) provides, at least on paper, a convenient 
framework to simulate fission in real time, since it does not require 
introducing collective variables or scission configurations. Assuming the 
original Runge-Gross theorem can be extended to the nuclear case, the (unknown) energy functional that gives the exact 
ground-state energy would only depend on the local, time-dependent, one-body 
density, and the Kohn-Sham scheme would reduce to solving Skyrme-like TDHF 
equations. 

All existing implementations of TDHF in computer codes are based on the 
coordinate-space formulation of the HF equations in a box using both 
absorption layers on the edges of the box and vanishing boundary conditions. 
The coordinate space approach is necessary because TDHF solutions at 
large time can have very extended geometries that cannot be accurately 
represented by a one-centre basis expansion such as the familiar HO basis. 
Because of the computational cost of the coordinate space approach, the 
first implementations of TDHF used several simplifications such as axial 
symmetry as in \cite{negele1978,berghammer1995}. The first application 
of a fully three-dimensional, Skyrme TDHF calculation was published in  
\cite{kim1997} in 1997 and was based on the adaptation of the {\ev}8 HF+BCS 
solver. Very recently, a full 3D implementation 
of the TDHF equations in coordinate space using the Fourier representation 
of spatial derivatives has been published \cite{maruhn2014}.

The simplest way to include pairing correlations in TDHF is the BCS 
approximation. Such a TDHF+BCS solver was developed in \cite{scamps2012} and 
applied to the study of fission of $^{258}$Fm in \cite{scamps2015-a}. The 
solver is also an extension of the {\ev} HF+BCS computer program and relies on 
the same underlying technology, in particularly the use of a 3D Cartesian 
mesh and a set of Lagrange functions, see the two previous sections 
\ref{subsubsec:mesh} and \ref{subsubsec:algorithms}. 

The full TDHFB equation is substantially more involved than the TDHF or 
TDHF+BCS equation because of the non-zero occupation of high-lying, 
de-localized quasiparticle states. The full TDHFB equation in nuclei was 
originally solved in spherical symmetry in \cite{avez2008}. For 
arbitrary deformations of the nuclear shape, the TDHFB equation has been 
implemented in the canonical basis, where the density matrix is diagonal 
and the pairing tensor has the canonical form, in \cite{ebata2010,
nakatsukasa2012,nakatsukasa2012-a}. The most advanced implementation of 
TDHFB today is by Bulgac, Roche and collaborators and was described in 
\cite{bulgac2008}. Their code, which is massively parallel and has been 
ported to GPU architectures, implements a local Kohn-Sham scheme for TDDFT 
dubbed the time-dependent superfluid local density approximation (equivalent
to a TDHFB theory with a functional of the local density $\rho(\gras{r})$ 
only) and was very recently applied for the first time to the description 
of neutron-induced fission in \cite{bulgac2016}. The code uses fast discrete 
Fourier transforms to evaluate derivatives on a three-dimensional Cartesian 
lattice, and a multi-step predictor-modifier-corrector method for the 
time evolution.

In contrast to real-time dynamics described by TDDFT methods, the description of fission dynamics 
as a large-amplitude collective motion driven by a small set of 
collective variables is especially suited to the calculation of the 
distributions of fission fragment properties. The time-dependent generator coordinate method (TDGCM) 
developed in the 1980ies at CEA Bruy{\`e}res-le-Ch{\^a}tel is currently 
the only microscopic theory capable of producing realistic fission fragment 
distributions. Until now, it has only been applied under the Gaussian 
overlap approximation. The very first applications of this approach to 
the dynamics of fission were reported in a series of paper by Berger 
and Gogny in \cite{berger1984,berger1989,berger1991}. The first actual 
calculation of fission fragment mass distributions with this technique 
were published in \cite{goutte2004,goutte2005}, with additional results 
reported in \cite{younes2012-a}. The implementation of the TDGCM was based on 
solving the collective Schr{\"o}dinger equation by using finite-differences 
for derivatives. This choice imposed the use of a regular grid of points 
that did not offer the possibility to discard regions of the collective 
space that were irrelevant to the dynamics (for example the points with a 
very high energy). In practice, applications were restricted to 
two-dimensional collective spaces, and large computational resources 
were needed to achieve good numerical accuracy. Very recently, the 
collaboration between CEA and LLNL developed a new program to solve  
the TDGCM+GOA equations for an arbitrary number of collective variables 
using finite element analysis \cite{regnier2016}. 


\subsubsection{Collective Inertia}
\label{subsubsec:inertia_num}

Computing collective inertia is usually accomplished by using the 
perturbative cranking formula (\ref{eq:ATDHFBMassPC}), which only 
requires moments of the collective operators in quasiparticle space and 
two-quasiparticle energies. However, as recalled in section \ref{subsubsec:atdhfb} (see also figure \ref{fig:corrections} page \pageref{fig:corrections}), 
this approximation falls too short and can underestimate the inertia 
by more than 40 \%. On the other hand, the exact evaluation of the ATDHFB 
inertia would require both inverting the linear response matrix $\mathcal{M}$ of (\ref{eq:qrpa_matrix}) and 
evaluating the partial derivatives of the generalized density matrix with 
respect to the collective variables, see (\ref{eq:ATDHFBMass}) page \pageref{eq:ATDHFBMass}. 
There exist analytical formulas giving the partial derivatives (\ref{eq:dRdq}) as a function of the matrix of the collective variables,
but this again requires inverting the linear response matrix as seen from (\ref{eq:P20}). A more convenient and
thrifty procedure consists in the numerical evaluation of the derivatives
using finite difference formulas such as the popular and accurate centred
difference formula. It requires the densities 
$\mathcal{R}(\gras{q} \pm \delta \gras{q})$ that can be obtained by any 
DFT solver by slightly modifying the value of the constraints 
$\gras{q}\rightarrow \gras{q} \pm \delta\gras{q}$. The choice of 
$\delta\gras{q}$ obviously depends on the collective variable and 
usually also on particle number (as $\gras{q}$ usually has such 
dependence). This approach has been taken in \cite{baran2011}.

Concerning the inversion of the linear response matrix, a possibility
used by the authors of \cite{raey} is to build the linear response matrix 
in a reduced subspace of two-quasiparticle excitations and compute the 
inverse there. This is a computationally intensive task, but the main 
advantage is that the convergence of the procedure can be easily tested 
by slightly increasing the size of the subspace and repeating the 
calculation. Other authors use the fact that the linear response matrix 
is diagonal dominant, with two-quasiparticle energies on the diagonal, 
to build an iterative method to compute the action of the inverse on a 
given vector (that is the only quantity required to compute the inertias). 
This method has only be used to compute the inertias associated to Goldstone 
modes in \cite{PhysRevC.86.034334}. Yet another approach aiming at the 
evaluation of the inertias associated to Goldstone modes (Thouless-Valatin 
moments of inertia) has been recently proposed in \cite{1507.00045}. It 
resorts to the ideas of the finite amplitude method (FAM) to compute the
action of the inverse of the linear response matrix on a given vector. 

%
%

\newpage
\section{Results}
\label{sec:results}

After discussing in details the theoretical framework used in the DFT 
approach to fission in sections \ref{sec:pes} and \ref{sec:dynamics} and 
its computational implementation in section \ref{sec:numerics}, we review  
here a selection of results obtained by various groups. The presentation 
is organized in four main themes: fission barriers in section 
\ref{subsec:barriers}, spontaneous fission half-lives in section 
\ref{subsec:spontaneous}, alternative fission modes in section 
\ref{subsec:other_modes} and induced fission in section \ref{subsec:distributions}. 

Since this article is about the microscopic theory of fission, we have not 
recalled the many results obtained with empirical or semi-microscopic 
methods. In many cases such as, e.g., the structure of fission barriers in 
actinides or superheavy elements, the evolution of these barriers with 
triaxiality, excitation energy or spin, DFT calculations confirm earlier 
predictions obtained with these approaches. Instead, the emphasis here is 
put both on the progress made toward a fully-fledged, rigorous implementation 
of DFT methods in fission studies, and on the universal predictions coming 
out of the DFT calculations. The reader interested in a presentation of 
results obtained with the macroscopic-microscopic method could consult the 
review articles of Brack and collaborators in \cite{brack1972} and that of 
Bj{\o}rnholm and Lynn in \cite{bjornholm1980} or the textbook by Nilsson and 
Ragnarsson \cite{nilsson1995}; for more recent results, see, 
e.g., the work by the Los Alamos and Berkeley collaboration in 
\cite{moeller1989,moeller2009,randrup2011,randrup2013} and references 
therein.


\subsection{Fission Barriers}
\label{subsec:barriers}

In principle, fission barrier heights $B_\mathrm{f}$ (both inner and outer) 
are not observable quantities since the ``experimental'' values are 
determined through a model-dependent analysis of various induced fission 
cross-sections. However, these quantities are used in various models aimed at 
describing, for instance, heavy-ion fusion reactions, the competition between 
neutron evaporation and fission in compound nucleus reactions and the cooling 
or fission recycling in the r-process. Fission barriers are thus important 
building blocks of modern nuclear reaction software suite such as EMPIRE of 
\cite{herman2007} or TALYS of \cite{koning2008}, which are extensively used 
in application of nuclear science in reactor technology.


%

Because of computational limitations, early DFT calculations of fission 
barriers were performed at the Hartree-Fock approximation with pairing
correlations typically treated within the BCS formalism. This is the case 
in \cite{flocard1974}, for instance, where the SIII parametrization of the 
Skyrme force and a seniority pairing force are used. In the nineteen 
eighties, the theory was extended by introducing the finite temperature 
formalism as, e.g., in \cite{campi1983,bartel1985}, to account for 
modifications of the fission barrier with excitation energy, and by adding 
constraints on the angular momentum \cite{nemeth1985}. In parallel, the 
first full HFB calculation of the fission barrier in $^{240}$Pu was reported 
by Berger and collaborators in \cite{berger1980}. This last work should be 
considered as an important milestone in the microscopic theory of fission 
since this was also the first example of a DFT calculation with a finite-range 
effective potential in both the particle-hole and particle-particle channel. 
In addition, the authors reported the first example of two-dimensional PES 
within this fully microscopic framework. Since then, progress has been made on three fronts 
(i) systematic calculations of fission barriers with DFT are now possible, 
see \cite{buervenich2004} for an early application; (ii) calculations involving 
more than one collective variable are commonplace, with several examples of 
three-dimensional adiabatic studies reported in the last few years in 
\cite{schunck2014,lu2012,younes2009,younes2013}, (iii) beyond mean-field 
correlations, beyond the subtraction of a zero point energy correction, 
have begun to be incorporated systematically in the calculations. For
instance in \cite{samyn2005}, the impact of variation after parity projection on 
the height of fission barriers was investigated. In \cite{bender2004}, a 
similar analysis was performed with respect to the role of 
simultaneous particle number and angular momentum projection. 

In spite of formal differences between non-relativistic and relativistic 
formulations of DFT, between zero- and finite-range nuclear potentials, and 
in the treatment of pairing correlations, the conclusions of DFT studies on 
fission barriers are remarkably similar qualitatively and in some cases even quantitatively.
First, as well known from earlier macroscopic-microscopic calculations, the 
fission barrier in actinide nuclei is in most of the cases double-humped: the ground-state is 
separated from a fission isomer by a first barrier, and the second barrier 
separates the fission isomer from the scission region. Quantitatively, the 
energy of the fission isomer and the height of the fission barriers depend 
on the functional, as reported in the comparative studies of 
\cite{buervenich2004,mcdonnell2013,baran2015}. 

\begin{figure}[!ht]
\begin{center}
\includegraphics[width=0.480\linewidth]{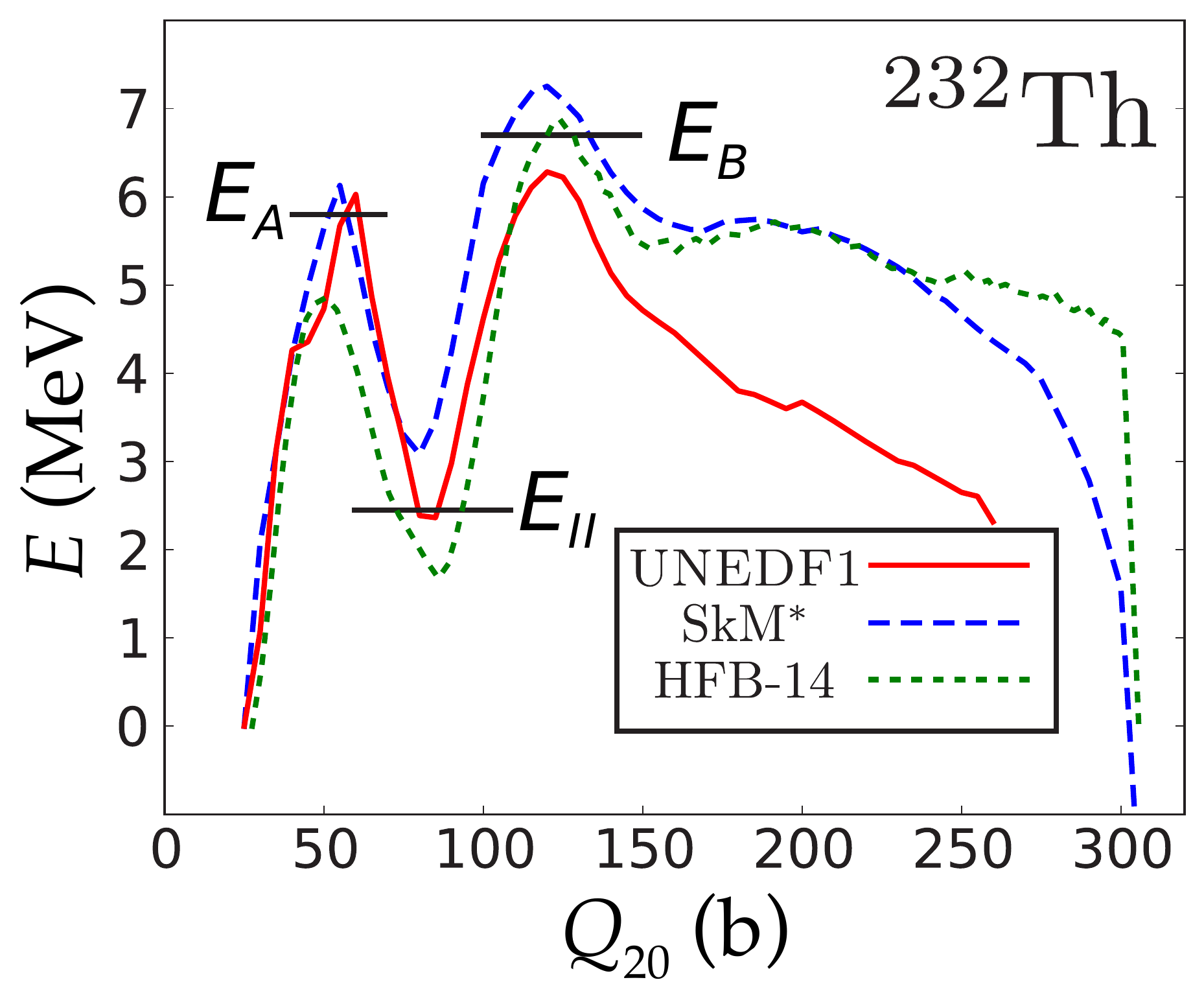}
\includegraphics[width=0.480\linewidth]{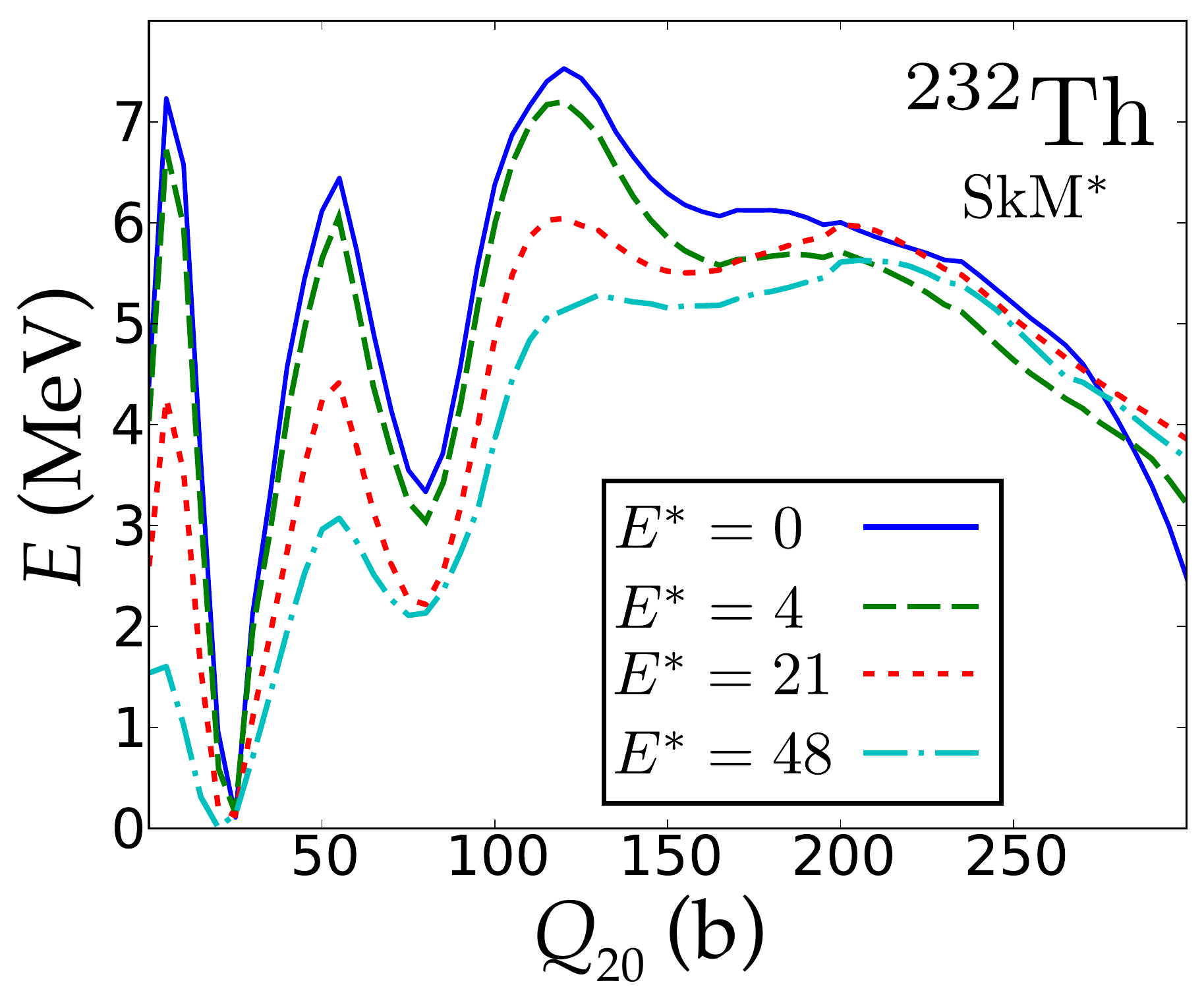}
\caption{Left: HFB deformation energy in $^{232}$Th computed for three 
different Skyrme energy functionals; Right: evolution of the HFB deformation 
energy with excitation energy $E^{*}$ for the SkM* functional.
Figure taken from \cite{mcdonnell2013}, courtesy of J. McDonnell; 
copyright 2013 by The American Physical Society.}
\label{fig:th232}
\end{center}
\end{figure}

Among the actinides, $^{232}$Th plays a special role, as it has been 
conjectured that it could have a triple-humped fission barrier; see 
\cite{mcdonnell2013} for references on experimental work. The 
early triaxial Gogny HFB calculations of \cite{berger1989} suggested the existence 
of the third barrier indeed, although predictions were model-dependent: 
the D1 parametrization of the Gogny potential did not exhibit any such 
minimum. Triaxial calculations with the Skyrme potential, either at the 
HF+BCS level with rotational correction, as in \cite{bonneau2006}, or 
at the HFB level as in \cite{mcdonnell2013}, suggest a very shallow minimum 
that becomes more pronounced for $N<142$. In \cite{mcdonnell2013}, additional 
finite temperature calculations pointed to the stabilizing effect of 
excitation energy through a reduction of pairing correlations. 

\begin{figure}[!ht]
\begin{center}
\includegraphics[width=0.80\linewidth]{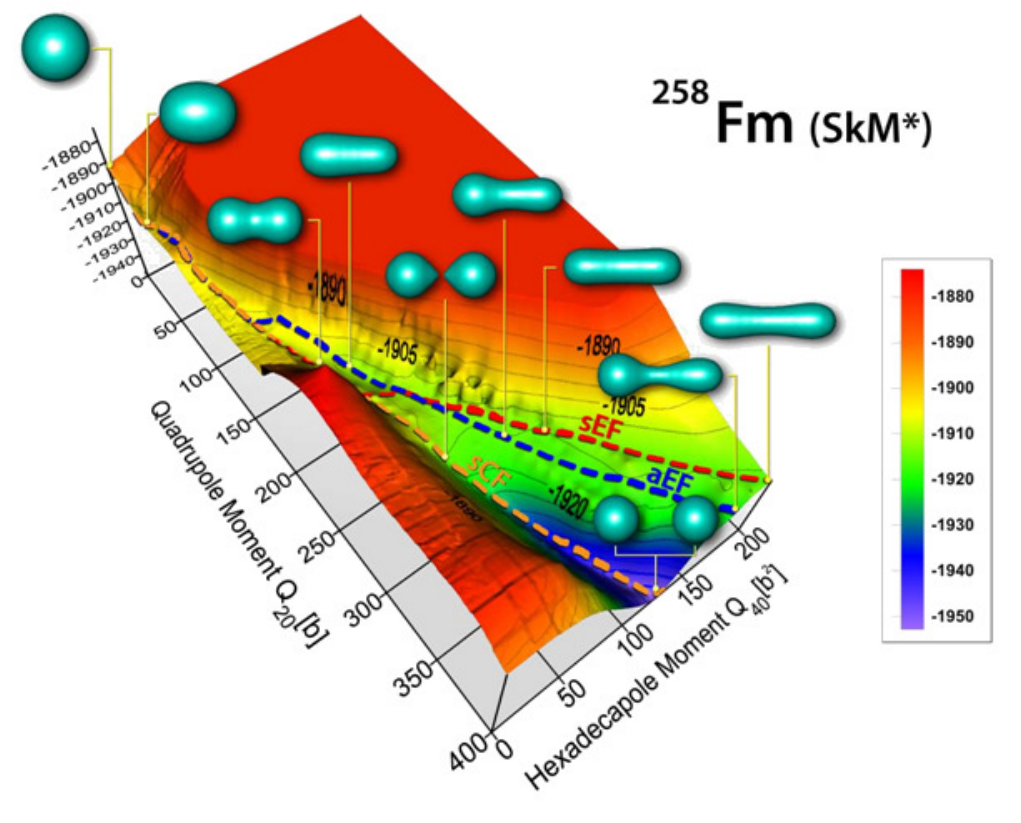}
\caption{Fission pathways in $^{258}$Fm computed in the Skyrme HFB approach 
with the SkM* parametrization. Figure courtesy of A. Staszczak.}
\label{fig:Fm_pathways}
\end{center}
\end{figure}

Fermium isotopes have also attracted considerable attention and are often 
used as test benches of theoretical approaches. DFT calculations were performed 
with Skyrme potentials both at the HF+BCS and HFB approximations in 
\cite{staszczak2005,staszczak2009,bonneau2006,bender1998} and with the Gogny 
D1S potential in \cite{warda2002,delaroche2006}. All these calculations predict 
a multimodal decay pattern in many of these isotopes, especially for 
$^{256,258}$Fm where several different fission pathways lead to significantly 
different geometries at scission (often labeled compact symmetric, compact asymmetric
and extended asymmetric). Figure \ref{fig:Fm_pathways} gives a visual representation 
of the nuclear shape in $^{258}$Fm along the different fission paths.

\begin{figure}[!ht]
\begin{center}
\includegraphics[width=0.80\linewidth]{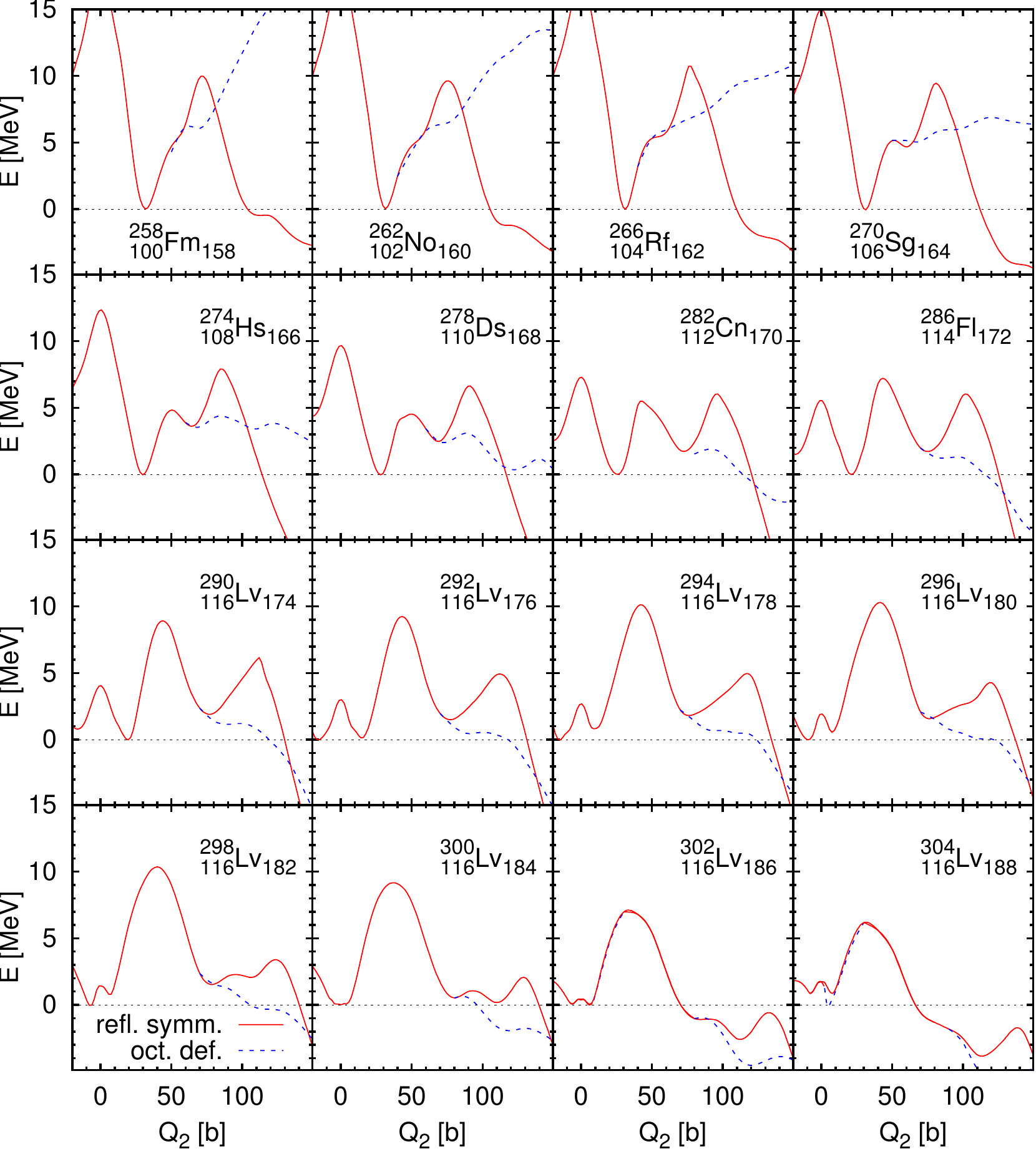}
\caption{Axial fission barriers for the Gogny D1S force. Solid (dashed) 
lines denote the reflection-symmetric (reflection-asymmetric) paths. 
Figure taken from \cite{baran2015}, courtesy of A. Baran; 
Copyright 2015, with permission from Elsevier}
\label{fig:SHbarriers}
\end{center}
\end{figure}

Another robust prediction of DFT models concerning fission barriers 
is the disappearance of the fission isomer in superheavy elements with 
$Z\geq 108$. This conclusion has been first obtained in \cite{bender1998} 
using Skyrme HF+BCS and confirmed in the systematic 
calculations of \cite{buervenich2004}. It was also found in studies based 
on the full HFB approach with the D1S parametrization of the Gogny force 
in \cite{warda2012,delaroche2006}. This general trend is illustrated in 
figure \ref{fig:SHbarriers}, which shows both reflection-symmetric and 
reflection-asymmetric fission barriers in superheavy elements for the D1S 
parametrization of the Gogny potential. 

It has been known since 1972 and the work of Larsson and collaborators in 
\cite{larsson1972} that breaking axial symmetry lowers the inner barrier 
in actinide and superheavy nuclei This result has been confirmed in all 
DFT calculations: in non-relativistic formulations with the Skyrme force 
at the HF+BCS level (\cite{staszczak2005,bender1998}) and at the HFB level 
(\cite{schunck2014,staszczak2009}), and with the Gogny potential at the 
HFB level (\cite{girod1983,warda2002}). 
This effect is typically of the order of 2 to 3 MeV. In actinide nuclei, 
there is almost no effect of triaxiality on the outer barrier. 

Another important component of the calculation affecting the barrier is 
the amount and nature of pairing correlations. Calculations reported in 
\cite{samyn2005} show a quantitative difference between the BCS and HFB 
approximations for pairing correlations: fission barriers at the HF+BCS 
level are typically larger than at the HFB level by about 0.5 MeV for Pu 
isotopes. However, this effect may be an indirect consequence of using a 
zero range pairing force (which involves introducing in quasiparticle space): 
in \cite{girod1983}, calculations with a finite-range Gogny force led to 
the exact opposite conclusion. The authors of \cite{samyn2005} also 
showed that reducing the strength of the pairing 
force results in an increase of fission barriers. The same conclusion was 
also obtained with Skyrme forces in \cite{schunck2014} and with the Gogny 
force and the BPCM functional in \cite{giuliani2013}. Finally, the impact 
of particle number projection was also investigated in \cite{reinhard1996,
samyn2005}, where it is shown that projection can reduce the barriers by 
about 0.5 MeV.

\begin{figure}[!ht]
\begin{center}
\includegraphics[width=0.70\linewidth]{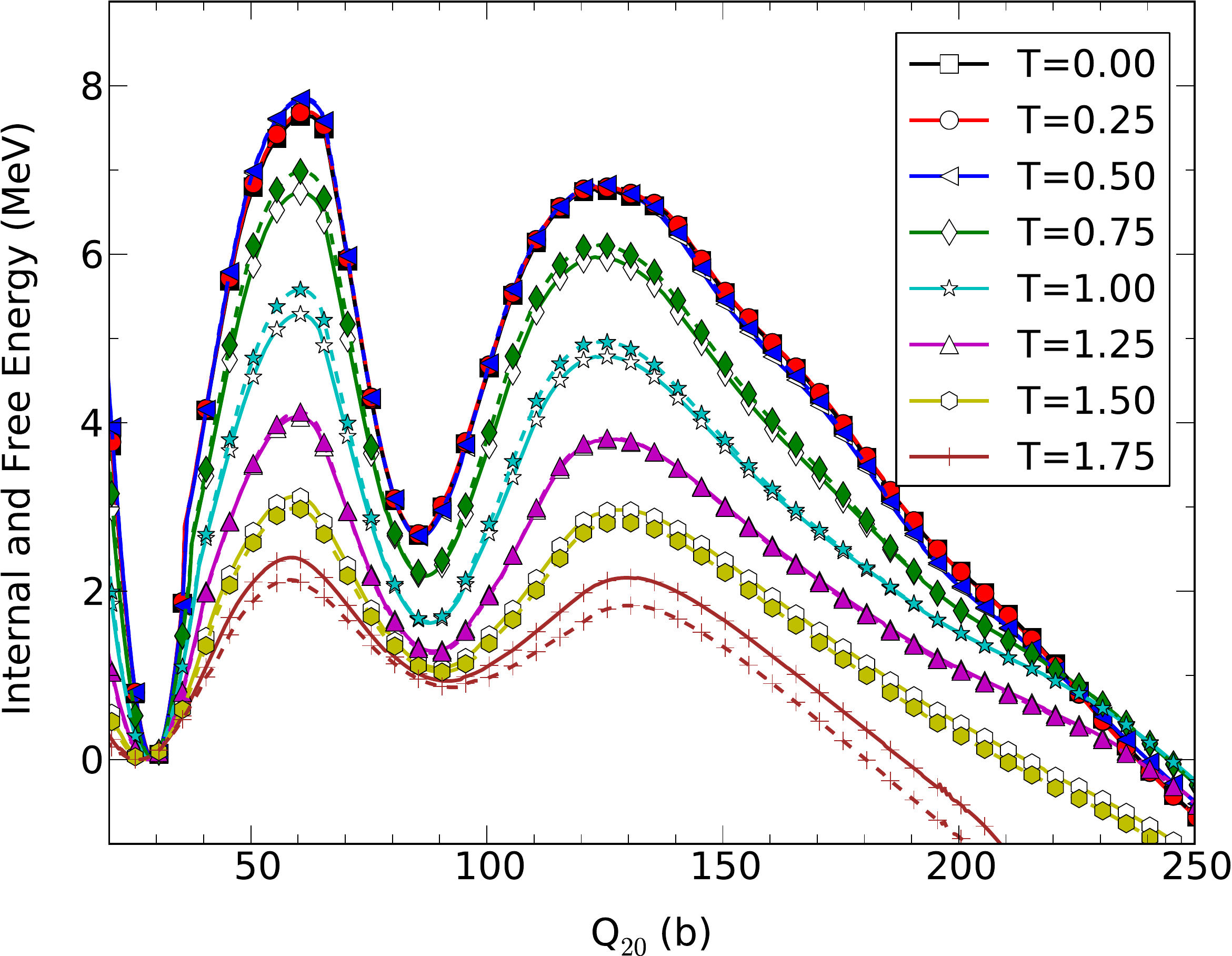}
\caption{Free energy (solid lines, open markers) and internal 
energy (dashed lines, plain markers) as a function the axial quadrupole 
moment in $^{240}$Pu for the Skyrme SkM* EDF. See \cite{schunck2015} 
for additional discussion.
Figure taken from Ref.\cite{schunck2015}, courtesy of N. Schunck; 
copyright 2015 by The American Physical Society.}
\label{fig:maxwell}
\end{center}
\end{figure}

The evolution of fission barriers with excitation energy is important for  
superheavy nuclei, since heavy elements are typically formed in cold- or 
hot-fusion heavy-ion reactions at an excitation energy that can reach up 
to 30 MeV, see \cite{hoffman2000} for a review. In induced fission, the 
compound nucleus is also at a non-zero excitation energy, and the evolution 
of fragment properties will depend on how the fission barriers change 
with that excitation energy. As recalled in section \ref{subsubsec:excitation}, 
finite temperature DFT is a convenient tool to model fission at $E^{*}>0$. 
Bartel and Quentin reported in \cite{bartel1985} the first example of a 
FT-HF calculation with the SkM* Skyrme force. They confirmed predictions 
from macroscopic-microscopic models that fission barriers decrease as $T$ 
increases, and have vanished for temperatures of the order of 3 MeV. This 
initial result was extended to the case of a FT-HFB calculation in 
\cite{pei2009,sheikh2009,martin2009}. The calculations in the low temperature 
regime of \cite{schunck2015} tell a slightly more nuanced story: for 
$0< T \leq 0.75$ MeV, the effect of temperature is to slightly increase 
fission barriers as a result of the dampening of pairing correlations; 
once the latter have completely vanished, around $T>1$ MeV, barriers 
monotonically decrease with $T$, as shown in figure \ref{fig:maxwell}.

Finally, fission barriers are also sensitive to angular momentum. There 
have been only two studies of this effect. In \cite{egido2000}, Egido 
and Robledo performed cranked HFB calculation with the Gogny force (D1S 
parametrization) and showed that the double-humped nature of the fission 
barrier in $^{254}$No persisted up to $I=60\hbar$. In addition, the 
energy of the fission isomer is pushed down so that it becomes the 
ground-state for $I>30\hbar$. In \cite{baran2014}, similar types of 
calculations, up to $I=16\hbar$, were performed with the SkM* 
parametrization of the Skyrme force and density-dependent pairing in Fm 
isotopes. The authors noticed again a gradual, weak decrease of the 
fission barrier, the magnitude of which depends on the number of neutrons.


\subsection{Spontaneous Fission Half-Lives}
\label{subsec:spontaneous}

Spontaneous fission half-lives $\tau_{1/2}^{\mathrm{SF}}$ are usually 
computed with the WKB formula, see section \ref{subsubsec:wkb}, combined 
with the least action principle to determine the most probable fission path, 
see \cite{brack1972} for a gentle introduction. Elements of the calculations 
that have been shown to play a major role are
\begin{enumerate}
\item {\bf Collective inertia}: As recalled in sections \ref{subsubsec:gcm} and 
\ref{subsubsec:atdhfb}, the collective inertia can be computed either from 
the GCM or ATDHFB formalism. Until now, the cranking approximation (where 
the residual interaction term of the QRPA matrix is neglected) was used in 
both cases. In addition, the perturbative version of it is also most commonly 
used: in the case of the GCM, it originates from the introduction of a local momentum 
operator in place of explicit derivatives with respect to $\qVec$ while in 
the case of ATDHFB, it implies expressing the derivatives $\partial\mathcal{R}/\partial q_{\alpha}$ 
in terms of the matrices of the operator $\hat{Q}_{\alpha}$. As recalled in 
\ref{subsubsec:atdhfb}, the perturbative and non-perturbative cranking formulas 
for the ATDHFB mass tensor leads to significant differences in fission half-lives. 
The most recent studies in \cite{sadhukhan2013,sadhukhan2014,sadhukhan2016,zhao2016} are therefore 
based on the non-perturbative expression. 
The collective inertia tensor is a function of the collective 
variables and depends sensitively on both the shell structure and pairing 
correlations.
\item {\bf Zero-point energy corrections}: One may distinguish two forms of 
zero-point energy corrections. On the one hand, any spontaneously broken symmetry can 
be associated with a collective variable. This leads to a zero-point energy correction 
if the resulting collective motion is sufficiently decoupled from the 
intrinsic motion. This is the case, for instance, for translational and 
rotational symmetry, as discussed in section \ref{subsubsec:beyond}. On 
the other hand, zero-point energy corrections also arise naturally as corrective 
terms to the collective Hamiltonian obtained after applying the GOA to the 
GCM equations, see section \ref{subsubsec:gcm}. 
\item {\bf Ground-state energy}: The energy $E_0$ is used to define the 
inner and outer turning point for the WKB formula. It is usually taken
as the quantal ground state energy obtained by adding the zero point energy correction of
the collective motion to the HFB potential energy -- see, for instance, \cite{staszczak2013}. 
Instead of the zero point energy many authors prefer to use a single 
constant value of the order of 1 MeV to  estimate this quantity.
\end{enumerate}

There is a rich literature on the semi-microscopic calculations of 
spontaneous fission half-lives, where the potential energy is computed 
in the macroscopic-microscopic framework and the inertia is computed from 
the Inglis cranking model or parametrized empirically. Again, since our 
goal is to discuss DFT predictions, we refer the reader interested to the 
reference calculations of \cite{baran1981} (actinide and transactinide
elements) and of \cite{smolaczuk1995} (superheavy elements). 

\begin{figure}[!ht]
\begin{center}
\includegraphics[width=0.68\linewidth]{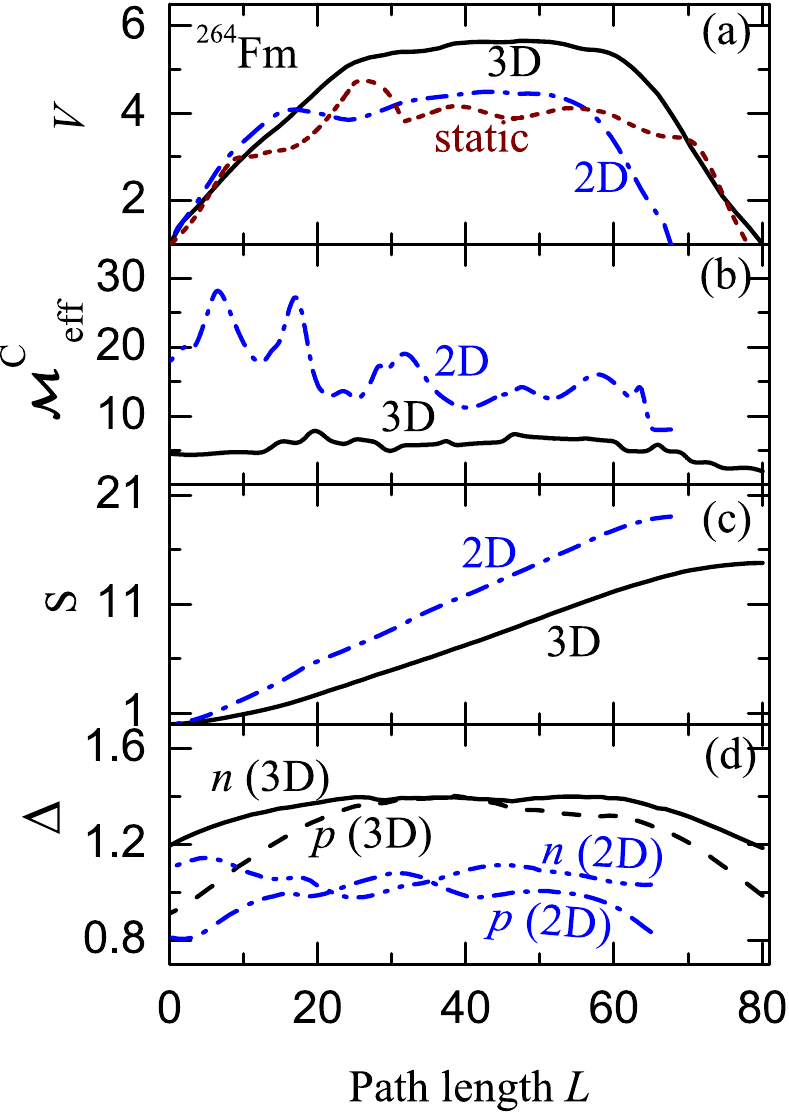}
\caption{(a) Potential V (in MeV), (b) effective inertia
$\mathcal{M}^{\mathrm{C}}_{\mathrm{eff}}$ (in $\hbar^{2}$ MeV$^{−1}$/1000),
(c) action S, and (d) average pairing gaps$\Delta_{n}$ and $\Delta_{p}$
(in MeV) plotted along the 2D (static pairing; dotted line) and 3D
(dynamic pairing; solid line) paths. The static fission barrier is
displayed for comparison in (a). Figure taken from Ref.
\cite{sadhukhan2014}, courtesy of J. Sadhukhan; copyright 2014 by The
American Physical Society.}
\label{fig:halflives_3D}
\end{center}
\end{figure}

Whatever the approach retained to compute the potential energy, half-lives 
calculations begin by determining the optimal fission path from the least 
action principle. The resulting dynamical path can be sensibly different 
from the static, least-energy path, leading to differences of several orders 
of magnitude for $\tau_{1/2}^{\mathrm{SF}}$ as reported in \cite{sadhukhan2013}; 
see also figure \ref{fig:cranking_perturbative} page \pageref{fig:cranking_perturbative} 
for an illustration.
This effect is amplified when the collective space includes pairing degrees of 
freedom, as illustrated in \cite{giuliani2013}, because of the approximate 
$1/\Delta^{2}$ dependence of collective inertia on the pairing gap discussed 
earlier in section \ref{subsubsec:other}. Similarly, the size of the 
collective space in which the fission path is determined can have a sizeable 
impact on the half-lives. The figure \ref{fig:halflives_3D} shows a comparison 
between one-dimensional trajectories through two- and three-dimensional collective 
spaces. At some points along the path, the 3D collective action can be lower 
by nearly a factor 2 than in the 2D scenario, which could result in huge 
differences for $\tau_{1/2}^{\mathrm{SF}}$ since the action is in the exponent, 
see (\ref{eq:tSF}). Note that in the case where the PES is determined by 1D 
constrained HFB calculations (typically involving the axial quadrupole moment), 
the fission path is automatically the least-energy path. This is the case, 
for instance, in the works of \cite{warda2012,berger2001,warda2006,schindzielorz2009,
erler2012}.

\begin{figure}[!ht]
\begin{center}
\includegraphics[width=0.65\linewidth]{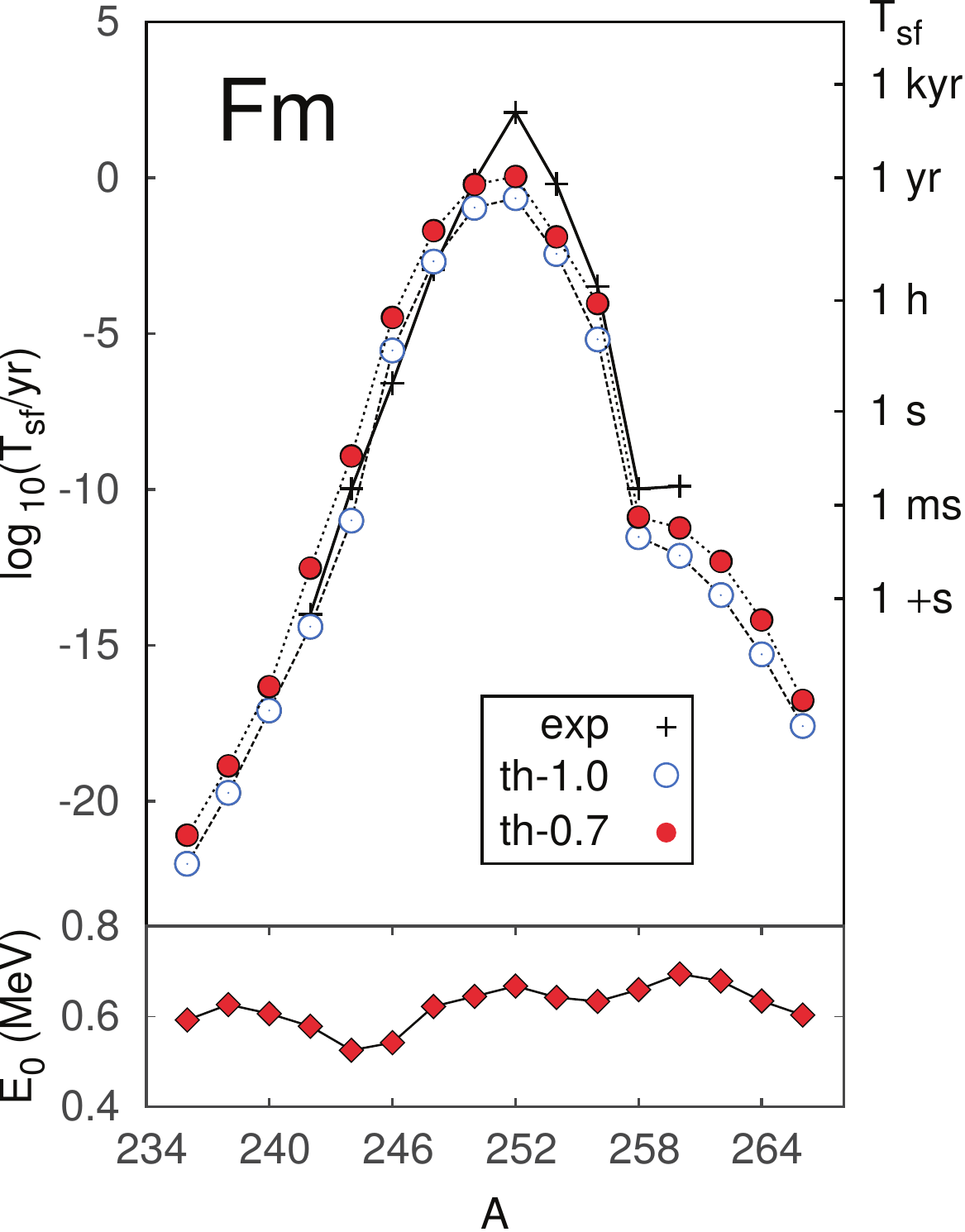}
\caption{Spontaneous fission half-lives of even-even Fm isotopes with
$236 \leq A \leq 266$, calculated with the SkM* nuclear EDF with initial
energy $E_{0} = 0.7\epsilon_{\mathrm{ZPE}}$, where $\epsilon_{\mathrm{ZPE}}$ 
is the zero-point energy (th-0.7) compared with experimental data. The
corresponding collective ground-state g.s. energies $E_0 = 0.7
\epsilon_{\mathrm{ZPE}}$ are shown in the lower panel. The results obtained
without scaling (th-1.0) are also shown. Figure taken from Ref.
\cite{staszczak2013}, courtesy of A. Staszczak; copyright 2013 by The
American Physical Society.}
\label{fig:Fm_halflives_SKMS}
\end{center}
\end{figure}

Owing to the variety of their fission modes, Fermium isotopes are excellent 
test bench of DFT calculations. Experimental half-lives are known in 10 different 
isotopes from $N=142$ to $N=160$ and cover about 15 orders of magnitude. In 
\cite{warda2012,warda2002,warda2006}, axially-symmetric HFB calculations of 
spontaneous fission half-lives with the Gogny force in one-dimensional collective 
space could reproduce qualitatively the evolution of the half-lives around $N=154$. 
In parallel, calculations breaking axial symmetry were performed along the entire 
Fm isotopic chain with the Skyrme functional (SkM* parametrization) in 
\cite{staszczak2005,staszczak2013} and reproduced quantitatively (within about 
1-2 orders of magnitude) the trend of $\tau_{1/2}^{\mathrm{SF}}$ as a function 
of neutron number. State-of-the-art results are summarized in figure 
\ref{fig:Fm_halflives_SKMS}.

\begin{figure}[!ht]
\begin{center}
\includegraphics[width=0.60\linewidth]{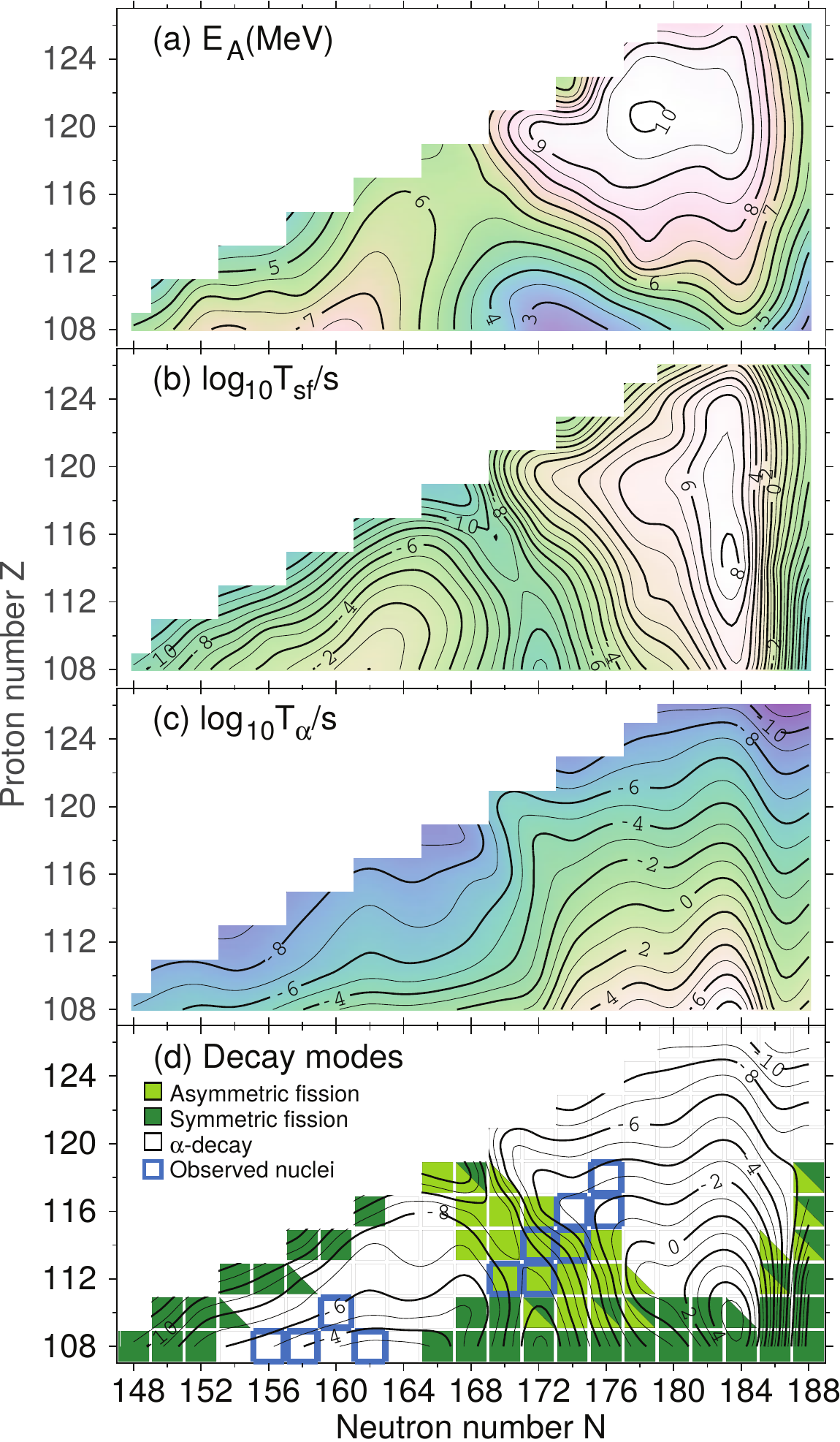}
\caption{Summary of results for even-even superheavy nuclei obtained with the 
SkM* nuclear EDF. (a) Inner fission barrier heights $E_A$ (in MeV); (b)
Spontaneous fission half lives log$_{10}\tau_{1/2}^{\mathrm{SF}}$ (in 
seconds); (c) $\alpha$ decay half-lives log$_{10}T_{\alpha}$ (in seconds); 
(d) Dominant decay modes. If two modes compete, this is marked by 
coexisting triangles. 
Figure taken from Ref.\cite{staszczak2013}, courtesy of A. Staszczak; 
copyright 2013 by The American Physical Society.}
\label{fig:SH_decay_SKMS}
\end{center}
\end{figure}

Since spontaneous fission is one of the major decay modes of superheavy elements, 
the accurate calculation of spontaneous fission half-lives is an important tool in the search for the next island of 
stability. The first systematic calculations of superheavy elements half-lives in the 
context of DFT were reported by Berger and collaborators in 
\cite{berger2001}. Calculations were performed at the HFB approximation 
along an axially-symmetric path in a 1D collective space using the Gogny 
D1S parametrization. In \cite{warda2012}, similar calculations were 
compared with the available data. As an illustration, the figure 
\ref{fig:SH_decay_SKMS}, which is taken from \cite{staszczak2013}, 
gives an overview of fission and $\alpha$ decay properties of superheavy  
elements in the particular case of the SkM* Skyrme functional. Overall, theoretical predictions 
overestimate the fission half-lives by a few orders of magnitude (recall 
that the experimental values are spread over about 15 orders of magnitude), 
but results pointed to the large variability of $\tau_{1/2}^{\mathrm{SF}}$ 
with respect to the symmetry breaking effects along the fission path. Using 
HF+BCS calculations, the authors of \cite{schindzielorz2009} showed the 
variability of the predictions with respect both to proton and neutron 
number, and to the parametrization of the functional. This variability 
can reach 10-20 orders of magnitude. These conclusions were confirmed and 
further discussed in \cite{baran2015}, where systematic comparisons of 
fission barriers and half-lives using the macroscopic-microscopic method, 
non-relativistic Skyrme and Gogny EDF and relativistic Lagrangian pointed 
to the huge uncertainties in the theory. As an example, the largest 
fission half-life for the SkM* Skyrme force is recorded for $Z=120$, 
$N=182$ and is of the order of $10^{11}$ s; for the D1S Gogny force, it 
is at $Z=124$, $N=184$ and in excess of $10^{20}$ s. In the same paper, 
uncertainty quantification methods based on the linear approximation of 
the covariance matrix showed that fission barriers can vary by $\pm 1$ 
MeV under even small changes in parameters. This result is consistent 
with the results of \cite{mcdonnell2015}, which used Bayesian inference 
techniques to propagate uncertainties in the parametrization of the EDF 
to model predictions for barriers. 

As mentioned in the introduction to this article, another important 
area of science where fission theory provides critical input is 
nucleosynthesis. There have been only two pioneer studies of spontaneous 
fission half-lives in very neutron-rich nuclei. In \cite{erler2012}, 
the same methodology as in \cite{schindzielorz2009} (Skyrme HF+BCS, 
one-dimensional collective space, zero-point quadrupole energy,
rotational correction) was employed to survey the spontaneous fission 
half-lives of superheavy elements up to the neutron drip line. In 
spite of the very large dependence on the parametrization of the EDF 
-- four different Skyrme forces are used -- mentioned earlier, there 
seems to be a consistently marked increase in spontaneous fission 
half-lives as a function of $N$ for $N/Z > 2$. A similar result was 
obtained in \cite{rodriguez-guzman2014,rodriguez-guzman2014-a}, where the focus was on the Uranium and 
Plutonium isotopic sequences, from the well-known $^{238}$U and $^{240}$Pu 
isotopes to the neutron drip line. Again, the paper highlights the extreme 
dependence of fission half-lives on the details of the microscopic 
calculation but identifies two robust trends: fission is favoured over 
$\alpha$ decay for $N>166$, and there is a marked increase of fission 
half-lives as $N$ increases for $N/Z > 1.8$, with a maximum around 
$N=184$. 


\subsection{Other fission modes}
\label{subsec:other_modes}

So far, we have focused our survey of DFT results on only two particular 
themes, spontaneous fission barriers and half-lives, and only presented 
results for even-even nuclei. Let us first note that odd mass nuclei pose 
a number of additional difficulties compared to even-even ones:
\begin{itemize}
\item The odd particle is typically handled in the HFB theory through the 
blocking approximation, where the HFB wave function for the odd nucleus is 
defined as a one-quasiparticle excitation of some reference even-even 
nucleus, see \cite{ring2000}. Such excitations break time-reversal symmetry 
internally, resulting in non-zero time-odd fields, see (\ref{eq:pes_Hodd}) 
for the form of these terms in the case of Skyrme functionals and 
\cite{rutz1998,duguet2001,schunck2010} for more general discussions. Of 
particular relevance for applications in potential energy surfaces for 
fission is the fact that time-odd fields can induce a small triaxial 
polarization of the nuclear shape as exemplified in \cite{schunck2010}. In 
practice, most large-scale calculations of odd mass nuclei such as, e.g., 
in \cite{bertsch2009}, are based on the equal filling approximation of the 
exact blocking prescription, which preserves time-reversal invariance and 
axial symmetry as discussed in \cite{perez-martin2008}.
\item When implementing the blocking approximation, whether exactly or 
through the equal filling approximation, it is not possible to know beforehand 
which quasiparticle excitation will yield the lowest energy for the odd 
mass system. Therefore, potential energy surfaces have to be computed for a 
number of different configurations -- in the case of axial symmetry, such 
configurations would typically be characterized by the projection $K$ of 
the angular momentum of the odd particle on the axis of symmetry. This 
can add substantially to the computational burden.
\item Both spontaneous and induced fission require the knowledge of the 
collective inertia tensor (\ref{gcm:masscrank}) or (\ref{eq:ATDHFBMass}) 
in section \ref{subsec:inertia}. While the extension of the GCM method to 
odd-mass nuclei has recently been published in \cite{bally2014}, the case 
of the ATDHFB theory remain open [Currently, the theory relies on the 
explicit hypothesis that the system is described by a time-even generalized 
density, see section \ref{subsubsec:atdhfb} page \pageref{subsubsec:atdhfb}]. 
In addition, the generalization of the Gaussian overlap approximation to 
odd nuclei published recently in \cite{rohoziski2015} is restricted to quadrupole 
collective variables, which is not sufficient to describe fission.
\end{itemize}
For all these reasons, fission in odd mass nuclei has been very rarely 
considered in DFT. The only exceptions can be found in \cite{hock2013}, 
where fission barriers for $^{235}$U and $^{239}$Pu are computed at the 
HF+BCS approximation, and in the earlier study of \cite{perez-martin2009}, 
where full HFB calculations using the equal filling approximation are 
performed in $^{235}$U.

Before we discuss 
neutron-induced fission, we should mention three other topics of interest
where DFT has been applied, albeit to various degrees of success: (i) 
$\gamma$-decay of fission isomers, (ii) $\beta$-delayed fission, (iii) 
cluster radioactivity:
\begin{itemize}
\item {\bf $\gamma$-decay of fission isomers - }
Although fission isomers decay predominantly by spontaneous fission (see 
\cite{bjornholm1980} for a comprehensive review), they could also 
$\gamma$-decay back to the ground-state. The rate of such a decay would, 
of course, affect the estimate of the fission isomer lifetime. There 
has been only few attempts in the literature at computing such quantities 
within a fully microscopic approach. In \cite{chinn1992}, the 
partial lifetime of the fission isomer in $^{238}$U, which decays via 
electric quadrupole radiation to the lowest $2^+$ state of the 
ground-state band, was estimated with the collective Hamiltonian derived 
from the GCM with the GOA approximation. Calculations were performed 
with the D1S parametrization of the Gogny force and were within 2 orders 
of magnitude of experimental data. Considering the restriction to 
axially-symmetric shapes for the HFB states used in the GCM, the authors 
concluded that the computation of the transition rates provided an upper 
limit on the $\gamma$-decay lifetime of the isomer. Simpler but more 
systematic studies of the $\gamma$ half-life $\tau_{1/2}^{\gamma}$ based 
on the WKB approximation were performed in \cite{delaroche2006}. The 
$\gamma$ half-life was computed in perfect analogy to fission half-lives 
(see section \ref{subsubsec:wkb} for details), only the trajectories 
relevant to $\gamma$ decay connect the fission isomer to the ground-state 
in the $(q_{20}, q_{22})$ space, instead of the ground-state to the outer 
turning point in the $(q_{20},q_{30})$ space. Results were typically within 
1--2 orders of magnitude of the experimental data.
\item {\bf$\beta$-delayed fission  - }
$\beta$-delayed fission (see \cite{RevModPhys.85.1541} for a review) is 
a mechanism relying on the fact that odd-odd nuclei may $\beta$-decay to  
an (highly) excited state of the even-even daughter. Assuming that the 
fission barrier does not depend much on the intrinsic configuration of 
the excited state, the extra excitation energy of the initial configuration 
in the even-even daughter leads to a decrease by the same amount of the 
effective fission barrier height. This qualitative mechanism explains why  
it is possible to observe fission in nuclei such as $^{180}$Hg, where the 
fission barrier for the ground-state would otherwise be too high. In this 
particular nucleus, recent experimental results showed that fission is 
asymmetric, contradicting simple arguments based on the symmetric split 
into two semi-magic $^{90}$Zr fragments. Calculations with the SkM* 
parametrization of the Skyrme functional and the D1S parametrization of 
the Gogny force, both at zero (\cite{warda2012}) and finite 
temperature (\cite{PhysRevC.90.021302}) in the mercury and polonium 
region have conclusively confirmed the reflection asymmetric nature of 
the fission paths in this region.
\item {\bf Cluster radioactivity  - }
The phenomenon denoted ``cluster radioactivity'' is a new kind of 
radioactivity where some atomic nuclei emit a light nuclei such as 
$^{14}$C and was first reported in \cite{rose1984}. The characteristics 
of the light fragment are strongly influenced by the magic character of 
the heavy fragment; doubly-magic heavy fragments such as $^{208}$Pb are 
common. This type of radioactivity shares features from both $\alpha$ 
emission and standard fission, and its properties can thus be described 
by invoking models developed in these two fields. From a DFT point of 
view, cluster radioactivity has been explained in \cite{egido2004} as 
a very asymmetric fission where the axial octupole moment $q_{30}$ is 
the driving coordinate. In figure \ref{fig:ClusterR_1}, we show the 
evolution of the density in the nucleus $^{224}$Ra from the ground-state 
to cluster emission as a function of the constraint on the octupole 
moment.

\begin{figure}[!ht]
\begin{center}
\includegraphics[width=0.65\linewidth]{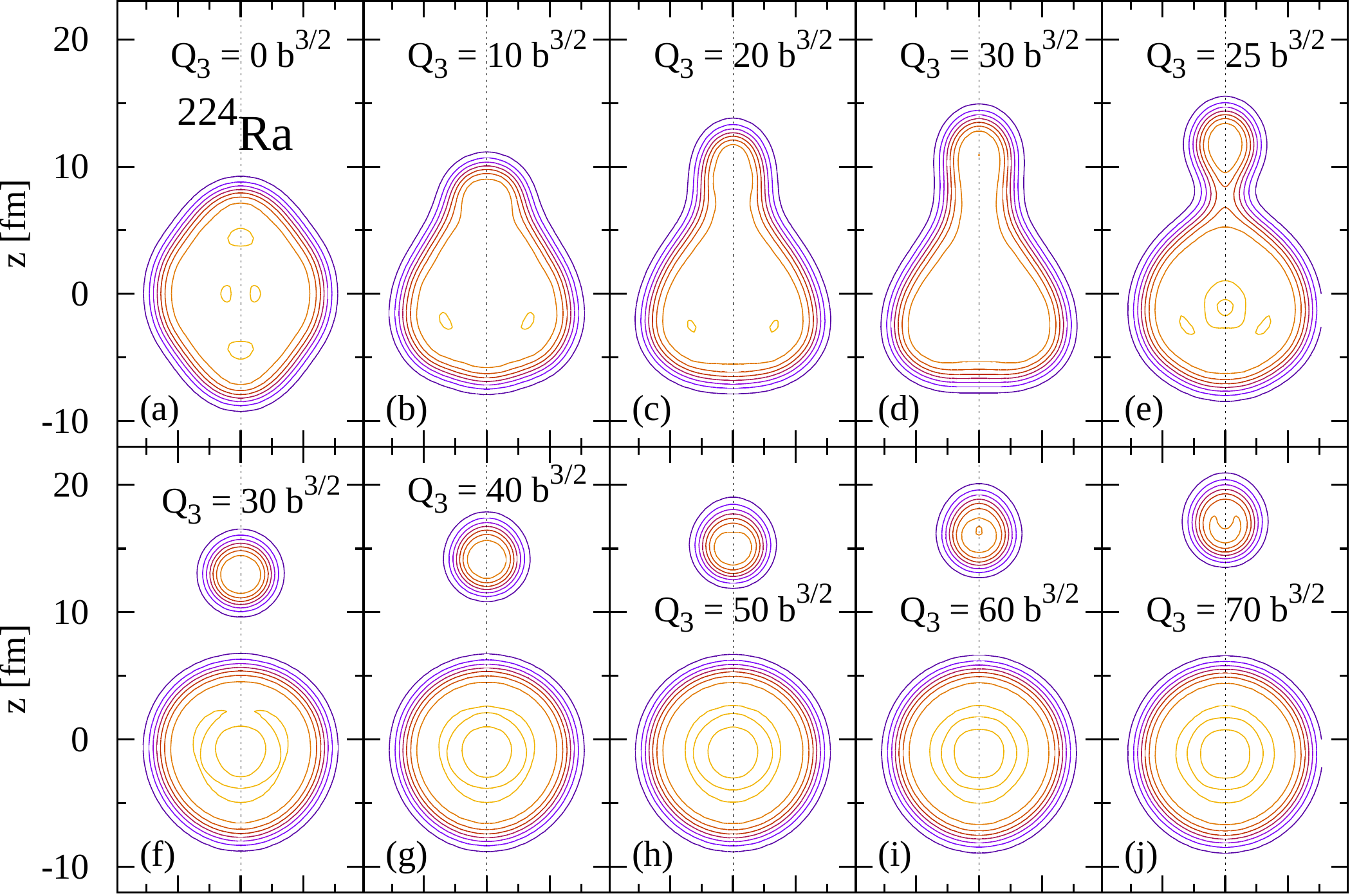}
\caption{Shape evolution of $^{224}$Ra as a function of the octupole 
moment $q_{30}$. Panels (a)–(d) correspond to the ascending (energy-wise) 
part of the fission path, panel (e) corresponds to the saddle region and 
panels (f)–(j) correspond to the descent from saddle to scission. 
Figure taken from \cite{warda2011}, courtesy of M. Warda; copyright 2011 
by The American Physical Society.}
\label{fig:ClusterR_1}
\end{center}
\end{figure}

In \cite{robledo2008,robledo2008-a,warda2011}, systematic calculations 
of cluster emission lifetimes using the same WKB framework as for 
fission have shown good agreement with experimental data in the actinides 
as well as in some neutron deficient Ba isotopes. Note that, like fission, 
cluster emission lifetimes cover a range of more than 15 orders of 
magnitude, thereby posing significant challenges to theory. In figure  
\ref{fig:ClusterR_2}, various theoretical estimates of spontaneous 
cluster emission lifetimes are compared to experimental data for several 
decay channels. In this particular case, theoretical values have been 
computed from the same one-dimensional WKB formula as in spontaneous 
fission by replacing the quadrupole inertia by the octupole one.

\begin{figure}[!ht]
\begin{center}
\includegraphics[width=0.65\linewidth]{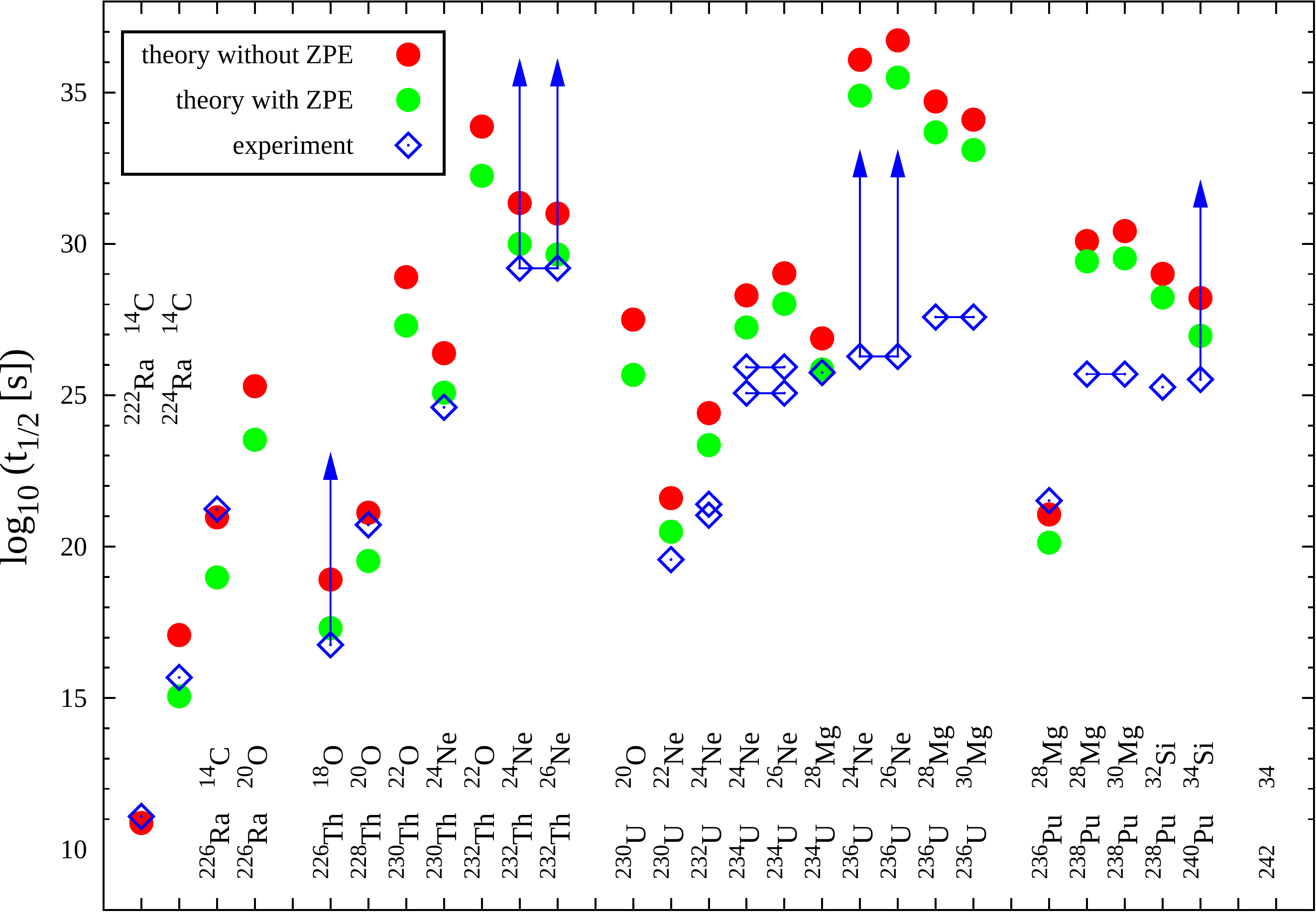}
\caption{Half-lives for cluster emission of various isotopes and various 
clusters. Blue diamonds show the experimental half-lives. Arrows indicate
the low experimental limit. Connected diamonds are for experimental values 
for two clusters. If experimental data from different experiments differ by 
more than 0.3, the extreme values are indicated.
Figure taken from Ref. \cite{warda2011}, courtesy of M. Warda; copyright 
2011 by The American Physical Society.}
\label{fig:ClusterR_2}
\end{center}
\end{figure}

\end{itemize}


\subsection{Neutron-Induced Fission}
\label{subsec:distributions}

\begin{figure}[!ht]
\begin{center}
\includegraphics[width=0.45\linewidth]{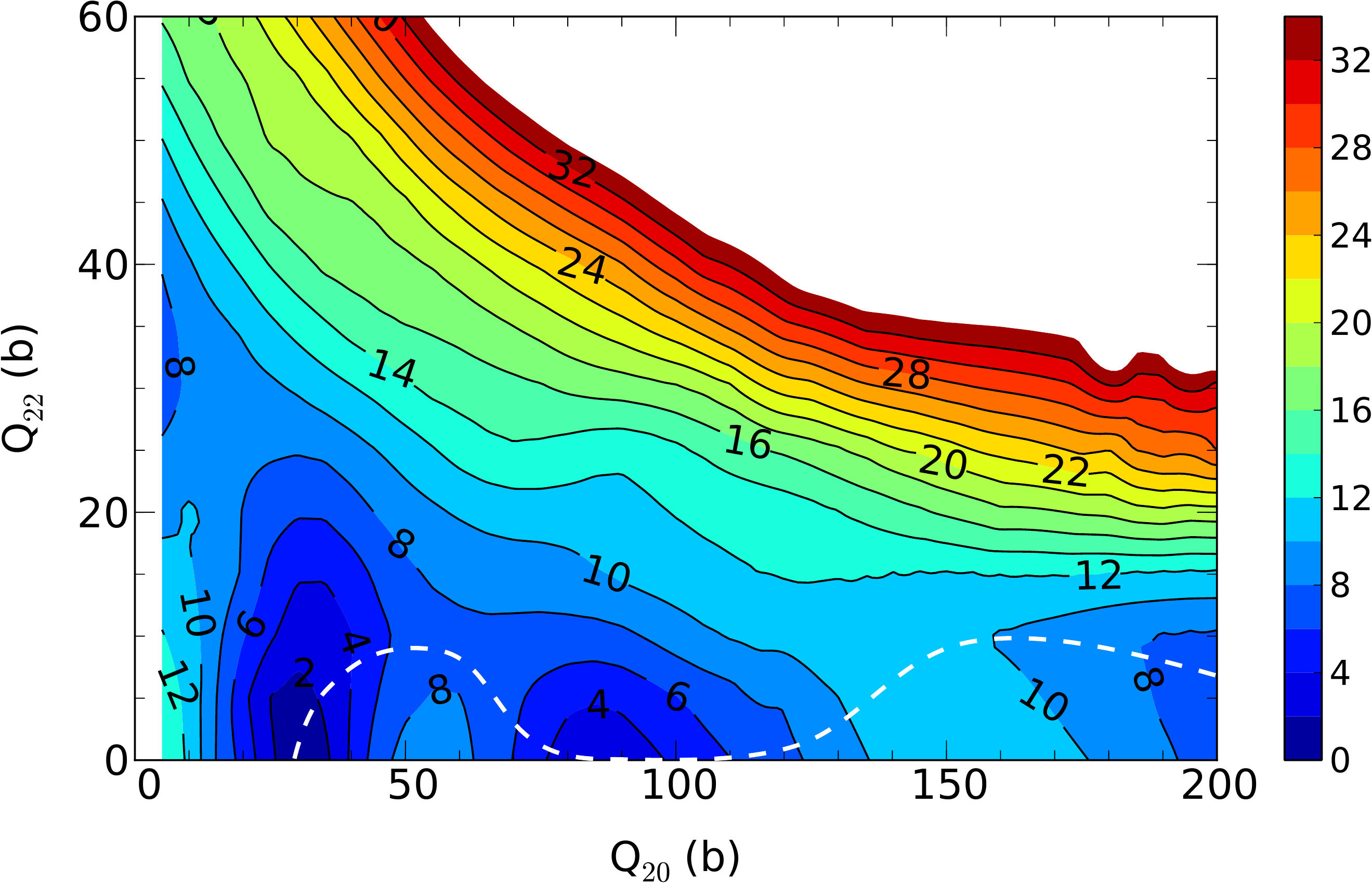}
\includegraphics[width=0.45\linewidth]{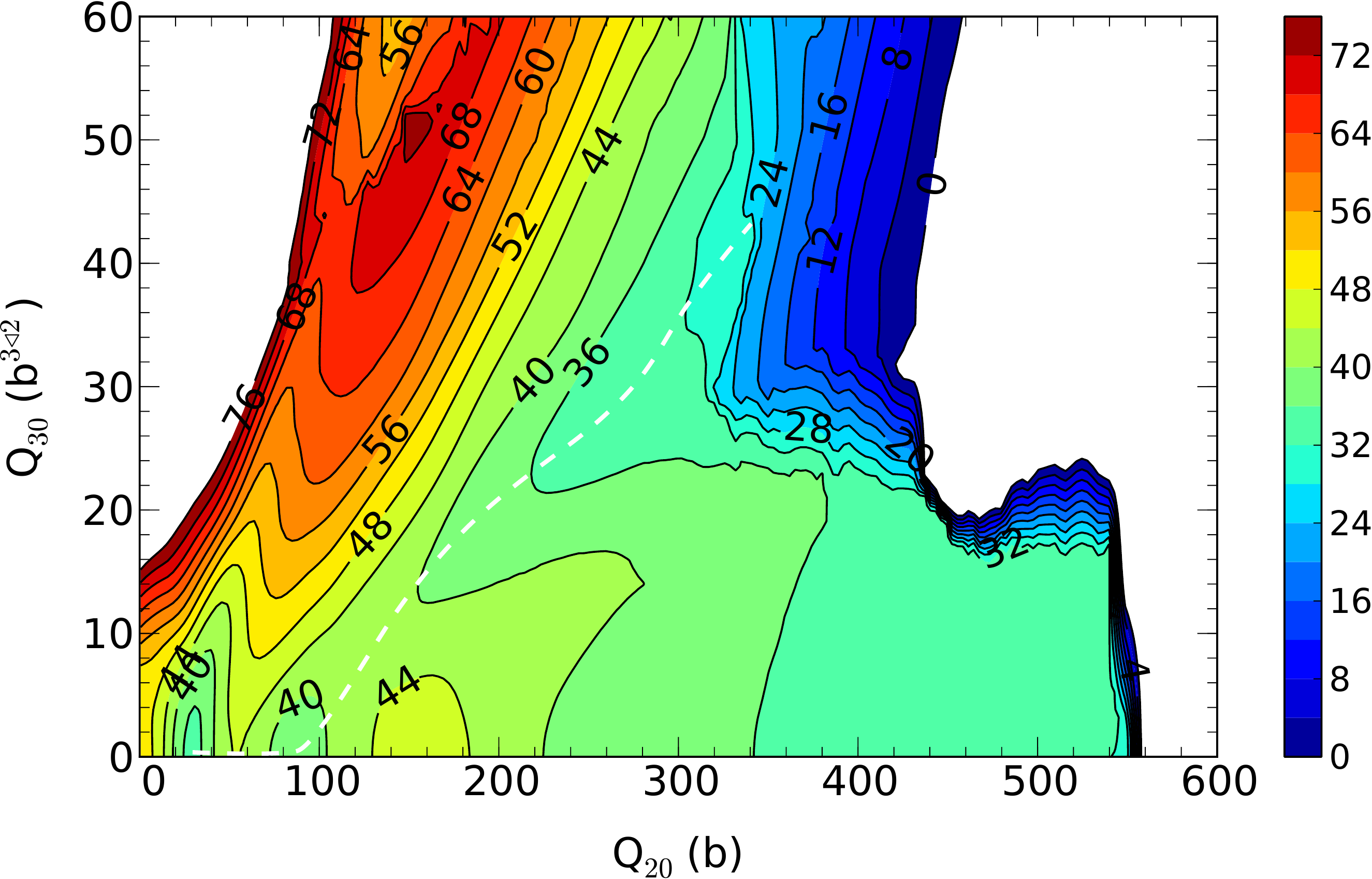}
\includegraphics[width=0.45\linewidth]{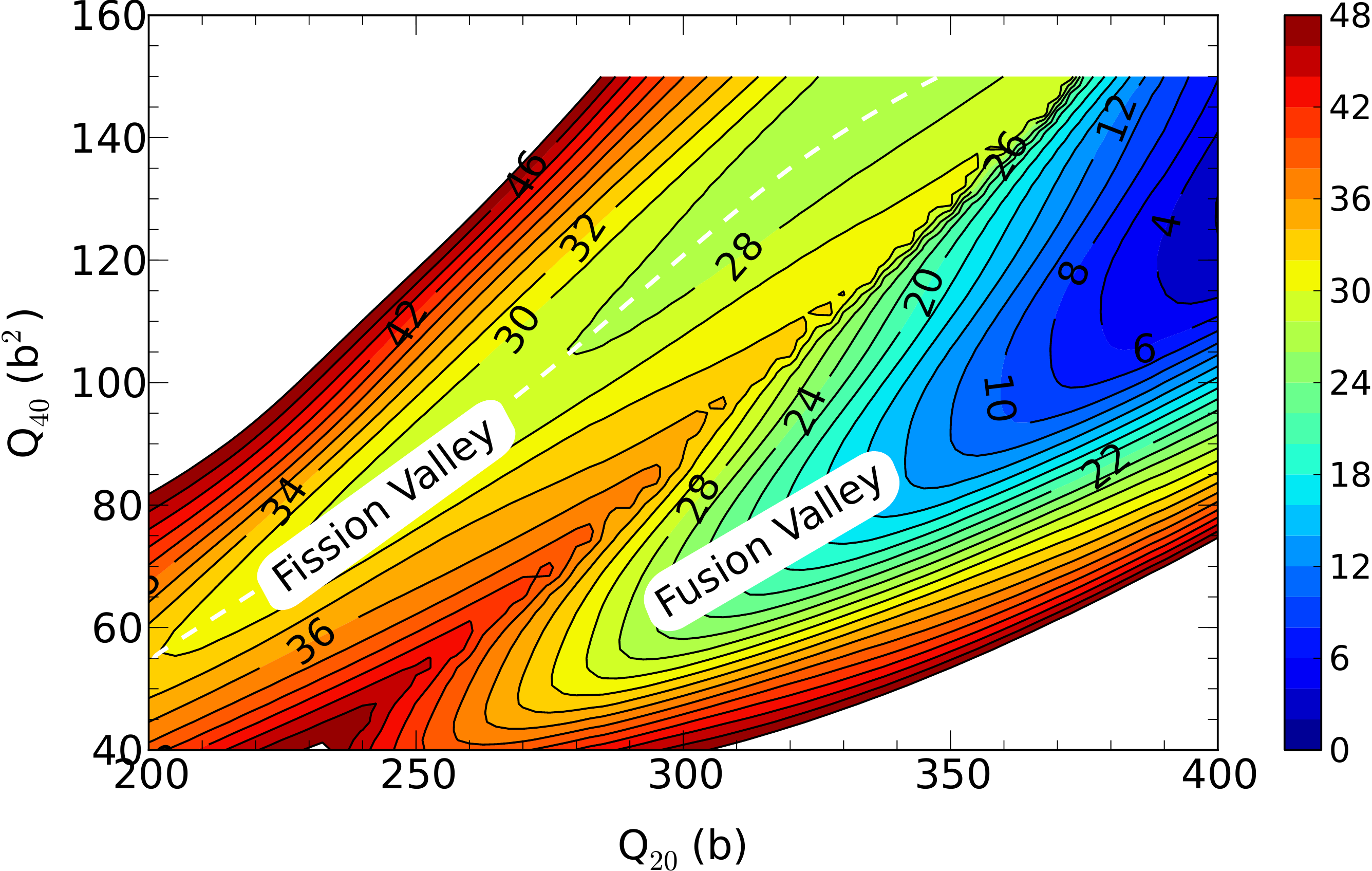}
\includegraphics[width=0.45\linewidth]{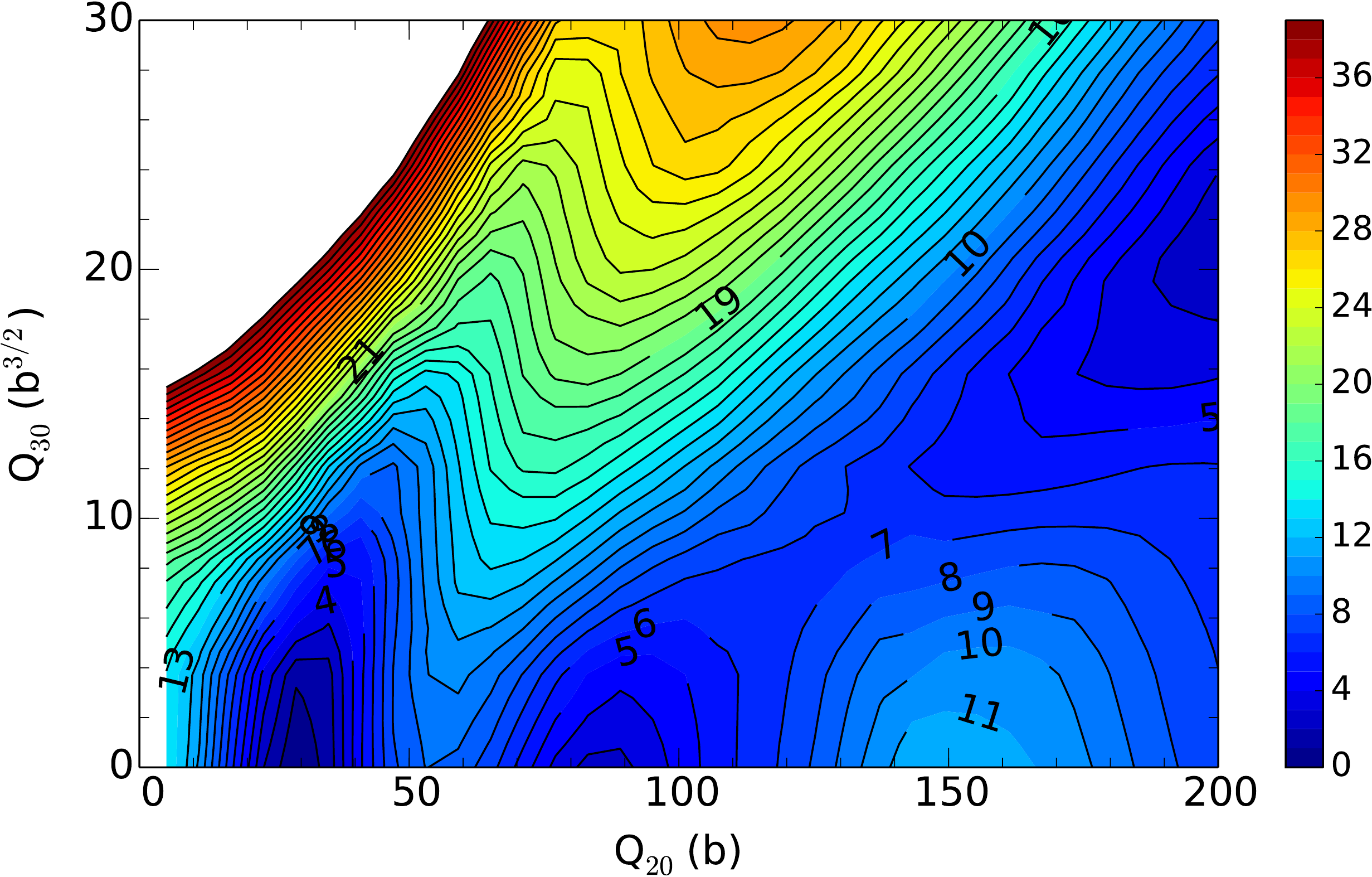}
\caption{Contour plots of the potential energy surface for $^{240}$Pu in the
$(q_{20},q_{22})$ collective space (top left), $(q_{20},q_{30})$ (top right), 
and $(q_{20},q_{40})$ (bottom left). The bottom right panel shows a close-up 
of the $(q_{20},q_{30})$ surface in the vicinity of the ground-state. 
Calculations were performed with the SkM* parametrization of the Skyrme energy 
density with surface-volume pairing according to the details given in 
\cite{schunck2014}.
Figure courtesy of N. Schunck from \cite{schunck2014}; 
copyright 2014 by the American Physical Society.}
\label{fig:pes_pu240}
\end{center}
\end{figure}

Most of the work on neutron-induced fission has been focused on actinide
nuclei, where there is a large amount of experimental data. As in spontaneous
fission, the essential ingredient is the calculation of potential energy
surfaces. Until now, only two-dimensional collective spaces have been
explicitly considered in the literature in the context of induced-fission within DFT.
In \cite{berger1984,berger1986}, Berger and Gogny from the
Bruy\`eres-le-Ch\^atel group were the first to compute fully microscopically
the PES in the $(q_{20},q_{30})$ and $(q_{20},q_{40})$ collective spaces at the
HFB level with a finite-range Gogny potential using a two-centre HO basis.
Similar calculations were performed later by Goutte and collaborators in
$^{238}$U in \cite{goutte2004,goutte2005}, and by Dubray and collaborators
in Th and Fm isotopes in \cite{dubray2008}. Recently, Younes and Gogny
provided additional information, such as the position of the scission line, energy 
for pre- and post-scission configurations, number of particles in the neck at scission, for the $(q_{20},q_{30})$ PES with the D1S
Gogny force in \cite{younes2009}. Schunck and collaborators provided a
similar analysis of the PES in $^{240}$Pu while including axial-symmetry
breaking within the Skyrme DFT framework in \cite{schunck2014}. In the
follow-up work presented in \cite{schunck2015}, the same authors also
calculated the evolution of the most likely fission path as a function of
nuclear temperature.

\begin{figure}[!ht]
\begin{center}
\includegraphics[width=0.70\linewidth]{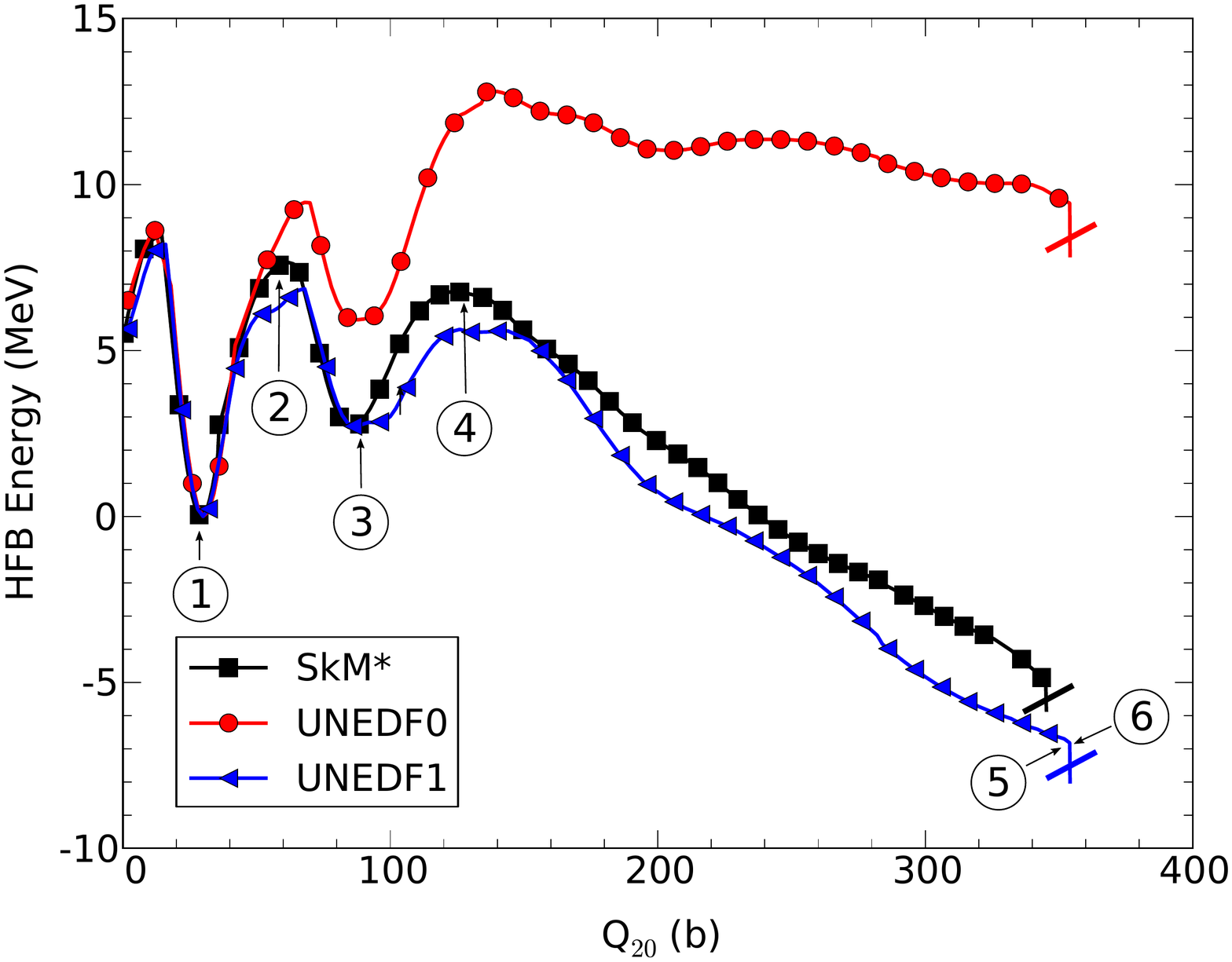}
\includegraphics[width=0.70\linewidth]{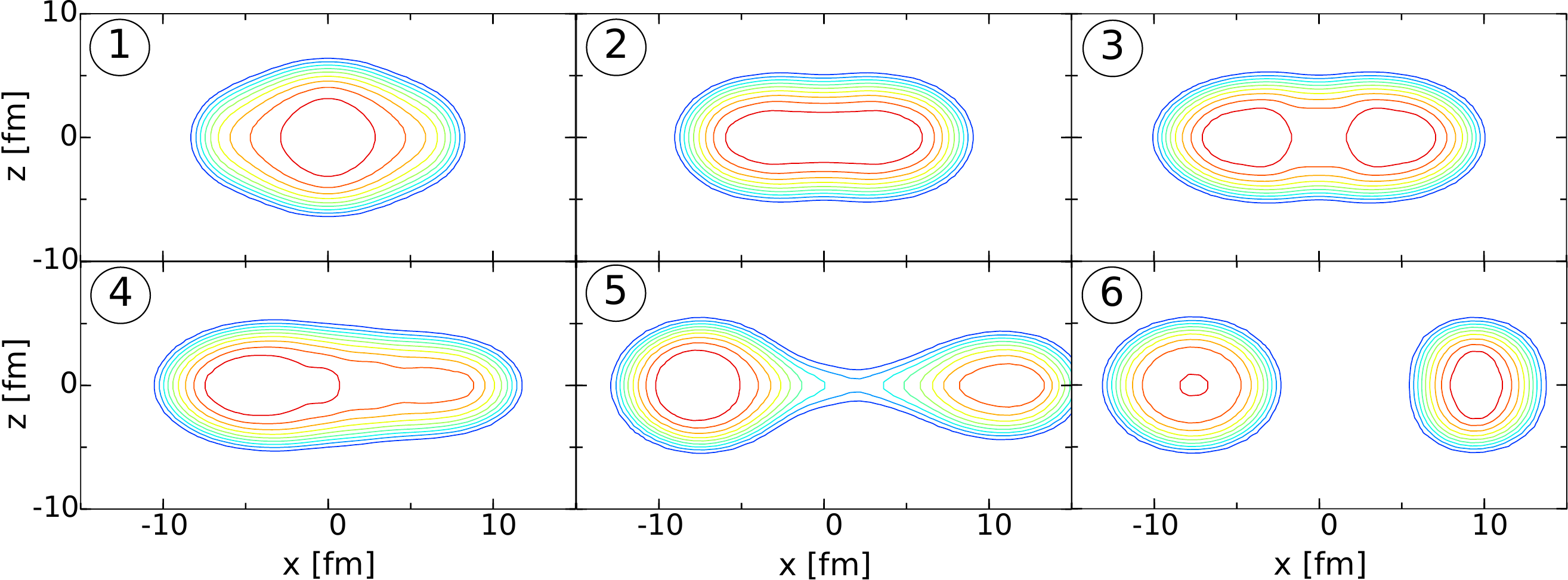}
\caption{Total energy along the most likely fission pathway in $^{240}$Pu. 
Calculations were performed with the SkM* parametrization of the Skyrme energy 
density with surface-volume pairing according to the details given in 
\cite{schunck2014}.
Figure courtesy of N. Schunck from \cite{schunck2014}; 
copyright 2014 by the American Physical Society.}
\label{fig:pes_pu240_1D}
\end{center}
\end{figure}

Calculations with Skyrme and Gogny show similar features for $^{240}$Pu: the
most likely fission pathway emerges into the fission valley (high $q_{40}$
values, see bottom left panel in figure \ref{fig:pes_pu240}) which coexist 
with the fusion valley (low $q_{40}$ values). This most likely fission path 
is clearly asymmetric, with values of $q_{30}$ at scission of the order of 45 
b$^{3/2}$, see top right and bottom right panels in figure \ref{fig:pes_pu240}. 
Another very asymmetric fission path exists at very high energy (cluster 
emission). Figure \ref{fig:pes_pu240_1D} shows the energy along the most likely 
fission pathway in that nucleus (top panel) along with illustrations of the 
density profiles at various points: ground-state, top of the first barrier, 
fission isomer, top of the second barrier, just before scission and just after 
scission.

As discussed in section \ref{subsubsec:multipole}, multipole moment operators
are not always the most adequate collective variables. In particular near
scission, two-dimensional potential energy surfaces obtained in the $(q_{20},q_{30})$ collective space show
marked discontinuities, reflecting the increasing role of additional degrees of freedom. As a consequence, 
the number of HFB iterations needed to reach convergence may increase substantially as the 
system has to explore a very large variational space characterized by changes in many different collective variables. 
The most important consequence of these discontinuities at scission is that some particular mass splits 
are missing along the scission line. Therefore, the sum of all probabilities $y(A)$ of having one of the 
fragment of mass $A$ is not equal to 1. Since fission product yields are normalized by this 
probability, recall (\ref{eq:yield_normalized}), the lack of given mass splits skews the actual distribution in 
an uncontrolled way.
In \cite{younes2012-a}, Younes and Gogny therefore suggested to use collective 
variables similar to the ones used in macroscopic-microscopic approaches and are 
directly related to the distances between the two pre-fragment and their mass asymmetry. We 
showed in figure \ref{fig:pes_Dxi} page \pageref{fig:pes_Dxi} an example of a PES in 
this new collective space.

\begin{figure}[!ht]
\begin{center}
\includegraphics[width=0.75\linewidth]{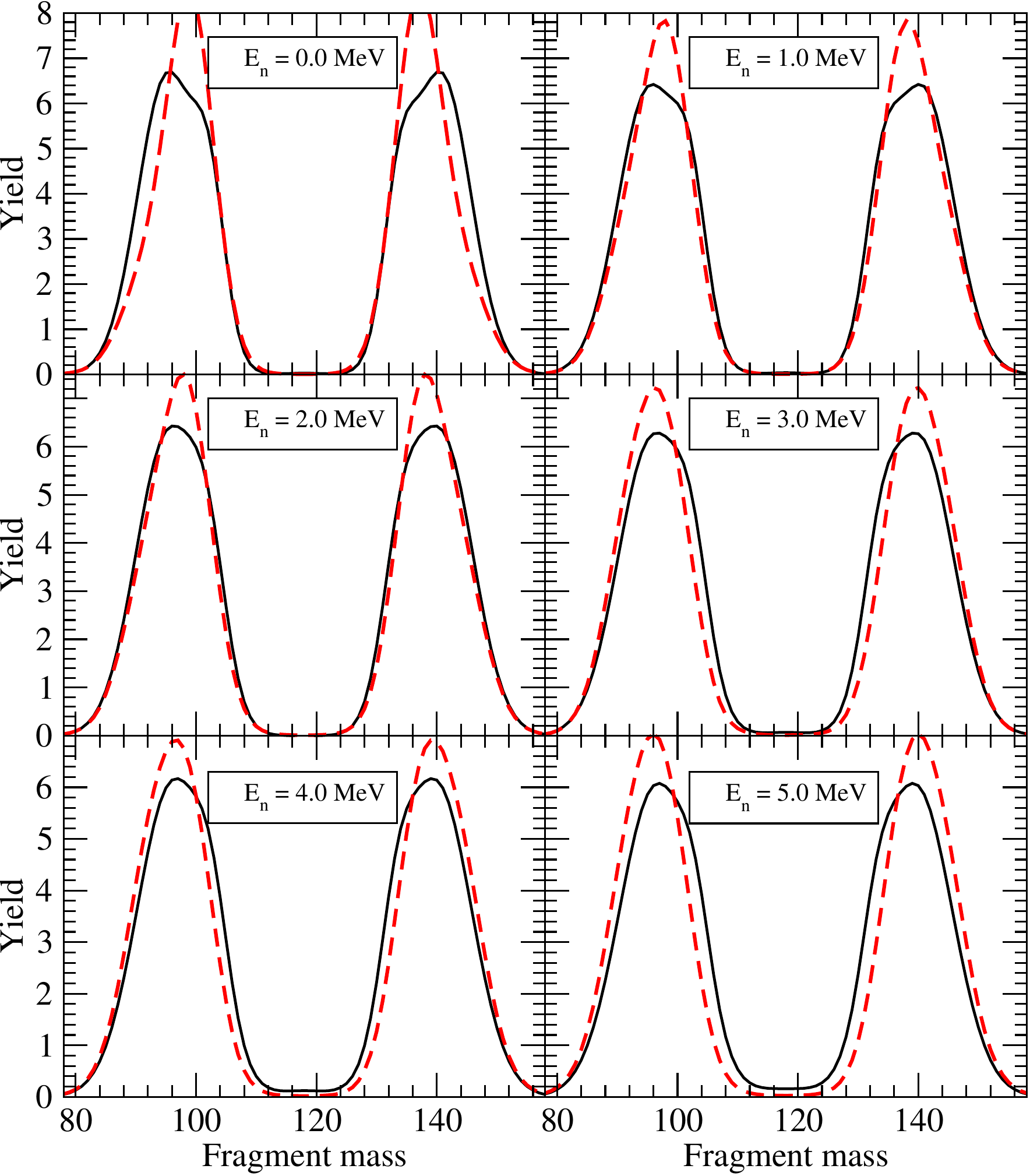}
\caption{Fission fragment mass distributions for $^{239}$Pu(n,f) as a
function of neutron incident energy given in \cite{younes2012-a}. The
results were obtained by evolving the nuclear collective wave-packet
with the TDGCM in the ($q_{20},q_{30})$ space up to scission. The
initial energy of the wave packet is defined by $E_{0} = E_{A} + E_{n}$,
where $E_{A}$ is the height off the first fission barrier. 
Figure courtesy of W. Younes from \cite{younes2012-a}.}
\label{fig:fpy}
\end{center}
\end{figure}

In the adiabatic approximation, the calculation of potential energy
surfaces and identification of scission configurations is the first
step required before the full calculation of fission fragment properties.
Among the latter, charge and mass distributions can be obtained by
solving the TDGCM equations as outlined in section \ref{subsubsec:tdgcm}.
The method was introduced by the French group at Bruy\`eres-le-Ch\^atel
in the nineteen eighties \cite{berger1984,berger1986,berger1989,berger1991}.
The first realistic calculations of fission fragment mass distributions
were reported by Goutte {\it et al.} in $^{237}$U(n,f) in \cite{goutte2004,
goutte2005}. The same overall approach was used by Younes and Gogny a few
years later, the main difference being the change of collective variables
discussed above $(q_{20},q_{30}) \rightarrow (D,\xi)$, and the use of the
neck degree of freedom to improve the description of the scission region.
The figure \ref{fig:fpy} taken from \cite{younes2012-a} shows the best
result obtained so far for the mass distributions $^{239}$Pu(n,f) using
the combination of DFT plus TDGCM. Fission product yields were convoluted
with a Gaussian folding function of width $\sigma=3.5$ to account both
for experimental uncertainties on pre-neutron emission mass yields (about
2-3 mass units) and theoretical uncertainties on particle number at scission
(between 2 and 5 mass units); see also discussion of open questions in
section \ref{subsec:open}.

\begin{figure}[!ht]
\begin{center}
\includegraphics[width=0.80\linewidth]{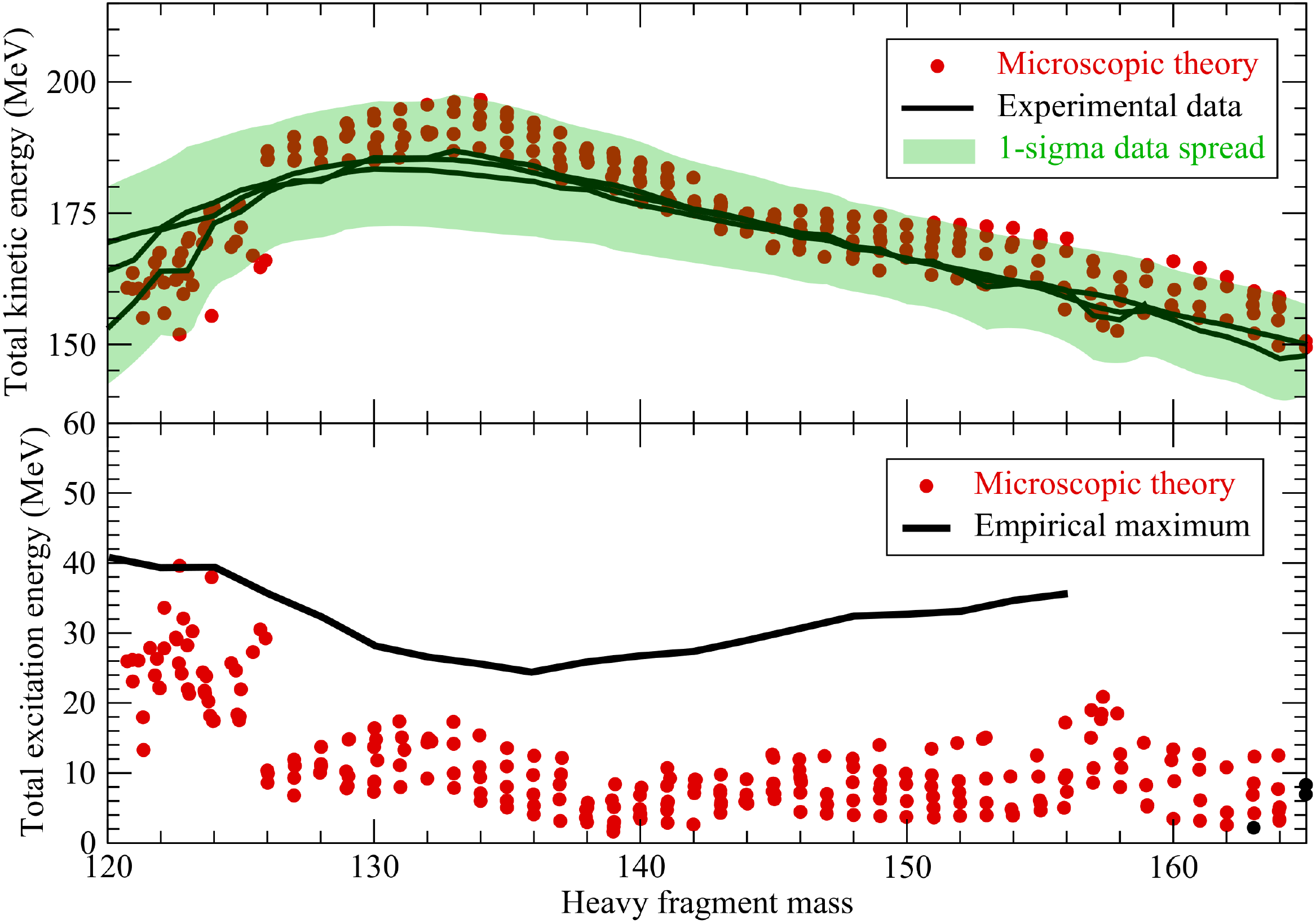}
\caption{Estimates of total kinetic energy (TKE, top panel) and total
excitation energy (TXE, bottom panel) for $^{239}$Pu(n,f). Calculations
were performed at the scission configurations of $^{240}$Pu with the
Gogny force in a three-dimensional collective space $(D,\xi, q_{N})$
(distance between the fragments and mass asymmetry, size of the neck).
Figure courtesy of W. Younes from \cite{younes2013}.}
\label{fig:tke}
\end{center}
\end{figure}

One of the most important quantities in induced-fission is the total
kinetic energy (TKE) carried out by the fission fragments. Early estimates  
of TKE in \cite{dubray2008,goutte2004} within the adiabatic approximation were based on assuming two spherical 
uniform charge distributions of $Z_1$ and $Z_2$ protons separated by a distance $d$, with 
the three quantities $Z_1$, $Z_2$ and $d$ extracted from the characteristics of the PES at scission. 
More realistic calculations of TKE involve precisely identifying the
scission region and disentangling the fragments by localizing them as discussed 
in section \ref{subsubsec:dynamical}. The TKE is then obtained by computing 
the direct Coulomb energy between the two charge distributions of the fragments -- which 
takes into account the deformation of the fragments. To our knowledge, this 
entire procedure was only applied by Younes and collaborators in \cite{younes2013} 
for the Gogny D1S force using a collective space composed of the collective 
variables $(D,\xi)$ (distance between the fragments and mass asymmetry). After 
the scission line has been obtained in that space, constraints on the size of 
the neck were added to remove the effect of discontinuities, and the two 
fragments were localized to reproduce asymptotic conditions. The results are 
shown in figure \ref{fig:tke} and show a very good agreement with experimental 
data.

\begin{figure}[!ht]
\begin{center}
\includegraphics[width=0.80\linewidth]{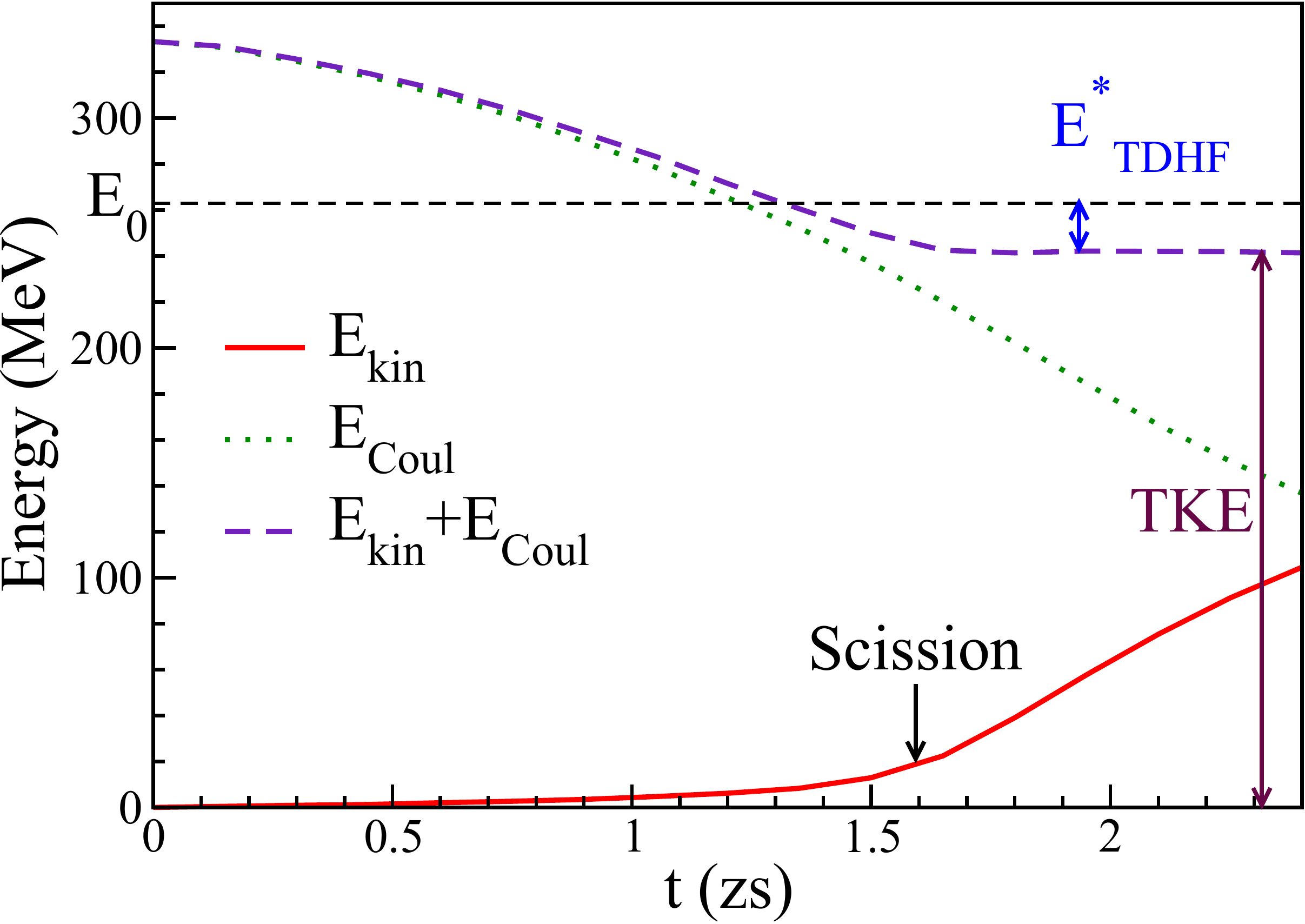}
\caption{Time evolution of the several energies during the fission of
$^{258}$Fm: $E_{0}$ is the mean TDHF energy (constant of motion),
$E_{\mathrm{Coul}}$ is the Coulomb repulsion energy between the two
fragments, while $E_{\mathrm{kin}}$ is the kinetic energy of both
fragments. By definition, $E^{*}_{\mathrm{TDHF}}$ represents the total
excitation energy of the fission fragments in the TDHF approach.
Figure taken from Ref. \cite{simenel2014}, courtesy of C. Simenel;
copyright 2014 by The American Physical Society.}
\label{fig:tdhf}
\end{center}
\end{figure}

As already discussed, the drawback of such an adiabatic calculation of TKE is
the dependence on the criterion used to define scission configurations.
Calculations of TKE are, therefore, more rigorously defined in the non-adiabatic
approach to fission dynamics based on TDDFT, since there is no need to
characterize scission (the nucleus ``automatically'' splits as a function of
time) and to disentangle the fragments. In TDDFT, the kinetic energy of the
fission fragments is calculable as a function of time, as is the direct
Coulomb energy, and the TKE is simply the sum of the two contributions.
Recently, Simenel and Umar have reported the first calculation of TKE in the
fission of $^{258}$Fm using the TDHF approximation to TDDFT. The results are shown in figure \ref{fig:tdhf}: the
clear advantage of TDHF over non-adiabatic methods is the possibility to
account for the transfer of energy between Coulomb repulsion and kinetic
energy of the fragments as a function of time owing to the conservation in energy. On the other hand, such
calculations remain expensive and can be performed only for a few cases: 
computing full TKE distributions (like mass or charge distributions) would 
require orders of magnitude increases in computing power.

Many uncertainties impacting the applications of nuclear fission come from the 
challenge of computing realistic estimates of fission fragment properties 
in a fully microscopic framework, especially at large excitation energy of 
the fissioning nucleus. In \cite{simenel2014} the {\it total} excitation 
energy TXE in the fission of $^{258}$Fm was extracted by again taking advantage of 
the conservation of energy in TDHF, as shown in figure \ref{fig:tdhf}: $E_{0} = 
\mathrm{TKE} + \mathrm{TXE}$. In principle, the TDHF framework could also 
provide a consistent framework to extract individual fragment properties, 
including their excitation energy. However, as discussed in \cite{goddard2014} 
in the particular case of $^{240}$Pu, one of the challenges is then to extract 
from the total TDHF excitation energy the contribution of the intrinsic 
excitations of the system only (be it of single-particle or collective nature). 
In the adiabatic approach to fission, the calculation of excitation energies is 
technically straightforward, but it depends very sensitively on the actual 
definition of scission configurations, on the localization of the fission 
fragments, and on the amount of dissipation of pre-scission energy into 
collective modes as discussed in \cite{younes2012}. Results reported in 
\cite{younes2011,younes2013} for $^{240}$Pu and shown in the bottom panel of 
figure \ref{fig:tke} do not reproduce experimental data very well, but come with large 
uncertainty bands. 

\begin{figure}[!ht]
\begin{center}
\includegraphics[width=0.80\linewidth]{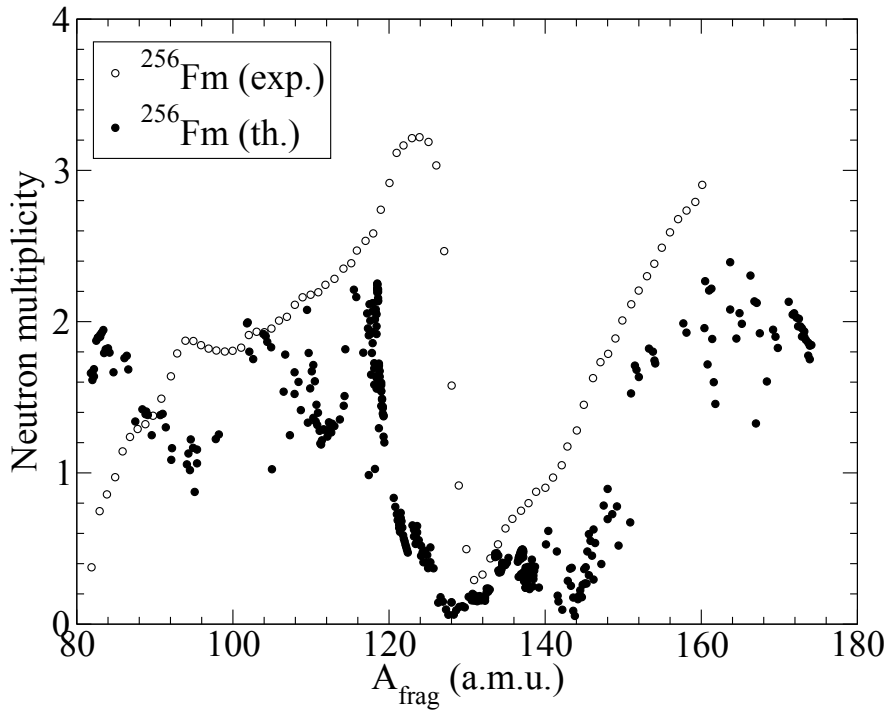}
\caption{Average prompt neutron multiplicity from fission fragments of
$^{258}$Fm calculated from (\ref{eq:results_nubar}) for the Gogny D1S
effective nuclear force.
Figure taken from Ref. \cite{dubray2008}, courtesy of N. Dubray;
copyright 2008 by The American Physical Society.}
\label{fig:nubar}
\end{center}
\end{figure}

Charge and mass distributions and TKE/TXE are natural outputs of DFT+TDGCM or 
TDFT calculations. In the present status of the theory, this is not the case for 
other important observables such as the multiplicity of the prompt neutrons 
emitted from the fragments. In this latter case, the average neutron multiplicity 
$\bar{\nu}$ can be obtained from a simple energy balance equation that defines 
how the total excitation energy available to the fragment can be distributed in 
various decay modes (neutrons, gamma, etc.). In \cite{dubray2008}, it was assumed 
that the excitation energy took the form of deformation energy, leading to
\begin{equation}
\bar{\nu}_{\mathrm{frag}} = 
\frac{E_{\mathrm{def}}}{\langle E_{\nu}\rangle + \bar{B}_{n}^{*}}
\label{eq:results_nubar}
\end{equation}
where $E_{\mathrm{def}}$ is the deformation energy of the fragment with
respect to its ground-state deformation, $\bar{B}_{n}^{*}$ is the (average) 
one-neutron separation energy in the fragment and $\langle E_{\nu}\rangle$ 
is an estimate of the mean energy of the emitted neutron (often taken from 
experimental data or evaluations). The figure \ref{fig:nubar} shows a comparison 
between theoretical predictions of the average neutron multiplicity, based on 
(\ref{eq:results_nubar}) and inputs from DFT calculations with the D1S 
parametrization of the Gogny force, and experimental data. The sawtooth feature 
of $\bar{\nu}$ is properly reproduced. Note that more realistic estimates of 
the neutron spectrum are typically obtained from reaction theory calculations, 
where fission fragment charge, mass, TKE and TXE distributions are important 
inputs; see for instance the code {\sc freya} of \cite{verbeke2015} and 
references therein.

In HFB calculations, particle number is only conserved 
on average. This point was already mentioned in section \ref{subsubsec:beyond} 
in the context of beyond mean-field corrections that can impact the potential 
energy surface. Particle number symmetry breaking has also consequences for 
fragment properties:
\begin{itemize}
\item In adiabatic approaches, fission fragments at scission are characterized 
by the functions $\rho_{\mathrm{f}}(\gras{r})$ with f=1,2 identifying the 
fragment, see (\ref{eq:densities_fragments}). Although these functions 
resemble the one-body density matrix of the HFB theory, they are not; in particular, 
the corresponding object
\begin{equation}
\mathcal{R}_{\mathrm{f}} = \left( \begin{array}{cc}
\rho_{\mathrm{f}} & \kappa_{\mathrm{f}} \\
-\kappa^{*}_{\mathrm{f}} & 1-\rho^{*}_{\mathrm{f}}
\end{array}\right)
\end{equation}
is not a projection operator ($\mathcal{R}_{\mathrm{f}}^{2} \neq \mathcal{R}_{\mathrm{f}}$). As a result, 
$\rho_{\mathrm{f}}$ and $\kappa_\mathrm{f}$ have been dubbed ``pseudodensities'' 
in \cite{schunck2014}. The charge and mass of the fission fragments along the 
scission configurations, however, are obtained by integration over space of 
these functions: as a result the charge and mass of fission fragments coming out 
of DFT calculations are often non-integer numbers.
\item This leads to uncertainties in the theoretical calculation of 
fission fragment yields in adiabatic approaches, such as those shown in figure \ref{fig:fpy}. First, the 
yield for each integer fragment mass $A$ is often obtained by summing all 
contributions from  all non-integer fragment masses $a$ such that 
$a\in[A-0.5,A+0.5]$. Then, the HFB wave function for mass $a$ is itself the 
superposition of several wave functions with good particle number, schematically 
$|a\rangle = \sum_{A} c_{A} |A\rangle$, Coefficients $c_{A}$ could be extracted by 
particle number projection, but the usual techniques are not easily applicable at 
scission because of the high degree of entanglement of the fragments, see below.

\begin{figure}[!ht]
\begin{center}
\includegraphics[width=0.90\linewidth]{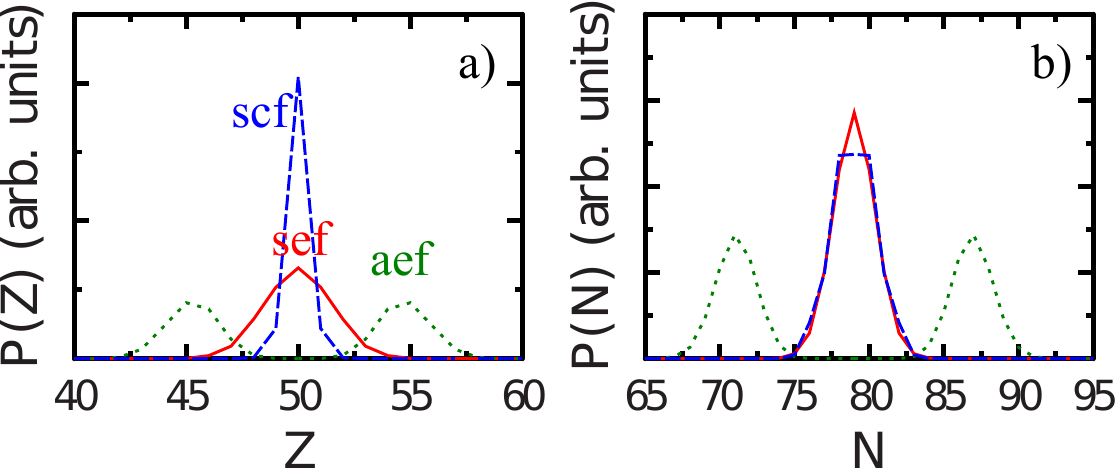}
\caption{Proton (a) and neutron (b) number distributions in the fission
fragments of $^{258}$Fm for the symmetric compact mode (scf, dashed blue),
symmetric extended mode (aef, plain red) and asymmetric extended mode
(aef, dotted green).
Figure taken from Ref. \cite{scamps2015-a}, courtesy of G. Scamps;
copyright 2015 by The American Physical Society.}
\label{fig:tdhfbcs_PNP}
\end{center}
\end{figure}

\item Particle number symmetry breaking also occurs in non-adiabatic approaches described by TDDFT as soon as 
pairing correlations are included. This is the case for the TDHF+BCS or TDHFB 
approximations to TDDFT. Just as 
in adiabatic approaches, fission fragments may have non-integer particle numbers. 
The advantage of TDHF+BCS or TDHFB, however, is that particle number projections techniques are 
readily applicable by simply extending the definition of the particle number 
following, e.g., the method proposed by Simenel in \cite{simenel2010}. Assuming 
the full space is partitioned in two regions, one defined by $r>0$ and the other by 
$r<0$, and that the fragment f is in the region $r>0$, then the particle number 
operator for isospin projection $\tau$ and for the fragment f reads
\begin{equation}
\hat{N}^{(\tau)}_{\mathrm{f}} 
= 
\sum_{\sigma}\int d^{3}\gras{r}\; \rho^{(\tau)}(\gras{r}\sigma,\gras{r}\sigma)H(r)
\end{equation}
where $H(r)$ is the Heaviside function, $H(r)=1, r>0$ and 0 elsewhere. $\rho^{(\tau)}(\gras{r}\sigma,\gras{r}\sigma)$ 
is the coordinate space representation of the one-body density matrix $ \rho_{ij}^{(\tau)}$ for the 
fissioning nucleus, which can be readily obtained by expanding (\ref{eq:pes_densities})
\begin{equation}
\rho^{(\tau)}(\gras{r}\sigma,\gras{r}\sigma)
=
\sum_{ij} \rho_{ij}^{(\tau)}\varphi^{*}_{i}(\gras{r}\sigma)\varphi_{j}(\gras{r}\sigma),
\end{equation}
where the $\varphi_{i}(\gras{r}\sigma)$ are the wave functions associated with the 
creation and annihilation operators $(c_{i},c_{i}^{\dagger})$ introduced in section 
\ref{subsubsec:hfb} page \pageref{subsubsec:hfb}. 
In \cite{scamps2015-a}, Scamps and 
collaborators used this technique to extract the decomposition of the fragment 
wave function into the sum of good-particle number components. The results are 
shown in figure \ref{fig:tdhfbcs_PNP} for the particular case of $^{258}$Fm. It 
is important to note that the spread in particle number for the 
fragment strongly depends on the fission pathway. 
\end{itemize}

%
%

\newpage
\section{Conclusions}
\label{sec:conclusions}


The purpose of this article was to review the current state of the 
microscopic methods used to describe nuclear fission. By ``microscopic'', 
we meant a theory centred on quantum many-body methods and the use of 
explicit nuclear forces, in contrast with semi-phenomenological approaches 
that are based on the liquid drop picture of the nucleus (with various 
corrective terms). At this point, a direct, {\it ab initio} approach to 
fission remains an utopia, and more effective methods must be used. Nuclear 
density functional theory, in its various formulations, provides the best 
microscopic framework to study fission. It is at the intersection of both 
phenomenological approaches -- since it is built upon the notion of 
independent particles in an independent mean field potential, and of the 
theory of nuclear forces and {\it ab initio} methods -- since EDF are 
often derived from effective nuclear forces. 


\subsection{Summary}
\label{subsec:summary}

In our discussion, we have tried to break down the DFT approach to 
fission into several elementary steps. We first emphasized in section 
\ref{sec:pes} the important hypothesis of adiabaticity, which remains one 
of the cornerstones of the theoretical view of fission and has its origin 
in phenomenological models. In the adiabatic approximation, it is assumed 
that fission is not driven by all of the nucleon degrees of freedom, 
but only by a small number $n$ of collective variables. Furthermore, it is 
also assumed, somewhat implicitly, that there is a perfect decoupling
between these $n$ collective degrees of freedom and the intrinsic 
excitations of the fissioning nucleus. These hypotheses were originally 
formulated by Bohr and Wheeler in their seminal paper \cite{bohr1939} on fission. 

In practice, the hypothesis of adiabaticity is implemented by choosing a 
small set of collective variables. There are few constraints on the type 
and nature that these collective variables must have, and choosing the 
right ones is something akin to an art. Fortunately, one can draw from the
large body of experience accumulated in phenomenological models of 
fission. Practitioners of the liquid drop model have developed very 
sophisticated parametrizations of the nuclear surface over the years, see 
section \ref{subsubsec:surface}, and have investigated the role of several 
degrees of freedom such as mass asymmetry, triaxiality, pairing. The 
same degrees of freedom can be explicitly introduced in DFT thanks to 
the mechanism of spontaneous symmetry breaking of the EDF and the use of 
constraint operators.

To make this clear, we have deliberately adopted a modern presentation of 
nuclear DFT, where the central object is a symmetry-breaking energy density 
functional treated at the Hartree-Fock-Bogoliubov approximation. We have 
thus recalled in section \ref{subsubsec:hfb} the basic features of the HFB 
approximation in the general case where the energy is only given as a 
functional of the density matrix $\rho$ and pairing tensor $\kappa$, which 
has the merit of highlighting the actual degrees of freedom of the theory.  
We have also summarized in section \ref{subsubsec:bcs} the BCS approximation to 
HFB, which remains popular in many applications. 
In practice, the EDF is often derived from an effective two-body nuclear 
potential, and we have recalled in section \ref{subsubsec:edf} the two most 
popular choices that have been used in fission, namely the Skyrme and Gogny 
force. The determination of a universal (=applicable to all nuclear 
properties) functional is currently a very active area of research, and we 
could only allude to some of the on-going work, whether on the form of the 
functional or on the determination of its parameters. We have also given a 
brief overview of the finite-temperature HFB theory in section 
\ref{subsubsec:excitation}, as it has been in practice one of the most 
popular tools to study the evolution of fission properties, both spontaneous 
and induced, with excitation energy. Symmetry-breaking manifests itself, 
among others, by quantum fluctuations that are associated with zero-point 
energy corrections, and section \ref{subsubsec:beyond} contains a discussion of some 
of the most important ones.

As already mentioned, the definition of the right collective variables is 
an important ingredient in the adiabatic view of fission. In section 
\ref{subsec:collective}, we have reviewed the various collective variables 
that have been identified over the years as essential. Most of them are 
``geometrical'', in the sense that they correspond to the actual 
distribution of mass in the intrinsic frame of the nucleus, see section 
\ref{subsubsec:multipole}, but collective variables associated with pairing 
correlations have recently been investigated and shown to be important. There 
are different ways to introduce pairing collective variables, which are 
briefly discussed in section \ref{subsubsec:other}.

One of the unpleasant consequences of adiabaticity is that the theory does 
not explicitly contain any scission mechanism. The importance of scission 
configurations had been recognized very early on the study of stability  
conditions for liquid drops. Extending these notions to DFT, one may define 
scission configurations based on some condition on the density of particles: 
if the density in the neck between the two pre-fragments is ``small enough'', 
one may decide that the system has scissioned. This is clearly unsatisfactory, 
and various other geometrical criteria have been explored and are discussed 
in section \ref{subsubsec:geometric}. These geometrical definitions are 
somewhat limited, though, and cannot really account for the complex physics 
of scission. For these reasons, it has been suggested recently to define 
scission quantum-mechanically by taking advantage of the invariance of the 
HFB solutions under a unitary transformation. This technique, briefly 
presented in \ref{subsubsec:dynamical}, gives a more realistic description 
of the fission fragments at scission.

Given a realistic nuclear EDF, a set of collective variables and a criterion 
to identify scission configurations, one then maps out the collective space in 
which fission will (adiabatically) take place by solving the HFB equations 
under a set of constraints. Confining the dynamics of fission in this 
collective space requires introducing a collective inertia tensor, which, 
roughly speaking, represents the overall mass of the nucleus as it 
evolves across the collective space. The collective inertia plays a major 
role both in spontaneous fission and in induced fission, and we give in  
section \ref{subsec:inertia} a comprehensive account of the two main 
techniques used to compute it. 
\begin{itemize}
\item In the generator coordinate method presented in section \ref{subsubsec:gcm}, 
the nuclear wave function is written as a linear superposition of HFB states 
corresponding to different values of the collective variables. The 
coefficients of this superposition can be determined by applying the 
variational principle. If one assumes a Gaussian form for the overlap 
kernel of the HFB states, it is possible to derive a collective Schr\"odinger-like 
equation that governs the dynamics of the system in the collective space. 
The collective inertia tensor appears in the kinetic energy part of this 
collective Hamiltonian and in the zero-point energy corrections to the collective potential. 
\item In the adiabatic time-dependent HFB theory outlined in section 
\ref{subsubsec:atdhfb}, the starting point is a low-momentum expansion of the 
full time-dependent HFB density. This expansion yields a set of coupled 
equations that drive the collective dynamics. Under the (most common) 
assumption that the collective space is predetermined as a set of 
constrained HFB calculations, it is also possible to extract a classical 
collective Hamiltonian, with an expression for the collective inertia that 
is more realistic than the GCM (for identical collective spaces).
\end{itemize}

The calculation of collective inertia, whether in the GCM or ATDHFB 
approximation, can be performed at each point of the collective space 
with only the knowledge of the HFB wave function or, equivalently, the 
HFB densities, at that point. In a sense, one could argue that the calculation 
of the PES and of collective inertia completes the static part of a DFT 
simulation of fission. For the dynamics itself, one has to distinguish 
between spontaneous and induced fission:
\begin{itemize}
\item
As pointed out in the introduction, spontaneous fission is essentially 
characterized by fission half-lives. The calculation of this observable
is presented in section \ref{subsec:t12}. It is done in analogy with the 
standard problem of tunnelling in quantum mechanics. Starting from a 
(multi-dimensional) potential energy surface -- the collective PES, one 
uses the WKB approximation to compute the action between inner and outer 
turning points. The half-life is then proportional to the exponential 
of the action corresponding to the least-action principle. The mass of 
the nucleus is represented by the collective inertia tensor. Section 
\ref{subsubsec:wkb} gives a quick overview of the various formulas 
involved. There have been early attempts to generalized this framework 
by using path integrals methods instead of the WKB approximation. For the sake of completeness, 
we outline this approach in section \ref{subsubsec:instantons}, although 
we should point out that no practical calculation has been performed so 
far in this framework.
\item
Induced fission is mostly concerned with the properties of the fission 
fragments, and it often involves an explicit time-dependent evolution of the 
system as discussed in section \ref{subsec:induced}. We chose to recall 
in section \ref{subsubsec:langevin} some of the classical methods used 
to describe to fission dynamics, based on the Langevin and Kramers equations, 
since these approaches can be coupled with DFT inputs on the potential 
energy and the collective inertia. Also, the Langevin equation can be 
viewed qualitatively as the classical analogue of time-dependent density 
functional theory, which is presented in section \ref{subsubsec:tddft}. 
TDDFT is the primary non-adiabatic theory that can be applied to studies 
of fission. While it is not suited to studies of spontaneous fission 
(TDDFT can not account for tunnelling), it can provide invaluable 
insights into the physics of scission, which is strongly non-adiabatic. 
On the other hand, the distribution of fission fragments is better 
formulated in the time-dependent extension of the GCM, which gives the 
time-evolution of the collective wave packet in the collective space 
up to the scission configurations. The TDGCM is briefly described in 
section \ref{subsubsec:tdgcm}, together with the techniques used to 
extract fission product yields.
\end{itemize}

In our presentation of results in section \ref{sec:results}, we 
have made the distinction between fission barriers, fission half-lives 
and results on neutron-induced fission.
\begin{itemize}
\item Fission barriers are not observables, but are very important 
inputs to models of the fission spectrum, of nuclear databases, of 
simulations of the r-process, etc. In section \ref{subsec:barriers}, we 
survey some of the results on fission barriers obtained in both relativistic 
and non-relativistic versions of DFT. Note that the most recent DFT 
calculations have a predictive power that is comparable with more 
phenomenological approaches.
\item Fission half-lives are very sensitive to the details of the DFT 
calculation, since they depend exponentially on both the HFB energy and 
the collective inertia. Most often, actinide nuclei are used as benchmarks 
to test the validity of DFT calculations. The real motivation for computing 
spontaneous fission half-lives, however, is in connection with superheavy 
and very neutron-rich nuclei involved in nucleosynthesis. A 
survey of the most recent results is given in section 
\ref{subsec:spontaneous}.
\item There have been comparatively few studies of neutron-induced fission 
in a microscopic setting. One possible reason is the much higher cost of 
performing the calculations, since it is necessary to compute the PES up 
to the scission point. By contrast, estimates of fission half-lives are 
based on the knowledge of the PES around the first fission barrier only. 
We summarize in section \ref{subsec:distributions} the few results published 
in the literature. Note that most of them are less than a decade old.
\end{itemize}


\subsection{Open Questions}
\label{subsec:open}

The only microscopic theory currently capable of predicting fission fragment 
distributions is the TDGCM (under the GOA approximation). As of today, the 
accuracy of fission product yields in actinides is of the order of 30\%. Can we reach 5\% accuracy 
without major changes to the theory, and can we quantify the associated 
uncertainty of such calculations? The latest results suggest that at least 3 
degrees of freedom corresponding to elongation, mass asymmetry and necking 
may be needed. Will constraints on pairing be also necessary to reach this 
level of accuracy? In terms of uncertainties, we already discussed in section  
\ref{subsec:scission} the dependence on the definition of scission configurations. 
In addition to these, the TDGCM+GOA approach is based on symmetry-breaking 
states and thereby inherits many of its limitations. In particular, symmetry 
restoration and beyond mean-field correlations, which were already discussed in 
section \ref{subsubsec:beyond} in the context of PES calculations, most likely 
have an effect on fission product yields. Currently, the flux of the collective wave function 
through a small surface element at the scission point $\gras{q}$ is 
associated with a fragment mass number $a$ (most often non integer as 
discussed in reference to figure \ref{fig:tdhfbcs_PNP}). Assuming the nascent 
fragments have been disentangled with the procedure outlined in 
\ref{subsubsec:dynamical}, how can we extract a realistic estimate of the 
yield of the true fragment with integer charge $Z$ and mass $A$? Will this 
change the current estimates of fission product yields significantly?

Individual properties of fission fragments, in particular their excitation 
energy and their spin, are currently very poorly known. Can a microscopic 
theory based on DFT provide a quantitative understanding of the energy 
sharing mechanism between the fragments? Although static DFT calculations 
combined with the localization procedure of section \ref{subsubsec:dynamical} 
are always feasible, TDDFT seems a more promising framework. Recently, 
pairing correlations have been implemented in TDDFT at the HFB approximation 
in \cite{bulgac2016}, and one may hope that progress in both 
algorithms and hardware will enable more systematic TDHFB calculations in the near future. 
At what accuracy can a full-blown TDHFB calculation predict the individual 
excitation energy of fission fragments in actinide nuclei for thermal 
neutrons?

Better modelling the excitation energy of the nucleus is another area where 
progress can be made. In the past few years, several groups have reported 
successful finite-temperature DFT calculations that reproduce well the 
trend of fission barriers with excitation energy; see, e.g. results in 
\cite{martin2009,sheikh2009,pei2009,mcdonnell2013,schunck2015}. However, 
most of these studies were restricted to barrier calculations. Only in 
\cite{martin2009} is there an attempt to generalize the formula for 
collective inertia to $T>0$. Also, the price to pay when using finite 
temperature is the increase of statistical fluctuations on observables, 
see some discussion of this effect in \cite{schunck2015}. In addition, 
many of the beyond mean-field techniques used to restore symmetry 
are not defined at $T>0$. In the low-energy regime, direct methods based 
on QRPA and/or the GCM at zero temperature could be more useful. For example, Bernard and 
collaborators introduced in \cite{bernard2011} techniques to derive a 
collective Hamiltonian from the GCM built on top of two-quasiparticle 
excitations. Dittrich and collaborators outlined a method to build a 
statistical density operator from GCM states in \cite{dietrich2010}. Most 
of these ideas are in their infancy and a lot more work is needed before 
they can become valuable options for applications. In addition, one 
should anticipate a tremendous need for computing power.

Most of the theoretical framework used in the microscopic theory of 
fission was invented and developed many decades ago. It is only in the 
past two decades that unprecedented gains in computing power have allowed 
scientists to test these theories without resorting to debilitating 
approximations. In some ways, this process of catching up with theory is 
still on-going, and, therefore, the jury is still out on the intrinsic 
predicting power of the current DFT framework. In the long term, there is 
little doubt that progress in the determination of reliable energy 
functionals better rooted into the theory of nuclear forces will become 
critical to increase accuracy. Processes that accompany, and compete with, 
fission such as particle evaporation (neutrons, protons, alpha particle) or 
gamma emission will also need to be incorporated within a unified 
theoretical framework. 

%
%

\ack

The authors thank G. Bertsch, K. Pomorski and D. Regnier for reading the manuscript 
and providing useful feedback.
This work was partly performed under the auspices of
the U.S.\ Department of Energy by Lawrence Livermore National Laboratory under
Contract DE-AC52-07NA27344. Computational resources were provided through
an INCITE award ``Computational Nuclear Structure'' by the National Center for
Computational Sciences (NCCS) and National Institute for Computational Sciences
(NICS) at Oak Ridge National Laboratory, and through an award by the Livermore
Computing Resource Center at Lawrence Livermore National Laboratory. The work
of LMR is supported in part by Spanish  MINECO grants Nos. FPA2012-34694 and FIS2012-34479
and by the Consolider-Ingenio 2010 program MULTIDARK CSD2009-00064.
\clearpage

%
%

\bibliographystyle{iopart-num}
\bibliography{zotero_output,books,FissionRPP}

\providecommand{\newblock}{}
\begin{thebibliography}{100}
\expandafter\ifx\csname url\endcsname\relax
  \def\url#1{{\tt #1}}\fi
\expandafter\ifx\csname urlprefix\endcsname\relax\def\urlprefix{URL }\fi
\providecommand{\eprint}[2][]{\url{#2}}

\bibitem{hahn1938}
Hahn O and Strassmann F 1938 {\em Naturwiss.\/} {\bf 26} 755

\bibitem{hahn1939}
Hahn O and Strassmann F 1939 {\em Naturwiss.\/} {\bf 27} 11

\bibitem{meitner1939}
Meitner L and Frisch O 1939 {\em Nature\/} {\bf 143} 239

\bibitem{bohr1936}
Bohr N 1936 {\em Nature\/} {\bf 137} 344

\bibitem{bohr1939}
Bohr N and Wheeler J~A 1939 {\em Phys. Rev.\/} {\bf 56} 426

\bibitem{bell1967}
Bell G~I 1967 {\em Phys. Rev.\/} {\bf 158} 1118

\bibitem{nix1969}
Nix J~R 1969 {\em Nucl. Phys. A\/} {\bf 130} 241

\bibitem{bolsterli1972}
Bolsterli M, Fiset E~O, Nix J~R and Norton J~L 1972 {\em Phys. Rev. C\/} {\bf
  5} 1050

\bibitem{brack1972}
Brack M, Damgaard J, Jensen A~S, Pauli H~C, Strutinsky V~M and Wong C~Y 1972
  {\em Rev. Mod. Phys.\/} {\bf 44} 320

\bibitem{hofmann2000}
Hofmann S and M\"{u}nzenberg G 2000 {\em Rev. Mod. Phys.\/} {\bf 72} 733

\bibitem{nilsson1995}
Nilsson S~G and Ragnarsson I 1995 {\em Shapes and Shells In Nuclear
  Structure\/} (Cambridge University Press)

\bibitem{krappe2012}
Krappe H~J and Pomorski K 2012 {\em Theory of Nuclear Fission\/} (Springer)

\bibitem{bender2000}
Bender M, Rutz K, Reinhard P~G and Maruhn J~A 2000 {\em Eur. Phys. J. A\/} {\bf
  7} 467

\bibitem{buervenich2004}
B\"{u}rvenich T, Bender M, Maruhn J and Reinhard P~G 2004 {\em Phys. Rev. C\/}
  {\bf 69} 014307

\bibitem{staszczak2005}
Staszczak A, Dobaczewski J and Nazarewicz W 2005 {Self-Consistent} {Study} of
  {Fission} {Barriers} of {Even-Even} {Superheavy} {Nuclei} {\em AIP Conference
  Proceedings\/} vol 798 p~93

\bibitem{pei2009}
Pei J, Nazarewicz W, Sheikh J and Kerman A 2009 {\em Phys. Rev. Lett.\/} {\bf
  102} 192501

\bibitem{sheikh2009}
Sheikh J~A, Nazarewicz W and Pei J~C 2009 {\em Phys. Rev. C\/} {\bf 80} 011302

\bibitem{abusara2012}
Abusara H, Afanasjev A~V and Ring P 2012 {\em Phys. Rev. C\/} {\bf 85} 024314

\bibitem{warda2012}
Warda M and Egido J~L 2012 {\em Phys. Rev. C\/} {\bf 86} 014322

\bibitem{staszczak2013}
Staszczak A, Baran A and Nazarewicz W 2013 {\em Phys. Rev. C\/} {\bf 87} 024320

\bibitem{goriely2011}
Goriely S, Bauswein A and Janka H~T 2011 {\em ApJ\/} {\bf 738} L32

\bibitem{korobkin2012}
Korobkin O, Rosswog S, Arcones A and Winteler C 2012 {\em MNRAS\/} {\bf 426}
  1940

\bibitem{just2015}
Just O, Bauswein A, Pulpillo R~A, Goriely S and Janka H~T 2015 {\em MNRAS\/}
  {\bf 448} 541

\bibitem{sneden2008}
Sneden C, Cowan J~J and Gallino R 2008 {\em ARA\&A\/} {\bf 46} 241

\bibitem{roederer2014}
Roederer I~U, Preston G~W, Thompson I~B, Shectman S~A and Sneden C 2014 {\em
  ApJ\/} {\bf 784} 158

\bibitem{nichols2008}
Nichols A, Aldama D and Verpelli M 2008 {Handbook} for nuclear data for
  safeguards: {Database} extensions, {August} 2008 Tech. Rep. INDC(NDS)-0534
  International Atomic Energy Agency Vienna

\bibitem{bogner2013}
Bogner S, Bulgac A, Carlson J, Engel J, Fann G, Furnstahl R~J, Gandolfi S,
  Hagen G, Horoi M and Johnson C 2013 {\em Comput. Phys. Comm.\/} {\bf 184}
  2235

\bibitem{ramayya1998}
Ramayya A~V, Hwang J~K, Hamilton J~H, Sandulescu A, Florescu A, Ter-Akopian
  G~M, Daniel A~V, Oganessian Y~T, Popeko G~S, Greiner W and Cole J~D 1998 {\em
  Phys. Rev. Lett.\/} {\bf 81} 947

\bibitem{bonneau2007}
Bonneau L, Quentin P and M\"{o}ller P 2007 {\em Phys. Rev. C\/} {\bf 76} 024320

\bibitem{capote2016}
Capote R, Chen Y~J, Hambsch F~J, Kornilov N, Lestone J, Litaize O, Morillon B,
  Neudecker D, Oberstedt S, Ohsawa T, Otuka N, Pronyaev V, Saxena A, Serot O,
  Shcherbakov O, Shu N~C, Smith D, Talou P, Trkov A, Tudora A, Vogt R and
  Vorobyev A 2016 {\em Nucl. Data Sheets\/} {\bf 131} 1

\bibitem{ring1996}
Ring P 1996 {\em Prog. Part. Nucl. Phys.\/} {\bf 37} 193

\bibitem{nikvsic2011}
Nik\v{s}i{\'c} T, Vretenar D and Ring P 2011 {\em Prog. Part. Nucl. Phys.\/}
  {\bf 66} 519

\bibitem{rutz1995}
Rutz K, Maruhn J~A, Reinhard P~G and Greiner W 1995 {\em Nucl. Phys. A\/} {\bf
  590} 680

\bibitem{bender1998}
Bender M, Rutz K, Reinhard P~G, Maruhn J~A and Greiner W 1998 {\em Phys. Rev.
  C\/} {\bf 58} 2126

\bibitem{karatzikos2010}
Karatzikos S, Afanasjev A~V, Lalazissis G~A and Ring P 2010 {\em Phys. Lett.
  B\/} {\bf 689} 72

\bibitem{abusara2010}
Abusara H, Afanasjev A~V and Ring P 2010 {\em Phys. Rev. C\/} {\bf 82} 044303

\bibitem{lu2012}
Lu B~N, Zhao E~G and Zhou S~G 2012 {\em Phys. Rev. C\/} {\bf 85} 011301

\bibitem{lu2014}
Lu B~N, Zhao J, Zhao E~G and Zhou S~G 2014 {\em Phys. Rev. C\/} {\bf 89} 014323

\bibitem{zhao2015}
Zhao J, Lu B~N, Vretenar D, Zhao E~G and Zhou S~G 2015 {\em Phys. Rev. C\/}
  {\bf 91} 014321

\bibitem{zhao2016}
Zhao J, Lu B~N, Nik\v{s}i\'{c} T, Vretenar D and Zhou S~G 2016 {\em Phys. Rev.
  C\/} {\bf 93} 044315

\bibitem{nix1965}
Nix J~R and Swiatecki W~J 1965 {\em Nucl. Phys.\/} {\bf 71} 1

\bibitem{wilkins1976}
Wilkins B~D, Steinberg E~P and Chasman R~R 1976 {\em Phys. Rev. C\/} {\bf 14}
  1832

\bibitem{moeller1995}
M\"{o}ller P, Nix J~R, Myers W~D and Swiatecki W~J 1995 {\em Atom. Data Nuc.
  Data Tab.\/} {\bf 59} 185

\bibitem{werner1995}
Werner T~R and Dudek J 1995 {\em Atom. Data Nuc. Data Tab.\/} {\bf 59} 1

\bibitem{moeller1997}
M\"{o}ller P, Nix J and Kratz K 1997 {\em Atom. Data Nuc. Data Tab.\/} {\bf 66}
  131

\bibitem{moeller2006}
M\"{o}ller P, Bengtsson R, Carlsson B~G, Olivius P and Ichikawa T 2006 {\em
  Phys. Rev. Lett.\/} {\bf 97} 162502

\bibitem{moeller1989}
M\"{o}ller P, Nix J~R and Swiatecki W~J 1989 {\em Nucl. Phys. A\/} {\bf 492}
  349

\bibitem{moeller2001}
M\"{o}ller P, Madland D, Sierk A~J and Iwamoto A 2001 {\em Nature\/} {\bf 409}
  785

\bibitem{moeller2009}
M\"{o}ller P, Sierk A~J, Ichikawa T, Iwamoto A, Bengtsson R, Uhrenholt H and
  Aberg S 2009 {\em Phys. Rev. C\/} {\bf 79} 064304

\bibitem{myers1966}
Myers W~D and Swiatecki W~J 1966 {\em Nucl. Phys.\/} {\bf 81} 1

\bibitem{myers1969}
Myers W~D and Swiatecki W~J 1969 {\em Ann. Phys.\/} {\bf 55} 395

\bibitem{myers1974}
Myers W~D and Swiatecki W~J 1974 {\em Ann. Phys.\/} {\bf 84} 186

\bibitem{dudek1988}
Dudek J, Herskind B, Nazarewicz W, Szymanski Z and Werner T~R 1988 {\em Phys.
  Rev. C\/} {\bf 38} 940

\bibitem{schunck2007}
Schunck N, Dudek J and Herskind B 2007 {\em Phys. Rev. C\/} {\bf 75} 054304

\bibitem{hasse1988}
Hasse R~W and Myers W~D 1988 {\em Geometrical Relashionships of Macroscopic
  Nuclear Physics\/} (Springer-Verlag)

\bibitem{bohr1975}
Bohr A and Mottelson B 1975 {\em Nuclear Structure\/} vol~II (New-York:
  Benjamin)

\bibitem{rohoziski1981}
Rohozi\'{n}ski S~G and Sobiczewski A 1981 {\em Acta Phys. Pol. B\/} {\bf 12}
  1001

\bibitem{rohoziski1997}
Rohozi\'{n}ski S~G 1997 {\em Phys. Rev. C\/} {\bf 56} 165

\bibitem{pashkevich1971}
Pashkevich V~V 1971 {\em Nucl. Phys. A\/} {\bf 169} 275

\bibitem{strutinsky1963}
Strutinsky V~M, Lyashchenko N~Y and Popov N~A 1963 {\em Nucl. Phys.\/} {\bf 46}
  639

\bibitem{ring2000}
Ring P and Schuck P 2000 {\em The Nuclear Many-Body Problem\/}
  (Springer-Verlag)

\bibitem{kowal2010}
Kowal M, Jachimowicz P and Sobiczewski A 2010 {\em Phys. Rev. C\/} {\bf 82}
  014303

\bibitem{randrup2011}
Randrup J and M\"{o}ller P 2011 {\em Phys. Rev. Lett.\/} {\bf 106} 132503

\bibitem{randrup2011-a}
Randrup J, M\"{o}ller P and Sierk A~J 2011 {\em Phys. Rev. C\/} {\bf 84} 034613

\bibitem{jachimowicz2012}
Jachimowicz P, Kowal M and Skalski J 2012 {\em Phys. Rev. C\/} {\bf 85} 034305

\bibitem{jachimowicz2013}
Jachimowicz P, Kowal M and Skalski J 2013 {\em Phys. Rev. C\/} {\bf 87} 044308

\bibitem{hohenberg1964}
Hohenberg P and Kohn W 1964 {\em Phys. Rev.\/} {\bf 136} B864

\bibitem{kohn1965}
Kohn W and Sham L~J 1965 {\em Phys. Rev.\/} {\bf 140} A1133

\bibitem{parr1989}
Parr R and Yang W 1989 {\em Density Functional Theory of Atoms and Molecules\/}
  (Oxford: Oxford University Press)

\bibitem{dreizler1990}
Dreizler R and Gross E 1990 {\em Density Functional Theory: An Approach to the
  Quantum Many-Body Problem\/} (Springer-Verlag)

\bibitem{eschrig1996}
Eschrig R 1996 {\em Fundamentals of Density Functional Theory\/} (Leipzig:
  Teubner)

\bibitem{drut2010}
Drut J, Furnstahl R and Platter L 2010 {\em Prog. Part. Nucl. Phys.\/} {\bf 64}
  120

\bibitem{engel2007}
Engel J 2007 {\em Phys. Rev. C\/} {\bf 75} 014306

\bibitem{barnea2007}
Barnea N 2007 {\em Phys. Rev. C\/} {\bf 76} 067302

\bibitem{giraud2008}
Giraud B~G, Jennings B~K and Barrett B~R 2008 {\em Phys. Rev. A\/} {\bf 78}
  032507

\bibitem{messud2009}
Messud J, Bender M and Suraud E 2009 {\em Phys. Rev. C\/} {\bf 80} 054314

\bibitem{messud2011}
Messud J 2011 {\em Phys. Rev. A\/} {\bf 84} 052113

\bibitem{lesinski2014}
Lesinski T 2014 {\em Phys. Rev. C\/} {\bf 89} 044305

\bibitem{bender2003}
Bender M, Heenen P~H and Reinhard P~G 2003 {\em Rev. Mod. Phys.\/} {\bf 75} 121

\bibitem{scheidenberger2014}
Duguet T 2014 {The} {Nuclear} {Energy} {Density} {Functional} {Formalism} {\em
  The Euroschool on Exotic Beams, Vol. IV\/} vol 879 ed Scheidenberger C and
  Pf\"{u}tzner M (Berlin, Heidelberg: Springer Berlin Heidelberg) p 293

\bibitem{blaizot1985}
Blaizot J~P and Ripka G 1985 {\em Quantum Theory of Finite Systems\/}
  (Cambridge: The MIT Press)

\bibitem{erler2010}
Erler J, Kl\"{u}pfel P and Reinhard P~G 2010 {\em J. Phys. G: Nucl. Part.
  Phys.\/} {\bf 37} 064001

\bibitem{chappert2015}
Chappert F, Pillet N, Girod M and Berger J~F 2015 {\em Phys. Rev. C\/} {\bf 91}
  034312

\bibitem{thouless1960}
Thouless D~J 1960 {\em Nucl. Phys.\/} {\bf 21} 225

\bibitem{perez-martin2008}
Perez-Martin S and Robledo L 2008 {\em Phys. Rev. C\/} {\bf 78} 014304

\bibitem{robledo2011}
Robledo L~M and Bertsch G~F 2011 {\em Phys. Rev. C\/} {\bf 84} 014312

\bibitem{girod1983}
Girod M and Grammaticos B 1983 {\em Phys. Rev. C\/} {\bf 27} 2317

\bibitem{samyn2005}
Samyn M, Goriely S and Pearson J 2005 {\em Phys. Rev. C\/} {\bf 72} 044316

\bibitem{dobaczewski1996-a}
Dobaczewski J, Nazarewicz W, Werner T~R, Berger J~F, Chinn C~R and Decharg\'{e}
  J 1996 {\em Phys. Rev. C\/} {\bf 53} 2809

\bibitem{skyrme1959}
Skyrme T~H~R 1959 {\em Nucl. Phys.\/} {\bf 9} 615

\bibitem{vautherin1972}
Vautherin D and Brink D~M 1972 {\em Phys. Rev. C\/} {\bf 5} 626

\bibitem{koehler1976}
K\"{o}hler H 1976 {\em Nucl. Phys. A\/} {\bf 258} 301

\bibitem{engel1975}
Engel Y~M, Brink D~M, Goeke K, Krieger S~J and Vautherin D 1975 {\em Nucl.
  Phys. A\/} {\bf 249} 215

\bibitem{lesinski2007}
Lesinski T, Bender M, Bennaceur K, Duguet T and Meyer J 2007 {\em Phys. Rev.
  C\/} {\bf 76} 014312

\bibitem{perliska2004}
Perli\'{n}ska E, Rohozi\'{n}ski S~G, Dobaczewski J and Nazarewicz W 2004 {\em
  Phys. Rev. C\/} {\bf 69} 014316

\bibitem{dobaczewski1996}
Dobaczewski J and Dudek J 1996 {\em Acta Phys. Pol. B\/} {\bf 27} 45

\bibitem{carlsson2008}
Carlsson B~G, Dobaczewski J and Kortelainen M 2008 {\em Phys. Rev. C\/} {\bf
  78} 044326

\bibitem{raimondi2011}
Raimondi F, Carlsson B~G, Dobaczewski J and Toivanen J 2011 {\em Phys. Rev.
  C\/} {\bf 84} 064303

\bibitem{raimondi2011-a}
Raimondi F, Carlsson B~G and Dobaczewski J 2011 {\em Phys. Rev. C\/} {\bf 83}
  054311

\bibitem{becker2015}
Becker P, Davesne D, Meyer J, Pastore A and Navarro J 2015 {\em J. Phys. G:
  Nucl. Part. Phys.\/} {\bf 42} 034001

\bibitem{sadoudi2013}
Sadoudi J, Duguet T, Meyer J and Bender M 2013 {\em Phys. Rev. C\/} {\bf 88}
  064326

\bibitem{gogny1975}
Gogny D 1975 {\em Nucl. Phys. A\/} {\bf 237} 399

\bibitem{bender2009}
Bender M, Bennaceur K, Duguet T, Heenen P~H, Lesinski T and Meyer J 2009 {\em
  Phys. Rev. C\/} {\bf 80} 064302

\bibitem{anguiano2012}
Anguiano M, Grasso M, Co' G, De~Donno V and Lallena A~M 2012 {\em Phys. Rev.
  C\/} {\bf 86} 054302

\bibitem{baldo2008}
Baldo M, Schuck P and Vi\~{n}as X 2008 {\em Phys. Lett. B\/} {\bf 663} 390

\bibitem{baldo2013}
Baldo M, Robledo L~M, Schuck P and Vi\~{n}as X 2013 {\em Phys. Rev. C\/} {\bf
  87} 064305

\bibitem{fayans1994}
Fayans S~A, Tolokonnikov S~V, Trykov E~L and Zawischa D 1994 {\em Phys. Lett.
  B\/} {\bf 338} 1

\bibitem{kroemer1995}
Kr\"{o}mer E, Tolokonnikov S~V, Fayans S~A and Zawischa D 1995 {\em Phys. Lett.
  B\/} {\bf 363} 12

\bibitem{anguiano2001}
Anguiano M, Egido J~L and Robledo L~M 2001 {\em Nucl. Phys. A\/} {\bf 683} 227

\bibitem{dobaczewski1984}
Dobaczewski J, Flocard H and Treiner J 1984 {\em Nucl. Phys. A\/} {\bf 422} 103

\bibitem{chasman1976}
Chasman R~R 1976 {\em Phys. Rev. C\/} {\bf 14} 1935

\bibitem{PhysRevC.60.064312}
Garrido E, Sarriguren P, Moya~de Guerra E and Schuck P 1999 {\em Phys. Rev.
  C\/} {\bf 60} 064312

\bibitem{kortelainen2010}
Kortelainen M, Lesinski T, Mor\'{e} J, Nazarewicz W, Sarich J, Schunck N,
  Stoitsov M~V and Wild S 2010 {\em Phys. Rev. C\/} {\bf 82} 024313

\bibitem{schunck2015-a}
Schunck N, McDonnell J~D, Sarich J, Wild S~M and Higdon D 2015 {\em J. Phys. G:
  Nucl. Part. Phys.\/} {\bf 42} 034024

\bibitem{schunck2015-b}
Schunck N, McDonnell J~D, Higdon D, Sarich J and Wild S 2015 {\em Nucl. Data
  Sheets\/} {\bf 123} 115

\bibitem{stone2007}
Stone J and Reinhard P~G 2007 {\em Prog. Part. Nucl. Phys.\/} {\bf 58} 587

\bibitem{bartel1982}
Bartel J, Quentin P, Brack M, Guet C and H{\aa}kansson H~B 1982 {\em Nucl.
  Phys. A\/} {\bf 386} 79

\bibitem{kortelainen2012}
Kortelainen M, McDonnell J, Nazarewicz W, Reinhard P~G, Sarich J, Schunck N,
  Stoitsov M~V and Wild S~M 2012 {\em Phys. Rev. C\/} {\bf 85} 024304

\bibitem{decharge1980}
Decharg\'{e} J and Gogny D 1980 {\em Phys. Rev. C\/} {\bf 21} 1568

\bibitem{berger1984}
Berger J~F, Girod M and Gogny D 1984 {\em Nucl. Phys. A\/} {\bf 428} 23

\bibitem{chappert2008}
Chappert F, Girod M and Hilaire S 2008 {\em Phys. Lett. B\/} {\bf 668} 420

\bibitem{goriely2009}
Goriely S, Hilaire S, Girod M and P\'{e}ru S 2009 {\em Phys. Rev. Lett.\/} {\bf
  102} 242501

\bibitem{berger1989}
Berger J~F, Girod M and Gogny D 1989 {\em Nucl. Phys. A\/} {\bf 502} 85

\bibitem{otsuka2006}
Otsuka T, Matsuo T and Abe D 2006 {\em Phys. Rev. Lett.\/} {\bf 97} 162501

\bibitem{hoffman2000}
Hoffman D, Ghiorso A and Seaborg G 2000 {\em The Transuranium People\/} (World
  Scientific, River Edge, NJ)

\bibitem{brack1974}
Brack M and Quentin P 1974 {\em Phys. Lett. B\/} {\bf 52} 159

\bibitem{bartel1985}
Bartel J, Brack M and Durand M 1985 {\em Nucl. Phys. A\/} {\bf 445} 263

\bibitem{martin2009}
Martin V and Robledo L~M 2009 {\em Int. J. Mod. Phys. E\/} {\bf 18} 861

\bibitem{schunck2015}
Schunck N, Duke D and Carr H 2015 {\em Phys. Rev. C\/} {\bf 91} 034327

\bibitem{descloizeaux1968}
des Cloizeaux J 1968 {Approximation} de {Hartree}-{Fock} et approximation de
  phase al\'eatoire à temp\'erature finie {\em Many-Body Physics\/} ed DeWitt
  C and Balian R (Gordon and Breach, Science Publishers, Inc.)

\bibitem{lee1979}
Lee H~C and Gupta S~D 1979 {\em Phys. Rev. C\/} {\bf 19} 2369

\bibitem{goodman1981}
Goodman A~L 1981 {\em Nucl. Phys. A\/} {\bf 352} 30

\bibitem{tanabe1981}
Tanabe K, Sugawara-Tanabe K and Mang H~J 1981 {\em Nucl. Phys. A\/} {\bf 357}
  20

\bibitem{egido1993}
Egido J and Ring P 1993 {\em J. Phys. G: Nucl. Part. Phys.\/} {\bf 19} 1

\bibitem{egido2000}
Egido J, Robledo L and Martin V 2000 {\em Phys. Rev. Lett.\/} {\bf 85} 26

\bibitem{martin2003}
Martin V, Egido J and Robledo L 2003 {\em Phys. Rev. C\/} {\bf 68} 034327

\bibitem{reichl1988}
Reichl L 1988 {\em A Modern Course in Statistical Physics\/} (John Wiley and
  Sons, Inc)

\bibitem{GAUDIN196089}
Gaudin M 1960 {\em Nucl. Phys.\/} {\bf 15} 89 -- 91

\bibitem{bonche1984}
Bonche P, Levit S and Vautherin D 1984 {\em Nucl. Phys. A\/} {\bf 427} 278

\bibitem{bonche1985}
Bonche P, Levit S and Vautherin D 1985 {\em Nucl. Phys. A\/} {\bf 436} 265

\bibitem{egido1988}
Egido J 1988 {\em Phys. Rev. Lett.\/} {\bf 61} 767

\bibitem{levit1984}
Levit S and Alhassid Y 1984 {\em Nucl. Phys. A\/} {\bf 413} 439

\bibitem{stoitsov2007}
Stoitsov M~V, Dobaczewski J, Kirchner R, Nazarewicz W and Terasaki J 2007 {\em
  Phys. Rev. C\/} {\bf 76} 014308

\bibitem{bender2009-a}
Bender M, Duguet T and Lacroix D 2009 {\em Phys. Rev. C\/} {\bf 79} 044319

\bibitem{duguet2009}
Duguet T and Lesinski T 2009 {Non-empirical} {Nuclear} {Energy} {Functionals},
  {Pairing} {Gaps} and {Odd-Even} {Mass} {Differences} ed Milin M, Niksic T,
  Szilner S and Vretenar D (AIP) p 243

\bibitem{lacroix2009}
Lacroix D, Duguet T and Bender M 2009 {\em Phys. Rev. C\/} {\bf 79} 044318

\bibitem{dobaczewski2009}
Dobaczewski J 2009 {\em J. Phys. G: Nucl. Part. Phys.\/} {\bf 36} 105105

\bibitem{hupin2011}
Hupin G 2011 {\em {Approches} fonctionnelles de la densit\'{e} pour les
  systemes finis appari\'{e}s\/} Ph.D. thesis Universit\'{e} de Caen Caen

\bibitem{hupin2012}
Hupin G and Lacroix D 2012 {\em Phys. Rev. C\/} {\bf 86} 024309

\bibitem{villars1971}
Villars F and Schmeing-Rogerson N 1971 {\em Ann. Phys.\/} {\bf 63} 443

\bibitem{PhysRevC.2.892}
Friedman W~A and Wilets L 1970 {\em Phys. Rev. C\/} {\bf 2} 892--902

\bibitem{PhysRevC.62.054319}
Rodr\'{i}guez-Guzm\'an R~R, Egido J~L and Robledo L~M 2000 {\em Phys. Rev. C\/}
  {\bf 62} 054319

\bibitem{peierls1957}
Peierls R~E and Yoccoz J 1957 {\em Proc. Phys. Soc. A\/} {\bf 70} 381

\bibitem{hao2012}
Hao T~V~N, Quentin P and Bonneau L 2012 {\em Phys. Rev. C\/} {\bf 86} 064307

\bibitem{Laft01}
Laftchiev H, Sams{\oe}n D, Quentin P and Piperova J 2001 {\em Eur. Phys. J.
  A\/} {\bf 12} 155

\bibitem{berger1980}
Berger J~F and Gogny D 1980 {\em Nucl. Phys. A\/} {\bf 333} 302

\bibitem{urbano1981}
Urbano J~N, Goeke K and Reinhard P~G 1981 {\em Nucl. Phys. A\/} {\bf 370} 329

\bibitem{goeke1983}
Goeke K, Gr\"{u}mmer F and Reinhard P~G 1983 {\em Ann. Phys.\/} {\bf 150} 504

\bibitem{skalski2007}
Skalski J 2007 {\em Phys. Rev. C\/} {\bf 76} 044603

\bibitem{dobaczewski2004}
Dobaczewski J and Olbratowski P 2004 {\em Comput. Phys. Comm.\/} {\bf 158} 158

\bibitem{nikolov2011}
Nikolov N, Schunck N, Nazarewicz W, Bender M and Pei J 2011 {\em Phys. Rev.
  C\/} {\bf 83} 034305

\bibitem{schunck2014}
Schunck N, Duke D, Carr H and Knoll A 2014 {\em Phys. Rev. C\/} {\bf 90} 054305

\bibitem{larsson1972}
Larsson S~E, Ragnarsson I and Nilsson S~G 1972 {\em Phys. Lett. B\/} {\bf 38}
  269

\bibitem{berger1986}
Berger J~F 1986 {Quantum} dynamics of wavepackets on two-dimensional potential
  energy surfaces governing nuclear fission {\em Dynamics of Wave Packets in
  Molecular and Nuclear Physics\/} ({\em Lecture Notes in Physics\/} no 256) ed
  Broeckhove J, Lathouwers L and Leuven P~v (Springer Berlin Heidelberg) p~21

\bibitem{dubray2012}
Dubray N and Regnier D 2012 {\em Comput. Phys. Comm.\/} {\bf 183} 2035

\bibitem{berger1990}
Berger J~F, Anderson J~D, Bonche P and Weiss M~S 1990 {\em Phys. Rev. C\/} {\bf
  41} R2483

\bibitem{egido1997}
Egido J~L, Robledo L~M and Chasman R~R 1997 {\em Phys. Lett. B\/} {\bf 393} 13

\bibitem{warda2002}
Warda M, Egido J, Robledo L and Pomorski K 2002 {\em Phys. Rev. C\/} {\bf 66}
  014310

\bibitem{younes2009-a}
Younes W and Gogny D 2009 {\em Phys. Rev. C\/} {\bf 80} 054313

\bibitem{scamps2015-a}
Scamps G, Simenel C and Lacroix D 2015 {\em Phys. Rev. C\/} {\bf 92} 011602

\bibitem{younes2012-a}
Younes W and Gogny D 2012 {Fragment} {Yields} {Calculated} in a
  {Time-Dependent} {Microscopic} {Theory} of {Fission} Tech. Rep.
  LLNL-TR-586678 Lawrence Livermore National Laboratory (LLNL), Livermore, CA

\bibitem{giuliani2013}
Giuliani S~A and Robledo L~M 2013 {\em Phys. Rev. C\/} {\bf 88} 054325

\bibitem{urin1966}
Urin M and Zaretsky D 1966 {\em Nucl. Phys.\/} {\bf 75} 101

\bibitem{moretto1974}
Moretto L~G and Babinet R~P 1974 {\em Phys. Lett. B\/} {\bf 49} 147

\bibitem{god1985-a}
G\'{o}\'{z}d\'{z} A 1985 {\em Phys. Lett. B\/} {\bf 152} 281

\bibitem{lazarev1987}
Lazarev Y~A 1987 {\em Phys. Scr.\/} {\bf 35} 255

\bibitem{PhysRevC.43.2200}
Bertsch G and Flocard H 1991 {\em Phys. Rev. C\/} {\bf 43}(5) 2200--2204

\bibitem{sadhukhan2014}
Sadhukhan J, Dobaczewski J, Nazarewicz W, Sheikh J~A and Baran A 2014 {\em
  Phys. Rev. C\/} {\bf 90} 061304

\bibitem{giuliani2014}
Giuliani S~A, Robledo L~M and Rodr\'{i}guez-Guzm\'{a}n R 2014 {\em Phys. Rev.
  C\/} {\bf 90} 054311

\bibitem{god1985}
G\'{o}\'{z}d\'{z} A, Pomorski K, Brack M and Werner E 1985 {\em Nucl. Phys.
  A\/} {\bf 442} 50

\bibitem{staszczak1989}
Staszczak A, Pi\l{}at S and Pomorski K 1989 {\em Nucl. Phys. A\/} {\bf 504} 589

\bibitem{bulgac2016}
Bulgac A, Magierski P, Roche K~J and Stetcu I 2016 {\em Phys. Rev. Lett.\/}
  {\bf 116} 122504

\bibitem{simenel2014}
Simenel C and Umar A~S 2014 {\em Phys. Rev. C\/} {\bf 89} 031601(R)

\bibitem{younes2011}
Younes W and Gogny D 2011 {\em Phys. Rev. Lett.\/} {\bf 107} 132501

\bibitem{younes2013}
Younes W, Gogny D and Schunck N 2013 {A} {Microscopic} {Theory} of {Low-Energy}
  {Fission:} {Fragment} {Properties} {\em Fission and Properties of
  Neutron-Rich Nuclei\/} (Sanibel Island, Florida, USA: World Scientific) p 605

\bibitem{staszczak2009}
Staszczak A, Baran A, Dobaczewski J and Nazarewicz W 2009 {\em Phys. Rev. C\/}
  {\bf 80} 014309

\bibitem{bonneau2006}
Bonneau L 2006 {\em Phys. Rev. C\/} {\bf 74} 014301

\bibitem{dubray2008}
Dubray N, Goutte H and Delaroche J~P 2008 {\em Phys. Rev. C\/} {\bf 77} 014310

\bibitem{duke2012}
Duke D, Carr H, Knoll A, Schunck N, Nam H~A and Staszczak A 2012 {\em IEEE
  Trans. Vis. Comp. Graph.\/} {\bf 18} 2033

\bibitem{davies1976}
Davies K 1976 {\em Phys. Rev. C\/} {\bf 13} 2385

\bibitem{gherghescu2008}
Gherghescu R, Poenaru D and Carjan N 2008 {\em Phys. Rev. C\/} {\bf 77} 044607

\bibitem{younes2009}
Younes W and Gogny D 2009 {The} microscopic theory of fission {\em AIP
  Conference Proceedings\/} vol 1175 (AIP Publishing) p~3

\bibitem{klein1991}
Klein A, Walet N~R and Do~Dang G 1991 {\em Ann. Phys.\/} {\bf 208} 90

\bibitem{feynman2012quantum}
Feynman R and Hibbs A 2012 {\em Quantum Mechanics and Path Integrals: Emended
  Edition\/} (Dover Publications, Incorporated)

\bibitem{COLEMAN1988178}
Coleman S 1988 {\em Nucl. Phys. B\/} {\bf 298} 178 -- 186

\bibitem{messiah1961}
Messiah A 1961 {\em Quantum Mechanics, Vol. 1\/} (North-Holland Publishing
  Company, Amsterdam)

\bibitem{landau2013quantum}
Landau L and Lifshitz E 2013 {\em Quantum Mechanics: Non-Relativistic Theory\/}
  (Elsevier Science)

\bibitem{brink1968}
Brink D~M and Weiguny A 1968 {\em Nucl. Phys. A\/} {\bf 120} 59

\bibitem{onishi1975}
Onishi N and Une T 1975 {\em Prog. Theor. Phys.\/} {\bf 53} 504

\bibitem{une1976}
Une T, Ikeda A and Onishi N 1976 {\em Prog. Theor. Phys.\/} {\bf 55} 498

\bibitem{REINHARD1975120}
Reinhard P 1975 {\em Nucl. Phys. A\/} {\bf 252} 120 -- 132

\bibitem{reinhard1987}
Reinhard P~G and Goeke K 1987 {\em Rep. Prog. Phys.\/} {\bf 50} 1

\bibitem{ring1977}
Ring P and Schuck P 1977 {\em Nucl. Phys. A\/} {\bf 292} 20

\bibitem{GOZDZ1985281}
G{\'o}{\'z}d{\'z} A 1985 {\em Phys. Lett. B\/} {\bf 152} 281

\bibitem{borisenko1968}
Borisenko A and Tarapov I 1968 {\em Vector and tensor analysis with
  applications\/} (Dover Publications, Inc.New York)

\bibitem{peierls1962}
Peierls R~E and Thouless D~J 1962 {\em Nucl. Phys.\/} {\bf 38} 154

\bibitem{goeke1980}
Goeke K and Reinhard P~G 1980 {\em Ann. Phys.\/} {\bf 124} 249

\bibitem{brink1976}
Brink D~M, Giannoni M~J and Veneroni M 1976 {\em Nucl. Phys. A\/} {\bf 258} 237

\bibitem{giannoni1976}
Giannoni M~J, Moreau F, Quentin P, Vautherin D, Veneroni M and Brink D~M 1976
  {\em Phys. Lett. B\/} {\bf 65} 305

\bibitem{villars1977}
Villars F 1977 {\em Nucl. Phys. A\/} {\bf 285} 269

\bibitem{baranger1978}
Baranger M and Veneroni M 1978 {\em Ann. Phys.\/} {\bf 114} 123

\bibitem{goeke1978}
Goeke K and Reinhard P~G 1978 {\em Ann. Phys.\/} {\bf 112} 328

\bibitem{giannoni1980}
Giannoni M~J and Quentin P 1980 {\em Phys. Rev. C\/} {\bf 21} 2076

\bibitem{giannoni1980-a}
Giannoni M~J and Quentin P 1980 {\em Phys. Rev. C\/} {\bf 21} 2060

\bibitem{giannoni1984}
Giannoni M~J 1984 {\em Nucl. Phys. A\/} {\bf 428} 63

\bibitem{yuldashbaeva1999}
Yuldashbaeva E~K, Libert J, Quentin P and Girod M 1999 {\em Phys. Lett. B\/}
  {\bf 461} 1

\bibitem{baran2011}
Baran A, Sheikh J~A, Dobaczewski J, Nazarewicz W and Staszczak A 2011 {\em
  Phys. Rev. C\/} {\bf 84} 054321

\bibitem{goeke1981}
Goeke K, Reinhard P~G and Rowe D~J 1981 {\em Nucl. Phys. A\/} {\bf 359} 408

\bibitem{mukherjee1981}
Mukherjee A~K and Pal M~K 1981 {\em Phys. Lett. B\/} {\bf 100} 457

\bibitem{mukherjee1982}
Mukherjee A~K and Pal M~K 1982 {\em Nucl. Phys. A\/} {\bf 373} 289

\bibitem{matsuo2000}
Matsuo M, Nakatsukasa T and Matsuyanagi K 2000 {\em Prog. Theor. Phys.\/} {\bf
  103} 959

\bibitem{hinohara2007}
Hinohara N, Nakatsukasa T, Matsuo M and Matsuyanagi K 2007 {\em Prog. Theor.
  Phys.\/} {\bf 117} 451

\bibitem{lechaftois2015}
Lechaftois F, Deloncle I and P\'{e}ru S 2015 {\em Phys. Rev. C\/} {\bf 92}
  034315

\bibitem{sadhukhan2013}
Sadhukhan J, Mazurek K, Baran A, Dobaczewski J, Nazarewicz W and Sheikh J 2013
  {\em Phys. Rev. C\/} {\bf 88} 064314

\bibitem{PhysRevC.13.229}
Randrup J, Larsson S~E, M\"oller P, Nilsson S~G, Pomorski K and Sobiczewski A
  1976 {\em Phys. Rev. C\/} {\bf 13}(1) 229--239

\bibitem{levit1980}
Levit S 1980 {\em Phys. Rev. C\/} {\bf 21} 1594

\bibitem{levit1980-a}
Levit S, Negele J~W and Paltiel Z 1980 {\em Phys. Rev. C\/} {\bf 22} 1979

\bibitem{negele1982}
Negele J~W 1982 {\em Rev. Mod. Phys.\/} {\bf 54} 913

\bibitem{puddu1987}
Puddu G and Negele J~W 1987 {\em Phys. Rev. C\/} {\bf 35} 1007

\bibitem{negele1989}
Negele J~W 1989 {\em Nucl. Phys. A\/} {\bf 502} 371

\bibitem{bonasera1997}
Bonasera A and Iwamoto A 1997 {\em Phys. Rev. Lett.\/} {\bf 78} 187

\bibitem{scamps2015}
Scamps G and Hagino K 2015 {\em Phys. Rev. C\/} {\bf 91} 044606

\bibitem{skalski2008}
Skalski J 2008 {\em Phys. Rev. C\/} {\bf 77} 064610

\bibitem{RING197850}
Ring P, Massmann H and Rasmussen J 1978 {\em Nucl. Phys. A\/} {\bf 296} 50 --
  76

\bibitem{abe1996}
Abe Y, Ayik S, Reinhard P~G and Suraud E 1996 {\em Phys. Rep.\/} {\bf 275} 49

\bibitem{nadtochy2007}
Nadtochy P~N, Keli\'{c} A and Schmidt K~H 2007 {\em Phys. Rev. C\/} {\bf 75}
  064614

\bibitem{sadhukhan2010}
Sadhukhan J and Pal S 2010 {\em Phys. Rev. C\/} {\bf 81} 031602(R)

\bibitem{grange1983}
Grang\'{e} P, Jun-Qing L and Weidenm\"{u}ller H~A 1983 {\em Phys. Rev. C\/}
  {\bf 27} 2063

\bibitem{jing-shang1983}
Jing-Shang Z and Weidenm\"{u}ller H~A 1983 {\em Phys. Rev. C\/} {\bf 28} 2190

\bibitem{weidenmueller1984}
Weidenm\"{u}ller H~A and Jing-Shang Z 1984 {\em J. Stat. Phys.\/} {\bf 34} 191

\bibitem{nadtochy2005}
Nadtochy P and Adeev G 2005 {\em Phys. Rev. C\/} {\bf 72} 054608

\bibitem{sadhukhan2011}
Sadhukhan J and Pal S 2011 {\em Phys. Rev. C\/} {\bf 84} 044610

\bibitem{nadtochy2012}
Nadtochy P~N, Ryabov E~G, Gegechkori A~E, Anischenko Y~A and Adeev G~D 2012
  {\em Phys. Rev. C\/} {\bf 85} 064619

\bibitem{randrup2013}
Randrup J and M\"{o}ller P 2013 {\em Phys. Rev. C\/} {\bf 88} 064606

\bibitem{sadhukhan2015}
Jhilam~Sadhukhan Witold~Nazarewicz N~S 2015 Microscopic modeling of mass and
  charge distributions in the spontaneous fission of $^{240}${Pu}
  arXiv:1510.08003 (\textit{Preprint} \eprint{arXiv:1510.08003})

\bibitem{fiolhais2003}
Fiolhais C, Nogueira F and Marques M~A (eds) 2003 {\em {A} {Primer} in
  {Density} {Functional} {Theory}\/} ({\em Lecture Notes in Physics\/} vol 649)
  (Springer-Verlag Berlin Heidelberg)

\bibitem{marques2004}
Marques M and Gross E 2004 {\em Annu. Rev. Phys. Chem.\/} {\bf 55} 427

\bibitem{runge1984}
Runge E and Gross E~K~U 1984 {\em Phys. Rev. Lett.\/} {\bf 52} 997

\bibitem{messud2009-a}
Messud J 2009 {\em Phys. Rev. C\/} {\bf 80} 054614

\bibitem{goddard2015}
Goddard P, Stevenson P and Rios A 2015 {\em Phys. Rev. C\/} {\bf 92} 054610

\bibitem{tanimura2015}
Tanimura Y, Lacroix D and Scamps G 2015 {\em Phys. Rev. C\/} {\bf 92} 034601

\bibitem{negele1978}
Negele J~W, Koonin S~E, M\"{o}ller P, Nix J~R and Sierk A~J 1978 {\em Phys.
  Rev. C\/} {\bf 17} 1098

\bibitem{avez2008}
Avez B, Simenel C and Chomaz P 2008 {\em Phys. Rev. C\/} {\bf 78} 044318

\bibitem{ebata2010}
Ebata S, Nakatsukasa T, Inakura T, Yoshida K, Hashimoto Y and Yabana K 2010
  {\em Phys. Rev. C\/} {\bf 82} 034306

\bibitem{nakatsukasa2012}
Nakatsukasa T, Ebata S, Avogadro P, Guo L, Inakura T and Yoshida K 2012 {\em J.
  Phys.: Conf. Ser.\/} {\bf 387} 012015

\bibitem{nakatsukasa2012-a}
Nakatsukasa T 2012 {\em Prog. Theor. Exp. Phys.\/} {\bf 2012} 1A207

\bibitem{scamps2012}
Scamps G, Lacroix D, Bertsch G~F and Washiyama K 2012 {\em Phys. Rev. C\/} {\bf
  85} 034328

\bibitem{berger1991}
Berger J~F, Girod M and Gogny D 1991 {\em Comput. Phys. Comm.\/} {\bf 63} 365

\bibitem{goutte2005}
Goutte H, Berger J~F, Casoli P and Gogny D 2005 {\em Phys. Rev. C\/} {\bf 71}
  024316

\bibitem{young2009}
Young G, Dean D and Savage M 2009 {Scientific} {Grand} {Challenges:}
  {Forefront} {Questions} in {Nuclear} {Science} and the {Role} of {Computing}
  at the {Extreme} {Scale} Tech. rep. U.S. Department of Energy Washington,
  D.C.

\bibitem{bishop2009}
Bishop A and Messina P 2009 {Scientific} {Grand} {Challenges} for {National}
  {Security:} {The} {Role} of {Computing} at the {Extreme} {Scale} Tech. rep.
  U.S. Department of Energy Washington, D.C.

\bibitem{talmi1952}
Talmi I 1952 {\em Helv. Phys. Acta\/} {\bf 25} 185

\bibitem{moshinsky1959}
Moshinsky M 1959 {\em Nucl. Phys. A\/} {\bf 13} 104

\bibitem{talman1970}
Talman J~D 1970 {\em Nucl. Phys. A\/} {\bf 141} 273

\bibitem{parrish2013}
Parrish R~M, Hohenstein E~G, Schunck N~F, Sherrill C~D and Mart\'{i}nez T~J
  2013 {\em Phys. Rev. Lett.\/} {\bf 111} 132505

\bibitem{younes2009-b}
Younes W 2009 {\em Comput. Phys. Comm.\/} {\bf 180} 1013

\bibitem{quentin1972}
Quentin P 1972 {\em J. Phys. France\/} {\bf 33} 457

\bibitem{dobaczewski2009-a}
Dobaczewski J, Satu\l{}a W, Carlsson B, Engel J, Olbratowski P, Powa\l{}owski
  P, Sadziak M, Sarich J, Schunck N, Staszczak A, Stoitsov M, Zalewski M and
  Zdu\'{n}czuk H 2009 {\em Comput. Phys. Comm.\/} {\bf 180} 2361

\bibitem{stoitsov2005}
Stoitsov M, Dobaczewski J, Nazarewicz W and Ring P 2005 {\em Comput. Phys.
  Comm.\/} {\bf 167} 43

\bibitem{stoitsov2013}
Stoitsov M, Schunck N, Kortelainen M, Michel N, Nam H, Olsen E, Sarich J and
  Wild S 2013 {\em Comput. Phys. Comm.\/} {\bf 184} 1592

\bibitem{dobaczewski1997}
Dobaczewski J and Dudek J 1997 {\em Comput. Phys. Comm.\/} {\bf 102} 166

\bibitem{dobaczewski1997-a}
Dobaczewski J and Dudek J 1997 {\em Comput. Phys. Comm.\/} {\bf 102} 183

\bibitem{dobaczewski2000}
Dobaczewski J and Dudek J 2000 {\em Comput. Phys. Comm.\/} {\bf 131} 164

\bibitem{dobaczewski2005}
Dobaczewski J and Olbratowski P 2005 {\em Comput. Phys. Comm.\/} {\bf 167} 214

\bibitem{schunck2012}
Schunck N, Dobaczewski J, McDonnell J, Satu\l{}a W, Sheikh J, Staszczak A,
  Stoitsov M and Toivanen P 2012 {\em Comput. Phys. Comm.\/} {\bf 183} 166

\bibitem{mcdonnell2013}
McDonnell J~D, Nazarewicz W and Sheikh J~A 2013 {\em Phys. Rev. C\/} {\bf 87}
  054327

\bibitem{schunck2013}
Schunck N 2013 {\em J. Phys.: Conf. Ser.\/} {\bf 436} 012058

\bibitem{schunck2013-a}
Schunck N 2013 {\em Acta Phys. Pol. B\/} {\bf 44} 263

\bibitem{warda2011}
Warda M and Robledo L~M 2011 {\em Phys. Rev. C\/} {\bf 84} 044608

\bibitem{rodriguez-guzman2014}
Rodr{\i}guez-Guzm\'{a}n R and Robledo L~M 2014 {\em Eur. Phys. J. A\/} {\bf 50}
  142

\bibitem{stoitsov1998}
Stoitsov M~V, Nazarewicz W and Pittel S 1998 {\em Phys. Rev. C\/} {\bf 58} 2092

\bibitem{schunck2008-a}
Schunck N and Egido J 2008 {\em Phys. Rev. C\/} {\bf 77} 011301(R)

\bibitem{grawe2007}
Grawe H, Langanke K and Mart\'{i}nez-Pinedo G 2007 {\em Rep. Prog. Phys.\/}
  {\bf 70} 1525

\bibitem{martinez-pinedo2007}
Mart\'{i}nez-Pinedo G, Mocelj D, Zinner N, Keli\'{c} A, Langanke K, Panov I,
  Pfeiffer B, Rauscher T, Schmidt K~H and Thielemann F~K 2007 {\em Prog. Part.
  Nucl. Phys.\/} {\bf 59} 199

\bibitem{goutte2004}
Goutte H, Casoli P and Berger J~F 2004 {\em Nucl. Phys. A\/} {\bf 734} 217

\bibitem{umar1991}
Umar A~S, Wu J, Strayer M~R and Bottcher C 1991 {\em J. Comput. Phys.\/} {\bf
  93} 426

\bibitem{kegleyjr.1996}
Kegley~Jr D~R, Oberacker V~E, Strayer M~R, Umar A~S and Wells J~C 1996 {\em J.
  Comput. Phys.\/} {\bf 128} 197

\bibitem{teran2003}
Ter\'{a}n E, Oberacker V~E and Umar A~S 2003 {\em Phys. Rev. C\/} {\bf 67}
  064314

\bibitem{blazkiewicz2005}
Blazkiewicz A, Oberacker V~E, Umar A~S and Stoitsov M 2005 {\em Phys. Rev. C\/}
  {\bf 71} 054321

\bibitem{pei2008}
Pei J, Stoitsov M, Fann G, Nazarewicz W, Schunck N and Xu F 2008 {\em Phys.
  Rev. C\/} {\bf 78} 064306

\bibitem{baye1986}
Baye D and Heenen P~H 1986 {\em J. Phys. A: Math. Gen.\/} {\bf 19} 2041

\bibitem{bonche2005}
Bonche P, Flocard H and Heenen P~H 2005 {\em Comput. Phys. Comm.\/} {\bf 171}
  49

\bibitem{ryssens2015}
Ryssens W, Heenen P~H and Bender M 2015 {\em Phys. Rev. C\/} {\bf 92} 064318

\bibitem{ryssens2015-a}
Ryssens W, Hellemans V, Bender M and Heenen P~H 2015 {\em Comput. Phys.
  Comm.\/} {\bf 187} 175

\bibitem{hashimoto2013}
Hashimoto Y 2013 {\em Phys. Rev. C\/} {\bf 88} 034307

\bibitem{bulgac2013}
Bulgac A and Forbes M~M 2013 {\em Phys. Rev. C\/} {\bf 87} 051301

\bibitem{maruhn2014}
Maruhn J~A, Reinhard P~G, Stevenson P~D and Umar A~S 2014 {\em Comput. Phys.
  Comm.\/} {\bf 185} 2195

\bibitem{poeschl1996}
P\"{o}schl W, Vretenar D and Ring P 1996 {\em Comput. Phys. Comm.\/} {\bf 99}
  128

\bibitem{poeschl1997}
P\"{o}schl W, Vretenar D, Lalazissis G~A and Ring P 1997 {\em Phys. Rev.
  Lett.\/} {\bf 79} 3841

\bibitem{poeschl1997-a}
P\"{o}schl W, Vretenar D, Rummel A and Ring P 1997 {\em Comput. Phys. Comm.\/}
  {\bf 101} 75

\bibitem{poeschl1998}
P\"{o}schl W 1998 {\em Comput. Phys. Comm.\/} {\bf 112} 42

\bibitem{pei2014}
Pei J~C, Fann G~I, Harrison R~J, Nazarewicz W, Shi Y and Thornton S 2014 {\em
  Phys. Rev. C\/} {\bf 90} 024317

\bibitem{baran2008}
Baran A, Bulgac A, Forbes M, Hagen G, Nazarewicz W, Schunck N and Stoitsov M
  2008 {\em Phys. Rev. C\/} {\bf 78} 014318

\bibitem{mang1976}
Mang H~J, Samadi B and Ring P 1976 {\em Z. Phys. A\/} {\bf 279} 325

\bibitem{egido1980}
Egido J, Mang H~J and Ring P 1980 {\em Nucl. Phys. A\/} {\bf 334} 1

\bibitem{egido1995}
Egido J~L, Lessing J, Martin V and Robledo L~M 1995 {\em Nucl. Phys. A\/} {\bf
  594} 70

\bibitem{Schmid1984205}
Schmid K, Gr{\"u}mmer F and Faessler A 1984 {\em Nucl. Phys. A\/} {\bf 431} 205

\bibitem{Reinhard1982418}
Reinhard P~G and Cusson R 1982 {\em Nucl. Phys. A\/} {\bf 378} 418 -- 442

\bibitem{davies1980}
Davies K, Flocard H, Krieger S and Weiss M 1980 {\em Nucl. Phys. A\/} {\bf 342}
  111

\bibitem{gall1994}
Gall B, Bonche P, Dobaczewski J, Flocard H and Heenen P~H 1994 {\em Z. Physik
  A\/} {\bf 348} 183--197

\bibitem{berghammer1995}
Berghammer H, Vretenar D and Ring P 1995 {\em Comput. Phys. Comm.\/} {\bf 88}
  293

\bibitem{kim1997}
Kim K~H, Otsuka T and Bonche P 1997 {\em J. Phys. G: Nucl. Part. Phys.\/} {\bf
  23} 1267

\bibitem{bulgac2008}
Bulgac A and Roche K~J 2008 {\em J. Phys.: Conf. Ser.\/} {\bf 125} 012064

\bibitem{regnier2016}
Regnier D, Verri\`{e}re M, Dubray N and Schunck N 2016 {\em Comput. Phys.
  Comm.\/} {\bf 200} 350

\bibitem{raey}
P{\'e}ru S and Martini M 2014 {\em Eur. Phys. J. A\/} {\bf 50}

\bibitem{PhysRevC.86.034334}
Li Z~P, Nik\ifmmode \check{s}\else \v{s}\fi{}i\ifmmode~\acute{c}\else
  \'{c}\fi{} T, Ring P, Vretenar D, Yao J~M and Meng J 2012 {\em Phys. Rev.
  C\/} {\bf 86} 034334

\bibitem{1507.00045}
Hinohara N 2015 Collective inertia of {Nambu-Goldstone} mode from linear
  response theory arXiv:1507.00045 (\textit{Preprint}
  \eprint{arXiv:1507.00045})

\bibitem{bjornholm1980}
Bj{\o}rnholm S and Lynn J~E 1980 {\em Rev. Mod. Phys.\/} {\bf 52} 725

\bibitem{herman2007}
Herman M, Capote R, Carlson B~V, Oblo\v{z}insk\'{y} P, Sin M, Trkov A, Wienke H
  and Zerkin V 2007 {\em Nucl. Data Sheets\/} {\bf 108} 2655

\bibitem{koning2008}
Koning A~J, Hilaire S and Duijvestijn M~C 2008 {TALYS-1.0} {\em Proceedings of
  the International Conference on Nuclear Data for Science and Technology\/}
  (Nice, France: EDP Sciences) p 211

\bibitem{flocard1974}
Flocard H, Quentin P, Vautherin D, Veneroni M and Kerman A~K 1974 {\em Nucl.
  Phys. A\/} {\bf 231} 176

\bibitem{campi1983}
Campi X and Stringari S 1983 {\em Z. Phys. A\/} {\bf 309} 239

\bibitem{nemeth1985}
Nemeth J, Dalili D and Ng\^{o} C 1985 {\em Phys. Lett. B\/} {\bf 154} 11

\bibitem{bender2004}
Bender M, Heenen P~H and Bonche P 2004 {\em Phys. Rev. C\/} {\bf 70} 054304

\bibitem{baran2015}
Baran A, Kowal M, Reinhard P~G, Robledo L~M, Staszczak A and Warda M 2015 {\em
  Nucl. Phys. A\/} {\bf 944} 442

\bibitem{delaroche2006}
Delaroche J~P, Girod M, Goutte H and Libert J 2006 {\em Nucl. Phys. A\/} {\bf
  771} 103

\bibitem{reinhard1996}
Reinhard P~G, Nazarewicz W, Bender M and Maruhn J~A 1996 {\em Phys. Rev. C\/}
  {\bf 53} 2776

\bibitem{baran2014}
Baran A and Staszczak A 2014 {\em Phys. Scr.\/} {\bf 89} 054002

\bibitem{sadhukhan2016}
Sadhukhan J, Nazarewicz W and Schunck N 2016 {\em Phys. Rev. C\/} {\bf 93}
  011304

\bibitem{baran1981}
Baran A, Pomorski K, Lukasiak A and Sobiczewski A 1981 {\em Nucl. Phys. A\/}
  {\bf 361} 83

\bibitem{smolaczuk1995}
Smola\'{n}czuk R, Skalski J and Sobiczewski A 1995 {\em Phys. Rev. C\/} {\bf
  52} 1871

\bibitem{berger2001}
Berger J~F, Bitaud L, Decharg\'{e} J, Girod M and Dietrich K 2001 {\em Nucl.
  Phys. A\/} {\bf 685} 1

\bibitem{warda2006}
Warda M, Egido J and Robledo L 2006 {\em Phys. Scr.\/} {\bf T125} 226

\bibitem{schindzielorz2009}
Schindzielorz N, Erler J, Kl\"{u}pfel P, Reinhard P~G and Hager G 2009 {\em
  Int. J. Mod. Phys. E\/} {\bf 18} 773

\bibitem{erler2012}
Erler J, Langanke K, Loens H~P, Mart\'{i}nez-Pinedo G and Reinhard P~G 2012
  {\em Phys. Rev. C\/} {\bf 85} 025802

\bibitem{mcdonnell2015}
McDonnell J~D, Schunck N, Higdon D, Sarich J, Wild S~M and Nazarewicz W 2015
  {\em Phys. Rev. Lett.\/} {\bf 114} 122501

\bibitem{rodriguez-guzman2014-a}
Rodr{\'i}guez-Guzm\'{a}n R and Robledo L~M 2014 {\em Phys. Rev. C\/} {\bf 89}
  054310

\bibitem{rutz1998}
Rutz K, Bender M, Reinhard P~G, Maruhn J~A and Greiner W 1998 {\em Nucl. Phys.
  A\/} {\bf 634} 67

\bibitem{duguet2001}
Duguet T, Bonche P, Heenen P~H and Meyer J 2001 {\em Phys. Rev. C\/} {\bf 65}
  014311

\bibitem{schunck2010}
Schunck N, Dobaczewski J, McDonnell J, Mor\'{e} J, Nazarewicz W, Sarich J and
  Stoitsov M~V 2010 {\em Phys. Rev. C\/} {\bf 81} 024316

\bibitem{bertsch2009}
Bertsch G, Bertulani C, Nazarewicz W, Schunck N and Stoitsov M 2009 {\em Phys.
  Rev. C\/} {\bf 79} 034306

\bibitem{bally2014}
Bally B 2014 {\em {Description} des noyaux impairs \`{a} l'aide d'une
  m\'{e}thode de fonctionnelle \'{e}nergie de la densit\'{e} \`{a} plusieurs
  \'{e}tats de r\'{e}f\'{e}rence\/} Ph.D. thesis Universit\'{e} de Bordeaux
  Bordeaux

\bibitem{rohoziski2015}
Rohozi\'{n}ski S~G 2015 {\em J. Phys. G: Nucl. Part. Phys.\/} {\bf 42} 025109

\bibitem{hock2013}
Hock K~M, Bonneau L, Quentin P and Wagiran H 2013 {Fission} barriers of
  odd-mass nuclei within the {HF-BCS} and {HTDA} approaches {\em EPJ Web of
  Conferences\/} vol~62 ed Chatillon A, Farget F, Faust H, Fioni G, Goutte D
  and Goutte H (EDP Sciences) p 04004

\bibitem{perez-martin2009}
Perez-Martin S and Robledo L~M 2009 {\em Int. J. Mod. Phys. E\/} {\bf 18} 788

\bibitem{chinn1992}
Chinn C~R, Berger J~F, Gogny D and Weiss M~S 1992 {\em Phys. Rev.\/} {\bf 45}
  1700

\bibitem{RevModPhys.85.1541}
Andreyev A~N, Huyse M and Van~Duppen P 2013 {\em Rev. Mod. Phys.\/} {\bf 85}(4)
  1541

\bibitem{PhysRevC.90.021302}
McDonnell J~D, Nazarewicz W, Sheikh J~A, Staszczak A and Warda M 2014 {\em
  Phys. Rev. C\/} {\bf 90} 021302

\bibitem{rose1984}
Rose H~J and Jones G~A 1984 {\em Nature\/} {\bf 307} 245

\bibitem{egido2004}
Egido J~L and Robledo L~M 2004 {\em Nucl. Phys. A\/} {\bf 738} 31

\bibitem{robledo2008}
Robledo L and Warda M 2008 {\em Int. J. Mod. Phys. E\/} {\bf 17} 2275

\bibitem{robledo2008-a}
Robledo L and Warda M 2008 {\em Int. J. Mod. Phys. E\/} {\bf 17} 204

\bibitem{goddard2014}
Goddard P~M 2014 {\em {A} {Microscopic} {Study} of {Nuclear} {Fission} using
  the {Time-Dependent} {Hartree-Fock} {Method}\/} Ph.D. thesis University of
  Surrey Guildford

\bibitem{younes2012}
Younes W and Gogny D 2012 {Collective} {Dissipation} from {Saddle} to
  {Scission} in a {Microscopic} {Approach} Tech. Rep. LLNL-TR-586694 Lawrence
  Livermore National Laboratory (LLNL), Livermore, CA

\bibitem{verbeke2015}
Verbeke J, Randrup J and Vogt R 2015 {\em Comput. Phys. Comm.\/} {\bf 191} 178

\bibitem{simenel2010}
Simenel C 2010 {\em Phys. Rev. Lett.\/} {\bf 105} 192701

\bibitem{bernard2011}
Bernard R, Goutte H, Gogny D and Younes W 2011 {\em Phys. Rev. C\/} {\bf 84}
  044308

\bibitem{dietrich2010}
Dietrich K, Niez J~J and Berger J~F 2010 {\em Nucl. Phys. A\/} {\bf 832} 249

\end{thebibliography}

\appendix

\end{document}